\documentclass[utf8,a4paper]{article}
\usepackage[top=2.5cm,bottom=2.5cm,left=2.5cm,right=2.5cm]{geometry}
\pdfoutput=1 
\usepackage{authblk}
\usepackage{amsmath}
\usepackage{amssymb}
\usepackage{amsthm}
\usepackage{color}
\usepackage{bm}
\usepackage[colorlinks,linkcolor=blue,anchorcolor=green,citecolor=red]{hyperref}
\usepackage{appendix}
\usepackage{graphicx}
\usepackage{subfigure}
\usepackage{booktabs}
\usepackage{epstopdf}

\allowdisplaybreaks[4]

\numberwithin{equation}{section}

\newtheorem{remark}{Remark}[section]
\usepackage{hyperref}
\newcommand{\RomanNumeralCaps}[1]
\linenumbers
\usepackage{booktabs}
\usepackage{lscape}

\title{ A phase field model for droplets suspended in viscous liquids under the influence of electric fields}
\author[a]{Yuzhe Qin}
\author[b,c,*]{Huaxiong Huang} 
\author[d]{Zilong Song}
\author[e,*]{Shixin Xu}
\affil[a]{Key Laboratory of Complex Systems and Data Science of Ministry of Education and School of Mathematical Sciences, Shanxi University, Taiyuan, 030006, China; }
\affil[b]{Advanced Institute of Natural Sciences, Beijing Normal University, Zhuhai, 519087, China}
\affil[c]{BNU-HKBU United International College, Zhuhai, 519087, China}
\affil[d]{Department of Mathematics and Statistics, Utah State University, Logan, UT, 84322, US}
\affil[e]{Duke Kunshan University, 8 Duke Ave, Kunshan, Jiangsu, China}

\affil[*]{Corresponding authors, shixin.xu@dukekunshan.edu.cn; hhuang@uic.edu.cn}
\date{}

\begin{document} 
\maketitle  
 	\begin{abstract}
 	In this paper, we propose a Poisson-Nernst-Planck-Navier-Stokes-Cahn-Hillard (PNP-NS-CH) model for an electrically charged droplet suspended in a viscous fluid subjected to an external electric field. Our model incorporates spatial variations of electric permittivity and diffusion constants, as well as interfacial capacitance. Based on a time scale analysis, we derive two approximations of the original model, namely a dynamic model for the net charge (by assuming conductance remains unchanged) and a leaky-dielectric model (by assuming both conductance and net charge remain unchanged). For the leaky-dielectric model,  we conduct a detailed asymptotic analysis to demonstrate the convergence of the diffusive-interface leaky-dielectric model to the sharp interface model as the interface thickness approaches zero. Numerical computations are performed to validate the asymptotic analysis and demonstrate the model's effectiveness in handling topology changes, such as electro-coalescence.
 	Our numerical results of these two approximation models reveal that the polarization force, which is induced by the spatial variation of electric permittivity in the direction perpendicular to the external electric field, consistently dominates the Lorentz force, which arises from the net charge. The equilibrium shape of droplets is determined by the interplay between these two forces along the direction of the electric field. Furthermore, in the presence of the interfacial capacitance, local variation of effective permittivity leads to an accumulation of counter-ions near the interface, resulting in a reduction in droplet deformation.
 	Our numerical solutions also confirm that the leaky-dielectric model serves as a reasonable approximation of the original PNP-NS-CH model when the electric relaxation time is sufficiently short. The Lorentz force and droplet deformation both decrease when diffusion of net charge is significant.
 \end{abstract}
	\section{Introduction}

%Electrohydrodynamics (EHD) is the study of liquid motion under the influence of an external electric field. 
When a charged droplet is suspended in a viscous fluid, it can exhibit complex behaviors, such as forming prolate or oblate shapes, pearling, and breaking up, under the influence of an externally applied electric field. Understanding and predicting droplet shape evolution are essential for many applications,  including microfluidics~(\cite{singh2020electrohydrodynamic}), printing~(\cite{zeng2004principles}), emulsion stability~(\cite{abbasi2020electro}), and biophysical systems~(\cite{hu2016vesicle,chakraborty2009electrohydrodynamics,wu2019electrohydrodynamics}). A more comprehensive review of this topic can be found in \cite{vlahovska2019electrohydrodynamics}.

Various mathematical models have been developed to investigate the dynamics of droplets under electric fields, ranging from simplified leaky-dielectric models to more sophisticated models that consider charge transport, fluid flow, and interfacial dynamics. One of the very first models was proposed by~\cite{Melcher1969Electrohydrodynamics}, known as the Melcher-Taylor (TM) model. It is a mathematical framework that describes the behavior of electrically charged fluids in the presence of electric fields, assuming electroneutrality, quasi-static electric field, leaky-dielectric fluids, weak deformation with no charge convection. The TM model has been experimentally validated and further extended by \cite{Saville1997electro}, \cite{lac2007axisymmetric}, and others to explore large drop deformations, stability, and breakup.
%TM model is  based on a few assumptions:  electroneutrality, a quasistatic electric field, leaky-dielectric fluids, weak deformation, and no charge convection. Specially, the dynamics of ions are neglected.   

Another class of models are based on the Poisson-Nernst-Planck (PNP) system  (\cite{xu2014energetic,bazant2009towards,bazant2015electrokinetics,wan2014self}), which has been applied successfully to problems involving ion transport.   Ryham et al. (\cite{Rolf2006PNP,ryham2006electro}) proposed a coupled PNP-Navier Stokes (PNP-NS) system to model the motion of electrolyte droplets.  Other researchers  (\cite{zholkovskij2002electrokinetic,schnitzer2015taylor,mori2018electrodiffusion}) have tried to build a link between PNP-NS and TM models.  Most theoretical models are limited to small deformation while under the influence of a strong electric field, however, droplets may exhibit large deformation and topology changes, resulting in droplets merge (electro-coalescence, \cite{Lin2012phasefield}) and breakup (\cite{karyappa2014breakup}).

Various numerical methods are proposed to  analyze  complex flows of droplets with large deformation.  \cite{zhang20052d}) proposed  a Lattice Boltzmann method to solve leaky-dielectric problem and reveal the flow pattern.   Boundary integral method is proposed by  \cite{lac2007axisymmetric})  to study droplet breakup. Most of the investigations can be classified into two main categories, sharp interface and diffusive interface methods. In sharp interface approach, there are level set method (\cite{bjorklund2009level,abbasi2017electro}), front tracking method (\cite{Hua2008Numerical}) and finite volume methods (\cite{lopez2011charge,cui2019numerical}), as well as hybrid immersed boundary (IB) and immersed interface method (IIM) \cite{HU2015Ahybrid} (for droplets) and \cite{hu2016vesicle} (for vesicles). Thanks to its ability to handle complex deformations and topological changes, diffusive interface models \cite{Lin2012phasefield,yang20133d,yang2017electrohydrodynamic} have also been widely used for droplets with certain types of interfacial forces/energies. On the other hand, the capacitance  and conductivity of droplet interface play important roles on their deformation and dynamics (\cite{aghdaei2008formation,boussoualem2014influence,hu2016vesicle}) since they change the continuity of electric potential and distribution of ions near the interface. How to integrate capacitance and conductance with spacial variation electric permittivity and diffusion consistently remains a challenge task. 

In this paper, we propose a Poisson-Nernst-Planck-Navier-Stokes-Cahn-Hillard (PNP-NS-CH) model for an electrically charged droplet suspended in a viscous fluid subjected to an external electric field, with the following specific features and objectives.
\begin{enumerate}
	\item We propose a thermodynamically consistent model for electrohydrodynamics (EHD) of two-phase flow, considering different electric conductance and permittivity, within the framework of the diffusive interface model. In this model, the capacitance of the interface will be taken into consideration;
	\item Based on an asymptotic analysis, we show that the obtained the diffusive-interface leaky dielectric model converges to the sharp interface one as the interface thickness tends to zero;
	\item We compare and analyze the differences between the leaky dielectric model (where both net charge and conductivity remain unchanged) and a net charge model where both diffusion and convection of net charge are considered. By examining these differences, a deeper understanding of the roles of various forces and processes in the system can be gained;
	\item We investigate the influence of electric forces on the equilibrium profile of the droplets. Particularly, we explore how variations in permittivity across the interface impact the distribution of ions and the resulting forces. This analysis provides insights into the mechanisms governing the behavior of the system under different conditions.
\end{enumerate}

The rest of the paper is organized as follows. 
In section \ref{sec: model}, we derive the phase field Poisson-Nernst-Planck-Navier-Stokes model based on the 
energy variational method. Then, the phase field leaky dielectric model is derived based on the time scale assumption. 
In section \ref{sec: sharp interface limit},  the sharp interface limit  is conducted to show our model could converge  to the former   sharp interface leaky dielectric model. 
In section \ref{sec: numerical results},  a series of numerical experiments are presented to verify 
the effectiveness of our phase field leaky dielectric model and study the capacitance 
effect at the interface.  The comparison of the results from leaky dielectric and net charge dynamic models are given in section \ref{sec:time scale}. 
Conclusions and the limitation of the current models and future directions are given in section \ref{sec: conclusions}.

	\section{Model derivation}\label{sec: model}
In this section, energy variation method is used to derive a thermal dynamically consistent phase field model for a droplet suspended in a viscous fluid under an external electric field. 
We definine two functionals for the total energy and dissipation of the system and introduce the corresponding kinematic equations based on physical laws of conservation. 
The specific forms of the flux and stress functions in the kinematic equations are obtained by taking the time derivative of the total energy functional and comparing with the dissipation functional. 
More details of this method can be found in \cite{shen2020energy}.

\subsection{Phase field model for a droplet suspended in electrolyte under electric field} 
Let $\Omega$ represent the computational domain that consists of the droplet and ambient fluid. $\psi$ is  an order parameter function which is equal to 1 in the droplet region and -1 in the ambient fluid region.   The interface between two domains can be described by the zero level set $\Gamma=\{\bm{x}:\psi(\bm{x},t)=0\}$.
Let $c_i, i=1\cdots N$ be the concentration of the $i_{th}$ ion, $\phi$ be the electric potential, and $\bm{u}$ be the fluid velocity. Based the the laws of conservation, we have the following kinematic assumptions   
\begin{subequations}\label{def: main}
	\begin{align} 
		& \frac{\partial \psi}{\partial t} 
		+ \nabla \cdot \left( \bm{u} \psi \right) 
		+ \nabla \cdot \bm{j}_{\psi} = 0, 
		& \mbox{in} \quad \Omega,
		\label{def: main psi} \\ 
		& \frac{\partial c_{i}}{\partial t} 
		+ \nabla \cdot \left( \bm{u} c_{i} \right) 
		+ \nabla \cdot \bm{j}_{i} = 0, 
		\qquad i = 1, \cdots, N, 
		& \mbox{in} \quad \Omega, 
		\label{def: main ci} \\
		& \nabla \cdot \bm{D} 
		= \sum_{i=1}^{N}z_{i}ec_{i}, 
		& \mbox{in} \quad \Omega, 
		\label{def: main D} \\ 
		& \bm{D} = \epsilon_{\it eff} \bm{E} 
		= - \epsilon_{\it eff} \nabla \phi, 
		& \mbox{in} \quad \Omega, 
		\label{def: main phi} \\ 
		& \rho \left( \frac{\partial \bm{u}}{\partial t} 
		+ \left( \bm{u} \cdot \nabla \right) \bm{u} \right) 
		= \nabla \cdot 	\bm{\sigma}_{\eta} 
		+ \nabla \cdot 	\bm{\sigma}_{e} +\nabla\cdot	\bm{\sigma}_{\psi}, 
		& \mbox{in} \quad \Omega,
		\label{def: main u} \\ 
		& \nabla \cdot \bm{u} = 0, 
		& \mbox{in} \quad \Omega, 
		\label{def: main nabla u}
	\end{align}
\end{subequations}
where  $z_{i}$ is the valency of $i$th ion,  $\bm{D}$ is the electric displacement,  $\bm{E}$ is the electric field intensity, $e$ is the elementary charge,  and $\rho$ is the density  of fluid. 
The first two equations are the conservation of each phase and ion with two unknown flux $\bm{j}_{\psi}$ and $\bm{j}_{i}$. The third and fourth equations are  based on Maxwell equation. The last two equations are the law of conservation of momentum for incompressible fluid with unknows  stresses induced by  
viscosity $\bm{\sigma}_{\eta}$, electricity $\bm{\sigma}_{e}$ and interface $\bm{\sigma}_{\psi}$, respectively. 
$\epsilon_{\it{eff}}$ is the dielectric coefficient, which is defined  by the harmonic average \cite{Qin2022111334}
\begin{equation}
	\epsilon_{\it{eff}}^{-1}
	= \frac{1-\psi}{2\epsilon^{-}}
	+ \frac{1+\psi}{2\epsilon^{+}} 
	+ \frac{\left(1-\psi^{2}\right)^{2}}{\delta C_{m}},
\end{equation} 
where $C_m$ is the  capacitance of the interface,  and $\epsilon^{\pm}$ is the dielectric constant of the droplet and ambient fluid, respectively. 

Here we consider a closed system during the derivation, i.e. the boundary condition is taken into account as 
\begin{equation}\label{def: bdc}
	\bm{u}|_{\partial \Omega} = 0, ~
	\bm{j}_{i}\cdot\bm{n}|_{\partial \Omega} = 0, ~
	\bm{j}_{\psi}\cdot\bm{n}|_{\partial \Omega} = 0, ~
	\phi|_{\partial \Omega} = 0,~
	\frac{\partial \psi}{\partial \bm{n}}|_{\partial \Omega} = 0,~
	\frac{\partial \mu_{\psi}}{\partial \bm{n}}|_{\partial \Omega} = 0.
\end{equation}

In the next, energy variational method is used to derive those unknown terms.  The total energy functional consists of kinetic energy, electrical static  energy, entropy and phase mixing energy, 
\begin{align}\label{def: total energy}
	E_{total} 
	= & E_{kin} + E_{es} + E_{ion} + E_{mix} 
	\nonumber \\
	= & \int_{\Omega} \left(\frac{\rho}{2}\left|\bm{u}\right|^{2} 
	+ \frac{1}{2}\bm{E}\cdot\bm{D} 
	+ k_{B}T\sum_{i=1}^{N}c_{i}\left(\ln\frac{c_{i}}{\tilde{c}}-1\right) 
	+ \lambda\left(\frac{\delta^{2}}{2}\left|\nabla\psi\right|^{2} 
	+ F\left(\psi\right) \right)\right)
	\mathrm{d}\bm{x}, 
\end{align} 
where $k_{B}$ and $T$ are Boltzmann constant and temperature respectively, 
$\tilde{c}$ is a reference concentration, 
$\lambda$ is the energy density for phase mixing energy,
$\delta$ is the thickness of diffuse interface and
$F\left(\psi\right)= \frac{1}{4}\left(1-\psi^{2}\right)^{2}$ is the double well potential.  

According to the total energy,  the chemical potentials are given by 
\begin{subequations}
	\begin{align}
		\mu_{i} = & z_{i} e \phi + k_{B}T \ln \frac{c_{i}}{\tilde{c}},~i =1\cdots N \\
		\mu_{\psi} = & \lambda \left( - \delta^{2} \nabla^{2} \psi + F^{\prime}(\psi) \right) 
		- \frac{1}{2} \frac{\partial \epsilon_{\it eff}}{\partial \psi} |\bm{E}|^{2}. 
	\end{align}
\end{subequations} 

The dissipation is mainly induced by viscosity of fluid in macroscale, invertible ions and  interface diffusion in microscale  
\begin{equation}\label{def: dissipation} 
	\Delta 
	= \int_{\Omega} 2 \eta \left|\bm{D}_{\eta}\right|^{2} \mathrm{d}\bm{x} 
	+ \int_{\Omega} \sum_{i=1}^{N} \frac{D_{i}c_{i}}{k_{B}T} 
	\left|\nabla\mu_{i}\right|^{2} \mathrm{d}\bm{x} 
	+ \int_{\Omega}\mathcal{M}\left|\nabla\mu_{\psi}\right|^{2}\mathrm{d}\bm{x}, 
\end{equation}
where $\eta$ is fluid viscosity, 
$\bm{D}_{\eta}=\frac{\nabla\bm{u}+(\nabla\bm{u})^{T}}{2}$ is the rate of strain, 
$\mu_{i}$ is the chemical potential
of $i$th ion, $\mathcal{M}$ and $\mu_{\psi}$ are the phenomenological 
mobility and chemical potential for $\psi$ respectively.  $D_i$ is the diffusion coefficient of $i_{th}$ ion
\begin{equation}
	D_{i}^{-1}
	= \frac{1-\psi}{2D_i^{-}}
	+ \frac{1+\psi}{2D_i^{+}}, 
\end{equation}
where $D_i^{\pm}$ is the diffusion coefficient  of $i_{th}$ ion in two regions, respectively.

For a closed system, the total energy satisfies the following energy dissipation law \cite{shen2020energy,xu2014energetic} 
\begin{align}\label{def: energy dissipation}
	\frac{dE_{total}}{dt}=& \frac{dE_{kin}}{dt} + \frac{dE_{es}}{dt}+ \frac{dE_{ion}}{dt}+ \frac{dE_{mix}}{dt}\nonumber\\
	=& I_1+I_2+I_3+I_4\nonumber\\
	=& -\Delta.  
\end{align}

%===================== Model Derivation =========================================

For the first term, using the Eqs.\eqref{def: main u}-\eqref{def: main nabla u} yields 
\begin{align}
	I_1 ~ 
	=& ~ \frac{\mathrm{d}}{\mathrm{d} t} \int_{\Omega} \frac{1}{2} \rho |\bm{u}|^{2} \mathrm{d}\bm{x} \nonumber \\
	=& ~  \frac{1}{2} \int_{\Omega} \left( \frac{\partial \rho}{\partial t} |\bm{u}|^{2} 
	+ 2 \rho \bm{u} \cdot \frac{\partial \bm{u}}{\partial t} \right) \mathrm{d}\bm{x} 
	\nonumber \\
	=& ~  \frac{1}{2} \int_{\Omega} \left( \frac{\partial \rho}{\partial t} |\bm{u}|^{2} 
	+ 2 \rho \bm{u} \cdot \left( \frac{\partial \bm{u}}{\partial t} + \bm{u} \cdot \nabla \bm{u} 
	- \bm{u} \cdot \nabla \bm{u} \right) \right) \mathrm{d}\bm{x} 
	\nonumber \\
	=& ~ \int_{\Omega} \rho \bm{u} \cdot \left( \frac{\partial \bm{u}}{\partial t} 
	+ \bm{u} \cdot \nabla \bm{u} \right) \mathrm{d}\bm{x}
	+ \frac{1}{2} \int_{\Omega} \left( \frac{\partial \rho}{\partial t} |\bm{u}|^{2} 
	- 2 \rho \bm{u} \cdot ( \bm{u} \cdot \nabla \bm{u}) \right) \mathrm{d}\bm{x} 
	\nonumber \\ 
	=& ~ \int_{\Omega} \bm{u} \cdot ( \nabla \cdot \sigma_{\eta} 
	+ \nabla \cdot \sigma_{e} +\nabla\cdot\sigma_{\psi}) \mathrm{d}\bm{x} 
	+ \frac{1}{2} \int_{\Omega} \left( \frac{\partial \rho}{\partial t} |\bm{u}|^{2} 
	- \rho \bm{u} \cdot \nabla |\bm{u}|^{2} \right) \mathrm{d}\bm{x} 
	\nonumber \\ 
	=& ~ - \int_{\Omega} \nabla \bm{u} : ( \sigma_{\eta} + \sigma_{e} +\sigma_{\psi}) \mathrm{d}\bm{x} 
	+ \frac{1}{2} \int_{\Omega} \left( \frac{\partial \rho}{\partial t} |\bm{u}|^{2} 
	+ \nabla \cdot (\rho \bm{u}) |\bm{u}|^{2} \right) \mathrm{d}\bm{x} 
	\nonumber \\
	& ~ - \frac{1}{2} \int_{\partial\Omega} \rho \bm{u} |\bm{u}|^{2} \cdot \bm{n} \mathrm{d}s 
	+ \int_{\partial \Omega} \bm{u} \cdot ( \sigma_{\eta} + \sigma_{e} +\sigma_{\psi}) \cdot \bm{n} \mathrm{d}s 
	\nonumber \\ 
	=& ~ - \int_{\Omega} \nabla \bm{u} : ( \sigma_{\eta} + \sigma_{e}  +\sigma_{\psi}) \mathrm{d}\bm{x} 
	- \int_{\Omega} p \nabla \cdot \bm{u} \mathrm{d}\bm{x} 
	\nonumber \\ 
	& ~ - \frac{1}{2} \int_{\partial\Omega} \rho \bm{u} |\bm{u}|^{2} \cdot \bm{n} \mathrm{d}s 
	+ \int_{\partial \Omega} \bm{u} \cdot ( \sigma_{\eta} + \sigma_{e}  +\sigma_{\psi}) \cdot \bm{n} \mathrm{d}s, 
\end{align}
where $\bm{n}$ is the unit outer normal vector of domain $\Omega$.

For electric potential energy, we have 
\begin{align}
	\frac{\mathrm{d} E_{es}}{\mathrm{d} t} ~ 
	= & ~ \frac{\mathrm{d}}{\mathrm{d} t} \int_{\Omega} \frac{1}{2} \bm{E} \cdot \bm{D} \mathrm{d}\bm{x} \nonumber \\
	= & ~ \frac{1}{2} \int_{\Omega} \left(\bm{D} \cdot \frac{\partial}{\partial t} \bm{E} 
	+ \bm{E} \cdot \frac{\partial}{\partial t} \bm{D}\right) \mathrm{d}\bm{x} \nonumber \\
	= & ~ \frac{1}{2} \int_{\Omega} \left(\bm{D} \cdot \frac{\partial}{\partial t} \bm{E} 
	+ \frac{\partial \epsilon_{\it eff}}{\partial \psi}\frac{\partial \psi}{\partial t} \bm{E} \cdot \bm{E} 
	+ \bm{D} \cdot \frac{\partial}{\partial t} \bm{E}  \right) \mathrm{d}\bm{x} \nonumber \\
	= & ~ \int_{\Omega} \bm{D} \cdot \frac{\partial}{\partial t} \bm{E} \mathrm{d}\bm{x}
	+ \frac{1}{2}\int_{\Omega} \frac{\partial \epsilon_{\it eff}}{\partial \psi} \frac{\partial \psi}{\partial t} \bm{E} \cdot \bm{E} \mathrm{d}\bm{x} \nonumber \\
	= & ~ \int_{\Omega} \bm{E} \cdot \epsilon_{\it eff} \frac{\partial}{\partial t} \bm{E} \mathrm{d}\bm{x}
	+ \frac{1}{2}\int_{\Omega} \frac{\partial \epsilon_{\it eff}}{\partial \psi}\frac{\partial \psi}{\partial t} \bm{E} \cdot \bm{E} \mathrm{d}\bm{x} \nonumber \\
	= & ~ \int_{\Omega} \bm{E} \cdot \left(\frac{\partial\bm{D}}{\partial t} 
	- \frac{\partial \epsilon_{\it eff}}{\partial \psi}
	\frac{\partial \psi}{\partial t} \bm{E} \right) \mathrm{d}\bm{x}
	+ \frac{1}{2}\int_{\Omega} \frac{\partial \epsilon_{\it eff}}{\partial \psi}\frac{\partial \psi}{\partial t} \bm{E} \cdot \bm{E} \mathrm{d}\bm{x} \nonumber \\
	= & ~ -\int_{\Omega} \nabla \phi \cdot \frac{\partial\bm{D}}{\partial t} \mathrm{d}\bm{x}
	- \frac{1}{2}\int_{\Omega} \frac{\partial \epsilon_{\it eff}}{\partial \psi}\frac{\partial \psi}{\partial t} \bm{E} \cdot \bm{E} \mathrm{d}\bm{x} \nonumber \\
	= & ~ -\int_{\partial \Omega} \phi \frac{\partial\bm{D}}{\partial t} \cdot \bm{n} \mathrm{d}s 
	+ \int_{\Omega} \phi \nabla \cdot \frac{\partial\bm{D}}{\partial t} \mathrm{d}\bm{x} 
	- \frac{1}{2}\int_{\Omega} \frac{\partial \epsilon_{\it eff}}{\partial \psi}\frac{\partial \psi}{\partial t} \bm{E} \cdot \bm{E} \mathrm{d}\bm{x} \nonumber \\
	= & ~ -\int_{\partial \Omega} \phi \frac{\partial\bm{D}}{\partial t} \cdot \bm{n} \mathrm{d}s 
	+ \int_{\Omega} \phi \left( \sum_{i=1}^{N} z_{i} e \frac{\partial c_{i}}{\partial t} \right) \mathrm{d}\bm{x}
	- \frac{1}{2}\int_{\Omega} \frac{\partial \epsilon_{\it eff}}{\partial \psi}\frac{\partial \psi}{\partial t} |\bm{E}|^{2} \mathrm{d}\bm{x} \nonumber \\
	= & ~ -\int_{\partial \Omega} \phi \frac{\partial\bm{D}}{\partial t} \cdot \bm{n} \mathrm{d}s 
	+ \int_{\Omega} \left( \sum_{i=1}^{N} z_{i} e \phi \frac{\partial c_{i}}{\partial t} \right) \mathrm{d}\bm{x}
	- \frac{1}{2}\int_{\Omega} \frac{\partial \epsilon_{\it eff}}{\partial \psi}\frac{\partial \psi}{\partial t} |\bm{E}|^{2} \mathrm{d}\bm{x}, 
\end{align}
where we used Eqs.\eqref{def: main ci}, \eqref{def: main phi}, and  
\begin{equation}\label{eqn: Dt}
	\frac{\partial\bm{D}}{\partial t} 
	= \frac{\partial \epsilon_{\it eff}}{\partial \psi}
	\frac{\partial \psi}{\partial t} \bm{E} 
	+ \epsilon_{\it eff}\frac{\partial \bm{E}}{\partial t}. 
\end{equation}

The derivative of  entropy is given by   
\begin{align}
	\frac{\mathrm{d} E_{ion}}{\mathrm{d} t} ~ 
	= & ~ \frac{\mathrm{d}}{\mathrm{d} t} \int_{\Omega} k_{B}T \sum_{i=1}^{N}c_{i} \left( \ln \frac{c_{i}}{\tilde{c}} - 1 \right) \mathrm{d}\bm{x} \nonumber \\
	= & ~ \int_{\Omega} k_{B}T \sum_{i=1}^{N} \frac{\partial c_{i}}{\partial t} \ln \frac{c_{i}}{\tilde{c}} \mathrm{d}\bm{x}. 
\end{align}
For the mixing energy,   the Eq. \eqref{def: main psi} yields 
\begin{align}
	\frac{\mathrm{d} E_{mix}}{\mathrm{d} t} ~ 
	= & ~ \frac{\mathrm{d}}{\mathrm{d} t} \int_{\Omega} \lambda \left(\frac{\delta^{2}}{2}|\nabla\psi|^{2} + F(\psi) \right) \mathrm{d}\bm{x} 
	\nonumber \\
	= & ~ \int_{\Omega} \lambda \frac{\partial }{\partial t} \left(\frac{\delta^{2}}{2}|\nabla\psi|^{2} + F(\psi) \right) \mathrm{d}\bm{x} 
	\nonumber \\ 
	= & ~ \int_{\Omega} \lambda \left( \delta^{2} \nabla \psi \cdot \frac{\partial \nabla \psi}{\partial t} + F^{\prime}(\psi) \frac{\partial \psi}{\partial t} \right) \mathrm{d}\bm{x} 
	\nonumber \\ 
	= & ~ \int_{\Omega} \lambda \left( \delta^{2} \nabla \psi \cdot \nabla \frac{\partial \psi}{\partial t} + F^{\prime}(\psi) \frac{\partial \psi}{\partial t} \right) \mathrm{d}\bm{x} 
	\nonumber \\ 
	= & ~ \int_{\Omega} \lambda \delta^{2} \nabla \psi \cdot \nabla \frac{\partial \psi}{\partial t} \mathrm{d}\bm{x} 
	+ \int_{\Omega} \lambda F^{\prime}(\psi) \frac{\partial \psi}{\partial t} \mathrm{d}\bm{x} 
	\nonumber \\ 
	= & ~ \int_{\partial \Omega} \lambda \delta^{2} \nabla \psi \frac{\partial \psi}{\partial t} \cdot \bm{n} \mathrm{d}s 
	- \int_{\Omega} \lambda \delta^{2} \nabla^{2} \psi \frac{\partial \psi}{\partial t} \mathrm{d}\bm{x} 
	+ \int_{\Omega} \lambda F^{\prime}(\psi) \frac{\partial \psi}{\partial t} \mathrm{d}\bm{x} 
	\nonumber \\ 
	= & ~ \int_{\partial \Omega} \lambda \delta^{2} \nabla \psi \frac{\partial \psi}{\partial t} \cdot \bm{n} \mathrm{d}s 
	+ \int_{\Omega} \lambda \left( - \delta^{2} \nabla^{2} \psi + F^{\prime}(\psi) \right) \frac{\partial \psi}{\partial t} \mathrm{d}\bm{x}. 
\end{align}
In summary, for the total energy, we have 
\begin{align}\label{DEDT}
	&\frac{\mathrm{d} E_{total}}{\mathrm{d} t} 	\nonumber \\ 
	= & I_1+I_2+I_3+I_4
	\nonumber \\ 
	= & ~ - \int_{\Omega} \nabla \bm{u} : ( \sigma_{\eta} + \sigma_{e}  +\sigma_{\psi}) \mathrm{d}\bm{x} 
	- \int_{\Omega} p \nabla \cdot \bm{u} \mathrm{d}\bm{x} 
	- \frac{1}{2} \int_{\partial\Omega} \rho \bm{u} |\bm{u}|^{2} \cdot \bm{n} \mathrm{d}s 
	+ \int_{\partial \Omega} \bm{u} \cdot ( \sigma_{\eta} + \sigma_{e} +\sigma_{\psi} ) \cdot \bm{n} \mathrm{d}s 
	\nonumber \\ 
	& ~ - \int_{\partial \Omega} \phi \frac{\partial\bm{D}}{\partial t} \cdot \bm{n} \mathrm{d}s 
	+ \int_{\Omega} \left( \sum_{i=1}^{N} z_{i} e \phi \frac{\partial c_{i}}{\partial t} \right) \mathrm{d}\bm{x} 
	- \frac{1}{2}\int_{\Omega} \frac{\partial \epsilon_{\it eff}}{\partial \psi}\frac{\partial \psi}{\partial t} |\bm{E}|^{2} \mathrm{d}\bm{x} 
	+ \int_{\Omega} k_{B}T \sum_{i=1}^{N} \frac{\partial c_{i}}{\partial t} \ln \frac{c_{i}}{\tilde{c}} \mathrm{d}\bm{x} 
	\nonumber \\ 
	& ~ + \int_{\partial \Omega} \lambda \delta^{2} \nabla \psi \frac{\partial \psi}{\partial t} \cdot \bm{n} \mathrm{d}s 
	+ \int_{\Omega} \lambda \left( - \delta^{2} \nabla^{2} \psi + F^{\prime}(\psi) \right) \frac{\partial \psi}{\partial t} \mathrm{d}\bm{x} 
	\nonumber \\ 
	= & ~ - \int_{\Omega} \nabla \bm{u} : ( \sigma_{\eta} + \sigma_{e}  + \sigma_{\psi}) \mathrm{d}\bm{x} 
	- \int_{\Omega} p \nabla \cdot \bm{u} \mathrm{d}\bm{x} 
	+ \int_{\Omega} \left( \sum_{i=1}^{N} z_{i} e \phi \frac{\partial c_{i}}{\partial t} \right) \mathrm{d}\bm{x} 
	+ \int_{\Omega} k_{B}T \sum_{i=1}^{N} \frac{\partial c_{i}}{\partial t} \ln \frac{c_{i}}{\tilde{c}} \mathrm{d}\bm{x} 
	\nonumber \\ 
	& ~ + \int_{\Omega} \lambda \left( - \delta^{2} \nabla^{2} \psi + F^{\prime}(\psi) \right) \frac{\partial \psi}{\partial t} \mathrm{d}\bm{x} 
	- \frac{1}{2}\int_{\Omega} \frac{\partial \epsilon_{\it eff}}{\partial \psi}\frac{\partial \psi}{\partial t} |\bm{E}|^{2} \mathrm{d}\bm{x} 
	\nonumber \\ 
	& ~ - \frac{1}{2} \int_{\partial\Omega} \rho \bm{u} |\bm{u}|^{2} \cdot \bm{n} \mathrm{d}s 
	+ \int_{\partial \Omega} \bm{u} \cdot ( \sigma_{\eta} + \sigma_{e} + \sigma_{\psi} ) \cdot \bm{n} \mathrm{d}s 
	- \int_{\partial \Omega} \phi \frac{\partial\bm{D}}{\partial t} \cdot \bm{n} \mathrm{d}s 
	+ \int_{\partial \Omega} \lambda \delta^{2} \nabla \psi \frac{\partial \psi}{\partial t} \cdot \bm{n} \mathrm{d}s 
	\nonumber \\ 
	= & ~ - \int_{\Omega} \nabla \bm{u} : ( \sigma_{\eta} + \sigma_{e}  + \sigma_{\psi}) \mathrm{d}\bm{x} 
	- \int_{\Omega} p \nabla \cdot \bm{u} \mathrm{d}\bm{x} 
	+ \int_{\Omega} \sum_{i=1}^{N} \left( \left( z_{i} e \phi + k_{B}T \ln \frac{c_{i}}{\tilde{c}} \right) \frac{\partial c_{i}}{\partial t} \right) \mathrm{d}\bm{x} 
	\nonumber \\ 
	& ~ + \int_{\Omega} \left(  \lambda \left( -\delta^{2} \nabla^{2} \psi + F^{\prime}(\psi) \right) 
	- \frac{1}{2} \frac{\partial \epsilon_{\it eff}}{\partial \psi} |\bm{E}|^{2} \right) \frac{\partial \psi}{\partial t} \mathrm{d}\bm{x} 
	\nonumber \\ 
	& ~ - \frac{1}{2} \int_{\partial\Omega} \rho \bm{u} |\bm{u}|^{2} \cdot \bm{n} \mathrm{d}s 
	+ \int_{\partial \Omega} \bm{u} \cdot ( \sigma_{\eta} + \sigma_{e}  + \sigma_{\psi}) \cdot \bm{n} \mathrm{d}s 
	- \int_{\partial \Omega} \phi \frac{\partial\bm{D}}{\partial t} \cdot \bm{n} \mathrm{d}s 
	+ \int_{\partial \Omega} \lambda \delta^{2} \nabla \psi \frac{\partial \psi}{\partial t} \cdot \bm{n} \mathrm{d}s 
	\nonumber \\ 
	= & ~ - \int_{\Omega} \nabla \bm{u} : ( \sigma_{\eta} + \sigma_{e}  + \sigma_{\psi}) \mathrm{d}\bm{x} 
	- \int_{\Omega} p \nabla \cdot \bm{u} \mathrm{d}\bm{x} 
	+ \int_{\Omega} \sum_{i=1}^{N} \mu_{i} \frac{\partial c_{i}}{\partial t} \mathrm{d}\bm{x} 
	+ \int_{\Omega} \mu_{\psi} \frac{\partial \psi}{\partial t} \mathrm{d}\bm{x} 
	\nonumber \\ 
	& ~ - \frac{1}{2} \int_{\partial\Omega} \rho \bm{u} |\bm{u}|^{2} \cdot \bm{n} \mathrm{d}s 
	+ \int_{\partial \Omega} \bm{u} \cdot ( \sigma_{\eta} + \sigma_{e} + \sigma_{\psi} ) \cdot \bm{n} \mathrm{d}s 
	- \int_{\partial \Omega} \phi \frac{\partial\bm{D}}{\partial t} \cdot \bm{n} \mathrm{d}s 
	+ \int_{\partial \Omega} \lambda \delta^{2} \nabla \psi \frac{\partial \psi}{\partial t} \cdot \bm{n} \mathrm{d}s 
	\nonumber \\ 
	= & ~ - \int_{\Omega} \nabla \bm{u} : ( \sigma_{\eta} + \sigma_{e} +\sigma_{\psi} ) \mathrm{d}\bm{x} 
	- \int_{\Omega} p \nabla \cdot \bm{u} \mathrm{d}\bm{x} \nonumber \\ 
	& ~ + \int_{\Omega} \sum_{i=1}^{N} \mu_{i} \left( -\nabla \cdot (\bm{u} c_{i}) - \nabla \cdot \bm{j}_{i} \right) \mathrm{d}\bm{x} 
	+ \int_{\Omega} \mu_{\psi} \left( -\nabla \cdot ( \bm{u} \psi ) - \nabla \cdot \bm{j}_{\psi} \right) \mathrm{d}\bm{x} 
	\nonumber \\ 
	& ~ - \frac{1}{2} \int_{\partial\Omega} \rho \bm{u} |\bm{u}|^{2} \cdot \bm{n} \mathrm{d}s 
	+ \int_{\partial \Omega} \bm{u} \cdot ( \sigma_{\eta} + \sigma_{e} + \sigma_{\psi} ) \cdot \bm{n} \mathrm{d}s 
	- \int_{\partial \Omega} \phi \frac{\partial\bm{D}}{\partial t} \cdot \bm{n} \mathrm{d}s 
	+ \int_{\partial \Omega} \lambda \delta^{2} \nabla \psi \frac{\partial \psi}{\partial t} \cdot \bm{n} \mathrm{d}s 
	\nonumber \\ 
	= & ~ - \int_{\Omega} \nabla \bm{u} : ( \sigma_{\eta} + \sigma_{e}  + \sigma_{\psi} ) \mathrm{d}\bm{x} 
	+ \int_{\Omega} p \nabla \cdot \bm{u} \mathrm{d}\bm{x} 
	\nonumber \\ 
	& ~ + \underbrace{\int_{\Omega} \sum_{i=1}^{N} c_{i} \bm{u} \cdot \nabla \mu_{i} \mathrm{d}\bm{x}}_{H_{1}} 
	+ \int_{\Omega} \sum_{i=1}^{N} \nabla \mu_{i} \cdot \bm{j}_{i} \mathrm{d}\bm{x} \quad ( := I_{1} + I_{2}) 
	\nonumber \\ 
	& ~ +\underbrace{ \int_{\Omega} \psi \bm{u} \cdot \nabla \mu_{\psi} \mathrm{d}\bm{x}}_{H_{2}} + \int_{\Omega} \nabla \mu_{\psi} \cdot \bm{j}_{\psi} \mathrm{d}\bm{x} \quad ( := I_{3} + I_{4})
	\nonumber \\ 
	& ~ - \int_{\partial \Omega} \sum_{i=1}^{N} c_{i} \mu_{i} \bm{u} \cdot \bm{n} \mathrm{d}s - \int_{\partial \Omega} \sum_{i=1}^{N} \mu_{i} \bm{j}_{i} \cdot \bm{n} \mathrm{d}s
	- \int_{\partial \Omega} \psi \mu_{\psi} \bm{u} \cdot \bm{n} \mathrm{d}s - \int_{\partial \Omega} \mu_{\psi} \bm{j}_{\psi} \cdot \bm{n} \mathrm{d}s
	\nonumber \\ 
	& ~ - \frac{1}{2} \int_{\partial\Omega} \rho \bm{u} |\bm{u}|^{2} \cdot \bm{n} \mathrm{d}s 
	+ \int_{\partial \Omega} \bm{u} \cdot ( \sigma_{\eta} + \sigma_{e}  + \sigma_{\psi} ) \cdot \bm{n} \mathrm{d}s 
	\nonumber \\ 
	& ~ - \int_{\partial \Omega} \phi \frac{\partial\bm{D}}{\partial t} \cdot \bm{n} \mathrm{d}s 
	+ \int_{\partial \Omega} \lambda \delta^{2} \nabla \psi \frac{\partial \psi}{\partial t} \cdot \bm{n} \mathrm{d}s. 
\end{align}
For $H_{1}$, combining the definition of chemical potential $\mu_i$
and Eq. \eqref{def: main phi} yields
\begin{align}\label{H1}
	& ~ \int_{\Omega} \sum_{i=1}^{N} c_{i} \bm{u} \cdot \nabla \mu_{i} \mathrm{d}\bm{x} 
	\nonumber \\ 
	= & ~ \int_{\Omega} \left( \sum_{i=1}^{N} z_{i} e c_{i} \right) \bm{u} \cdot \nabla \phi \mathrm{d}\bm{x} 
	+ \int_{\Omega} \sum_{i=1}^{N} k_{B} T \bm{u} \cdot \nabla c_{i} \mathrm{d}\bm{x} 
	\nonumber \\ 
	= & ~ \int_{\Omega} \left( \nabla \cdot \bm{D} \right) \bm{u} \cdot \nabla \phi \mathrm{d}\bm{x} 
	+ \int_{\Omega} \sum_{i=1}^{N} k_{B} T \bm{u} \cdot \nabla c_{i} \mathrm{d}\bm{x} 
	\nonumber \\ 
	= & ~ - \int_{\Omega} \bm{D} \cdot \nabla ( \bm{u} \cdot \nabla \phi ) \mathrm{d}\bm{x} 
	+ \int_{\partial \Omega} \sum_{i=1}^{N} k_{B} T c_{i} \bm{u} \cdot \bm{n} \mathrm{d}s 
	+ \int_{\partial \Omega} ( \bm{u} \cdot \nabla \phi ) \bm{D} \cdot \bm{n} \mathrm{d}s 
	\nonumber \\ 
	= & ~ \int_{\Omega} \epsilon_{\it eff} \nabla \phi \cdot \nabla \bm{u} \cdot \nabla \phi \mathrm{d}\bm{x} 
	+ \int_{\Omega} \epsilon_{\it eff} \nabla \phi \cdot \nabla \nabla \phi \cdot \bm{u} \mathrm{d}\bm{x} 
	+ \int_{\partial \Omega} \sum_{i=1}^{N} k_{B} T c_{i} \bm{u} \cdot \bm{n} \mathrm{d}s 
	+ \int_{\partial \Omega} ( \bm{u} \cdot \nabla \phi ) \bm{D} \cdot \bm{n} \mathrm{d}s 
	\nonumber \\ 
	= & ~ \int_{\Omega} \epsilon_{\it eff} \nabla \phi \cdot \nabla \bm{u} \cdot \nabla \phi \mathrm{d}\bm{x} 
	+ \int_{\Omega} \epsilon_{\it eff} \frac{1}{2} \nabla | \nabla \phi |^{2} \cdot \bm{u} \mathrm{d}\bm{x} 
	+ \int_{\partial \Omega} \sum_{i=1}^{N} k_{B} T c_{i} \bm{u} \cdot \bm{n} \mathrm{d}s 
	+ \int_{\partial \Omega} ( \bm{u} \cdot \nabla \phi ) \bm{D} \cdot \bm{n} \mathrm{d}s 
	\nonumber \\ 
	= & ~ \int_{\Omega} \epsilon_{\it eff} \nabla \phi \cdot \nabla \bm{u} \cdot \nabla \phi \mathrm{d}\bm{x} 
	- \int_{\Omega} \frac{1}{2} | \nabla \phi |^{2} \nabla \cdot ( \epsilon_{\it eff} \bm{u} ) \mathrm{d}\bm{x} 
	\nonumber \\ 
	& ~ + \int_{\partial \Omega} \sum_{i=1}^{N} k_{B} T c_{i} \bm{u} \cdot \bm{n} \mathrm{d}s 
	+ \int_{\partial \Omega} ( \bm{u} \cdot \nabla \phi ) \bm{D} \cdot \bm{n} \mathrm{d}s 
	+ \int_{\partial \Omega} \epsilon_{\it eff} \frac{1}{2} | \nabla \phi |^{2} \bm{u} \cdot \bm{n} \mathrm{d}s
	\nonumber \\
	= & ~ \int_{\Omega} \epsilon_{\it eff} \nabla \phi \cdot \nabla \bm{u} \cdot \nabla \phi \mathrm{d}\bm{x} 
	- \int_{\Omega} \frac{1}{2} | \nabla \phi |^{2} \nabla \epsilon_{\it eff} \cdot \bm{u} \mathrm{d}\bm{x} 
	- \int_{\Omega} \frac{1}{2} | \nabla \phi |^{2} \epsilon_{\it eff} \nabla \cdot \bm{u} \mathrm{d}\bm{x} 
	\nonumber \\ 
	& ~ + \int_{\partial \Omega} \sum_{i=1}^{N} k_{B} T c_{i} \bm{u} \cdot \bm{n} \mathrm{d}s 
	+ \int_{\partial \Omega} ( \bm{u} \cdot \nabla \phi ) \bm{D} \cdot \bm{n} \mathrm{d}s 
	+ \int_{\partial \Omega} \epsilon_{\it eff} \frac{1}{2} | \nabla \phi |^{2} \bm{u} \cdot \bm{n} \mathrm{d}s 
	\nonumber \\ 
	= & ~ \int_{\Omega} \epsilon_{\it eff} \left( \nabla \phi \otimes \nabla \phi -\frac{1}{2} \left| \nabla \phi \right|^{2} \textbf{I} \right) : \nabla \bm{u} \mathrm{d}\bm{x} 
	- \int_{\Omega} \frac{1}{2} | \nabla \phi |^{2} \nabla \epsilon_{\it eff} \cdot \bm{u} \mathrm{d}\bm{x} 
	\nonumber \\ 
	& ~ + \int_{\partial \Omega} \sum_{i=1}^{N} k_{B} T c_{i} \bm{u} \cdot \bm{n} \mathrm{d}s 
	+ \int_{\partial \Omega} ( \bm{u} \cdot \nabla \phi ) \bm{D} \cdot \bm{n} \mathrm{d}s 
	+ \int_{\partial \Omega} \epsilon_{\it eff} \frac{1}{2} | \nabla \phi |^{2} \bm{u} \cdot \bm{n} \mathrm{d}s. 
\end{align}
Similarly, for $H_2$ terms, the chemical potential of $\mu_{\psi}$ gives   
\begin{align}\label{H2}
	& ~ \int_{\Omega} \psi \bm{u} \cdot \nabla \mu_{\psi} \mathrm{d}\bm{x} 
	\nonumber \\ 
	= & ~ \int_{\Omega} \psi \bm{u} \cdot \nabla \left( \lambda \left( - \delta^{2} \nabla^{2} \psi + F^{\prime}(\psi) \right) 
	- \frac{1}{2} \frac{\partial \epsilon_{\it eff}}{\partial \psi} |\bm{E}|^{2} \right) \mathrm{d}\bm{x} 
	\nonumber \\ 
	= & ~ - \int_{\Omega} \nabla \cdot ( \psi \bm{u} ) \left( \lambda \left( - \delta^{2} \nabla^{2} \psi + F^{\prime}(\psi) \right) 
	- \frac{1}{2} \frac{\partial \epsilon_{\it eff}}{\partial \psi} |\bm{E}|^{2} \right) \mathrm{d}\bm{x} 
	\nonumber \\ 
	& ~ + \int_{\partial \Omega} \left( \lambda \left( - \delta^{2} \nabla^{2} \psi + F^{\prime}(\psi) \right) 
	- \frac{1}{2} \frac{\partial \epsilon_{\it eff}}{\partial \psi} |\bm{E}|^{2} \right) \psi \bm{u} \cdot \bm{n} \mathrm{d}s 
	\nonumber \\ 
	= & ~ - \int_{\Omega} \bm{u} \cdot \nabla \psi \left( \lambda \left( - \delta^{2} \nabla^{2} \psi + F^{\prime}(\psi) \right) 
	- \frac{1}{2} \frac{\partial \epsilon_{\it eff}}{\partial \psi} |\bm{E}|^{2} \right) \mathrm{d}\bm{x} 
	\nonumber \\ 
	& ~ + \int_{\partial \Omega} \left( \lambda \left( - \delta^{2} \nabla^{2} \psi + F^{\prime}(\psi) \right) 
	- \frac{1}{2} \frac{\partial \epsilon_{\it eff}}{\partial \psi} |\bm{E}|^{2} \right) \psi \bm{u} \cdot \bm{n} \mathrm{d}s 
	\nonumber \\ 
	= & ~ - \int_{\Omega} \left( \lambda \left( - \delta^{2} \bm{u} \cdot \nabla \psi \nabla^{2} \psi 
	+ \bm{u} \cdot \nabla F(\psi) \right) 
	- \frac{1}{2} |\bm{E}|^{2} \bm{u} \cdot \nabla \epsilon_{\it eff} \right) \mathrm{d}\bm{x} 
	\nonumber \\ 
	& ~ + \int_{\partial \Omega} \left( \lambda \left( - \delta^{2} \nabla^{2} \psi + F^{\prime}(\psi) \right) 
	- \frac{1}{2} \frac{\partial \epsilon_{\it eff}}{\partial \psi} |\bm{E}|^{2} \right) \psi \bm{u} \cdot \bm{n} \mathrm{d}s 
	\nonumber \\ 
	= & ~ - \int_{\Omega} \left( - \lambda \delta^{2} \bm{u} \cdot \left( \nabla \cdot ( \nabla \psi \otimes \nabla \psi ) -\frac{1}{2} \nabla | \nabla \psi |^{2} \right) 
	+ \lambda \bm{u} \cdot \nabla F(\psi) 
	- \frac{1}{2} |\bm{E}|^{2} \bm{u} \cdot \nabla \epsilon_{\it eff} \right) \mathrm{d}\bm{x} 
	\nonumber \\ 
	& ~ + \int_{\partial \Omega} \left( \lambda \left( - \delta^{2} \nabla^{2} \psi + F^{\prime}(\psi) \right) 
	- \frac{1}{2} \frac{\partial \epsilon_{\it eff}}{\partial \psi} |\bm{E}|^{2} \right) \psi \bm{u} \cdot \bm{n} \mathrm{d}s 
	\nonumber \\ 
	= & ~ \int_{\Omega} \lambda \delta^{2} \bm{u} \cdot \nabla \cdot ( \nabla \psi \otimes \nabla \psi ) \mathrm{d}\bm{x} 
	- \int_{\Omega} \lambda \bm{u} \cdot \nabla \left( \frac{\delta^{2}}{2} | \nabla \psi |^{2} - F(\psi) \right) \mathrm{d}\bm{x} 
	+ \int_{\Omega} \frac{1}{2} |\bm{E}|^{2} \bm{u} \cdot \nabla \epsilon_{\it eff} \mathrm{d}\bm{x} 
	\nonumber \\ 
	& ~ + \int_{\partial \Omega} \left( \lambda \left( - \delta^{2} \nabla^{2} \psi + F^{\prime}(\psi) \right) 
	- \frac{1}{2} \frac{\partial \epsilon_{\it eff}}{\partial \psi} |\bm{E}|^{2} \right) \psi \bm{u} \cdot \bm{n} \mathrm{d}s 
	\nonumber \\ 
	= & ~ - \int_{\Omega} \nabla \bm{u} : ( \lambda \delta^{2} \nabla \psi \otimes \nabla \psi ) \mathrm{d}\bm{x} 
	+ \int_{\Omega} \frac{1}{2} |\bm{E}|^{2} \bm{u} \cdot \nabla \epsilon_{\it eff} \mathrm{d}\bm{x} 
	- \int_{\partial \Omega} \lambda \left( \frac{\delta^{2}}{2} | \nabla \psi |^{2} - F(\psi) \right) \bm{u} \cdot \bm{n} \mathrm{d}s 
	\nonumber \\ 
	& ~ + \int_{\partial \Omega} \left( \lambda \left( - \delta^{2} \nabla^{2} \psi + F^{\prime}(\psi) \right) 
	- \frac{1}{2} \frac{\partial \epsilon_{\it eff}}{\partial \psi} |\bm{E}|^{2} \right) \psi \bm{u} \cdot \bm{n} \mathrm{d}s 
	+ \int_{\partial \Omega} \lambda \delta^{2} \bm{u} \cdot \nabla \psi \otimes \nabla \psi \cdot \bm{n} \mathrm{d}s. 
\end{align}

Substituting Eqs. \eqref{H1} and \eqref{H2} into Eq. \eqref{DEDT} yields
\begin{align} 
	&\frac{\mathrm{d} E_{total}}{\mathrm{d} t} \nonumber\\
	= & ~ - \int_{\Omega} \nabla \bm{u} : ( \sigma_{\eta} + \sigma_{\psi} + \sigma_{e} ) \mathrm{d}\bm{x} 
	- \int_{\Omega} p \nabla \cdot \bm{u} \mathrm{d}\bm{x} 
	+ \int_{\Omega} \sum_{i=1}^{N} \nabla \mu_{i} \cdot \bm{j}_{i} \mathrm{d}\bm{x} 
	+ \int_{\Omega} \nabla \mu_{\psi} \cdot \bm{j}_{\psi} \mathrm{d}\bm{x} 
	\nonumber \\ 
	& ~ + \int_{\Omega} \epsilon_{\it eff} \left( \nabla \phi \otimes \nabla \phi -\frac{1}{2} \left| \nabla \phi \right|^{2} \textbf{I} \right) : \nabla \bm{u} \mathrm{d}\bm{x} 
	- \int_{\Omega} \frac{1}{2} | \nabla \phi |^{2} \nabla \epsilon_{\it eff} \cdot \bm{u} \mathrm{d}\bm{x} 
	\nonumber \\ 
	& ~ - \int_{\Omega} \nabla \bm{u} : ( \lambda \delta^{2} \nabla \psi \otimes \nabla \psi ) \mathrm{d}\bm{x} 
	+ \int_{\Omega} \frac{1}{2} |\bm{E}|^{2} \bm{u} \cdot \nabla \epsilon_{\it eff} \mathrm{d}\bm{x} 
	+ \int_{\partial \Omega} \lambda \left( \frac{\delta^{2}}{2} | \nabla \psi |^{2} - F(\psi) \right) \bm{u} \cdot \bm{n} \mathrm{d}s 
	\nonumber \\ 
	& ~ + \int_{\partial \Omega} \left( \lambda \left( -\delta^{2} \nabla^{2} \psi + F^{\prime}(\psi) \right) 
	- \frac{1}{2} \frac{\partial \epsilon_{\it eff}}{\partial \psi} |\bm{E}|^{2} \right) \psi \bm{u} \cdot \bm{n} \mathrm{d}s 
	+ \int_{\partial \Omega} \lambda \delta^{2} \bm{u} \cdot \nabla \psi \otimes \nabla \psi \cdot \bm{n} \mathrm{d}s 
	\nonumber \\ 
	& ~ + \int_{\partial \Omega} \sum_{i=1}^{N} k_{B} T c_{i} \bm{u} \cdot \bm{n} \mathrm{d}s 
	+ \int_{\partial \Omega} ( \bm{u} \cdot \nabla \phi ) \bm{D} \cdot \bm{n} \mathrm{d}s 
	+ \int_{\partial \Omega} \epsilon_{\it eff} \frac{1}{2} | \nabla \phi |^{2} \bm{u} \cdot \bm{n} \mathrm{d}s 
	\nonumber \\ 
	& ~ - \int_{\partial \Omega} \sum_{i=1}^{N} c_{i} \mu_{i} \bm{u} \cdot \bm{n} \mathrm{d}s 
	- \int_{\partial \Omega} \sum_{i=1}^{N} \mu_{i} \bm{j}_{i} \cdot \bm{n} \mathrm{d}s
	- \int_{\partial \Omega} \psi \mu_{\psi} \bm{u} \cdot \bm{n} \mathrm{d}s - \int_{\partial \Omega} \mu_{\psi} \bm{j}_{\psi} \cdot \bm{n} \mathrm{d}s
	\nonumber \\ 
	& ~ - \frac{1}{2} \int_{\partial\Omega} \rho \bm{u} |\bm{u}|^{2} \cdot \bm{n} \mathrm{d}s 
	+ \int_{\partial \Omega} \bm{u} \cdot ( \sigma_{\eta} + \sigma_{e} ) \cdot \bm{n} \mathrm{d}s 
	- \int_{\partial \Omega} \phi \frac{\partial\bm{D}}{\partial t} \cdot \bm{n} \mathrm{d}s 
	+ \int_{\partial \Omega} \lambda \delta^{2} \nabla \psi \frac{\partial \psi}{\partial t} \cdot \bm{n} \mathrm{d}s, 
	\nonumber \\ 
	= & ~ - \int_{\Omega} \nabla \bm{u} : ( \sigma_{\eta} + \sigma_{\psi} + \sigma_{e} ) \mathrm{d}\bm{x} 
	- \int_{\Omega} p \textbf{I} : \nabla \bm{u} \mathrm{d}\bm{x} 
	+ \int_{\Omega} \sum_{i=1}^{N} \nabla \mu_{i} \cdot \bm{j}_{i} \mathrm{d}\bm{x} 
	+ \int_{\Omega} \nabla \mu_{\psi} \cdot \bm{j}_{\psi} \mathrm{d}\bm{x} 
	\nonumber \\ 
	& ~ + \int_{\Omega} \epsilon_{\it eff} \left( \nabla \phi \otimes \nabla \phi -\frac{1}{2} \left| \nabla \phi \right|^{2} \textbf{I} \right) : \nabla \bm{u} \mathrm{d}\bm{x} 
	- \int_{\Omega} \nabla \bm{u} : ( \lambda \delta^{2} \nabla \psi \otimes \nabla \psi ) \mathrm{d}\bm{x},
	%	& ~ + \int_{\partial \Omega} \lambda \left( \frac{\delta^{2}}{2} | \nabla \psi |^{2} - F(\psi) \right) \bm{u} \cdot \bm{n} \mathrm{d}s 
	%	\nonumber \\ 
	%	& ~ + \int_{\partial \Omega} \left( \lambda \left( -\delta^{2} \nabla^{2} \psi + F^{\prime}(\psi) \right) 
	%	- \frac{1}{2} \frac{\partial \epsilon_{\it eff}}{\partial \psi} |\bm{E}|^{2} \right) \psi \bm{u} \cdot \bm{n} \mathrm{d}s 
	%	+ \int_{\partial \Omega} \lambda \delta^{2} \bm{u} \cdot \nabla \psi \otimes \nabla \psi \cdot \bm{n} \mathrm{d}s,
	%	& ~ + \int_{\partial \Omega} \sum_{i=1}^{N} k_{B} T c_{i} \bm{u} \cdot \bm{n} \mathrm{d}s 
	%	+ \int_{\partial \Omega} ( \bm{u} \cdot \nabla \phi ) \bm{D} \cdot \bm{n} \mathrm{d}s 
	%	+ \int_{\partial \Omega} \epsilon_{\it eff} \frac{1}{2} | \nabla \phi |^{2} \bm{u} \cdot \bm{n} \mathrm{d}s 
	%	\nonumber \\ 
	%	& ~ - \int_{\partial \Omega} \sum_{i=1}^{N} c_{i} \mu_{i} \bm{u} \cdot \bm{n} \mathrm{d}s 
	%	- \int_{\partial \Omega} \sum_{i=1}^{N} \mu_{i} \bm{j}_{i} \cdot \bm{n} \mathrm{d}s
	%	- \int_{\partial \Omega} \psi \mu_{\psi} \bm{u} \cdot \bm{n} \mathrm{d}s 
	%	- \int_{\partial \Omega} \mu_{\psi} \bm{j}_{\psi} \cdot \bm{n} \mathrm{d}s
	%	\nonumber \\ 
	%	& ~ - \frac{1}{2} \int_{\partial\Omega} \rho \bm{u} |\bm{u}|^{2} \cdot \bm{n} \mathrm{d}s 
	%	+ \int_{\partial \Omega} \bm{u} \cdot ( \sigma_{\eta} + \sigma_{e} ) \cdot \bm{n} \mathrm{d}s 
	%	\nonumber \\ 
	%	& ~ - \int_{\partial \Omega} \phi \frac{\partial\bm{D}}{\partial t} \cdot \bm{n} \mathrm{d}s 
	%	+ \int_{\partial \Omega} \lambda \delta^{2} \nabla \psi \frac{\partial \psi}{\partial t} \cdot \bm{n} \mathrm{d}s, 
\end{align}
where the close boundary conditions \eqref{def: bdc} are used. 

Comparing with the predefined dissipation functional \eqref{def: dissipation}, we obtain the expression for those unknowns variables, 
\begin{subequations}
	\begin{align}
		\bm{j}_{i} ~ = & ~ -\frac{D_{i}c_{i}}{k_{B}T}\nabla\mu_{i},  \\ 
		\bm{j}_{\psi} ~ = & ~ -\mathcal{M}\nabla\mu_{\psi},  \\ 
		\bm{\sigma}_{\eta} ~ = & ~ 2 \eta \bm{D}_{\eta} - p \textbf{I},  \\ 
		\bm{\sigma}_{\psi} ~=&~  -\lambda \delta^{2} \nabla \psi \otimes \nabla \psi, \\
		\bm{\sigma}_{e} ~ = & ~  \epsilon_{\it eff} \left( \nabla \phi \otimes \nabla \phi -\frac{1}{2} \left| \nabla \phi \right|^{2} \textbf{I} \right). 
	\end{align}
\end{subequations}

Therefore,  the Poisson-Nernst-Planck-Navier-Stokes-Cahn-Hilliard system for a droplet in electrolyte under electric fields  is summarized as follows
\begin{subequations}\label{eqn: dimensional eqns}
	\begin{align}
		&\frac{\partial \psi}{\partial t} 
		+ \nabla \cdot \left( \bm{u} \psi \right) 
		= \nabla \cdot(\mathcal{M} \nabla \mu_{\psi}), 
		& \mbox{in} \quad \Omega, 
		\label{def: dimensional psi}\\ 
		%		&\bm{j}_{\psi} = -\mathcal{M} \nabla \mu_{\psi},  
		%		& \mbox{in} \quad \Omega, 
		%		\label{def: dimensional j psi}\\ 
		&\mu_{\psi} = \lambda \left( - \delta^{2} \nabla^{2} \psi 
		+ F^{\prime}(\psi) \right) 
		- \frac{1}{2} \frac{\partial \epsilon_{\it eff}}{\partial \psi} |\bm{E}|^{2}, 
		& \mbox{in} \quad \Omega, 
		\label{def: dimensional mu psi}\\
		&	\frac{\partial c_{i}}{\partial t} 
		+ \nabla \cdot \left(\bm{u} c_{i}\right) 
		= \nabla\cdot\left(\frac{D_{i}c_{i}}{k_{B}T}z_{i}e\nabla\phi 
		+ D_{i}\nabla c_{i}\right),  \qquad i=1,\cdots,N, 
		& \mbox{in} \quad \Omega, 
		\label{def: dimensional ci}\\
		%		&\bm{j}_{i} = -\frac{D_{i}c_{i}}{k_{B}T} \nabla \mu_{i}, 
		%		& \mbox{in} \quad \Omega, 
		%		\label{def: dimensional ji}\\ 
		%		&\mu_{i} = z_{i} e \phi + k_{B}T \ln \frac{c_{i}}{\tilde{c}}, 
		%		& \mbox{in} \quad \Omega, 
		%		\label{def: dimensional mui}\\
		&-\nabla \cdot (\epsilon_{\it eff} \nabla \phi) = \sum_{i=1}^{N}z_{i}ec_{i}, 
		& \mbox{in} \quad \Omega, 
		\label{def: dimensional E}\\ 
		%		&\bm{D} = \epsilon_{\it eff} \bm{E} = - \epsilon_{\it eff} \nabla \phi, 
		%		& \mbox{in} \quad \Omega, 
		%		\label{def: dimensional phi} \\
		&\rho \left( \frac{\partial \bm{u}}{\partial t} 
		+ \left( \bm{u} \cdot \nabla \right) \bm{u} \right) 
		= \nabla \cdot\bm{\sigma}_{\eta} 
		+ \nabla \cdot \bm{\sigma}_{\psi} 
		+ \nabla \cdot \bm{\sigma}_{e}, 
		& \mbox{in} \quad \Omega, 
		\label{def: dimensional u}\\
		& \nabla \cdot \bm{u} = 0, & \mbox{in} \quad \Omega, 
		\label{def: dimensional nabla u} \\ 
		&\bm{\sigma}_{\eta} = 2 \eta \bm{D}_{\eta}  - p \textbf{I},  
		& \mbox{in} \quad \Omega, 
		\label{def: dimensional sigma eta}\\
		&\bm{\sigma}_{\psi} = - \lambda \delta^{2} \nabla \psi \otimes \nabla \psi,  
		& \mbox{in} \quad \Omega, 
		\label{def: dimensional sigma psi}\\
		&\bm{\sigma}_{e} = \epsilon_{\it eff} \left( \nabla \phi \otimes \nabla \phi 
		-\frac{1}{2} \left| \nabla \phi \right|^{2} \textbf{I} \right), 
		& \mbox{in} \quad \Omega, 
		\label{def: dimensional sigma e}\\
		&\epsilon_{\it{eff}}^{-1} 
		= \frac{1-\psi}{2\epsilon^{-}} 
		+ \frac{1+\psi}{2\epsilon^{+}}
		+ \frac{\left(1-\psi^{2}\right)^{2}}{\delta C_{m}}, 
		& \mbox{in} \quad \Omega, 
		\label{def: dimensional epsilon}%\\
		%        &D_{i}^{-1} 
		%        = \frac{(1-\psi^{2})^{2}}{A_{2} \delta \frac{g_{i}}{\left(z_{i}e\right)^{2}}
			%        \frac{\partial\mu_{i}}{\partial c_{i}}} 
		%        + \frac{1-\psi}{2D_{i}^{-}} 
		%        + \frac{1+\psi}{2D_{i}^{+}}, 
		%        & \mbox{in} \quad \Omega,  
		%        \label{def: dimensional D}
	\end{align}
\end{subequations}
with boundary condition \eqref{def: bdc}. 

\subsection{Time scale analysis and approximation models}
Following previous work \cite{HU2015Ahybrid}, if $R$ is the radius of the initial drop,
the characteristic length scale, velocity,  time, pressure  are set to be  $L=R$,  $\tilde{u}=\sqrt{\frac{\lambda\delta}{\rho R}}$,  $\tilde{t}=\sqrt{\frac{\rho R^{3}}{\lambda\delta}}$,  and $\tilde{p} = \frac{\lambda \delta}{L} = \frac{\sigma_{s}}{L}, $ respectively. Here we define $\sigma_{s}$ as the surface tension.   The characteristic dielectric constant is set to be the one in the outer region $\epsilon^{-}$.  And the characteristic electric field intensity is defined as $E_{\infty}$.

The dimensionless Navier-Stokes equations are given by 
\begin{align}\label{eqn: non-d-2}
	\left( \frac{\partial \bm{u}^{\prime}}{\partial t^{\prime}} 
	+ \left( \bm{u}^{\prime} \cdot \nabla \right) \bm{u}^{\prime} \right) 
	+ \nabla p^{\prime} 
	= &\frac{1}{Re}\nabla\cdot(\eta^{\prime}(\nabla\bm{u}^{\prime}+(\nabla\bm{u}^{\prime})^T)\\
	&	- \delta^{\prime} \nabla \cdot \left( \nabla \psi \otimes \nabla \psi\right) 
	+ Ca_{E} \nabla \cdot \epsilon_{\it eff}^{\prime} \left( \bm{E}^{\prime} \otimes \bm{E}^{\prime} 
	- \frac{1}{2} \left| \bm{E}^{\prime} \right|^{2} \textbf{I} \right).\nonumber
\end{align} 
Here  $Re = \frac{\rho\tilde{u}L}{\tilde{\eta}}$ is the Reynolds number, 
$\delta^{\prime} = \frac{\delta}{L}$ is the non-dimensional thickness of diffuse interface,  $Ca_{E}$ is the  electrical capillary number
defined as $Ca_{E} = \epsilon^{-} R \left|E_{\infty}\right|^{2}/\lambda \delta$, 
which measures the strength of the electric field relative 
to the surface tension force.

%  
%Firstly, we substitute \eqref{def: sigma eta}, \eqref{def: sigma psi} and 
%\eqref{def: sigma e} into \eqref{def: u}, and then we have 
%\begin{align}\label{eqn: non-d-1}
%	& \frac{\rho \tilde{u}}{\tilde{t}} \left( \frac{\partial \bm{u}^{\prime}}{\partial t^{\prime}} 
%	+ \left( \bm{u}^{\prime} \cdot \nabla \right) \bm{u}^{\prime} \right) 
%	+\frac{\tilde{p}}{L}\nabla p^{\prime} \nonumber\\ 
%	= & \frac{\eta \tilde{u}}{L^{2}}\nabla^{2}\bm{u}^{\prime} 
%	- \frac{\lambda \delta^{2}}{L^{3}} \nabla \cdot \left( \nabla \psi \otimes \nabla \psi\right) 
%	+ \frac{\epsilon^{-}\left|E_{\infty}\right|^{2}}{L}
%	\nabla \cdot \epsilon_{\it eff}^{\prime} \left( \bm{E}^{\prime} \otimes \bm{E}^{\prime}
%	- \frac{1}{2} \left| \bm{E}^{\prime} \right|^{2} \textbf{I} \right), 
%\end{align}
%where we take 
%\begin{equation} 
%	\tilde{p} = \frac{\lambda \delta}{L} = \frac{\sigma_{s}}{L}, 
%\end{equation}
%and here we define $\sigma_{s}$ as the surface tension. 
%In addition, an electrical capillary number $Ca_{E}$, 
%defined as $Ca_{E} = \epsilon^{-} R \left|E_{\infty}\right|^{2}/\lambda \delta$, 
%is introduced here to measures the strength of the electric field relative 
%to the surface tension force, 
%which means equation \eqref{eqn: non-d-1} will be transformed into 

The non-dimensional effective dielectric coefficient is as follows, 
\begin{equation}
	\epsilon^{\prime}_{\it eff} 
	= \frac{1}{\frac{\left(1-\psi^{2}\right)^{2}}{\delta^{\prime} C_{m}^{\prime}}
		+ \frac{1-\psi}{2}
		+ \frac{1+\psi}{2\epsilon_{r}}}			
	\left(\mbox{where}\quad \epsilon_{r} 
	= \frac{\epsilon^{+}}{\epsilon^{-}} \mbox{ and }
	C_{m}^{\prime}=\frac{C_{m}R}{\epsilon^{-}}\sim o\left(\frac{1}{\delta^{\prime}}\right)\right). 
\end{equation}
%If  we  define the characteristic  electric potential $\tilde{\phi} = E_{\infty} L$, the equation of electric potential could be written as 
%\begin{equation}
%	\nabla \cdot \left( \sigma_{\it{eff}}^{\prime} \nabla \phi \right) = 0, 
%\end{equation}
%where the non-dimensional effective conductivity $\sigma_{\it{eff}}$ 
%is defined as 
%\begin{equation}
%	\sigma^{\prime}_{\it eff} = \frac{1}{ \frac{1-\psi}{2} + \frac{1+\psi}{2\sigma_{r}}}			
%	\left(\mbox{where}\quad \sigma_{r}=\frac{\sigma^{+}}{\sigma^{-}}\right). 
%\end{equation}
%It is pointed that $\epsilon_{r}$ and $\sigma_{r}$ are the ratio of 
%dielectric coefficient and conductivity between inner and outer for the interface. 
%The non-dimensional current density is acquired directly when we define 
%characteristic current density $\tilde{\bm{J}} = \sigma^{-}E_{\infty}$, 
%\begin{equation}
%	\bm{J}^{\prime} = \sigma_{\it{eff}}^{\prime}\nabla\phi^{\prime}. 
%\end{equation}
For the Cahn-Hilliard equation, if let  $\tilde{\mu}=\frac{\sigma_{s}}{L}$ be the characteristic chemical potential, we have 
%\begin{align}
%	& \frac{1}{\tilde{t}}\frac{\partial \psi}{\partial t^{\prime}} 
%	+ \frac{\tilde{u}}{L}\nabla \cdot \left( \bm{u}^{\prime} \psi \right) 
%	= \frac{\mathcal{M}\tilde{\mu}}{L^{2}} \nabla^{2} \mu_{\psi}^{\prime}, 
%	\\
%	& \tilde{\mu} \mu_{\psi}^{\prime} 
%	= \lambda \left( - \frac{\delta^{2}}{L^{2}} \nabla^{2} \psi 
%	+ F^{\prime}\left(\psi\right) \right)
%	- \frac{\epsilon^{-}E_{\infty}^{2}}{2} 
%	\frac{\partial \epsilon_{\it eff}^{\prime}}{\partial \psi} 
%	\left|\bm{E}^{\prime}\right|^{2}, 
%\end{align}
%when we take $\tilde{\mu}=\frac{\sigma_{s}}{L}$, we have 
%\begin{equation}
%	\mu_{\psi}^{\prime} 
%	= - \delta^{\prime} \nabla^{2} \psi 
%	+ \frac{1}{\delta^{\prime}}F^{\prime}\left( \psi \right)
%	- \frac{Ca_{E}}{2} 
%	\frac{\partial \epsilon_{\it eff}^{\prime}}{\partial \psi} 
%	\left|\bm{E}^{\prime}\right|^{2}. 
%\end{equation}
the Cahn-Hilliard equations as follows, 
\begin{align}
	& \frac{\partial \psi}{\partial t^{\prime}} 
	+ \nabla \cdot \left( \bm{u}^{\prime} \psi \right) 
	= M^{\prime} \nabla^{2} \mu_{\psi}^{\prime}, 
	\\
	& \mu_{\psi}^{\prime} 
	= - \delta^{\prime} \nabla^{2} \psi 
	+ \frac{1}{\delta^{\prime}}F^{\prime}\left(\psi\right)
	- \frac{Ca_{E}}{2} 
	\frac{\partial \epsilon_{\it eff}^{\prime}}{\partial \psi} 
	\left|\bm{E}^{\prime}\right|^{2},
\end{align}
where  $M^{\prime} = \frac{\mathcal{M}\tilde{\mu}}{L \tilde{u}}$.

We define net charge as $\rho_e = \sum_{i=1}^N z_ic_ie$ and the characteristic electric potential as $\tilde{\phi} =E_{\infty}L$,  the dimensionless of Poisson equation is given by 
\begin{equation}
	-\zeta^2\nabla\cdot(\epsilon_{\it eff}^{\prime}\nabla\phi') = \rho_e'
\end{equation}
where $\zeta = \frac{\lambda_D}{L}$ and $\lambda_D = \sqrt{\frac{\epsilon^-E_{\infty}L}{\tilde{c}e}}$ is the Debye length.

For the ion concentration, multiplying above equation \eqref{def: dimensional ci} by   $z_{i}e$ 
respectively and summing up from $1$ to $N$ give
\begin{equation}
	\frac{\partial \left(\sum\limits_{i=1}^{N}z_{i}c_{i}e\right)}{\partial t} 
	+ \nabla \cdot \left(\bm{u}\left(\sum\limits_{i=1}^{N}z_{i}c_{i}e\right)\right) 
	= \nabla\cdot\left(\frac{\left(\sum\limits_{i=1}^{N}D_iz_{i}^{2}c_{i}\right)e^2}{k_{B}T}\nabla\phi 
	+ \sum\limits_{i=1}^{N}D_{i}\nabla \left(z_{i}c_{i}e\right)\right). 
\end{equation}
For simplicity, we assume that $D_i=D$ for all ionic species for the rest of the paper and the case of different $D_i$ will be considered in a follow up paper. The equation above could be rewritten as 
\begin{equation}\label{def:rhoe}
	\frac{\partial \rho_e}{\partial t} 
	+ \nabla \cdot \left(\bm{u}\rho_e\right) 
	= \nabla\cdot\left(\sigma_{c}\nabla\phi 
	+ D\nabla \rho_e\right). 
\end{equation}
where $\sigma_{c} = \frac{D\sum_{i=1}^Nz_i^2c_ie^2}{k_BT}$ is the effective conductivity. 
If we define the diffusion time $\tilde{t}_D = \frac{L^2}{D}$ and electric relaxation time $\tilde{t}_E = \frac{\epsilon^-}{\sigma^*}$, then the dimensionless of charge density is 
\begin{equation}\label{eqn: charge}
	\frac{1}{\tilde{t}}\left(\frac{\partial \rho'_e}{\partial t} 
	+ \nabla \cdot \left(\bm{u}\rho'_e\right) \right)
	= \frac{\zeta^2}{\tilde{t}_E}\nabla\cdot\left(\sigma_{c}^{\prime} \nabla\phi' \right)
	+\frac{1}{\tilde{t}_D}\nabla\cdot\left( D\nabla \rho'_e\right). 
\end{equation}

When the electric relaxation time is much shorter than the diffusion time and macro time scale, i.e. $\tilde{t}_E<<\tilde{t}_D$ and $\tilde{t}_E<<\tilde{t}$ , the ions in   bulk attain steady state simultaneously and conductivity  could be treated as constants  $\sigma^{\pm}$ in the bulk region. 

\subsubsection{Leaky-dielectric model}

In the leaky-dielectric model, we assume that the conductivities $\sigma^\pm$ remains as constants, and the conductivity in phase field frame work could be defined as follows 
\begin{equation}
	\sigma^{\prime}_{\it eff} = \frac{1-\psi}{2} + \frac{1+\psi}{2\sigma^+/\sigma^-} =  \frac{1-\psi}{2} + \frac{1+\psi}{2\sigma_r}. 
\end{equation}
Equation (\ref{eqn: charge}) can be approximated by 
\begin{equation}
	\nabla\cdot\left(\sigma_{\it eff} \nabla\phi' \right) = 0. 
\end{equation}
Replacing (\ref{eqn: charge}) by the equation above, we obtain the non-dimensional leaky-dielectric  phase field system as follows, 
%\begin{subequations}
%	\begin{align}
	%		&\left( \frac{\partial \bm{u}^{\prime}}{\partial t^{\prime}} 
	%		+ \left( \bm{u}^{\prime} \cdot \nabla \right) \bm{u}^{\prime} \right) 
	%		+ \nabla p^{\prime} 
	%		= \frac{1}{Re}\nabla^{2}\bm{u}^{\prime} 
	%		- \delta^{\prime}\nabla \cdot \left( \nabla \psi \otimes \nabla \psi\right) 
	%		\nonumber \\
	%		&\qquad \qquad \qquad \qquad \qquad \qquad 
	%		+ Ca_{E} \nabla \cdot \epsilon_{\it eff}^{\prime} \left( \bm{E}^{\prime} \otimes \bm{E}^{\prime} 
	%		- \frac{1}{2} \left| \bm{E}^{\prime} \right|^{2} \textbf{I} \right),
	%		\\
	%		& \nabla \cdot \bm{u}^{\prime} = 0, 
	%		\\ 
	%		&\nabla \cdot \left( \sigma_{\it eff}^{\prime} \nabla \phi^{\prime} \right) = 0, 
	%		\\ 
	%		&\bm{J}^{\prime} = \sigma_{\it{eff}}^{\prime}\nabla\phi^{\prime}, 
	%		\\
	%		&\frac{\partial \psi}{\partial t^{\prime}} 
	%		+ \nabla \cdot \left( \bm{u}^{\prime} \psi \right) 
	%		= M^{\prime} \nabla^{2} \mu_{\psi}^{\prime}, 
	%		\\
	%		&\mu_{\psi}^{\prime} = - \delta^{\prime} \nabla^{2} \psi 
	%		+ \frac{1}{\delta^{\prime}}F^{\prime}\left(\psi\right)
	%		- \frac{Ca_{E}}{2} 
	%		\frac{\partial \epsilon_{\it eff}^{\prime}}{\partial \psi} 
	%		\left|\bm{E}^{\prime}\right|^{2}.
	%	\end{align}
%\end{subequations}
\begin{subequations}\label{eq:leakydielectricsystem}
	\begin{align}
		& \left( \frac{\partial \bm{u}}{\partial t} 
		+ \left( \bm{u} \cdot \nabla \right) \bm{u} \right) 
		+ \nabla p 
		= \frac{1}{Re}\nabla^{2}\bm{u} 
		- \delta\nabla \cdot \left( \nabla \psi \otimes \nabla \psi\right) 
		\nonumber \\
		&\qquad\qquad\qquad\qquad\qquad+ Ca_{E} \nabla \cdot \epsilon_{\it eff} \left( \bm{E} \otimes \bm{E} 
		- \frac{1}{2} \left| \bm{E} \right|^{2} \textbf{I} \right),
		\\
		& \nabla \cdot \bm{u} = 0,
		\\
		& \nabla \cdot \left( \sigma_{\it eff} \nabla \phi \right) = 0, 
		\\ 
		%		& \bm{J} = \sigma_{\it{eff}} \nabla \phi, 
		%		\\
		& \frac{\partial \psi}{\partial t} 
		+ \nabla \cdot \left( \bm{u} \psi \right) 
		= M \nabla^{2} \mu_{\psi}, 
		\\
		& \mu_{\psi} = - \delta \nabla^{2} \psi 
		+ \frac{1}{\delta}F^{\prime}\left(\psi\right)
		- \frac{Ca_{E}}{2} 
		\frac{\partial \epsilon_{\it eff}}{\partial \psi} 
		\left|\bm{E}\right|^{2}. 
	\end{align}
\end{subequations}
where   we omit superscript $\prime$ for simplicity's sake.

Note that the following fact, 
\begin{equation}
	\nabla \cdot \left(\nabla \psi \otimes \nabla \psi \right) 
	= \nabla^{2} \psi \nabla \psi + \frac{1}{2} \nabla \left| \nabla \psi \right|^{2} 
	= -\frac{1}{\delta} \mu \nabla \psi 
	+ \nabla \left( \frac{1}{2} \left| \nabla \psi \right|^{2} + \frac{1}{\delta^{2}} F\left(\psi\right)\right). 
\end{equation}
Then, if we define a new variable 
\begin{equation}
	\mu = - \delta \nabla^{2} \psi + \frac{1}{\delta}F\left(\psi\right), 
	\label{def: mu}
\end{equation}
and let  
\begin{equation}
	\hat{p} = p +\frac{1}{\delta}F\left(\psi\right) + \frac{\delta}{2}|\nabla\psi|^{2}, 
\end{equation}
we can rewrite the above system as follows, 
\begin{subequations}\label{eqn: main}
	\begin{align}
		& \left( \frac{\partial \bm{u}}{\partial t} 
		+ \left( \bm{u} \cdot \nabla \right) \bm{u} \right) 
		+ \nabla p 
		= \frac{1}{Re}\nabla^{2}\bm{u} 
		+ \mu \nabla \psi 
		+ Ca_{E} \nabla \cdot \epsilon_{\it eff} \left( \bm{E} \otimes \bm{E} 
		- \frac{1}{2} \left| \bm{E} \right|^{2} \textbf{I} \right),
		\label{eqn: u}\\
		& \nabla \cdot \bm{u} = 0,
		\label{eqn: nabla u}\\
		& \nabla \cdot \bm{J} = 0, 
		\label{eqn: sigma phi}\\ 
		& \bm{J} = \sigma_{\it{eff}} \nabla \phi, 
		\label{eqn: J}\\
		& \frac{\partial \psi}{\partial t} 
		+ \nabla \cdot \left( \bm{u} \psi \right) 
		= M \nabla^{2} \mu_{\psi}, 
		\label{eqn: psi}\\
		& \mu_{\psi} = - \delta \nabla^{2} \psi 
		+ \frac{1}{\delta}F^{\prime}\left(\psi\right)
		- \frac{Ca_{E}}{2} 
		\frac{\partial \epsilon_{\it eff}}{\partial \psi} 
		\left|\bm{E}\right|^{2}, 
		\label{eqn: mu_psi}\\
		& \mu = - \delta \nabla^{2} \psi + \frac{1}{\delta}F\left(\psi\right),
		\label{eqn: mu}
	\end{align}
\end{subequations}  
where  $\hat{\cdot}$ is omitted for simplicity. 

\subsubsection{Net charge model}
In this model, we also assume that the conductivities $\sigma^{\pm}$ remain as constants and the effective conductivity given by 
\begin{equation}
	\sigma_{\it eff} = \frac{1-\psi}{2} + \frac{1+\psi}{2\sigma^+/\sigma^-} =  \frac{1-\psi}{2} + \frac{1+\psi}{2\sigma_r}. 
\end{equation}
However, we keep all the terms in (\ref{eqn: charge}) and use it to computer $\rho_e$ instead of $\phi$. The electric potential can be computed with the Poisson equation, which was ignored in the leaky-dielectric model. The phase field system including the evolution of net charge is given as follows, 
\begin{subequations}\label{eqn: main2}
	\begin{align}
		& \left( \frac{\partial \bm{u}}{\partial t} 
		+ \left( \bm{u} \cdot \nabla \right) \bm{u} \right) 
		+ \nabla p 
		= \frac{1}{Re}\nabla^{2}\bm{u} 
		+ \mu \nabla \psi 
		+ Ca_{E} \nabla \cdot \epsilon_{\it eff} \left( \bm{E} \otimes \bm{E} 
		- \frac{1}{2} \left| \bm{E} \right|^{2} \textbf{I} \right),
		\label{eqn: u2}\\
		& \nabla \cdot \bm{u} = 0,
		\label{eqn: nabla u2}\\
		& \frac{1}{\tilde{t}}\left(\frac{\partial \rho_e}{\partial t} 
		+ \nabla \cdot \left(\bm{u}\rho_e\right) \right)
		= \frac{\zeta^2}{\tilde{t}_E}\nabla\cdot\left(\sigma_{\it eff} \nabla\phi \right)
		+\frac{1}{\tilde{t}_D}\nabla\cdot\left( D\nabla \rho_e\right),
		\label{eqn: sigma rhoe2}\\ 
		& -\zeta^2\nabla\cdot(\epsilon_{\it eff}\nabla\phi) = \rho_e
		\label{eq: phi2}\\
		& \frac{\partial \psi}{\partial t} 
		+ \nabla \cdot \left( \bm{u} \psi \right) 
		= M \nabla^{2} \mu_{\psi}, 
		\label{eqn: psi2}\\
		& \mu_{\psi} = - \delta \nabla^{2} \psi 
		+ \frac{1}{\delta}F^{\prime}\left(\psi\right)
		- \frac{Ca_{E}}{2} 
		\frac{\partial \epsilon_{\it eff}}{\partial \psi} 
		\left|\bm{E}\right|^{2}, 
		\label{eqn: mu_psi2}\\
		& \mu = - \delta \nabla^{2} \psi + \frac{1}{\delta}F\left(\psi\right).
		\label{eqn: mu2}
	\end{align}
\end{subequations}  
%where  $\hat{\cdot}$ is omitted for simplicity. 

\begin{remark} \label{remarkontheelectricforce}
	Here in Eq. \eqref{eqn: u} or  Eq. \eqref{eqn: u2}, the electric force term $\nabla\cdot\bm{\sigma_e}$ could be written as follows
	\begin{align}\label{eqn: equivalent form of nabla sigma_e}
		& \nabla \cdot \epsilon_{\it eff} 
		\left( \nabla \phi \otimes \nabla \phi 
		-\frac{1}{2} \left| \nabla \phi \right|^{2} \textbf{I} \right) 
		\nonumber \\
		= & \nabla \cdot 
		\left( \epsilon_{\it eff} \nabla \phi \otimes \nabla \phi \right) 
		- \nabla \cdot 
		\left( \frac{\epsilon_{\it eff}}{2} 
		\left| \nabla \phi \right|^{2} \textbf{I} \right) 
		\nonumber \\ 
		= & \nabla \cdot \left( \epsilon_{\it eff} \nabla \phi \right) 
		\nabla \phi 
		+ \frac{\epsilon_{\it eff}}{2} \nabla \left| \nabla \phi \right|^{2} 
		- \frac{1}{2} \left|\nabla \phi \right|^{2} \nabla \epsilon_{\it eff} 
		- \frac{\epsilon_{\it eff}}{2} \nabla \left| \nabla \phi \right|^{2} 
		\nonumber \\ 
		= & \nabla \cdot \left( \epsilon_{\it eff} \nabla \phi \right) \nabla \phi 
		- \frac{1}{2} \left|\nabla \phi \right|^{2} \nabla \epsilon_{\it eff} 
		\nonumber \\
		= & \nabla \cdot \left( \epsilon_{\it eff} \nabla \phi \right) \nabla \phi 
		- \frac{1}{2} \left|\nabla \phi \right|^{2} 
		\frac{\partial \epsilon_{\it{eff}}}{\partial \psi} \nabla \psi\nonumber\\
		= & - \frac{1}{\zeta^{2}}\rho_{e} \nabla \phi 
		- \frac{1}{2} \left|\nabla \phi \right|^{2} 
		\frac{\partial \epsilon_{\it{eff}}}{\partial \psi} \nabla \psi\nonumber\\
		=	& \bm{F}_L +\bm{F}_p
	\end{align} 
	where the first term  $\bm{F}_L$ is the Lorentz force  due to the interaction of net charges with electric field and the second term $\bm{F}_p$ is due to the polarization stress. 
	Then Eq.\eqref{eqn: u}  or  Eq. \eqref{eqn: u2} could be rewritten as 
	\begin{equation}\label{eqn：	urevised}
		\left( \frac{\partial \bm{u}}{\partial t} 
		+ \left( \bm{u} \cdot \nabla \right) \bm{u} \right) 
		+ \nabla p 
		= \frac{1}{Re}\nabla^{2}\bm{u} 
		+ \mu_{\psi} \nabla \psi 
		- \frac{Ca_{E}}{\zeta^2} \rho_{e} \nabla \phi. 
	\end{equation}
\end{remark}

	\section{Sharp interface limit of the diffuse leaky-dielectric interface model}
In this section, a detailed asymptotic analysis is presented to show that as $\delta\rightarrow 0$, 
the limit of the obtained system \eqref{eqn: main} is consistent with the sharp interface Taylor–Melcher model \cite{Saville1997electro,Melcher1969Electrohydrodynamics,HU2015Ahybrid}.  
Here the mobility is assumed as $M = \alpha_1 \delta^{2}$ and capacitance $C_m = \alpha_2\delta^{-1}$, 
where $\alpha_1$ and $\alpha_2$ are constants independent of $\delta$. In the following, 
$ \left[\![f]\!\right] = f^+-f^-$ denotes the jump across the interface.
\label{sec: sharp interface limit}
%In this section, we will use abbreviation h.o.t. to instead of high order terms. 
\subsection{Outer expansions}
In the outer region for the bulk fluids away from the interface defined by $\Gamma=\{\bm{x}:\psi(\bm{x},t)=0\}$, 
we consider the sharp interface limit by taking $\delta\to 0$, and using the expansions as follows, 
\begin{subequations}\label{eq:electricforce}
	\begin{align}
		&\psi^{\pm} = \psi_{0}^{\pm} + \delta \psi_{1}^{\pm} 
		+ \delta^{2} \psi_{2}^{\pm} + o\left(\delta^{2}\right), 
		\label{out: psi}\\ 
		&\mu_{\psi}^{\pm} = \delta^{-1}\mu_{\psi 0}^{\pm} 
		+ \mu_{\psi 1}^{\pm} + \delta \mu_{\psi 2}^{\pm} 
		+ \delta^{2} \mu_{\psi 3}^{\pm} + o\left(\delta^{2}\right), 
		\label{out: mu_psi}\\
		&\mu^{\pm} = \delta^{-1}\mu_{0}^{\pm} 
		+ \mu_{1}^{\pm} + \delta \mu_{2}^{\pm} 
		+ \delta^{2} \mu_{3}^{\pm} + o\left(\delta^{2}\right), 
		\label{out: mu}\\
		&\phi^{\pm} = \phi_{0}^{\pm} + \delta \phi_{1}^{\pm} 
		+ \delta^{2} \phi_{2}^{\pm} + o\left(\delta^{2}\right), 
		\label{out: phi}\\ 
		&\bm{u}^{\pm} = \bm{u}_{0}^{\pm} + \delta \bm{u}_{1}^{\pm} 
		+ \delta^{2} \bm{u}_{2}^{\pm} + o\left(\delta^{2}\right), 
		\label{out: u}\\ 
		&p^{\pm} = p_{0}^{\pm} + \delta p_{1}^{\pm} + \delta^{2} p_{2}^{\pm} 
		+ o\left(\delta^{2}\right), 
		\label{out: p}\\ 
		&\bm{\sigma}_{e}^{\pm} = \bm{\sigma}_{e0}^{\pm} + \delta \bm{\sigma}_{e1}^{\pm} 
		+ \delta^{2} \bm{\sigma}_{e2}^{\pm} + o\left(\delta^{2}\right), 
		\label{out: sigmae}\\
		&\bm{J}^{\pm} = \bm{J}_{0}^{\pm} + \delta \bm{J}_{1}^{\pm} + \delta^{2} \bm{J}_{2}^{\pm} 
		+ o\left(\delta^{2}\right), 
		\label{out: p} 
	\end{align}
\end{subequations}

For the Cahn-Hilliard equation, we first consider the chemical potential of $\psi$. 
Substituting \eqref{out: psi} into \eqref{eqn: mu_psi} and \eqref{eqn: mu}  yields 
\begin{subequations}
	\begin{align} 
		& \delta^{-1}\mu_{0}^{\pm} + \mu_{1}^{\pm} + \delta \mu_{2}^{\pm} 
		+ \delta^{2} \mu_{3}^{\pm} + o\left(\delta^{2}\right) 
		= - \delta \nabla^{2} \left(\psi_{0}^{\pm} + \delta \psi_{1}^{\pm} 
		+ \delta^{2} \psi_{2}^{\pm} + o\left(\delta^{2}\right)\right) 
		\nonumber \\
		& + \delta^{-1} \left(\psi_{0}^{\pm} + \delta \psi_{1}^{\pm} 
		+ \delta^{2} \psi_{2}^{\pm} + o\left(\delta^{2}\right)\right) 
		\left(\left(\psi_{0}^{\pm} + \delta \psi_{1}^{\pm} + \delta^{2} \psi_{2}^{\pm} 
		+ o\left(\delta^{2}\right)\right)^{2}-1\right),\label{eq:muout} 
		\\
		& \delta^{-1}\mu_{\psi 0}^{\pm} + \mu_{\psi 1}^{\pm} + \delta \mu_{\psi 2}^{\pm} 
		+ \delta^{2} \mu_{\psi 3}^{\pm} + o\left(\delta^{2}\right) 
		= \delta^{-1}\mu_{0}^{\pm} + \mu_{1}^{\pm} + \delta \mu_{2}^{\pm} 
		+ \delta^{2} \mu_{3}^{\pm} + o\left(\delta^{2}\right) 
		\nonumber \\
		& - \frac{Ca_{E}}{2} \frac{\partial \epsilon_{\it eff}}{\partial \psi} 
		\left|\nabla \left(\phi_{0}^{\pm} + \delta \phi_{1}^{\pm} + \delta^{2} \phi_{2}^{\pm} 
		+ o\left(\delta^{2}\right)\right)\right|^{2}.\label{eq: mupsiout}
	\end{align}
\end{subequations}
The leading order of Eq. \eqref{eq:muout} yields 
\begin{equation}
	\mu^{\pm}_0 = \psi_0^{\pm}((\psi_0^{\pm})^2-1).
\end{equation}
According to the definition of $\epsilon_{\it eff}$, we have
\begin{equation}
	\frac{\partial \epsilon_{\it{eff}}}{\partial \psi} 
	= \frac{\frac{1}{2}-\frac{1}{2\epsilon_{r}}
		+\frac{4\psi\left(1-\psi^{2}\right)}{\delta C_{m}}}
	{\left(\frac{1-\psi}{2}
		+ \frac{1+\psi}{2\epsilon_{r}}
		+ \frac{\left(1-\psi^{2}\right)^{2}}{\delta C_{m}}\right)^{2}}, 
\end{equation}
which means 
\begin{align}\label{out eqn: mupsi} 
	&\left(\frac{1-\psi}{2}
	+ \frac{1+\psi}{2\epsilon_{r}}
	+ \frac{\left(1-\psi^{2}\right)^{2}}{\delta C_{m}}\right)^{2}
	\left(\delta^{-1}\mu_{\psi 0}^{\pm} + \mu_{\psi 1}^{\pm} 
	+ \delta \mu_{\psi 2}^{\pm} + \delta^{2} \mu_{\psi 3}^{\pm} 
	+ o\left(\delta^{2}\right) \right) 
	\nonumber\\ 
	=&\left(\frac{1-\psi}{2}
	+ \frac{1+\psi}{2\epsilon_{r}}
	+ \frac{\left(1-\psi^{2}\right)^{2}}{\delta C_{m}}\right)^{2}
	\left(\delta^{-1}\mu_{0}^{\pm} + \mu_{1}^{\pm} 
	+ \delta \mu_{2}^{\pm} + \delta^{2} \mu_{3}^{\pm} 
	+ o\left(\delta^{2}\right) \right) 
	\nonumber \\ 
	&- \frac{Ca_{E}}{2}\left(\frac{1}{2}-\frac{1}{2\epsilon_{r}}
	+ \frac{4\psi\left(1-\psi^{2}\right)}{\delta C_{m}}\right) 
	\left|\nabla \left(\phi_{0}^{\pm} + \delta \phi_{1}^{\pm} 
	+ \delta^{2} \phi_{2}^{\pm} + o\left(\delta^{2}\right)\right)\right|^{2}. 
\end{align}
The leading order of  above equation \eqref{out eqn: mupsi} gives  
\begin{equation}\label{eq:psi0mu0out}
	\mu_{\psi 0}^{\pm} 
	= \mu_{0}^{\pm} 
	= \psi_{0}^{\pm}\left(\left(\psi_{0}^{\pm}\right)^{2}-1\right). 
\end{equation}

For Navier-Stokes equations, when substituting the outer expansion of 
$\bm{u}$, $p$ and $\mu$ into momentum equation \eqref{eqn: u} and incompressibility \eqref{eqn: nabla u}, 
we have 
\begin{align*}
	& \frac{\partial\left(\bm{u}_{0}^{\pm} + \delta \bm{u}_{1}^{\pm} 
		+ \delta^{2} \bm{u}_{2}^{\pm} + o\left(\delta^{2}\right)\right)}{\partial t} 
	+ \left( \left(\bm{u}_{0}^{\pm} + \delta \bm{u}_{1}^{\pm} + \delta^{2} \bm{u}_{2}^{\pm} 
	+ o\left(\delta^{2}\right)\right)\cdot \nabla \right) 
	\left(\bm{u}_{0}^{\pm} + \delta \bm{u}_{1}^{\pm} + \delta^{2} \bm{u}_{2}^{\pm} 
	+ o\left(\delta^{2}\right)\right)  
	\\
	&+ \nabla \left(p_{0}^{\pm} + \delta p_{1}^{\pm} + \delta^{2} p_{2}^{\pm} 
	+ o\left(\delta^{2}\right)\right) 
	\\
	= &\frac{1}{Re} \nabla^{2} \left(\bm{u}_{0}^{\pm} + \delta \bm{u}_{1}^{\pm} 
	+ \delta^{2} \bm{u}_{2}^{\pm} + o\left(\delta^{2}\right) \right)\nonumber\\
	&+ \left( \delta^{-1}\mu_{0}^{\pm} + \mu_{1}^{\pm} 
	+ \delta \mu_{2}^{\pm} + \delta^{2} \mu_{3}^{\pm} 
	+ o\left(\delta^{2}\right) \right) 
	\nabla \left( \psi_{0}^{\pm} + \delta \psi_{1}^{\pm} + \delta^{2} \psi_{2}^{\pm} 
	+ o\left(\delta^{2}\right) \right) 
	\\
	& + Ca_{E} \nabla \cdot \left( \bm{\sigma}_{e0}^{\pm} + \delta \bm{\sigma}_{e1}^{\pm} 
	+ \delta^{2} \bm{\sigma}_{e2}^{\pm} + o\left(\delta^{2}\right)\right). 
\end{align*}
and 
\begin{align}
	\nabla\cdot\left(\bm{u}_{0}^{\pm} + \delta \bm{u}_{1}^{\pm} 
	+ \delta^{2} \bm{u}_{2}^{\pm} + o\left(\delta^{2}\right)\right)=0. 
\end{align}
The leading order yields
\begin{equation} 
	\mu_{0}^{\pm}\nabla\psi_{0}^{\pm}=0, 
\end{equation}
which means 
\begin{equation} 
	\nabla\left(1-\left(\psi_{0}^{\pm}\right)^{2}\right)^{2}=0, 
\end{equation}
i.e. $\psi_{0}^{\pm}=C_0^{\pm}$ and $\mu_{\psi 0}^{\pm} = \mu_0^{\pm} = (C_0^{\pm})^3 - C_0^{\pm}$.  
The next order is 
\begin{align*}
	\frac{\partial\bm{u}_0^{\pm}}{\partial t} 
	+ \bm{u}_0^{\pm}\cdot\nabla\bm{u}_0^{\pm}+\nabla p_0^{\pm} 
	= \frac{1}{Re}\nabla^2\bm{u}_0^{\pm} 
	+ \mu_0^{\pm}\nabla\psi_1^{\pm}+Ca_E\nabla\cdot\bm{\sigma}_{e0}^{\pm}. 
\end{align*}

For equation \eqref{eqn: sigma phi} and \eqref{eqn: J}, 
the leading order term gives us that 
\begin{equation}\label{out eqn: phi}
	\nabla \cdot\bm{J}_{0}^{\pm} = 0, \nonumber
\end{equation}
and 
\begin{equation}\label{out eqn: J}
	\bm{J}_{0}^{\pm} = 
	\frac{1}{ \frac{1-C_0^{\pm}}{2} + \frac{1+C_0^{\pm}}{2\sigma_{r}}}\nabla \phi^{\pm}_{0}. \nonumber
\end{equation}

For the  Maxwell stress, we have  
\begin{align}
	&\left(\frac{1-\psi}{2}
	+ \frac{1+\psi}{2\epsilon_{r}}
	+ \frac{\left(1-(\psi)^{2}\right)^{2}}{\delta C_{m}}\right)\left(	\bm{\sigma}^{\pm}_{e,0}+\delta\bm{\sigma}^{\pm}_{e,0}+o\left(\delta^{2}\right)\right )\nonumber\\
	&=\nabla \left(\phi^{\pm}_0+\delta\phi^{\pm}_1+o\left(\delta^{2}\right)\right)\otimes\nabla\left(\phi^{\pm}_0+\delta\phi^{\pm}_1+o\left(\delta^{2}\right)\right) -\frac 1 2 \left|\nabla\left(\phi^{\pm}_0+\delta\phi^{\pm}_1+o\left(\delta^{2}\right)\right)\right|^2\textbf{I}. \nonumber
\end{align}
Since $C_m = \alpha_2 \delta^{-1}$, the leading order of above equation is  
\begin{align*}
	\left(\frac{1-\psi^{\pm}}{2}
	+ \frac{1+\psi^{\pm}}{2\epsilon_{r}}
	+ \frac{\left(1-(\psi^{\pm})^{2}\right)^{2}}{\alpha_2}\right) 	\bm{\sigma}^{\pm}_{e,0} = \nabla\phi^{\pm}_0\otimes\nabla\phi^{\pm}_0-\frac 1 2|\nabla\phi^{\pm}_0|^2\textbf{I}. 
\end{align*}

In summary, for the outer region, $(\bm{u}_0^{\pm},p_0^{\pm},\mu_0^{\pm},\psi_0^{\pm},\phi_0^{\pm})$ satisfy 
\begin{subequations}
	\begin{align}
		\frac{\partial\bm{u}_0^{\pm}}{\partial t} +\bm{u}_0^{\pm}\cdot\nabla\bm{u}_0^{\pm}+\nabla p_0^{\pm} = \frac{1}{Re}\nabla^2\bm{u}_0^{\pm}+\mu_0^{\pm}\nabla\psi_1^{\pm}+Ca_E\nabla\cdot\bm{\sigma}_{e0}^{\pm},\\
		\nabla\cdot(\bm{u}_0^{\pm})=0,\\
		\frac{\partial\psi_0^{\pm}}{\partial t} +\nabla(\bm{u}_0^{\pm}\psi_0^{\pm})=0,\\
		\psi_{0}^{\pm}=C_0^{\pm}, ~\mu_{\psi 0}^{\pm} = \mu_0^{\pm} = (C_0^{\pm})^3 - C_0^{\pm},\\
		\nabla \cdot \left( 
		\frac{1}{ \frac{1-C_0^{\pm}}{2} + \frac{1+C_0^{\pm}}{2\sigma_{r}}} 
		\nabla \phi^{\pm}_{0} \right) = 0,\\
		\left(\frac{1-\psi^{\pm}}{2}
		+ \frac{1+\psi^{\pm}}{2\epsilon_{r}}
		+ \frac{\left(1-(\psi^{\pm})^{2}\right)^{2}}{\alpha_2}\right) 	\bm{\sigma}^{\pm}_{e,0} = \nabla\phi^{\pm}_0\otimes\nabla\phi^{\pm}_0-\frac 1 2|\nabla\phi^{\pm}_0|^2\textbf{I}.
	\end{align}
\end{subequations}

\subsection{Inner expansions}
Firstly, we introduce the signed distance function $d\left(\bm{x}\right)$ to the interface $\Gamma$. 
Immediately, we have $\nabla d=\bm{n}$. 
After defining a new rescaled variable 
\begin{equation}
	\xi=\frac{d\left(\bm{x}\right)}{\delta}, 
\end{equation}
for any scalar function $f\left(\bm{x}\right)$, we can rewrite it as 
\begin{equation}
	f\left(\bm{x}\right)=\tilde{f}\left(\bm{x},\xi\right), 
\end{equation}
and the relevant operators are
\begin{subequations}
	\begin{align}
		&\nabla f\left(\bm{x}\right) = 
		\nabla_{\bm{x}}\tilde{f}+\delta^{-1}\partial_{\xi}\tilde{f}\bm{n},
		\\
		&\nabla^{2} f\left(\bm{x}\right)
		=\nabla^{2}_{\bm{x}}\tilde{f}
		+\delta^{-1}\partial_{\xi}\tilde{f}\kappa
		+2\delta^{-1}\left(\bm{n}\cdot\nabla_{\bm{x}}\right)
		\partial_{\xi}\tilde{f}
		+\delta^{-2}\partial_{\xi\xi}\tilde{f},
		\\
		&\partial_{t}f=\partial_{t}\tilde{f}+\delta^{-1}\partial_{\xi}\tilde{f}\partial_{t}d, 
	\end{align}
\end{subequations}
and for a vector function $\bm{g}\left(\bm{x}\right)$, we have 
\begin{equation}
	\nabla\cdot\tilde{\bm{g}}\left(\bm{x}\right)=
	\nabla_{\bm{x}}\cdot\tilde{\bm{g}}
	+\delta^{-1}\partial_{\xi}\tilde{\bm{g}}\cdot\bm{n}. 
\end{equation}

Here the $\nabla_{\bm{x}}$ and $\nabla^{2}_{\bm{x}}$ 
stand for the gradient and Laplace 
with respect to $\bm{x}$, respectively. 
And we use the fact that $\nabla_{\bm{x}}\cdot\bm{n}=\kappa$. 
$\kappa\left(\bm{x}\right)$ for $\bm{x}\in\Gamma\left(t\right)$ is 
the mean curvature of the interface and 
is positive if the domain $\Omega_{-}$ is convex near $\bm{x}$. 
In the inner region, we assume that 
\begin{subequations}\label{eqn: inner expansions}
	\begin{align}
		&\tilde{\psi}
		=\tilde{\psi}_{0}+\delta\tilde{\psi}_{1}
		+\delta^{2}\tilde{\psi}_{2}+o\left(\delta^{2}\right),
		\label{inner: psi}\\
		&\tilde{\mu}_{\psi} 
		= \delta^{-1}\tilde{\mu}_{\psi0}
		+ \tilde{\mu}_{\psi1}
		+ \delta\tilde{\mu}_{\psi2}
		+ \delta^{2}\tilde{\mu}_{\psi3}
		+ o\left(\delta^{2}\right),
		\label{inner: mu_psi}\\
		&\tilde{\mu} 
		=\delta^{-1}\tilde{\mu}_{0}+\tilde{\mu}_{1}
		+\delta\tilde{\mu}_{2}+\delta^{2}\tilde{\mu}_{3}
		+o\left(\delta^{2}\right),
		\label{inner: mu}\\ 
		&\tilde{\phi}
		=\tilde{\phi}_{0}+\delta\tilde{\phi}_{1}
		+\delta^{2}\tilde{\phi}_{2}+o\left(\delta^{2}\right),
		\label{inner: phi}\\
		&\tilde{\bm{u}}
		=\tilde{\bm{u}}_{0}+\delta\tilde{\bm{u}}_{1}
		+\delta^{2}\tilde{\bm{u}}_{2}+o\left(\delta^{2}\right), 
		\label{inner: u}\\ 
		&\tilde{p}
		=\tilde{p}_{0}+\delta\tilde{p}_{1}
		+\delta^{2}\tilde{p}_{2}+o\left(\delta^{2}\right), 
		\label{inner: p}\\
		&\tilde{\bm{J}}
		=\tilde{\bm{J}}_{0}+\delta\tilde{\bm{J}}_{1}
		+\delta^{2}\tilde{\bm{J}}_{2}+o\left(\delta^{2}\right), 
		\label{inner: J}\\
		&\tilde{\bm{\sigma}}_{e}
		=\tilde{\bm{\sigma}}_{e0}+\delta\tilde{\bm{\sigma}}_{e1}
		+\delta^{2}\tilde{\bm{\sigma}}_{e2} 
		+o\left(\delta^{2}\right), 
		\label{inner: sigma}
	\end{align}
\end{subequations}
and  the following matching conditions for inner and outer expansions: 
\begin{subequations}
	\begin{align}
		&\lim_{\xi\to\pm\infty}\tilde{f}_{i}\left(\bm{x},\xi\right)
		=f_{i}^{\pm}\left(\bm{x}\right), \\
		&\lim_{\xi\to\pm\infty}\left(\nabla_{\bm{x}}\tilde{f}_{i}\left(\bm{x},\xi\right)
		+\partial_{\xi}\tilde{f}_{i+1}\left(\bm{x},\xi\right)\bm{n}\right)
		=\nabla f_{i}^{\pm}\left(\bm{x}\right). 
	\end{align}
\end{subequations}

Firstly, for the electric potential, we have 
\begin{subequations}
	\begin{align}
		&	\nabla_{\bm{x}} \cdot\left(\tilde{\bm{J}}_{0}+\delta\tilde{\bm{J}}_{1}
		+\delta^{2}\tilde{\bm{J}}_{2}+o\left(\delta^{2}\right)\right)+\delta^{-1}\partial_{\xi} \left(\tilde{\bm{J}}_{0}+\delta\tilde{\bm{J}}_{1}
		+\delta^{2}\tilde{\bm{J}}_{2}+o\left(\delta^{2}\right)\right)\cdot \bm{n}=0,\label{in eqn: J}  \\
		&\left(\frac{1-\left(\tilde{\psi}_{0}
			+ \delta \tilde{\psi}_{1}
			+ \delta^{2} \tilde{\psi}_{2}
			+ o\left(\delta^{2}\right)\right)}{2}
		+ \frac{1+\tilde{\psi}_{0}
			+ \delta \tilde{\psi}_{1}
			+ \delta^{2} \tilde{\psi}_{2}
			+ o\left(\delta^{2}\right)}{2\sigma_{r}}\right) 
		\left( \tilde{\bm{J}}_{0} 
		+ \delta \tilde{\bm{J}}_{1} 
		+ \delta^{2} \tilde{\bm{J}}_{2} 
		+ o\left(\delta^{2}\right) \right) 
		\nonumber \\
		& 
		= \nabla_{\bm{x}} \left(\tilde{\phi}_{0} 
		+ \delta\tilde{\phi}_{1} 
		+ \delta^{2}\tilde{\phi}_{2} 
		+ o\left(\delta^{2}\right)\right) 
		+ \delta^{-1} \partial_{\xi} \left(\tilde{\phi}_{0} 
		+ \delta\tilde{\phi}_{1} 
		+ \delta^{2}\tilde{\phi}_{2} 
		+ o\left(\delta^{2}\right)\right)\bm{n}, 
		\label{in eqn: D and phi}  
	\end{align}
\end{subequations}
where we used the definition of $\sigma_{\it eff}$.

The leading order of two equations are 
\begin{subequations}
	\begin{align}
		&\partial_{\xi} \tilde{\bm{J}}_0\cdot \bm{n} =0 \\
		&	\partial_{\xi}  \tilde{\phi}_{0} = 0,
	\end{align}	
\end{subequations}
which yields $\tilde{\phi}_0$ does not depends on $\xi$ in the inner region
\begin{align}
	\tilde{\phi}_0(\xi,x) = \tilde{\phi}_0(x),
\end{align}
and the continuity of  electric potential and flux 
\begin{equation}
	\left[\!\left[\phi_{0}\right]\!\right] = 0,~~	\left[\!\left[\bm{J}_{0}\cdot \bm{n}\right]\!\right] = 0. \label{eq: Jphicontinuous}
\end{equation}

%%where we suppose $\tilde{\phi}_{0}$ is a constant. 
%%We should explain the origin of formula \eqref{inner: sigma}. 

We have 
\begin{equation}
	\nabla_{\bm{x}}\tilde{\phi} + \delta^{-1}\partial_{\xi}\tilde{\phi}\bm{n} 
	= -\tilde{\epsilon}_{\it{eff}} \tilde{\bm{D}} 
	= -\left(\frac{1-\tilde{\psi}}{2}+\frac{1+\tilde{\psi}}{2\epsilon_{r}}\right)
	\tilde{\bm{D}}. 
\end{equation}
So $\tilde{\bm{D}}$ have the following expansion 
\begin{equation}
	\tilde{\bm{D}} = \tilde{\bm{D}}_{0} + \delta \tilde{\bm{D}}_{1}
	+ \delta^{2} \tilde{\bm{D}}_{2} + o\left(\delta^{2}\right). 
\end{equation} 
And due to the fact 
\begin{equation}
	\epsilon_{\it{eff}} \sigma_{e} = \bm{D} \otimes \bm{D} 
	- \frac{1}{2} \left| \bm{D} \right|^{2}\textbf{I}, 
\end{equation}
we have 
\begin{equation}
	\sigma_{e} = \epsilon_{\it{eff}}^{-1} \left(\bm{D} \otimes \bm{D} 
	- \frac{1}{2} \left| \bm{D} \right|^{2}\textbf{I}\right), 
\end{equation}
and the expansion \eqref{inner: sigma} is apparent. 
It is the same reason why \eqref{inner: J} is established. 

For the chemical potential,  substituting expansions \eqref{inner: mu}, \eqref{inner: psi}, 
\eqref{inner: phi} into \eqref{eqn: mu_psi} and \eqref{eqn: mu} gives 
\begin{subequations}
	\begin{align}
		& \delta^{-1} \tilde{\mu}_{0} 
		+ \tilde{\mu}_{1} 
		+ \delta \tilde{\mu}_{2} 
		+ \delta^{2} \tilde{\mu}_{3} 
		+ o\left(\delta^{2}\right) 
		= - \delta \nabla^{2} \left(\tilde{\psi}_{0}
		+ \delta \tilde{\psi}_{1} 
		+ \delta^{2} \tilde{\psi}_{2} 
		+ o\left(\delta^{2}\right)\right) 
		\nonumber \\ 
		& \qquad \qquad 
		+ \delta^{-1} \left(\tilde{\psi}_{0} 
		+ \delta \tilde{\psi}_{1} 
		+ \delta^{2} \tilde{\psi}_{2} 
		+ o\left(\delta^{2}\right)\right) 
		\left(\left(\tilde{\psi}_{0} 
		+ \delta \tilde{\psi}_{1} 
		+ \delta^{2} \tilde{\psi}_{2} 
		+ o\left(\delta^{2}\right)\right)^{2}-1\right), 
		\label{in eqn: mu} 
		\\
		& \Bigg(\frac{1-\left(\tilde{\psi}_{0}
			+ \delta\tilde{\psi}_{1}
			+ \delta^{2}\tilde{\psi}_{2}
			+ o\left(\delta^{2}\right)\right)}{2}
		+ \frac{1+\left(\tilde{\psi}_{0}
			+ \delta\tilde{\psi}_{1}
			+ \delta^{2}\tilde{\psi}_{2}
			+ o\left(\delta^{2}\right)\right)}{2\epsilon_{r}}\nonumber\\
		&+ \frac{\left(1-\left(\tilde{\psi}_{0}
			+ \delta\tilde{\psi}_{1}
			+ \delta^{2}\tilde{\psi}_{2}
			+ o\left(\delta^{2}\right)\right)^{2}\right)^{2}}{\delta C_{m}}\Bigg)^{2} \times \left(\delta^{-1}\mu_{\psi 0} 
		+ \tilde{\mu}_{\psi 1} 
		+ \delta \tilde{\mu}_{\psi 2} 
		+ \delta^{2} \tilde{\mu}_{\psi 3} 
		+ o\left(\delta^{2}\right) \right) 
		\nonumber \\ 
		= & \Bigg(\frac{1-\left(\tilde{\psi}_{0}
			+ \delta\tilde{\psi}_{1}
			+ \delta^{2}\tilde{\psi}_{2}
			+ o\left(\delta^{2}\right)\right)}{2}
		+ \frac{1+\left(\tilde{\psi}_{0}
			+ \delta\tilde{\psi}_{1}
			+ \delta^{2}\tilde{\psi}_{2}
			+ o\left(\delta^{2}\right)\right)}{2\epsilon_{r}}\nonumber\\
		&	+ \frac{\left(1-\left(\tilde{\psi}_{0}
			+ \delta\tilde{\psi}_{1}
			+ \delta^{2}\tilde{\psi}_{2}
			+ o\left(\delta^{2}\right)\right)^{2}\right)^{2}}{\delta C_{m}}\Bigg)^{2}\times \left(\delta^{-1} \tilde{\mu}_{0} 
		+ \tilde{\mu}_{1} 
		+ \delta \tilde{\mu}_{2} 
		+ \delta^{2} \tilde{\mu}_{3} 
		+ o\left(\delta^{2}\right) \right) 
		\nonumber \\
		& + \left(\frac{1}{2}-\frac{1}{2\epsilon_{r}}
		+ \frac{4\left(\tilde{\psi}_{0}
			+ \delta\tilde{\psi}_{1}
			+ \delta^{2}\tilde{\psi}_{2}
			+ o\left(\delta^{2}\right)\right)\left(1-\left(\tilde{\psi}_{0}
			+ \delta\tilde{\psi}_{1}
			+ \delta^{2}\tilde{\psi}_{2}
			+ o\left(\delta^{2}\right)\right)^{2}\right)}{\delta C_{m}}\right) 
		\nonumber \\ 
		& \times 
		\left|\nabla_{\bm{x}} \left(\tilde{\phi}_{0} 
		+ \delta \tilde{\phi}_{1} 
		+ \delta^{2} \tilde{\phi}_{2} 
		+ o\left(\delta^{2}\right)\right)
		+ \delta^{-1}\partial_{\xi}\left(\tilde{\phi}_{0} 
		+ \delta \tilde{\phi}_{1} 
		+ \delta^{2} \tilde{\phi}_{2} 
		+ o\left(\delta^{2}\right)\right)\bm{n}\right|^{2}.
		\label{in eqn: mu_psi} 
	\end{align}
\end{subequations}

The leading order $\delta^{-1}$ of above two equations   gives 

\begin{align}
	\tilde{\mu}_{\psi 0} = \tilde{\mu}_{0} 
	= & -\partial_{\xi\xi} \tilde{\psi}_{0} 
	+ \tilde{\psi}_{0}\left(\tilde{\psi}_{0}^{2}-1\right), 
	\label{in: leading mu}
\end{align}
and the next order $\delta^0$ of Eq. \eqref{in eqn: mu} is 
\begin{align}
	\tilde{\mu}_{ 1} 
	= &-\partial_{\xi} \tilde{\psi}_{0}\kappa
	-2 \left(\bm{n}\cdot\nabla_{\bm{x}}\right)\partial_{\xi} \tilde{\psi}_{0}
	-\partial_{\xi\xi} \tilde{\psi}_{1}+3\tilde{\psi}_{0}^{2}\tilde{\psi}_{1}-\tilde{\psi}_{1}.
	\label{in: next mu}
\end{align}
For the Cahn-Hilliard equation, we substitute expansions \eqref{inner: psi}, 
\eqref{inner: mu_psi} and \eqref{inner: u} into \eqref{eqn: psi}, 
then we have 
\begin{align} 
	&\partial_{t}\left(\tilde{\psi}_{0}+\delta\tilde{\psi}_{1} 
	+\delta^{2}\tilde{\psi}_{2}+o\left(\delta^{2}\right)\right) 
	+ \delta^{-1}\partial_{\xi}\left(\tilde{\psi}_{0}+\delta\tilde{\psi}_{1} 
	+\delta^{2}\tilde{\psi}_{2}+o\left(\delta^{2}\right)\right) \partial_{t}d 
	\nonumber \\
	+& \left(\tilde{\bm{u}}_{0}+\delta\tilde{\bm{u}}_{1} 
	+\delta^{2}\tilde{\bm{u}}_{2}+o\left(\delta^{2}\right)\right) \cdot\nabla_{\bm{x}}  \left( 
	\left(\tilde{\psi}_{0}+\delta\tilde{\psi}_{1} +\delta^{2}\tilde{\psi}_{2} 
	+o\left(\delta^{2}\right)\right) \right) 
	\nonumber \\ 
	+&\delta^{-1}\left(\tilde{\bm{u}}_{0}+\delta\tilde{\bm{u}}_{1} 
	+\delta^{2}\tilde{\bm{u}}_{2}+o\left(\delta^{2}\right)\right) \partial_{\xi}\left( 
	\left(\tilde{\psi}_{0}+\delta\tilde{\psi}_{1} +\delta^{2}\tilde{\psi}_{2}
	+o\left(\delta^{2}\right)\right) \right) \cdot \bm{n}
	\nonumber \\
	-& M \left[\nabla_{\bm{x}}^{2} \left(\delta^{-1}\tilde{\mu}_{\psi0}
	+\tilde{\mu}_{\psi 1} +\delta\tilde{\mu}_{\psi 2}
	+\delta^{2}\tilde{\mu}_{\psi 3}+o\left(\delta^{2}\right)\right)\right.  \nonumber\\
	&+ \delta^{-1}\partial_{\xi}\left(\delta^{-1}\tilde{\mu}_{\psi0}
	+\tilde{\mu}_{\psi 1} +\delta\tilde{\mu}_{\psi 2} 
	+\delta^{2}\tilde{\mu}_{\psi 3}+o\left(\delta^{2}\right)\right) \kappa
	\nonumber \\
	&\left. 
	+2\delta^{-1}\left(\bm{n}\cdot\nabla_{\bm{x}}\right)\partial_{\xi}
	\left(\delta^{-1}\tilde{\mu}_{\psi0}+\tilde{\mu}_{\psi1} 
	+\delta\tilde{\mu}_{\psi2}+\delta^{2}\tilde{\mu}_{3}
	+o\left(\delta^{2}\right)\right)\right.
	\nonumber \\
	&\left.
	+\delta^{-2}\partial_{\xi\xi}\left(\delta^{-1}\tilde{\mu}_{\psi0}
	+\tilde{\mu}_{\psi1} +\delta\tilde{\mu}_{\psi2}+\delta^{2}\tilde{\mu}_{3}
	+o\left(\delta^{2}\right)\right)\right] = 0. 
	\label{in: psi} 
\end{align}

Since $M=\alpha_1\delta^{2}$, 
then the leading order of above equation   is 
\begin{equation}
	\partial_{\xi} \tilde{\psi}_{0} \partial_{t}d 
	+ \tilde{\bm{u}}_{0}\cdot \bm{n}\partial_{\xi}\left(\tilde{\psi}_{0} \right) 
	-\alpha_1\partial_{\xi\xi} \tilde{\mu}_{\psi0}=0, 
	\label{in: leading psi} 
\end{equation} 
If We integrate \eqref{in: leading psi} with respect to $\xi$ in $\left(-\infty,+\infty\right)$, 
and use the matching condition 
$\lim\limits_{\xi\to\pm\infty}\partial_{\xi}\tilde{\mu}_{0}=0$, 
then above equation yields
\begin{equation}
	\partial_{t}d +\tilde{\bm{u}}_{0}\cdot \bm{n}=0. 
\end{equation}
This implies that the normal velocity of the interface $\Gamma$ is 
\begin{equation}
	V_{n}=\bm{u}_{0}\cdot\bm{n}, 
\end{equation} 
and $\tilde{\mu}_{\psi 0}$ is a linear function about $\xi$, 
it is $\tilde{\mu}_{0}=c_{1} \xi + c_{0}$. 
Because of the matching condition
\begin{equation}
	\lim_{\xi\to\pm\infty} \tilde{\mu}_{0} = \mu_{0}^{\pm}, 
\end{equation}
$\tilde{\mu}_{0} = c_{0}$ is obvious. 

For the Navier-Stokes equations, when we substitute \eqref{inner: u}, 
\eqref{inner: p}, \eqref{inner: psi}, \eqref{inner: mu}, 
\eqref{inner: phi} and \eqref{inner: sigma} 
into \eqref{eqn: u} and \eqref{eqn: nabla u}, we have 
\begin{subequations}
	\begin{align}
		& \partial_{t}\left(\tilde{\bm{u}}_{0} 
		+ \delta\tilde{\bm{u}}_{1} 
		+ \delta^{2}\tilde{\bm{u}}_{2} + 
		o\left(\delta^{2}\right)\right) 
		+ \delta^{-1}\partial_{\xi}\left(\tilde{\bm{u}}_{0} 
		+ \delta\tilde{\bm{u}}_{1} 
		+ \delta^{2}\tilde{\bm{u}}_{2} 
		+ o\left(\delta^{2}\right)\right) \partial_{t}d 
		\nonumber \\
		& + \left(\tilde{\bm{u}}_{0} 
		+ \delta\tilde{\bm{u}}_{1} 
		+ \delta^{2}\tilde{\bm{u}}_{2} 
		+ o\left(\delta^{2}\right)\right) \cdot \nabla_{\bm{x}}
		\left(\tilde{\bm{u}}_{0} 
		+ \delta\tilde{\bm{u}}_{1} 
		+ \delta^{2}\tilde{\bm{u}}_{2} 
		+ o\left(\delta^{2}\right)\right) 
		\nonumber \\
		& + \left( \tilde{\bm{u}}_{0} 
		+ \delta\tilde{\bm{u}}_{1} 
		+ \delta^{2}\tilde{\bm{u}}_{2} 
		+ o\left(\delta^{2}\right)\right)\cdot\delta^{-1}\partial_{\xi}
		\left(\tilde{\bm{u}}_{0} 
		+ \delta\tilde{\bm{u}}_{1} 
		+ \delta^{2}\tilde{\bm{u}}_{2} 
		+ o\left(\delta^{2}\right)\right)\bm{n} 
		\nonumber \\
		& + \nabla_{\bm{x}}\left(\tilde{p}_{0} 
		+ \delta\tilde{p}_{1} 
		+ \delta^{2}\tilde{p}_{2} 
		+ o\left(\delta^{2}\right)\right)
		+ \delta^{-1}\partial_{\xi}\left(\tilde{p}_{0} 
		+ \delta\tilde{p}_{1} 
		+ \delta^{2}\tilde{p}_{2} 
		+ o\left(\delta^{2}\right)\right)\bm{n} 
		\nonumber \\
		= & \frac{1}{Re} \left[ \nabla_{\bm{x}}^{2} \left(\tilde{\bm{u}}_{0} 
		+ \delta\tilde{\bm{u}}_{1} 
		+ \delta^{2}\tilde{\bm{u}}_{2} 
		+ o\left(\delta^{2}\right)\right) 
		+ \delta^{-1}\partial_{\xi}\left(\tilde{\bm{u}}_{0} 
		+ \delta\tilde{\bm{u}}_{1} 
		+ \delta^{2}\tilde{\bm{u}}_{2} 
		+ o\left(\delta^{2}\right)\right) \kappa 
		\right. 
		\nonumber \\
		& \left. + 2\delta^{-1} \left( \bm{n} \cdot \nabla_{\bm{x}} \right) 
		\partial_{\xi} \left(\tilde{\bm{u}}_{0} 
		+ \delta\tilde{\bm{u}}_{1} 
		+ \delta^{2}\tilde{\bm{u}}_{2} 
		+ o\left(\delta^{2}\right)\right) 
		+ \delta^{-2}\partial_{\xi\xi}\left(\tilde{\bm{u}}_{0} 
		+ \delta\tilde{\bm{u}}_{1} 
		+ \delta^{2}\tilde{\bm{u}}_{2} 
		+ o\left(\delta^{2}\right)\right) \right] 
		\nonumber \\
		& + \left(\delta^{-1}\tilde{\mu}_{0}
		+ \tilde{\mu}_{1} 
		+ \delta\tilde{\mu}_{2}
		+ \delta^{2}\tilde{\mu}_{3} 
		+ o\left(\delta^{2}\right)\right)
		\nabla_{\bm{x}}\left(\tilde{\psi}_{0} 
		+ \delta\tilde{\psi}_{1} 
		+ \delta^{2}\tilde{\psi}_{2} 
		+ o\left(\delta^{2}\right)\right)
		\nonumber \\
		& + \left(\delta^{-1}\tilde{\mu}_{ 0}
		+ \tilde{\mu}_{ 1} 
		+ \delta\tilde{\mu}_{ 2}
		+ \delta^{2}\tilde{\mu}_{3} 
		+ o\left(\delta^{2}\right)\right)
		\delta^{-1}\partial_{\xi}\left(\tilde{\psi}_{0} 
		+ \delta\tilde{\psi}_{1} 
		+ \delta^{2}\tilde{\psi}_{2} 
		+ o\left(\delta^{2}\right)\right)\bm{n} 
		\nonumber \\
		& + Ca_{E} \left(\nabla_{\bm{x}}\cdot\left( \tilde{\sigma}_{e0} 
		+ \delta \tilde{\sigma}_{e1} 
		+ \delta^{2} \tilde{\sigma}_{e2} 
		+ o\left( \delta^{2} \right)\right)
		+ \delta^{-1}\partial_{\xi}\left( \tilde{\sigma}_{e0} 
		+ \delta \tilde{\sigma}_{e1} 
		+ \delta^{2} \tilde{\sigma}_{e2} 
		+ o\left( \delta^{2} \right)\right)\cdot\bm{n}\right), 
		\label{in: u}\\
		& \nabla_{\bm{x}}\cdot\left(\tilde{\bm{u}}_{0} 
		+ \delta\tilde{\bm{u}}_{1} 
		+ \delta^{2}\tilde{\bm{u}}_{2} 
		+ o\left(\delta^{2}\right)\right)
		+ \delta^{-1}\partial_{\xi}\left(\tilde{\bm{u}}_{0} 
		+ \delta\tilde{\bm{u}}_{1} 
		+ \delta^{2}\tilde{\bm{u}}_{2} 
		+ o\left(\delta^{2}\right)\right)\cdot\bm{n} = 0. 
		\label{in: nablau}
	\end{align}
\end{subequations}
The leading order of \eqref{in: nablau} is 
\begin{equation}
	\partial_{\xi} \tilde{\bm{u}}_{0}\cdot\bm{n}=0, \label{in: leading nablau}
\end{equation}
which means $ \tilde{\bm{u}}_{0}\cdot\bm{n}$ is a constant with respect to $\xi$. 
We integrate above equation \eqref{in: leading nablau} in$\left(-\infty,+\infty\right)$, the following result can be obtained directely, 
\begin{equation}
	\int_{-\infty}^{+\infty}\partial_{\xi} \tilde{\bm{u}}_{0}\cdot\bm{n}\mathrm{d}\xi=0, 
\end{equation}
which means 
\begin{equation}\label{in:ucontinuous}
	[\![\bm{u}_{0}\cdot\bm{n}]\!]=0. 
\end{equation} 
So the velocity is continuous across the interface. 

The leading $(\delta^{-2})$ and next order $(\delta^{-1})$ of equation \eqref{in: u} are 
\begin{subequations}
	\begin{align}
		& \partial_{\xi\xi} \tilde{\bm{u}}_{0} 
		+ \tilde{\mu}_{ 0}\partial_{\xi} \tilde{\psi}_{0}\bm{n} = 0, 
		\label{in: leading u}\\ 
		& \partial_{\xi}\tilde{\bm{u}}_{0}\partial_{t}d 
		+ \tilde{\bm{u}}_{0}\cdot\partial_{\xi}\tilde{\bm{u}}_{0}\bm{n} 
		+ \partial_{\xi} \tilde{p}_{0}\bm{n} 
		= \frac{1}{Re} \left( \partial_{\xi}\tilde{\bm{u}}_{0}\kappa 
		+ 2 \left(\bm{n}\cdot\nabla_{\bm{x}}\right)\partial_{\xi}\tilde{\bm{u}}_{0} 
		+ \partial_{\xi\xi} \tilde{\bm{u}}_{1} \right) 
		\nonumber \\
		& \qquad \qquad \qquad 
		+ \left(\tilde{\mu}_{ 0}\nabla_{\bm{x}} \tilde{\psi}_{0}
		+ \tilde{\mu}_{ 0}\partial_{\xi} \tilde{\psi}_{1}\bm{n} 
		+ \tilde{\mu}_{ 1}\partial_{\xi} \tilde{\psi}_{0}\bm{n}\right)
		+ Ca_{E} \partial_{\xi}\tilde{\sigma}_{e0}\cdot\bm{n}. 
		\label{in: next u} 
	\end{align}
\end{subequations}
Integrate equation \eqref{in: leading u} in $(-\infty,+\infty)$  yields
\begin{equation}
	\partial_{\xi}\tilde{\bm{u}}_{0}|_{-\infty}^{\infty} 
	+ c_{0} \tilde{\psi}_{0}\bm{n}|_{-\infty}^{+\infty} = 0,
\end{equation}
where we used the result that $\tilde{\mu}_{0} = c_{0}$ is independence with $\xi$.

Multiplying $\bm{n}$ on both sides of the above equation, we have  
\begin{equation}
	c_{0} \tilde{\psi}_{0}|_{-\infty}^{+\infty} = 0, 
\end{equation}
which means $c_{0} \left(\psi_{0}^{+}-\psi_{0}^{-}\right)=0$, 
and hence $c_{0} = 0$, it is $\tilde{\mu}_{0} = 0$.

Then  equation \eqref{in: leading mu} gives  
\begin{equation}
	\tilde{\psi}_{0}=\tanh(\xi/\sqrt{2}),
\end{equation}
and as $\xi\rightarrow\pm\infty$, the matching condition yields 
\begin{subequations}\label{in:psi0}
	\begin{align}
		\psi_0^+ = \lim_{\xi\rightarrow\infty}\tilde{\psi}_0 =1,\\
		\psi_0^- = \lim_{\xi\rightarrow-\infty}\tilde{\psi}_0 =-1.
	\end{align}
\end{subequations}
If we integrate equation \eqref{in: next u} in $(-\infty,+\infty)$ 
and use the results in \cite{Xu2018Sharp}, 
then we obtain
\begin{equation}\label{in:interfacecon}
	\left[\!\left[-p_{0}\bm{n} 
	+ \frac 1 {Re}\left(\bm{n}\cdot\nabla\bm{u}_{0} \right) 
	+ Ca_{E} \sigma_{e0}\cdot \bm{n}\right]\!\right] 
	= \sigma \kappa \bm{n}. 
\end{equation}
%\begin{align}
%	& \left(\frac{1-\left(\tilde{\psi}_{0}
	%		+ \delta \tilde{\psi}_{1}
	%		+ \delta^{2} \tilde{\psi}_{2}
	%		+ o\left(\delta^{2}\right)\right)}{2}
%	+ \frac{1+\tilde{\psi}_{0}
	%		+ \delta \tilde{\psi}_{1}
	%		+ \delta^{2} \tilde{\psi}_{2}
	%		+ o\left(\delta^{2}\right)}{2\sigma_{r}}\right) 
%	\left( \tilde{\bm{J}}_{0} 
%	+ \delta \tilde{\bm{J}}_{1} 
%	+ \delta^{2} \tilde{\bm{J}}_{2} 
%	+ o\left(\delta^{2}\right) \right) 
%	\nonumber \\
%	& 
%	= \nabla_{\bm{x}} \left(\tilde{\phi}_{0} 
%	+ \delta\tilde{\phi}_{1} 
%	+ \delta^{2}\tilde{\phi}_{2} 
%	+ o\left(\delta^{2}\right)\right) 
%	+ \delta^{-1} \partial_{\xi} \left(\tilde{\phi}_{0} 
%	+ \delta\tilde{\phi}_{1} 
%	+ \delta^{2}\tilde{\phi}_{2} 
%	+ o\left(\delta^{2}\right)\right), 
%	\label{in eqn: D and phi}  
%\end{align}
%and the leading order term of the above equation gives us that 
%\begin{equation}
%	\partial_{\xi} \tilde{\phi}_{0} = 0, 
%\end{equation}
%which means 
%\begin{equation}
%	\left[\!\left[ \phi_{0} \right]\!\right] = 0. 
%\end{equation}
%The leading order term of \eqref{eqn: J} gives us that 
%\begin{equation}
%	\partial_{\xi} \tilde{\bm{J}}_{0} = 0, 
%\end{equation}
%which means $\left[\!\left[\bm{J}_{0}\right]\!\right] = 0$. 
Using the results \eqref{eq: Jphicontinuous}, \eqref{in:ucontinuous},  \eqref{in:psi0} and \eqref{in:interfacecon},  we obtain the sharp interface limit of system \eqref{eqn: main} 
\begin{subequations}
	\begin{align}
		& \frac{\partial \bm{u}_{0}^{\pm}}{\partial t} 
		+ \left( \bm{u}_{0}^{\pm} \cdot \nabla \right) \bm{u}_{0}^{\pm} 
		+ \nabla p_{0}^{\pm} 
		= \frac{1}{Re} \nabla^{2} \bm{u}_{0}^{\pm} 
		+ Ca_{E} \nabla \cdot \sigma_{e0}^{\pm}, 
		\\
		&\nabla\cdot\bm{u}_{0}^{\pm} = 0, \\
		&\bm{\sigma}_{e,0}^+ =\epsilon_r\left(\nabla\phi_0^+\otimes\nabla\phi_0^+-\frac 1 2 |\nabla\phi_0^+|^2I\right), ~~\bm{\sigma}_{e,0}^- =\left(\nabla\phi_0^-\otimes\nabla\phi_0^--\frac 1 2 |\nabla\phi_0^-|^2I\right),\\ 
		&\nabla\cdot\left(\sigma_{r}\nabla\phi^{+}\right) = 0, ~~\nabla^{2}\phi^{-} = 0, 
		\\
		& \left[\!\left[\bm{u}_{0}\cdot\bm{n}\right]\!\right]=0, 
		\\
		&	\left[\!\left[-p_{0}\bm{n} 
		+ \frac 1 {Re}\left(\bm{n}\cdot\nabla\bm{u}_{0} \right) 
		+ Ca_{E} \sigma_{e0}\cdot \bm{n}\right]\!\right] 
		= \sigma \kappa \bm{n}, 
		\\ 
		& \left[\!\left[ \phi_{0} \right]\!\right] = 0, 
		\\ 
		& \left[\!\left[\bm{J}_{0}\right]\!\right] = 0, 
	\end{align} 
\end{subequations}
which is the consistent with the leaky-dielectric  model in \cite{Hua2008Numerical,HU2015Ahybrid}. 

	\section{Numerical results}\label{sec: numerical results}
In this section, we conduct a series of numerical experiments 
to illustrate the validity of the estabilished diffuse interface model. 
%To make the computation more simple, 
%
%
%Then the Navier-Stokes equation can be written as 
%\begin{equation}
%	\left( \frac{\partial \bm{u}}{\partial t} 
%	+ \left( \bm{u} \cdot \nabla \right) \bm{u} \right) 
%	+ \nabla p 
%	= \frac{1}{Re}\nabla^{2}\bm{u} 
%	%+ \left(\mu - \frac{Ca_{E}}{2} \left|\nabla \phi \right|^{2} 
%	%\frac{\partial \epsilon_{\it{eff}}}{\partial \psi}\right) \nabla \psi 
%	+ \mu_{\psi} \nabla \psi 
%	+ Ca_{E} \nabla \cdot \left( \epsilon_{\it eff} \nabla \phi \right) \nabla \phi.  
%\end{equation}
%We will construct numerical scheme for the system with the above new 
%above form of Navier-Stokes equation. 
%Because the numerical scheme is not the focus of our this paper, 
The traditional semi-implicit numerical scheme is adopted. Specifically, we add a linear stabilization factor 
to solve the Cahn-Hilliard equations, 
and the classical pressure correction method to 
deal with the Navier-Stokes equations. 
We use the Mark and Cell(MAC) finite difference to discrete 
the space variables, which means the scalar variables are located at 
the center of every mesh, however the vector is located at the center of edge.  
The specific semi-discrete numerical scheme in time is shown as follows, 

{\bf Step 1.} We solve the Cahn-Hilliard equation firstly with the help of 
stabilization method, 
\begin{subequations}
	\begin{align}
		& \frac{\psi^{n+1} - \psi^{n}}{\Delta t} 
		+ \nabla \cdot \left( \bm{u}^{n} \psi^{n}\right) 
		= M \nabla^{2} \mu_{\psi}^{n+1}, 
		\\
		& \mu_{\psi}^{n+1} 
		= -\delta \nabla^{2} \psi^{n+1} 
		+ \frac{s}{\delta} \left( \psi^{n+1} - \psi^{n} \right)
		+ \frac{1}{\delta}F^{\prime}\left(\psi^{n}\right)
		- \frac{Ca_{E}}{2} \frac{\partial \epsilon_{\it{eff}^{n}}}{\partial \psi^{n}}
		\left|\nabla \phi^{n}\right|^{2}, 
	\end{align}
\end{subequations}
where $s$ is the so-called stabilization factor 
to keep the numerical scheme more stable. 

{\bf Step 2.} The electric potential $\phi^{n+1}$ is obtain immediately 
when $\psi^{n+1}$ is known by the following Poisson equation, 
\begin{subequations}
	\begin{align}
		& \nabla \cdot \left( \sigma_{\it{eff}}^{n+1} \nabla \phi^{n+1} \right) = 0, 
		\\
		& \frac{1}{\sigma_{\it{eff}}^{n+1}} 
		= \frac{1-\psi^{n+1}}{2}+\frac{1+\psi^{n+1}}{2\sigma_{r}}. 
	\end{align}
\end{subequations}

{\bf Step 3.} We use the pressure correction method to decoupled 
velocity $\bm{u}$ and pressure $p$ in Navier-Stokes equation \eqref{eqn：	urevised}, 
which means we solve an intermediate variable velocity $\tilde{\bm{u}}$ 
shown as follows, 
\begin{subequations}
	\begin{align}
		& \frac{\tilde{\bm{u}}^{n+1} - \bm{u}^{n}}{\Delta t} 
		+ \left(\bm{u}^{n} \cdot \nabla \tilde{\bm{u}}^{n+1}\right)
		+ \nabla p^{n} 
		= \frac{1}{Re} \nabla^{2} \tilde{\bm{u}}^{n+1} 
		+ \mu_{\psi}^{n+1} \nabla \psi^{n+1} 
		%		+ Ca_{E} \nabla \cdot 
		%		\left( \epsilon_{\it{eff}}^{n+1} \nabla \phi^{n+1} \right) 
		%		\nabla \phi^{n+1}, 
		-\frac{Ca_{E}}{\zeta^{2}} \rho_{e}^{n+1} 
		\nabla \phi^{n+1}, 
		\\ 
		%		& \mu_{\psi}^{n+1}  
		%		= \left(\mu^{n+1}- \frac{Ca_{E}}{2} \left|\nabla \phi^{n+1} \right|^{2} 
		%		\frac{\epsilon_{\it eff}^{n+1}}{\partial \psi^{n+1}}\right), 
		%		\\ 
		& \frac{1}{\epsilon_{\it eff}^{n+1}} 
		= \frac{\left(1-\left(\psi^{n+1}\right)^{2}\right)^{2}}{\delta C_{m}} 
		+ \frac{1-\psi^{n+1}}{2} 
		+ \frac{1+\psi^{n+1}}{2\epsilon_{r}}.
	\end{align}
\end{subequations}

{\bf Step 4.} We get the pressure $p^{n+1}$ and velocity $\tilde{\bm{u}}^{n+1}$, 
\begin{subequations}
	\begin{align}
		& \frac{\bm{u}^{n+1} - \tilde{\bm{u}}^{n+1}}{\Delta t} 
		+ \nabla\left(p^{n+1} - p^{n}\right) 
		= 0, 
		\\ 
		& \nabla \cdot \bm{u}^{n+1} = 0. 
	\end{align}
\end{subequations}
Firstly, a convergence study is carried out to varify the correctness of our codes. 
Then we check the sharp interface limit in Section \ref{sec: sharp} by choosing a relatively small values 
of thickness $\delta$ and observe the result. 
In the next, to compare with the benchmark solution, we choose the numerical solutions in \cite{HU2015Ahybrid}. 
Additionally, we study the merge phenomenon with two drops located in different directions. 
Finally, we study the influence with the capacitance by some numerical examples.

\subsection{Convergence test}\label{sec: convergence}
%In this section, we perform some numerical experiments to support the theoretical results. 
In this section, we perform a convergence study to support the efficiency of our inhouse code. 
We use a uniform Cartesian grid to discretize a square domain $\Omega=(0,1)^{2}\subset \mathbb{R}^{2}$, 
namely  the mesh size $N_{x}=N_{y}=N$ is adopted here, and if not specified, uniform mesh is always tenable. 

The initial condition is chosen as follows, 
\begin{subequations}\label{eqn: convergence}
	\begin{align}
		&\psi(\bm{x},0)=0.2+0.5\cos\left(2\pi x\right)\cos\left(2\pi y\right),\\
		&\phi(\bm{x},0)=y,\\
		&u(\bm{x},0)=-0.25\sin^{2}\left(\pi x\right)\cos\left(2\pi y\right),\\
		&v(\bm{x},0)=0.25\sin^{2}\left(\pi y\right)\cos\left(2\pi x\right). 
	\end{align}
\end{subequations} 
The parameters in model are set as 
\begin{equation}
	\delta = 0.1,\quad Re = 1,\quad Ca_{E} = 1,\quad 
	\sigma_{r} = 2,\quad \epsilon_{r} = 1.
\end{equation} 

The Cauchy error in \cite{Wise2007Solving} is used to test the convergence rate. 
In this test method, error between two different spacial mesh sizes $h$ and $h/2$ is calculated by 
$\left\|e_{\theta}\right\|=\left\|\theta_{h}-\theta_{h/2}\right\|$, 
where $\theta$ is the function to be solved. 
The mesh sizes are set to  be $h=1/16,~1/32, ~1/64, ~ 1/128$ 
and time step is fixed as $\delta t=10^{-4}$. 
The $L^{2}$ and $L^{\infty}$ numerical errors and convergence rate at chosen time $T= 0.1$ 
are displayed in Table $\ref{tab: L2 space convergence}$ 
and Table $\ref{tab: Linfty space convergence}$, respectively. 
The second order spatial accuracy is apparently observed for all the variables. 
\begin{table}
	\centering
	\caption{The discrete $L^{2}$ error and convergence rate at $t = 0.1$ 
		with initial data $\eqref{eqn: convergence}$ and the given parameters.}\label{tab: L2 space convergence}
	\vskip 0.2cm
	\begin{tabular}{lcccccccccc}
		\toprule
		Grid sizes    &Error$(\psi)$ & Rate & Error$(\phi)$ &Rate&Error$(u)$ & Rate & Error$(v)$ &Rate& Error$(p)$ &Rate\\
		\midrule
		$16\times 16$    & 2.3377e-02 &  --  & 8.8527e-04 &  --  & 2.4760e-04 &  --  & 4.1204e-04 &  --  & 1.7182e-02 &  --  \\
		$32\times 32$    & 5.7835e-03 & 2.02 & 2.2773e-04 & 1.96 & 7.7879e-05 & 1.67 & 1.2048e-04 & 1.77 & 4.5426e-03 & 1.92 \\
		$64\times 64$    & 1.4428e-03 & 2.00 & 5.7333e-05 & 1.99 & 2.2016e-05 & 1.82 & 3.3117e-05 & 1.86 & 1.1686e-03 & 1.96 \\
		$128\times 128$  & 3.6066e-04 & 2.00 & 1.5150e-05 & 1.92 & 5.8081e-06 & 1.92 & 8.6179e-06 & 1.94 & 2.9555e-04 & 1.98 \\
		\bottomrule
	\end{tabular}
\end{table}

\begin{figure}
	\begin{center}
		\includegraphics[width=1.0\textwidth]{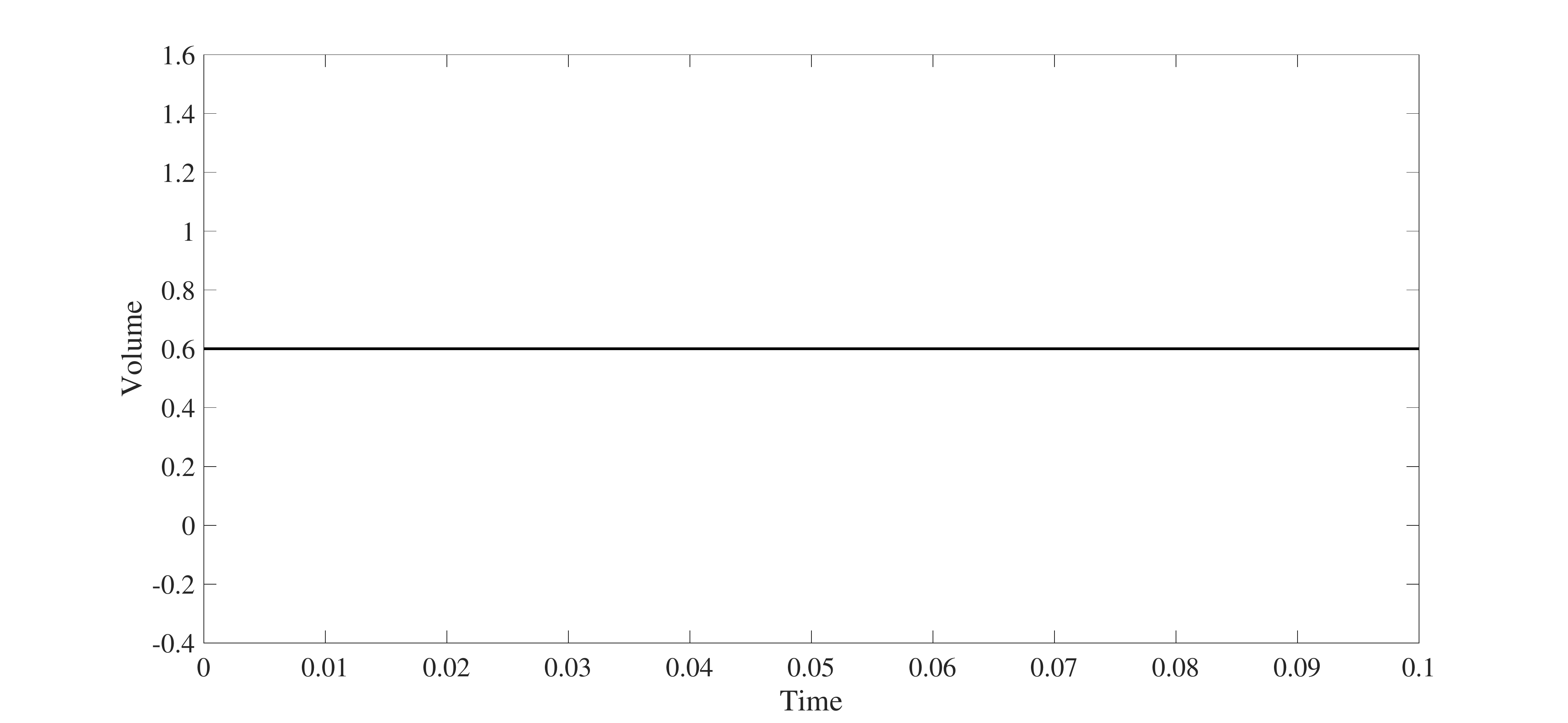}
	\end{center}
	\caption{Conservation of volume of the numerical scheme.}
	\label{fig: conservation}
\end{figure}

\begin{table}
	\centering
	\caption{The discrete $L^{\infty}$ error and convergence rate at $t = 0.1$ 
		with initial data $\eqref{eqn: convergence}$ and the given parameters.}\label{tab: Linfty space convergence}
	\vskip 0.2cm
	\begin{tabular}{lcccccccccc}
		\toprule
		Grid sizes    &Error$(\psi)$ & Rate & Error$(\phi)$ &Rate&Error$(u)$ & Rate & Error$(v)$ &Rate& Error$(p)$ &Rate\\
		\midrule
		$16\times 16$    & 5.3663e-02 &  --  & 1.8178e-03 &  --  & 7.1054e-04 &  --  & 1.0068e-03 &  --  & 1.9289e-02 &  --  \\
		$32\times 32$    & 1.3440e-02 & 2.00 & 4.7406e-04 & 1.94 & 2.0492e-04 & 1.79 & 2.9178e-04 & 1.79 & 7.1935e-03 & 1.42 \\
		$64\times 64$    & 3.3507e-03 & 2.00 & 1.2254e-04 & 1.95 & 7.8757e-05 & 1.38 & 7.9105e-05 & 1.88 & 2.2039e-03 & 1.71 \\
		$128\times 128$  & 8.3845e-04 & 2.00 & 3.5675e-05 & 1.78 & 2.5285e-05 & 1.64 & 2.0504e-05 & 1.95 & 6.1615e-04 & 1.84 \\
		\bottomrule
	\end{tabular}
\end{table}
We utilize the result to give the conservation of volume about our numerical scheme for $\psi$ 
as illustrated in Fig. $\ref{fig: conservation}$. 
It demonstrates the volume of $\psi$ doesn't change over time. 

\subsection{Sharp interface limit test in one dimension}\label{sec: sharp}
\begin{figure}
	\begin{center}
		\includegraphics[width=1.0\textwidth]{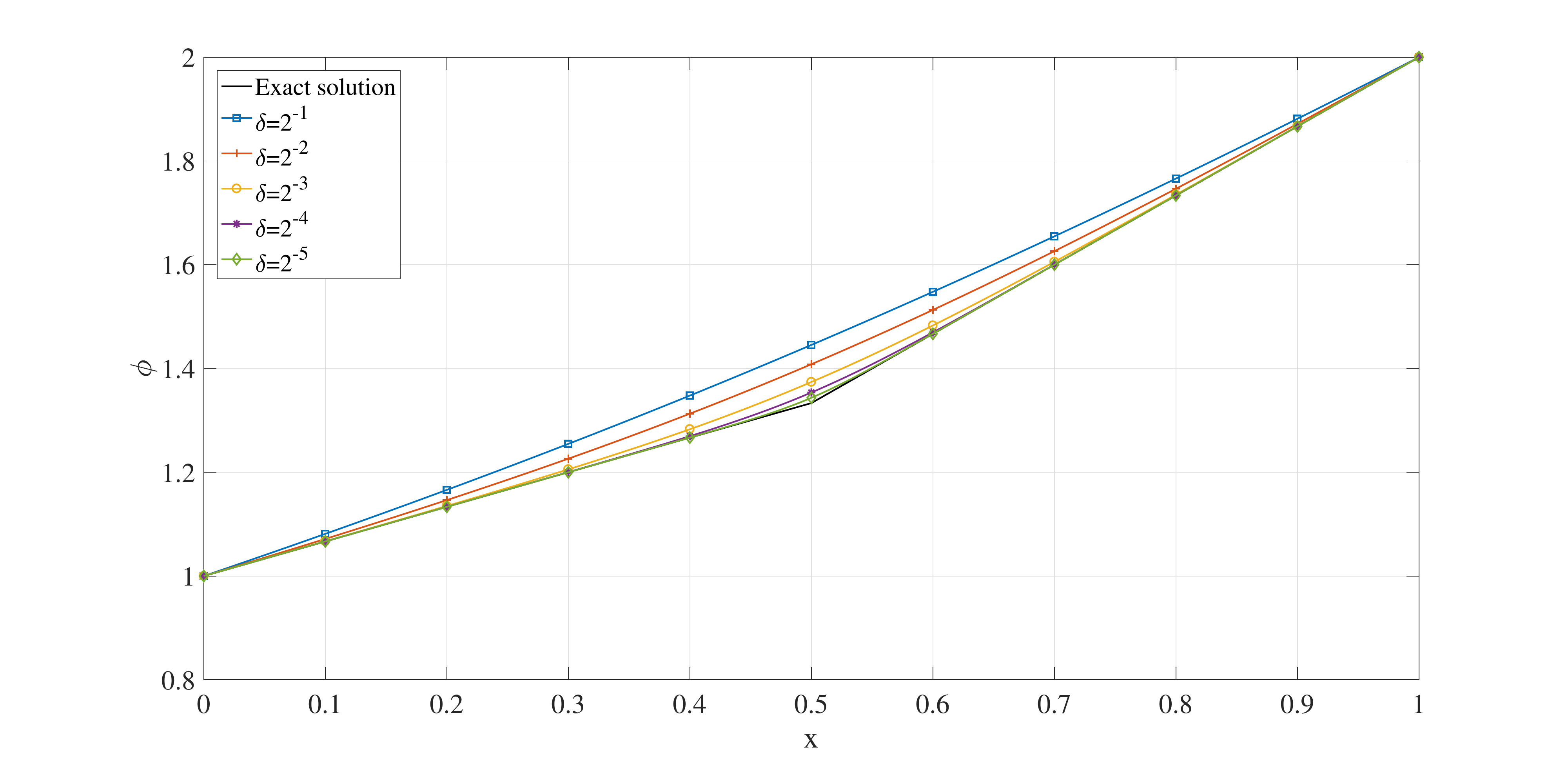}
	\end{center}
	\caption{Sharp interface limit test of electric potential in one dimension, 
		the fixed interface is chosen as $x_{0} = 0.5$.
		The left interval is assumed as the inner region nevertheless 
		the right interval is assumed as the outer region of the interface.}
	\label{fig: sharp interface}
\end{figure}

In this example, we consider the steady state solution in one dimensional situation 
to verify sharp interface limit of electric potential $\phi$ in our proposed model. 
For simplicity, we fix the interface and assume the left interval $(0,0.5)$ is 
the inner region and the right interval $(0.5,1)$ is the outer region of the interface. 
The electric potential equation \eqref{eqn: sigma phi} is used to illustrate the sharp interface limit results. 
The ration of two region conductivities is set to be  $\sigma_{r} = 2$. 
Besides, the Dirichlet boundary condition $\phi(x=0) = 1,~\phi(x=1) = 2$ is used. 
With the help of continuity condition $\left[\!\left[\phi\right]\!\right] = 0$ 
and $\left[\!\left[\sigma\nabla\phi\right]\!\right] = 0$ 
in the sharp interface situation, 
it is easy to get the exact solution of sharp interface model is a 
piecewise linear function 
\begin{equation}\label{eqn: sharp exact solution}
	\phi=\left\{
	\begin{aligned}
		& \frac{2}{3}x+1, && x<x_{0}, \\
		& \frac{4}{3}x+\frac{2}{3}, && x\geq x_{0}.
	\end{aligned}
	\right.
\end{equation}
In Figure $\ref{fig: sharp interface}$, 
the exact solution \eqref{eqn: sharp exact solution} is shown 
in black solid line without any marker and the colored lines with different markers 
are the solutions of Eq. \eqref{eqn: sigma phi} 
where the phase field function is chosen as 
$\psi=\tanh\left(\frac{x-x_{0}}{\sqrt{2}\delta}\right)$ 
with different interface thickness $\delta$. 
In the bulk region, solutions of two methods fit very well. 
As $\delta\rightarrow 0$, the proposed diffusive model solutions change 
much sharper near the interface and converge to the sharp interface solution, 
which is consistent with our analysis in Section \ref{sec: sharp interface limit}. 

\subsection{Comparison with the sharp interface model}\label{sec: comparison with sharp model}

In this section, we compare  our diffusive interface model results with sharp interface model results conducted in \cite{HU2015Ahybrid} 
where a hybrid immersed boundary and immersed interface method is used to model the deformation of a leaky dielectric droplet.
The 2D computational domain is set as $\Omega = [-4,4]\times[-4,4]$ and the mesh size is set to  be $h=8/256$. 
The time step is chosen as $\Delta t = 10^{-4}$ and the interface thickness is set to be $\delta = 0.1$. 
We assume the initial profile of droplet is 
\begin{equation}\label{eqn: initial single drop}
	\psi\left(x,y,0\right) = \tanh\frac{1-\sqrt{x^{2}+y^{2}}}{\sqrt{2}\delta}. 
\end{equation}
Nonflux boundary for $\psi$ and nonslip boundary for velocity are adopted here.  
For the electric potential, Dirichlet boundary condition $\phi(x,4) = 4$ and $\phi(x,-4)=-4$ is used on the top and bottom boundaries; 
the homogeneous Neumann boundary  is used on the left and right boundaries. 
As in \cite{HU2015Ahybrid}, the dielectric coefficients and electric capillary number are chosen as 
$\epsilon_{r}=3.5$ and $Ca_{E}=1$, respectively. 
And there is no capacitance on the interface, i.e. $C_m = \infty$ in Eq. \eqref{def: dimensional epsilon}. 

\begin{figure}
	\begin{center}
		\includegraphics[width=0.32\textwidth]{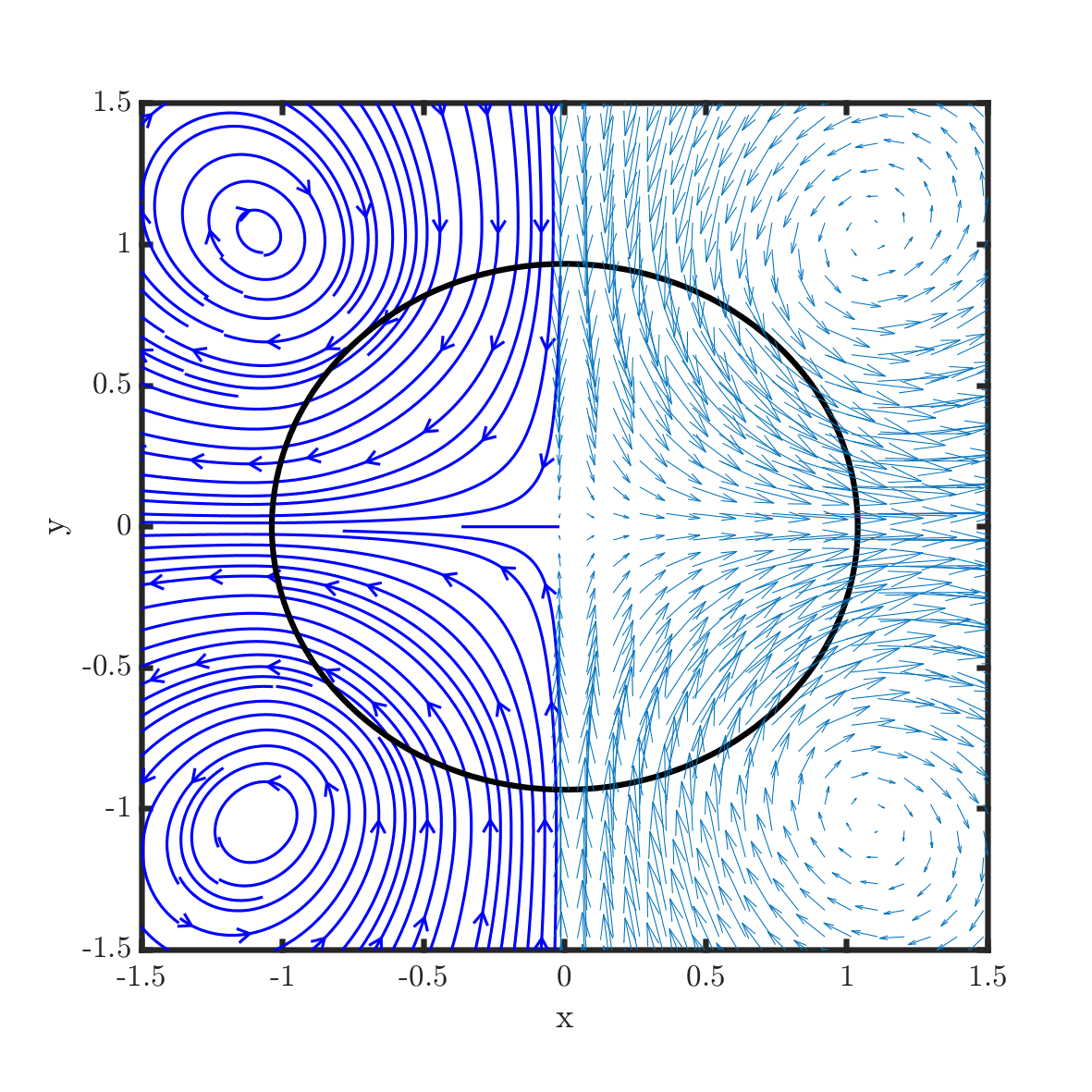}
		\includegraphics[width=0.32\textwidth]{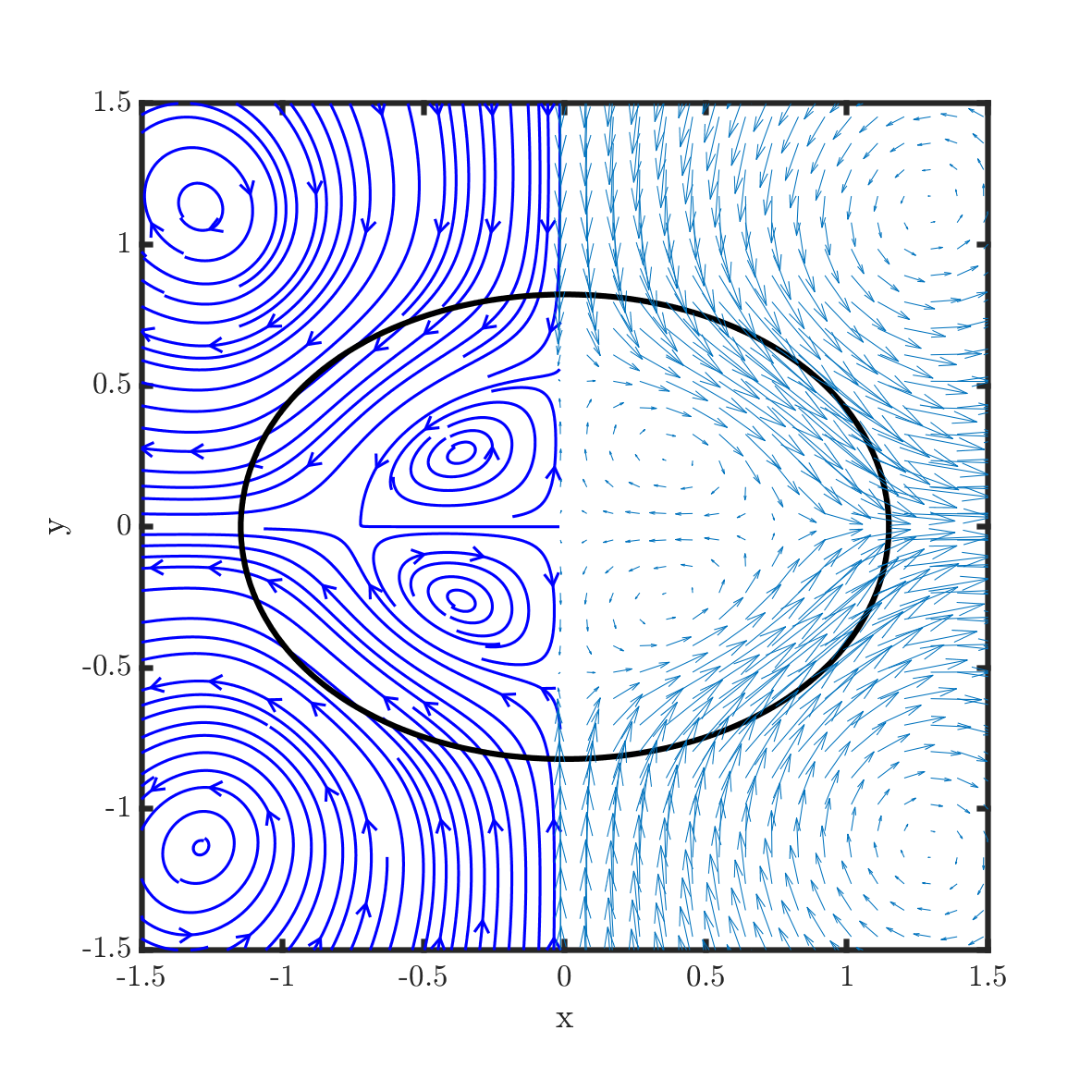}
		\includegraphics[width=0.32\textwidth]{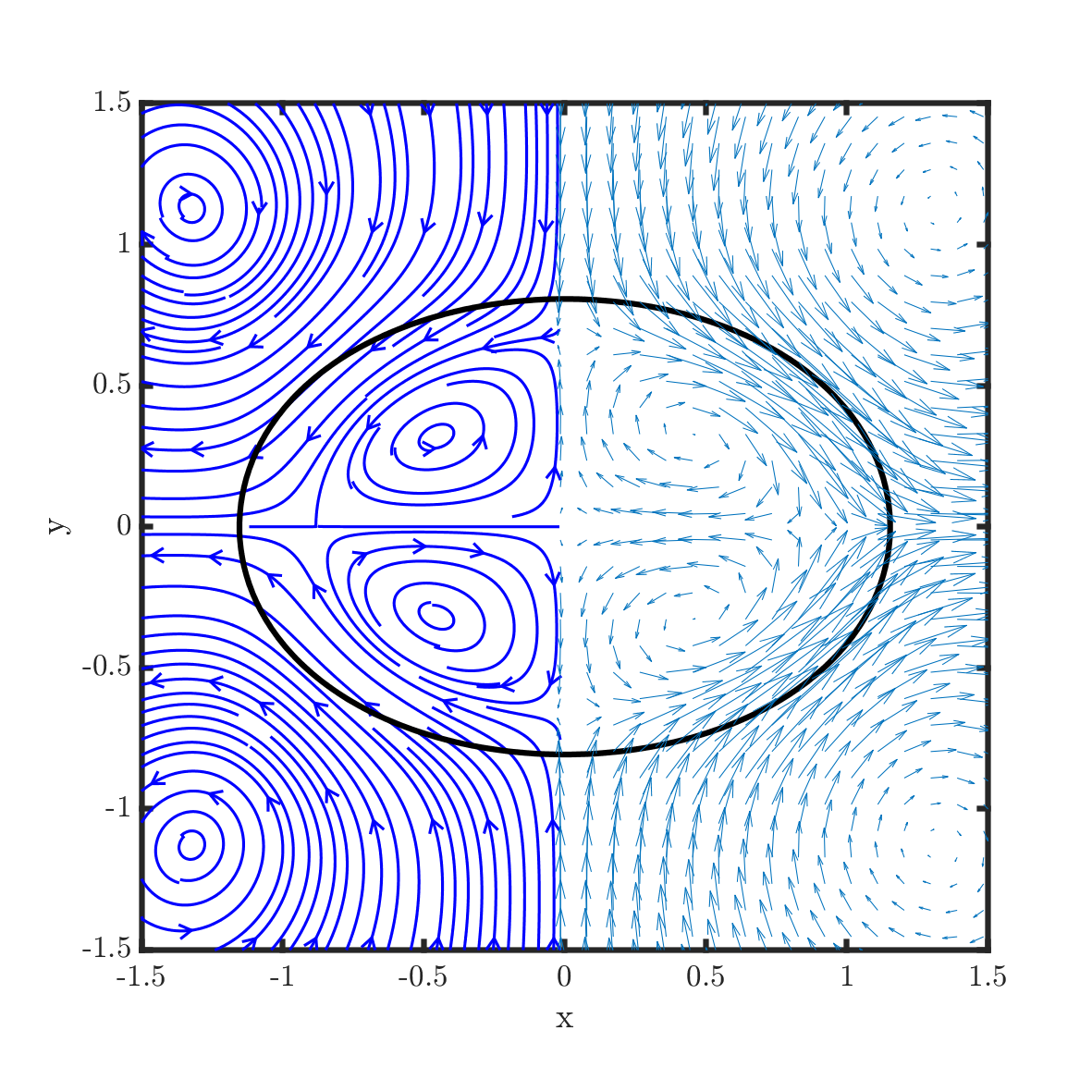}
		\includegraphics[width=0.32\textwidth]{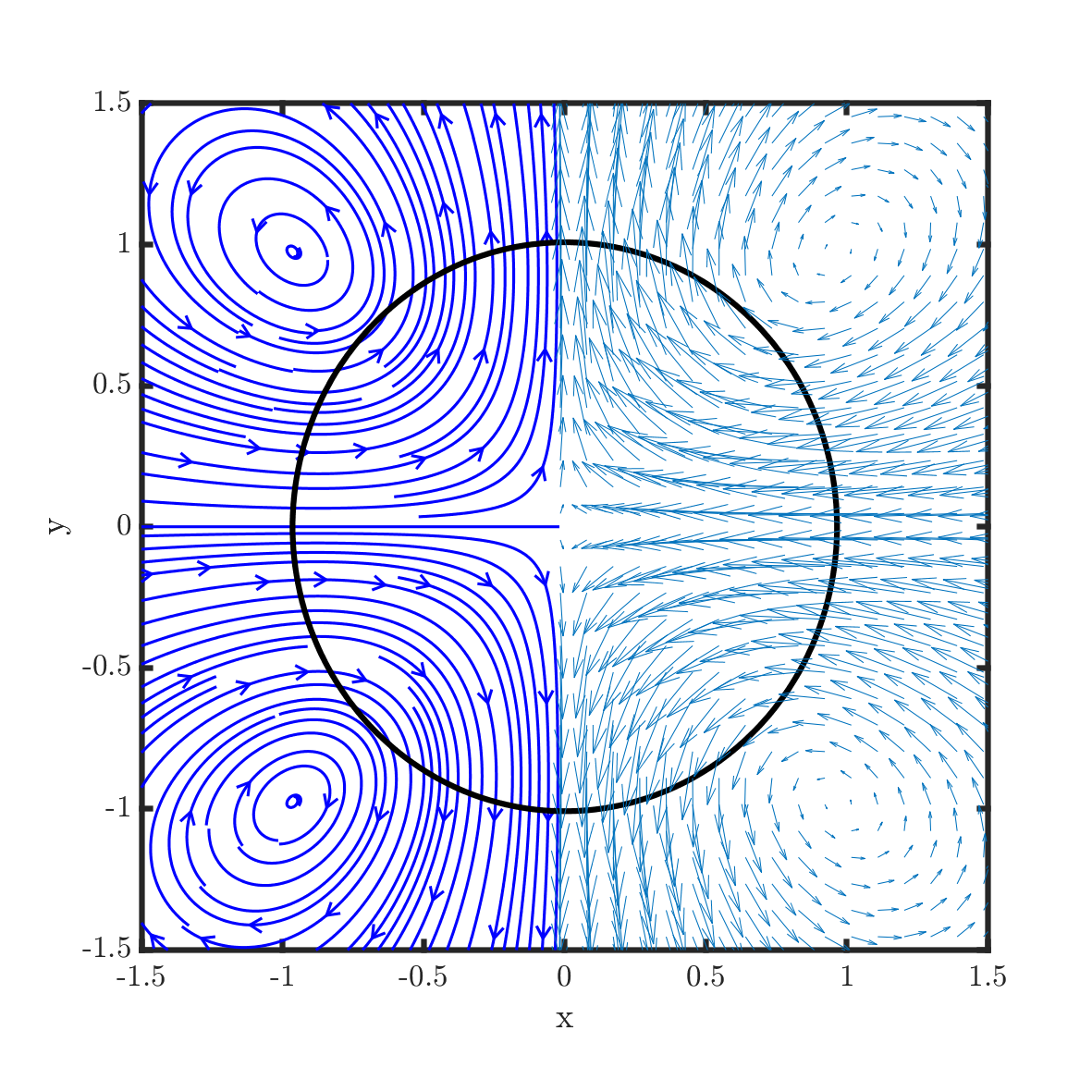}
		\includegraphics[width=0.32\textwidth]{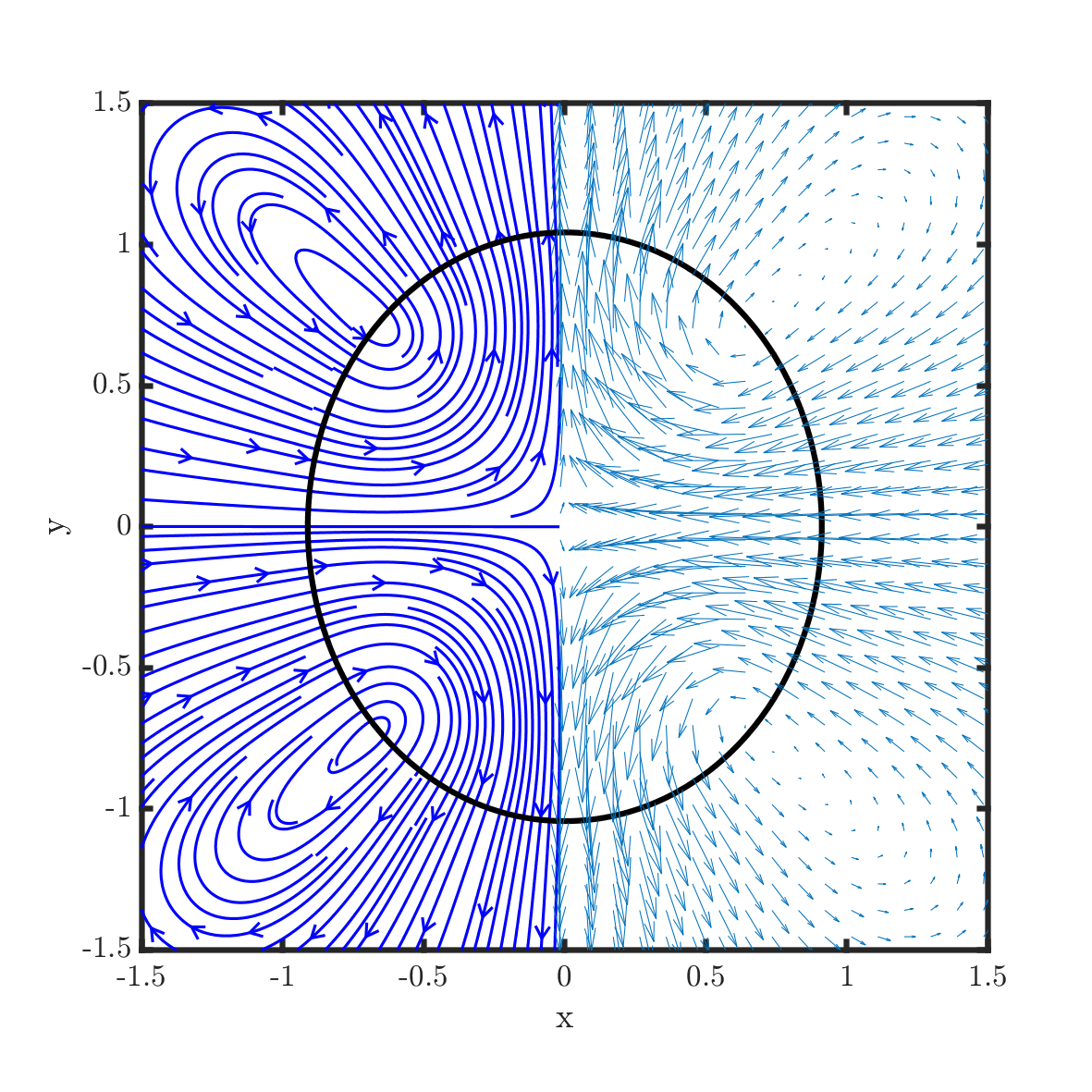}
		\includegraphics[width=0.32\textwidth]{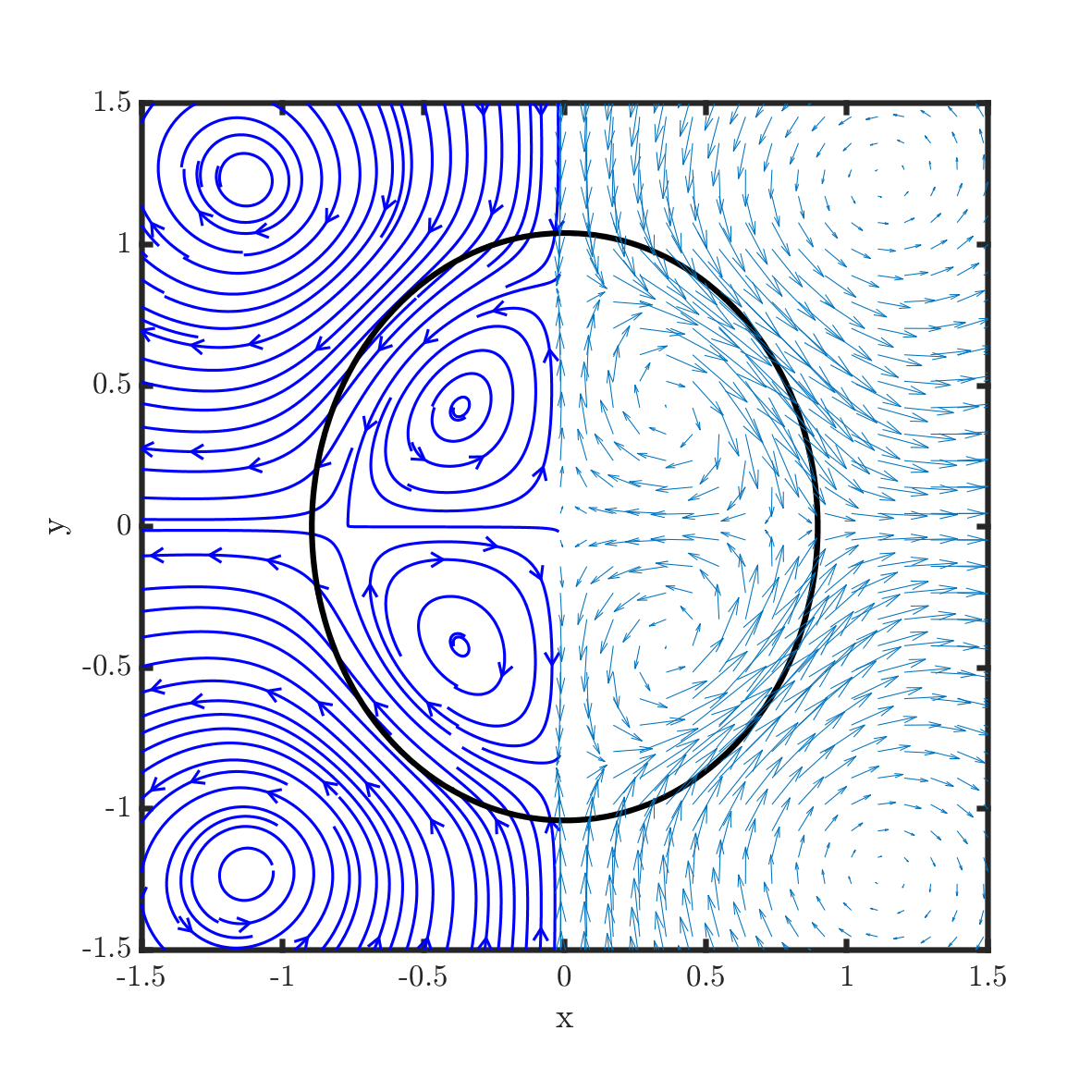}
		\includegraphics[width=0.32\textwidth]{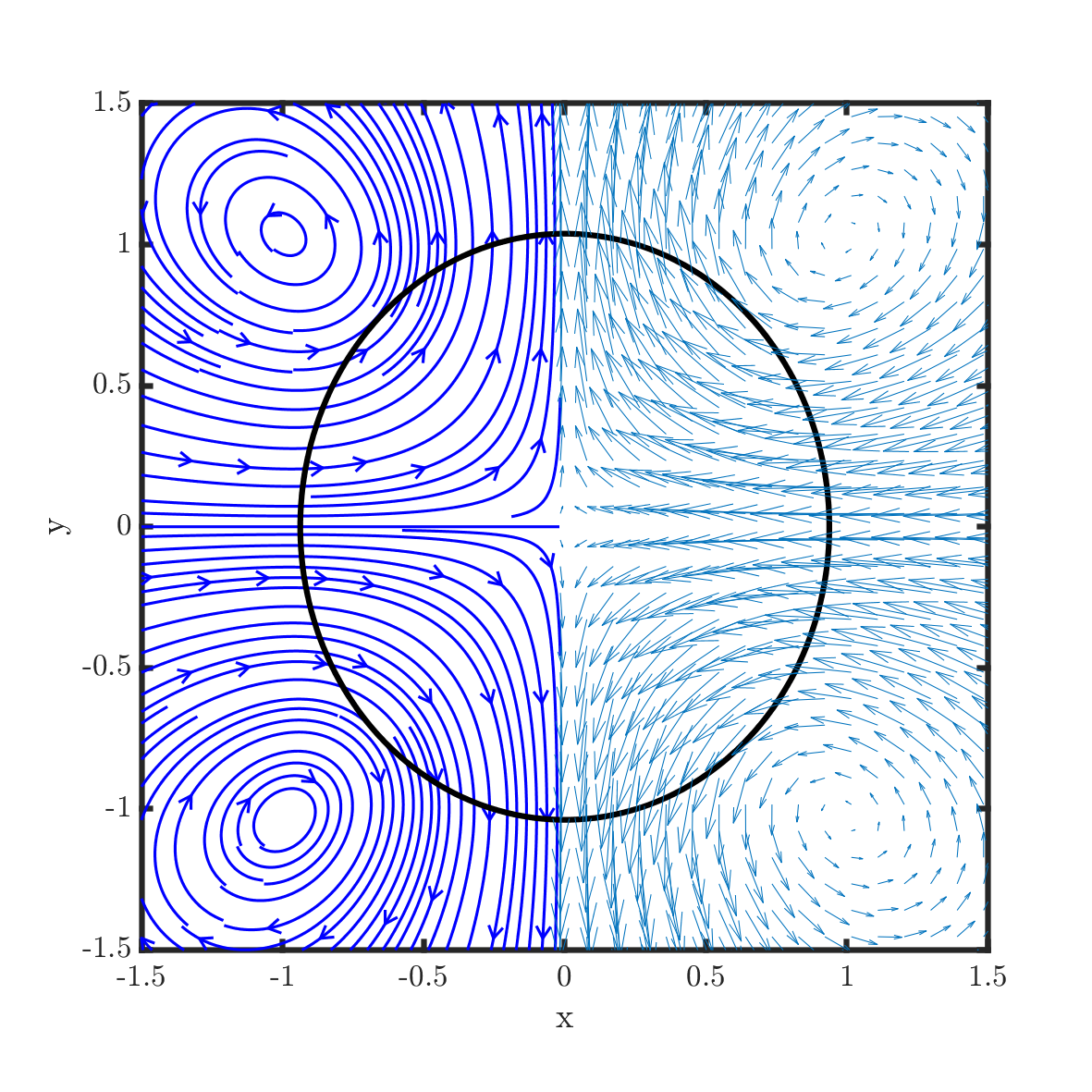}
		\includegraphics[width=0.32\textwidth]{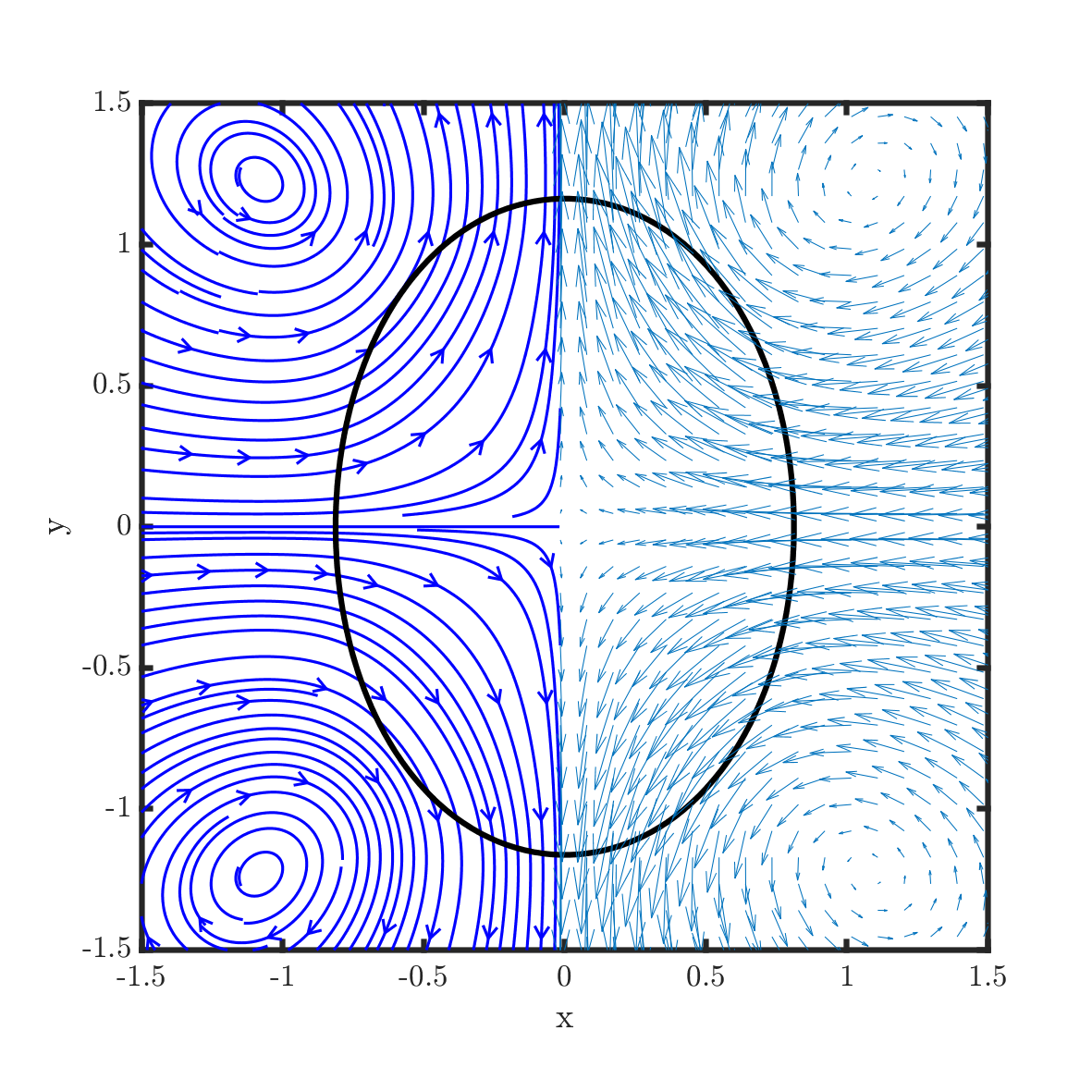}
		\includegraphics[width=0.32\textwidth]{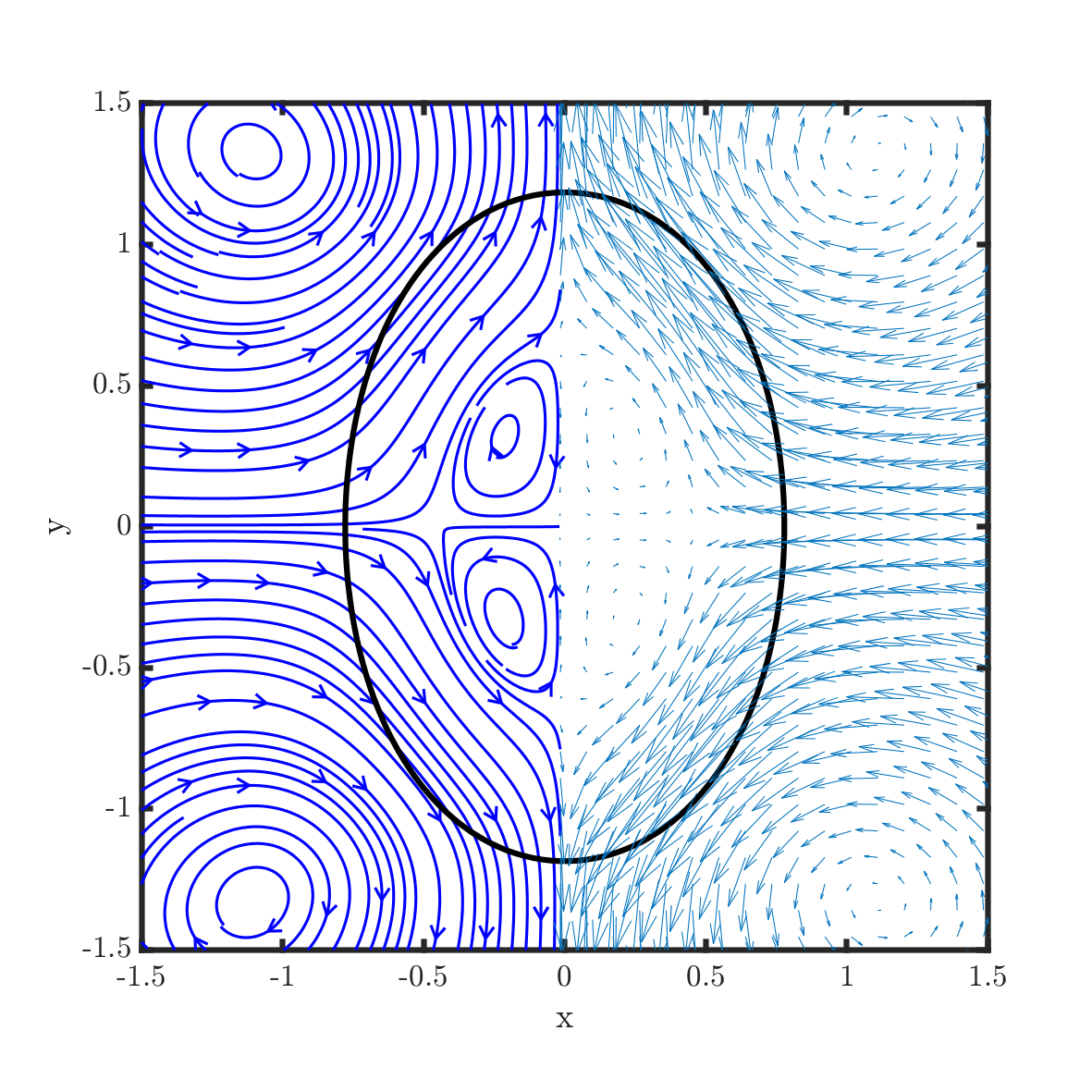}
	\end{center}
	%\vskip -2.0cm
	\caption{Drop shapes and flow patterns for $\sigma_{r}=1.75$ (top), 
		$\sigma_{r}=3.25$ (middle) and $\sigma_{r}=4.75$ (bottom)  
		at $t=1$ (left), $t=5$ (middle) and $t=10$ (right) for example \ref{eqn: initial single drop} 
		in section \ref{sec: comparison with sharp model}. 
		The dielectric coefficient ratio is set to $\epsilon_{r}=3.5$. 
		The electric capiliary number is chosen as $Ca_{E} = 1$.
		We choose the coefficients are all the same as \cite{HU2015Ahybrid} 
		for comparison. In each subpplot, the black solid line shows the zero level 
		set of $\psi$ to describe the position of interface, the velocity quivers 
		are depicted in the right half while the corresponding stream lines 
		are shown in the left.}
	\label{fig: comparison with Lai}
\end{figure}

\begin{figure}
	\begin{center}
		\includegraphics[width=0.9\textwidth]{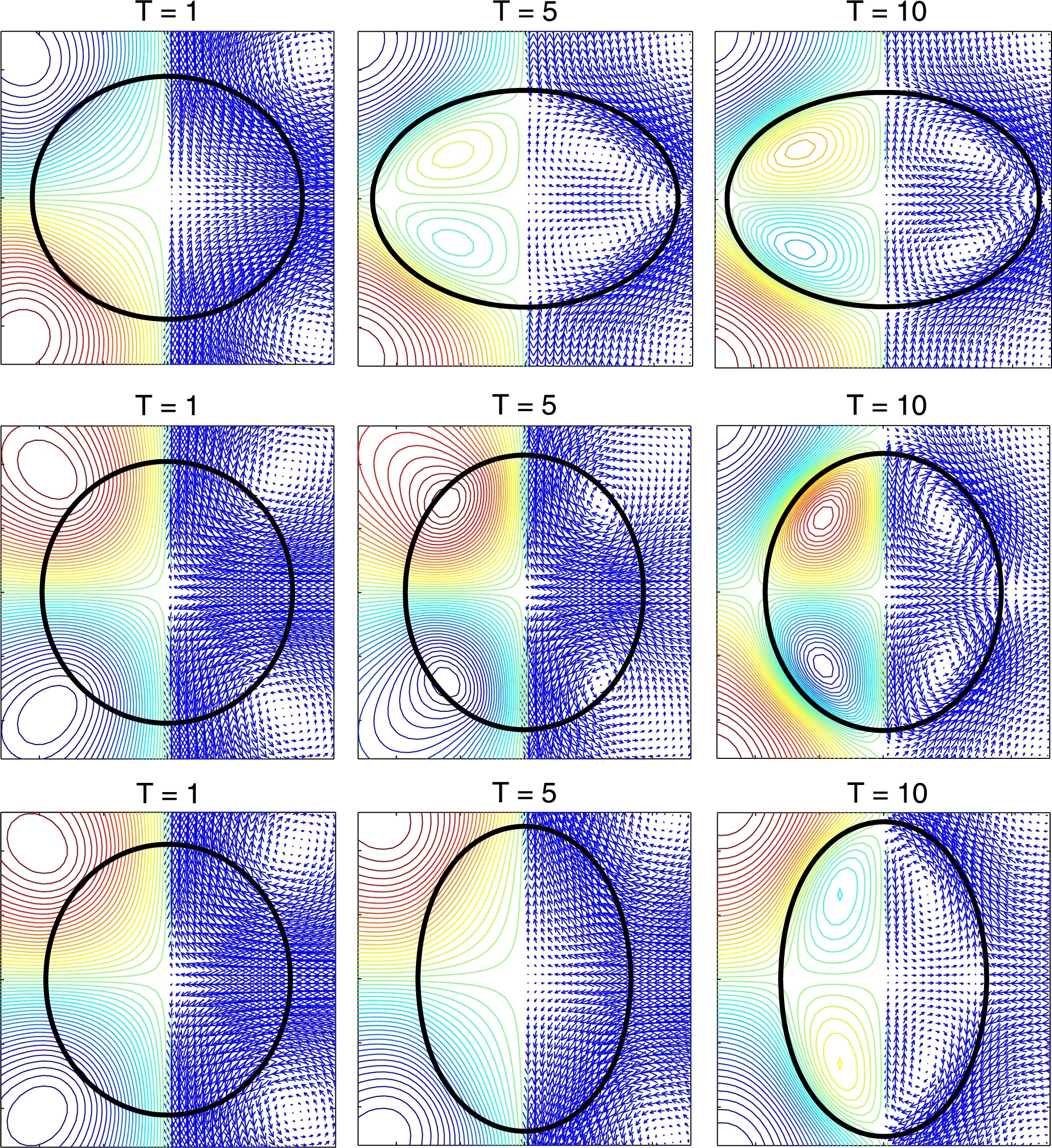}
	\end{center}
	\caption{The numerical results obtained by a hybrid immersed boundary 
		and immersed interface method in \cite{HU2015Ahybrid}. 
		The black solid line is the location of interface, and left part is 
		the stream lines and the right part is the velocity quivers. 
		The different conductivity ratios are set as $\sigma_{r}=1.75$ (top), 
		$\sigma_{r}=3.25$ (middle), $\sigma_{r}=4.75$ (bottom). 
		Besides, the dielectric coefficient ratio and electric capiliary are 
		$\epsilon_{r}=3.5$ and $Ca_{E} = 1$, respectively.}
	\label{fig: Lai_DC_field}
\end{figure}

Fig. \ref{fig: comparison with Lai} shows snapshots of the droplets and flow patterns at 
different time $t = 1$ (left), $t = 5$ (middle) and $t = 10$ (right) 
with three different conductivity ratio cases 
$\sigma_{r}=1.75$ (top), $3.25$ (middle) and $4.75$ (bottom) which are compared with the 
previous work in \cite{HU2015Ahybrid}. 
In each subpplot, the black solid line shows the zero level set to describe the position of interface. 
The velocity quivers are depicted in the right half while the corresponding stream lines 
are shown in the left.  When  conductivity ratio is small $\sigma_r=1.75, $ the drop shape 
is oblate (top) and  the induced circulatory flow inside the first quadrant 
is clockwise (from the pole to the equator). When ratio become larger, the droplet changes to be  the prolate shape.
The induced circulatory flow can be clockwise (middle) and 
counterclockwise (from the equator to the pole, bottom). 
The numerical results obtained by a hybrid immersed boundary 
and immersed interface method in \cite{HU2015Ahybrid} is shown in Fig. \ref{fig: Lai_DC_field}. 
We can see that these flow patterns are in good agreement with those from the previous work. 
Slight difference may be observed from the comparison due to the thickness of diffuse interface. 

\begin{figure}
	\begin{center}
		\includegraphics[width=0.325\textwidth]{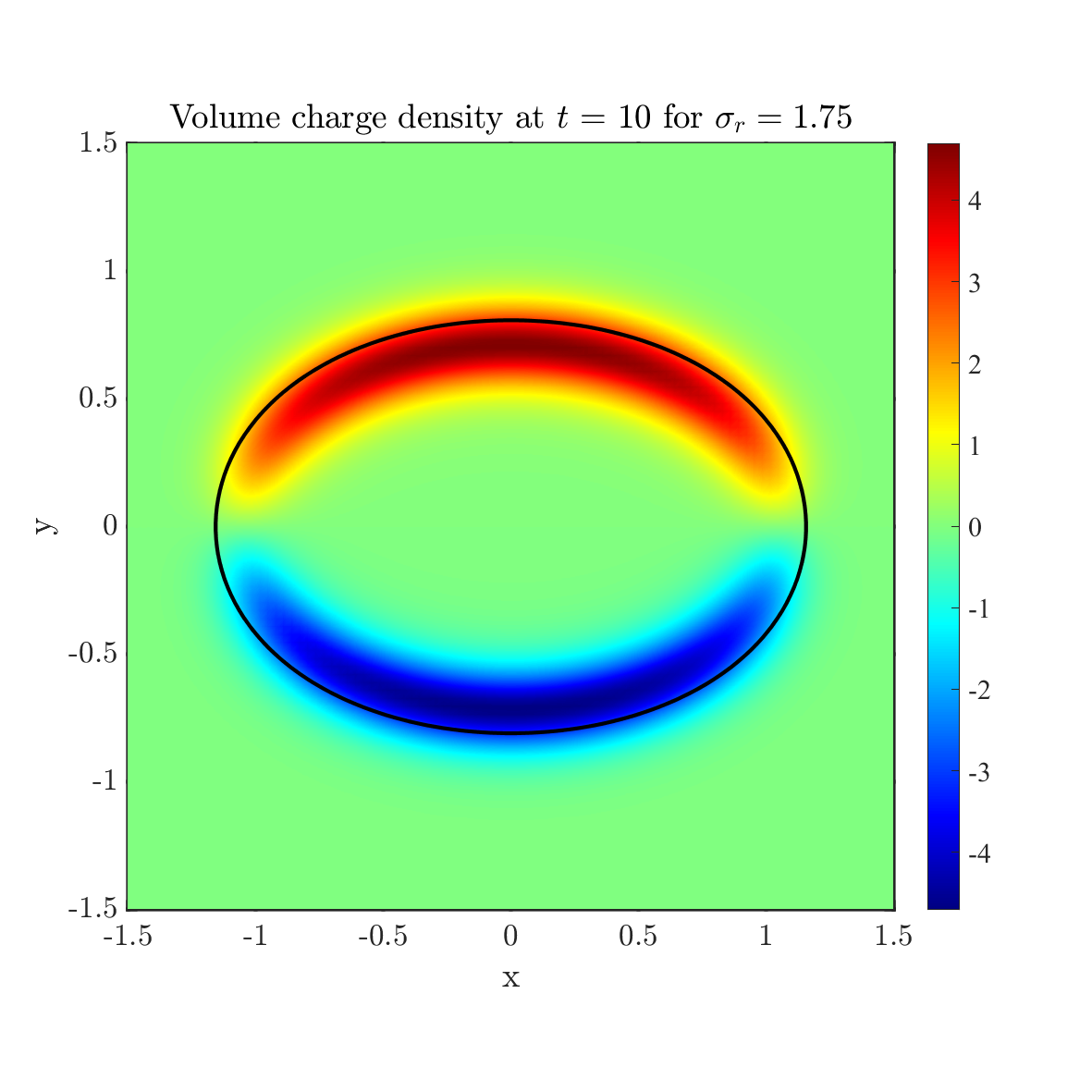}
		\includegraphics[width=0.325\textwidth]{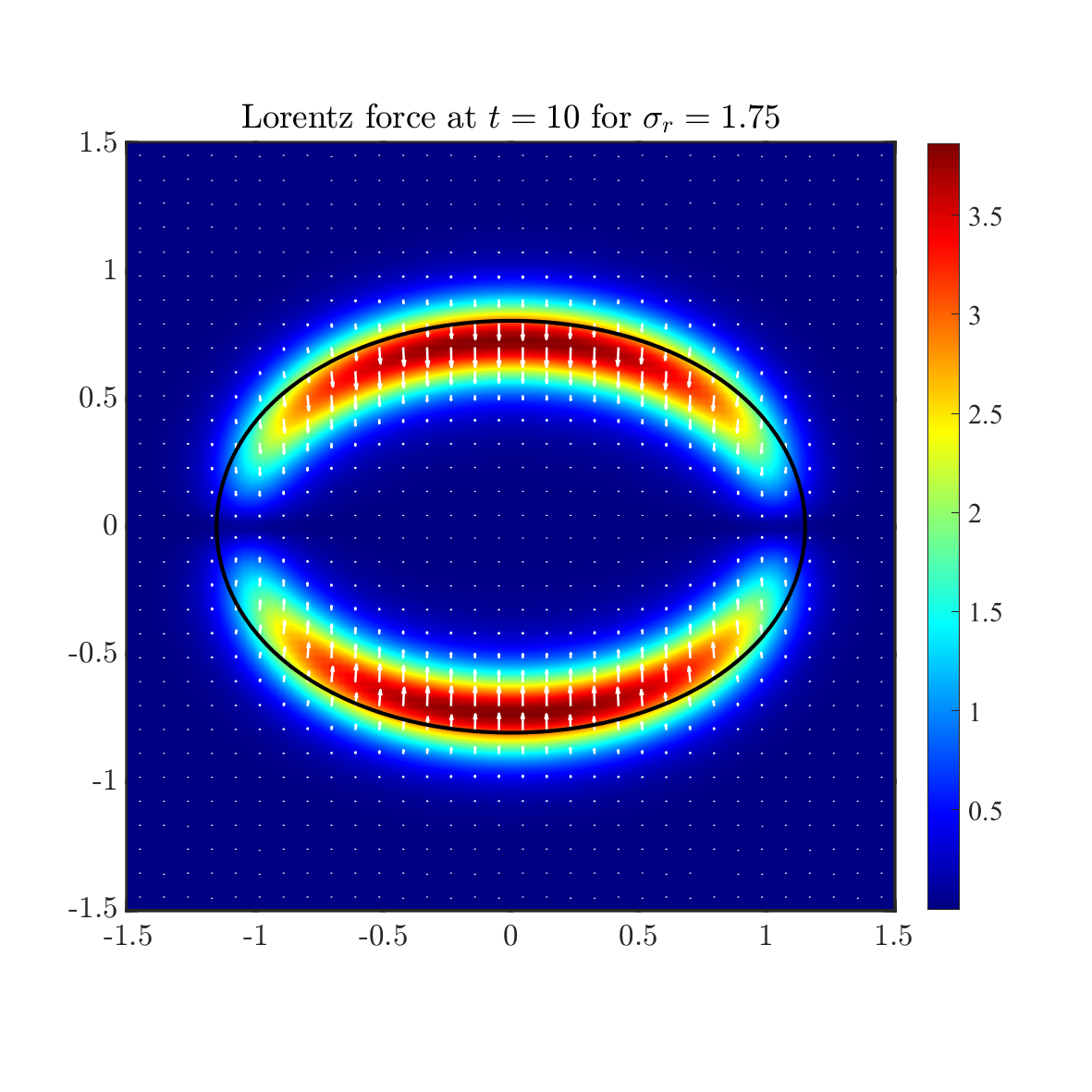}
		\includegraphics[width=0.325\textwidth]{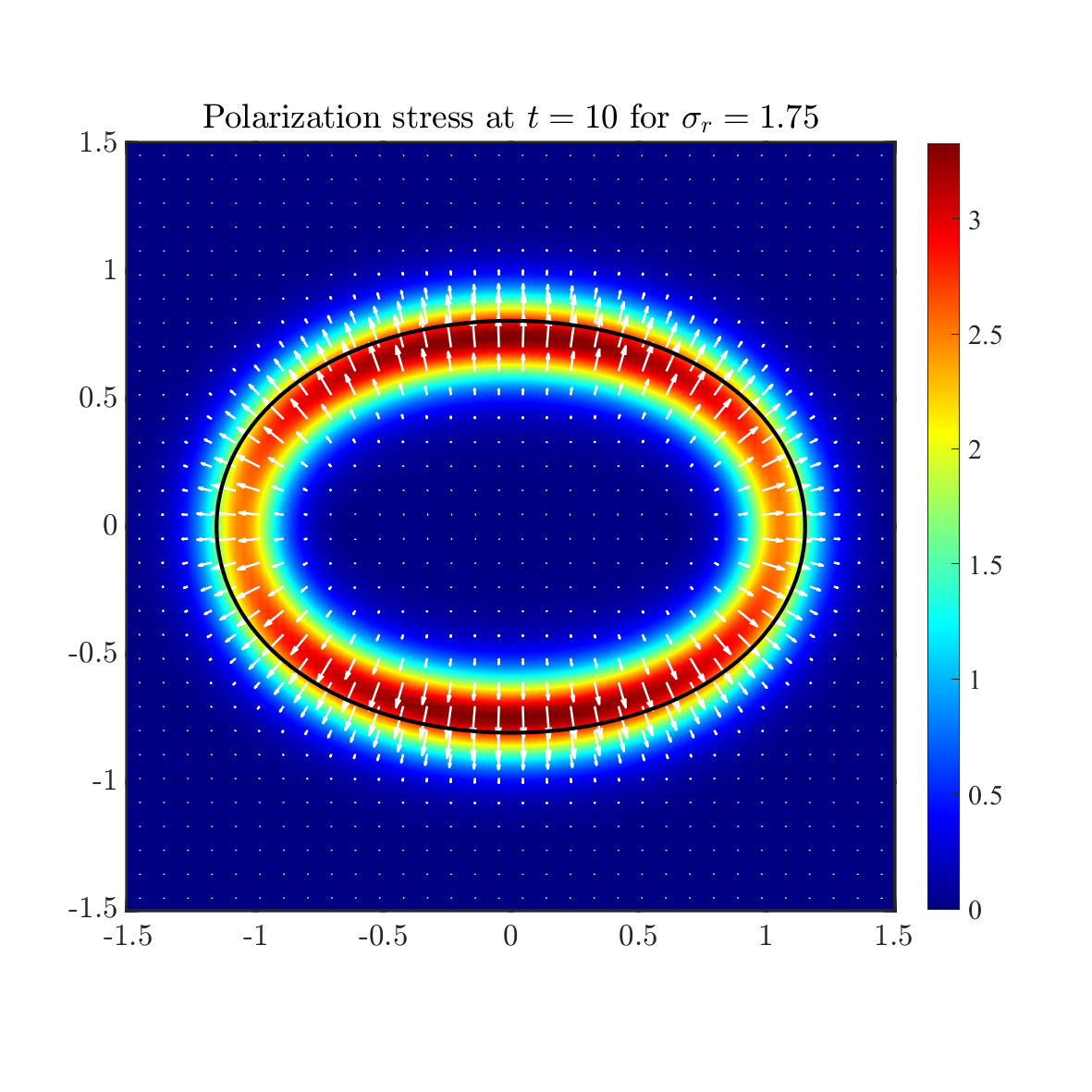}
		\includegraphics[width=0.325\textwidth]{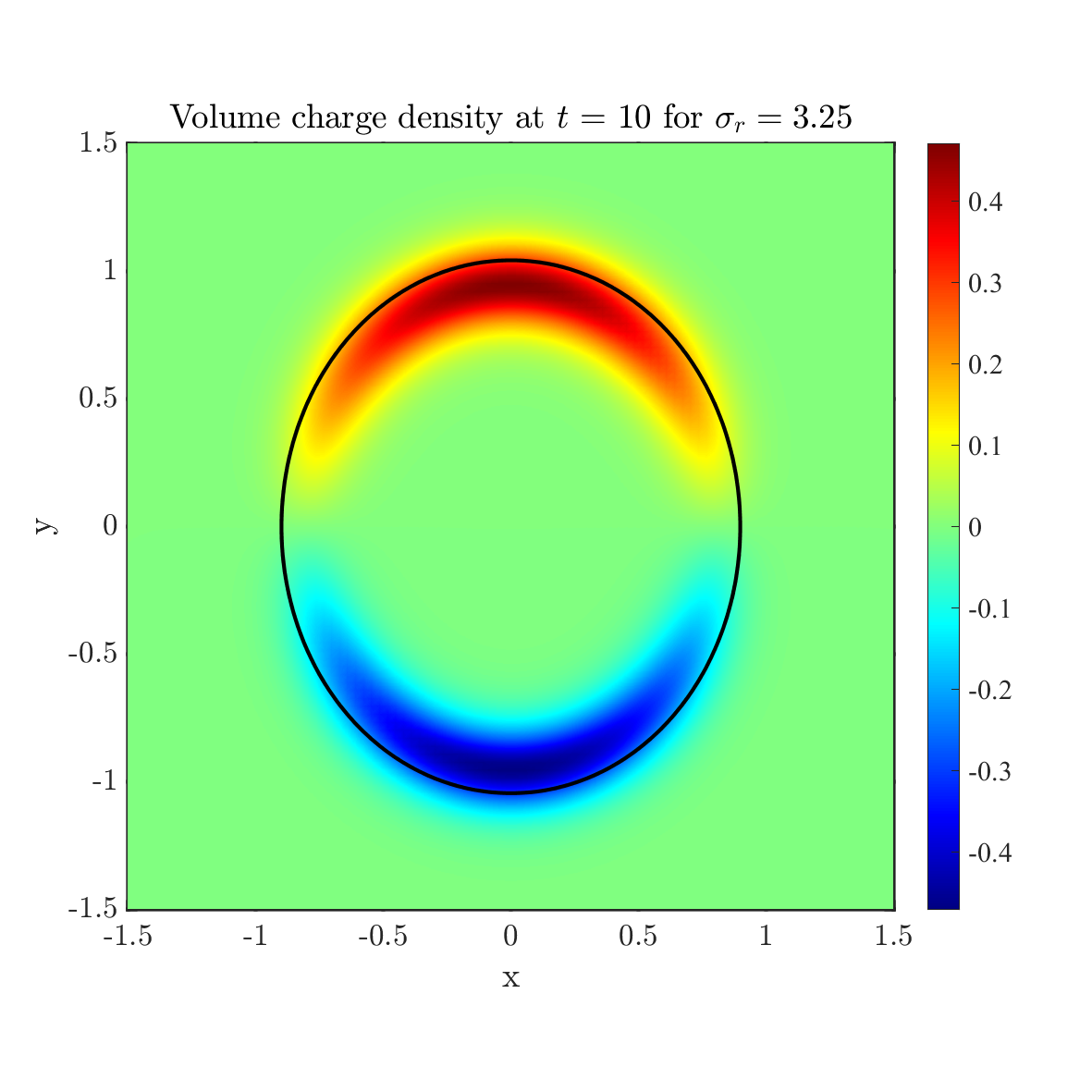}
		\includegraphics[width=0.325\textwidth]{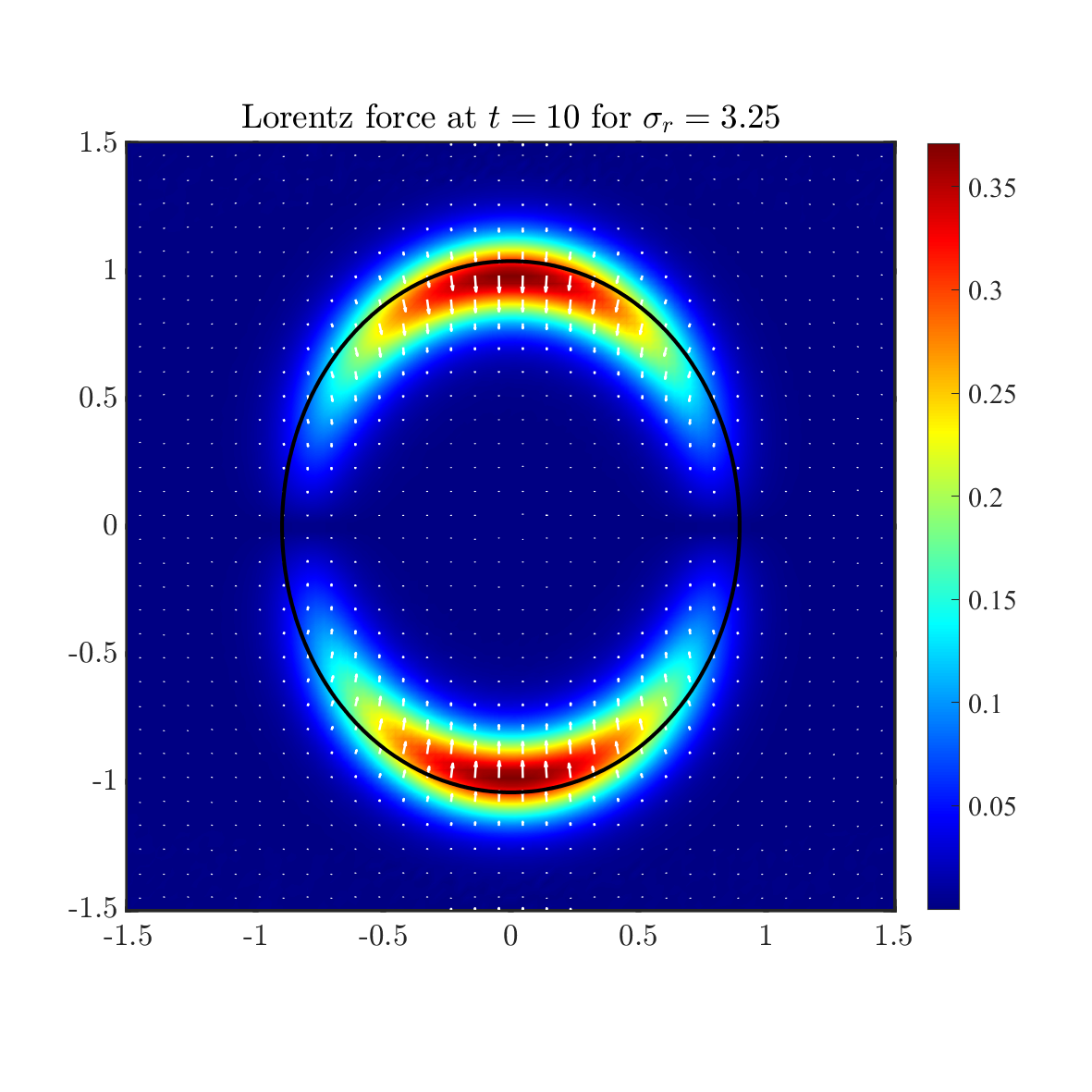}
		\includegraphics[width=0.325\textwidth]{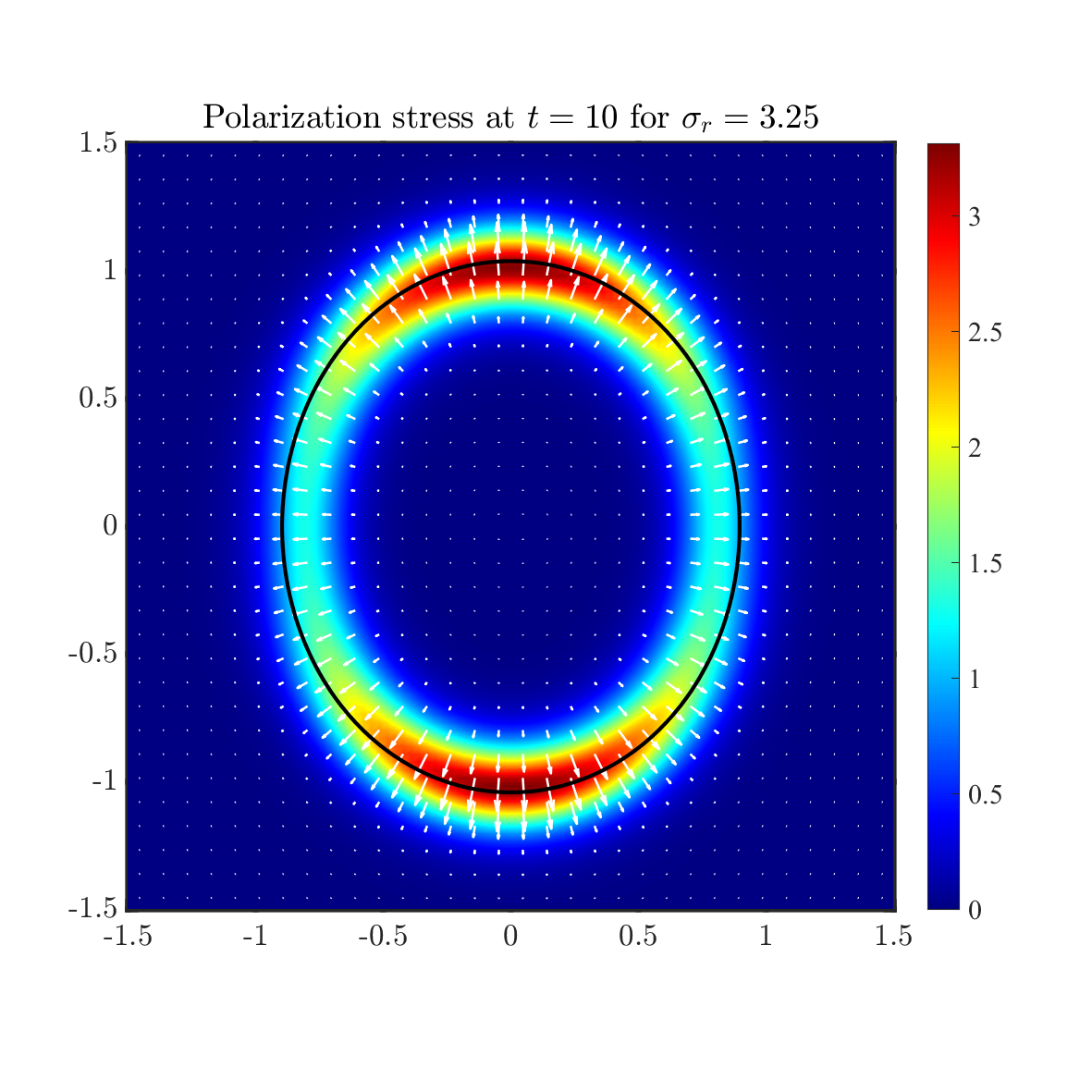}
		\includegraphics[width=0.325\textwidth]{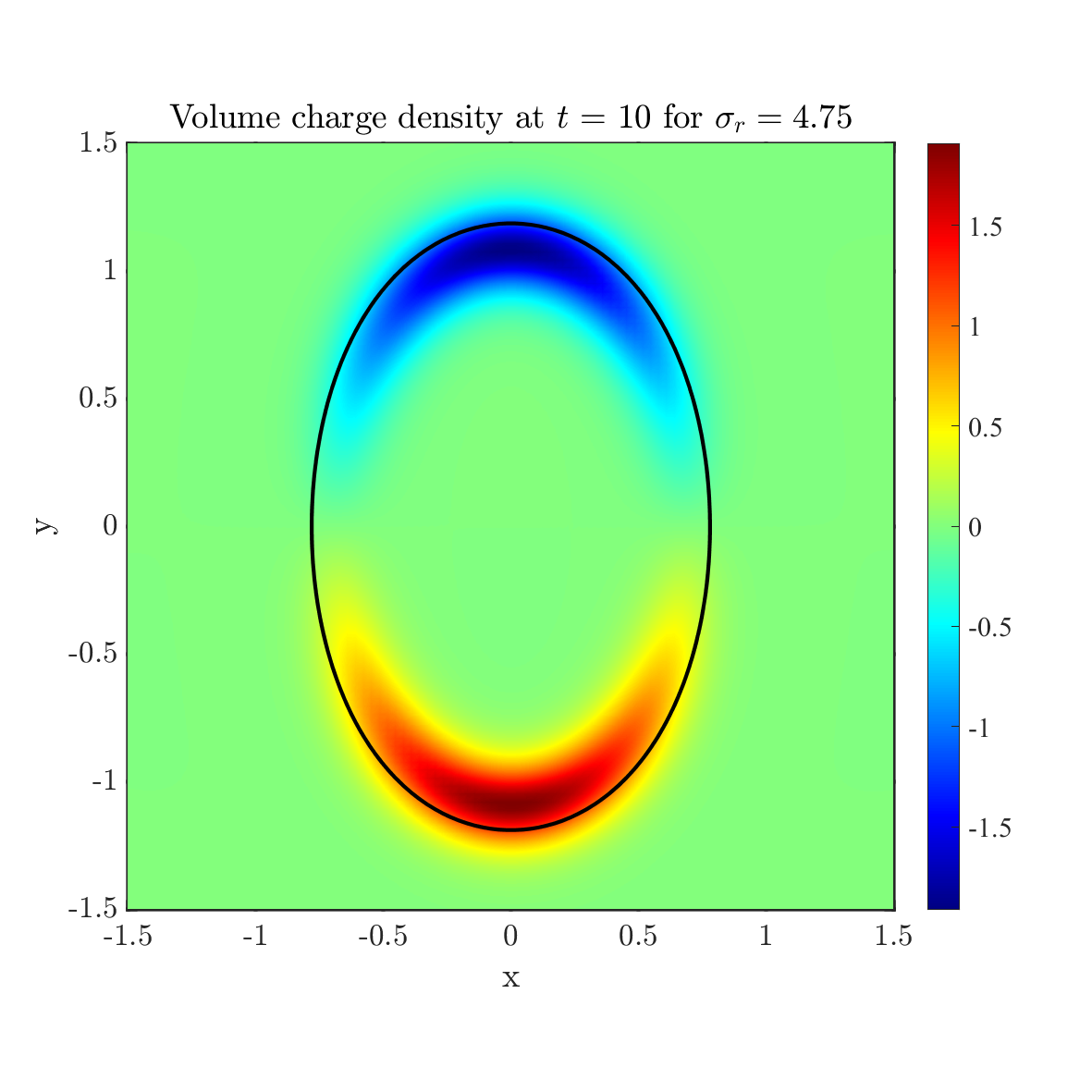}
		\includegraphics[width=0.325\textwidth]{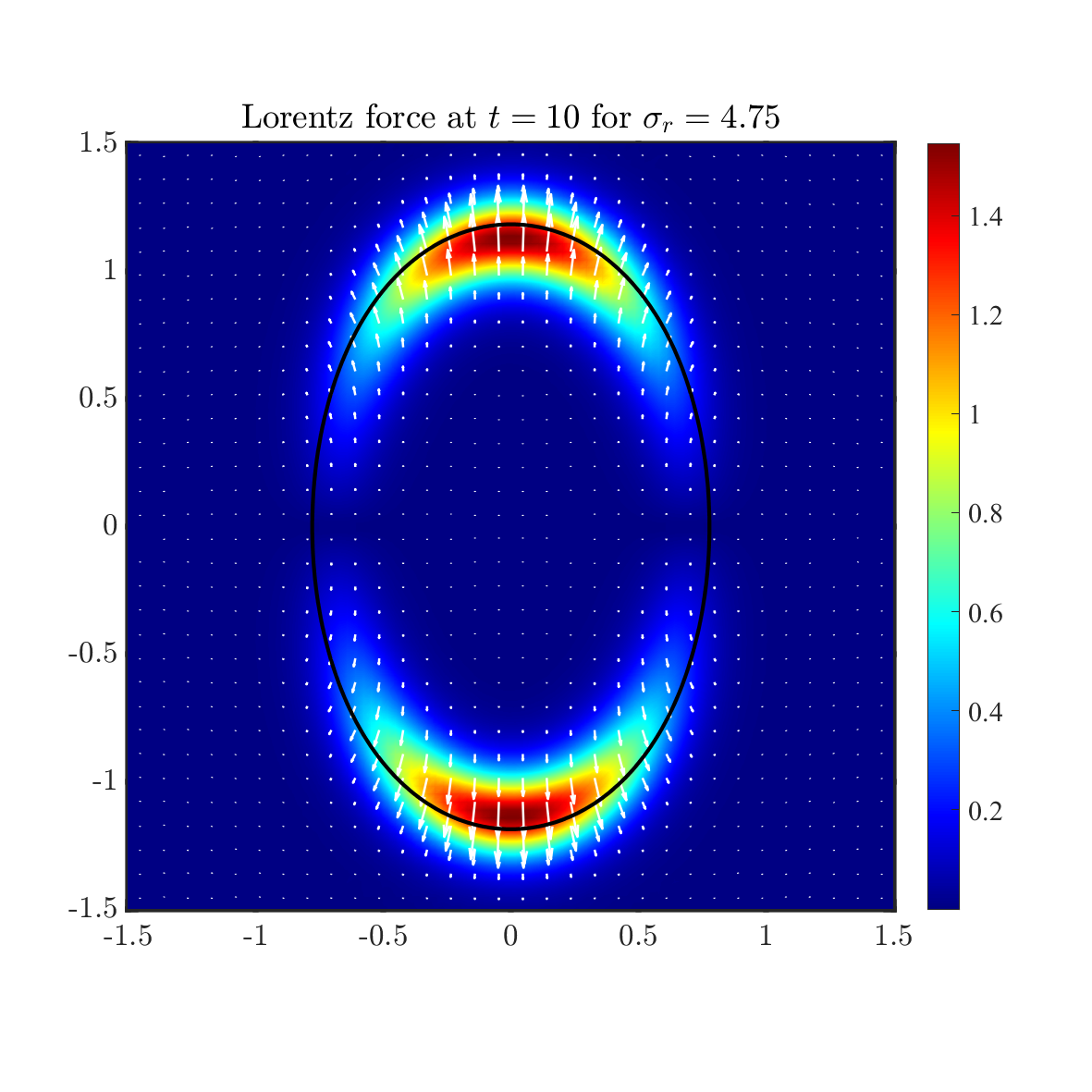}
		\includegraphics[width=0.325\textwidth]{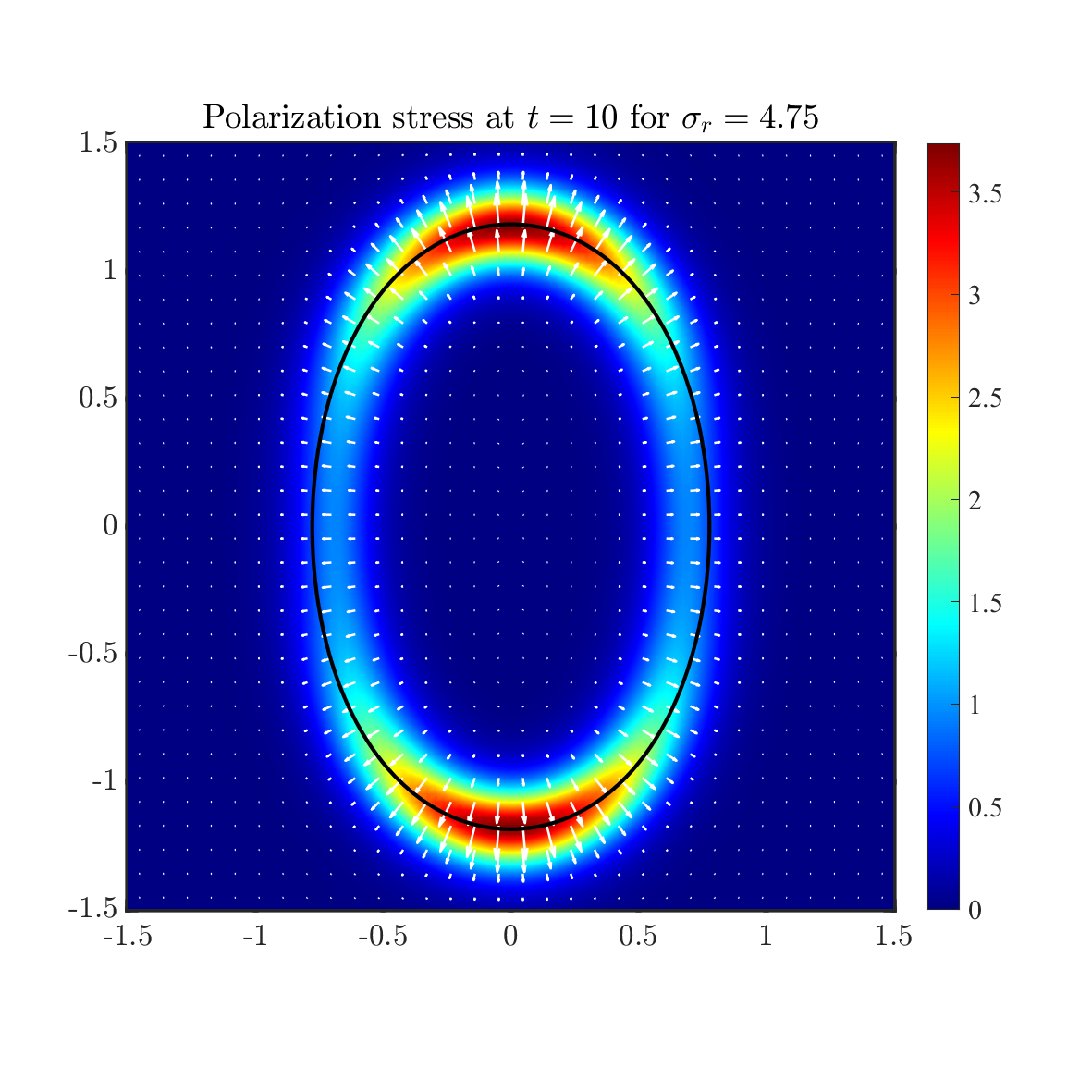}
	\end{center}
	%\vskip -2.0cm
	\caption{Distributions of charge density $\rho_{e}$ (left), Lorentz electric force $\bm{F}_L$ (middle) 
		and polarization force $\bm{F}_p$ (right) with different conductivity ratios 
		$\sigma_{r}=1.75$ (top), $\sigma_{r}=3.25$ (middle), $\sigma_{r}=4.75$ (bottom) at $t = 10$ 
		for example \ref{eqn: initial single drop} in section \ref{sec: comparison with sharp model}.
		The positive ions accumulate at the top of the interface when $\sigma_{r} = 1.75$ 
		and $\sigma_{r} = 3.25$, whereas the negative ions on the contrary. 
		The positive ions accumulate at the bottom of the interface when $\sigma_{r} = 4.75$. 
		The dielectric coefficient ratio and electric capiliary are $\epsilon_{r}=3.5$ and $Ca_{E} = 1$, respectively.}
	\label{fig: some other information}
\end{figure}

In Fig. \ref{fig: some other information}, the charge densities $\rho_e$ with different $\sigma_r$ at equilibrium states are presented. In the sharp interface limit case, the continuity of current across the boundary $-\sigma^-\nabla\phi^- \cdot\bm{n}= -\sigma^+\nabla\phi\cdot \bm{n}$ yields $\nabla\phi^+\cdot\bm{n}= \frac{1}{\sigma_r} \nabla\phi\cdot\bm{n}$. Then, the net charge density could be calculated by $\rho_e = -\epsilon^-\nabla\phi^-\cdot\bm{n}+\epsilon^+\nabla\phi^+\cdot\bm{n} = (\frac{\epsilon_r}{\sigma_r}-1)\nabla\phi^-\cdot\bm{n}$, as in \cite{Saville1997electro,Melcher1969Electrohydrodynamics}. Therefore, when $\sigma_r<\epsilon_r$, positive net charges accumulate on the top of the droplet, and negative charges are on the bottom. This can be observed in the first and second rows of Fig. \ref{fig: some other information} for the cases $\sigma_r=1.75, 3.25$. When $\sigma_r>\epsilon_r$, the opposite is true (third row). Also, the closer $\sigma_r$ is to $\epsilon_r$, the smaller the net charge is (second row of Fig. \ref{fig: some other information}). The different profiles of the droplets are induced by the electric force on the droplet. As mentioned in Remark \ref{remarkontheelectricforce}, there are two forces induced by Maxwell stress: the first term is the Lorentz force $\bm{F}L=-\frac{1}{\zeta^2}\rho_e\nabla\phi$ due to the interaction of net charges with the electric field, and the second term $\bm{F}p=- \frac{1}{2} \left|\nabla \phi \right|^{2} \frac{\partial \epsilon{\it{eff}}}{\partial \psi} \nabla \psi$ is due to the polarization stress. The distributions of these two forces are shown in the second and third column of Fig. \ref{fig: some other information}, where the color represents the magnitude, and the arrow represents the direction of the force. It shows that the force $\bm{F}p$ is always pointing outside of the droplet since $-\nabla\psi = \bm{n}$ near the interface, and the magnitude is almost the same in all cases. On the left and right sides, it becomes weaker because of the decrease of the electric field magnitude and the conservation of current across the interface (See Fig. \ref{fig: electric field intensity} in Appendix). The Lorentz forces point to the inner side of the droplets with $\sigma_r=1.75,~3.25$, and the outer side of the droplet with $\sigma_r = 4.75$. On the left and right sides, the magnitude of $\bm{F}L$ is almost zero since there is no accumulated net charge. 

\begin{figure}
	\begin{center} 
		\includegraphics[width=0.325\textwidth]{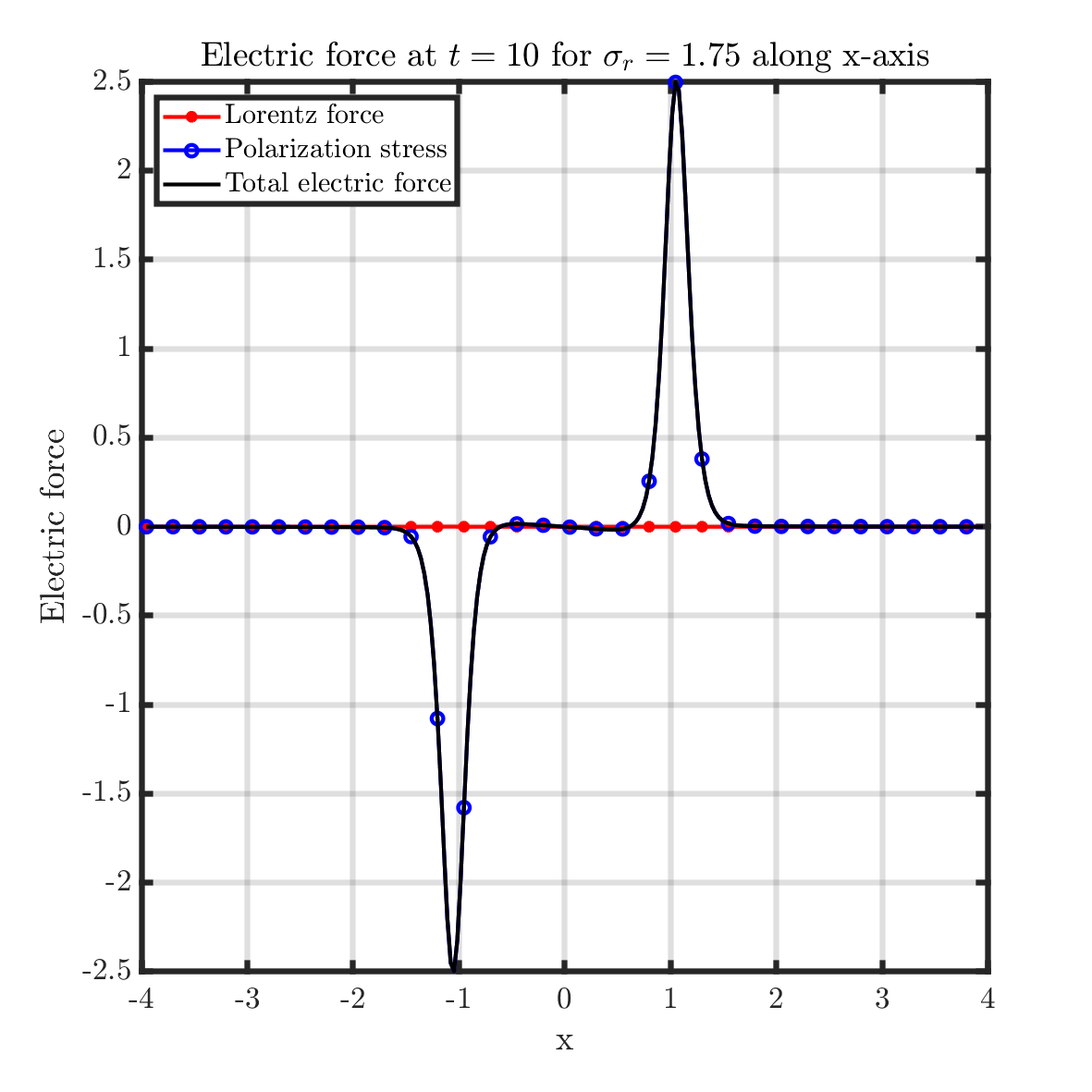}
		\includegraphics[width=0.325\textwidth]{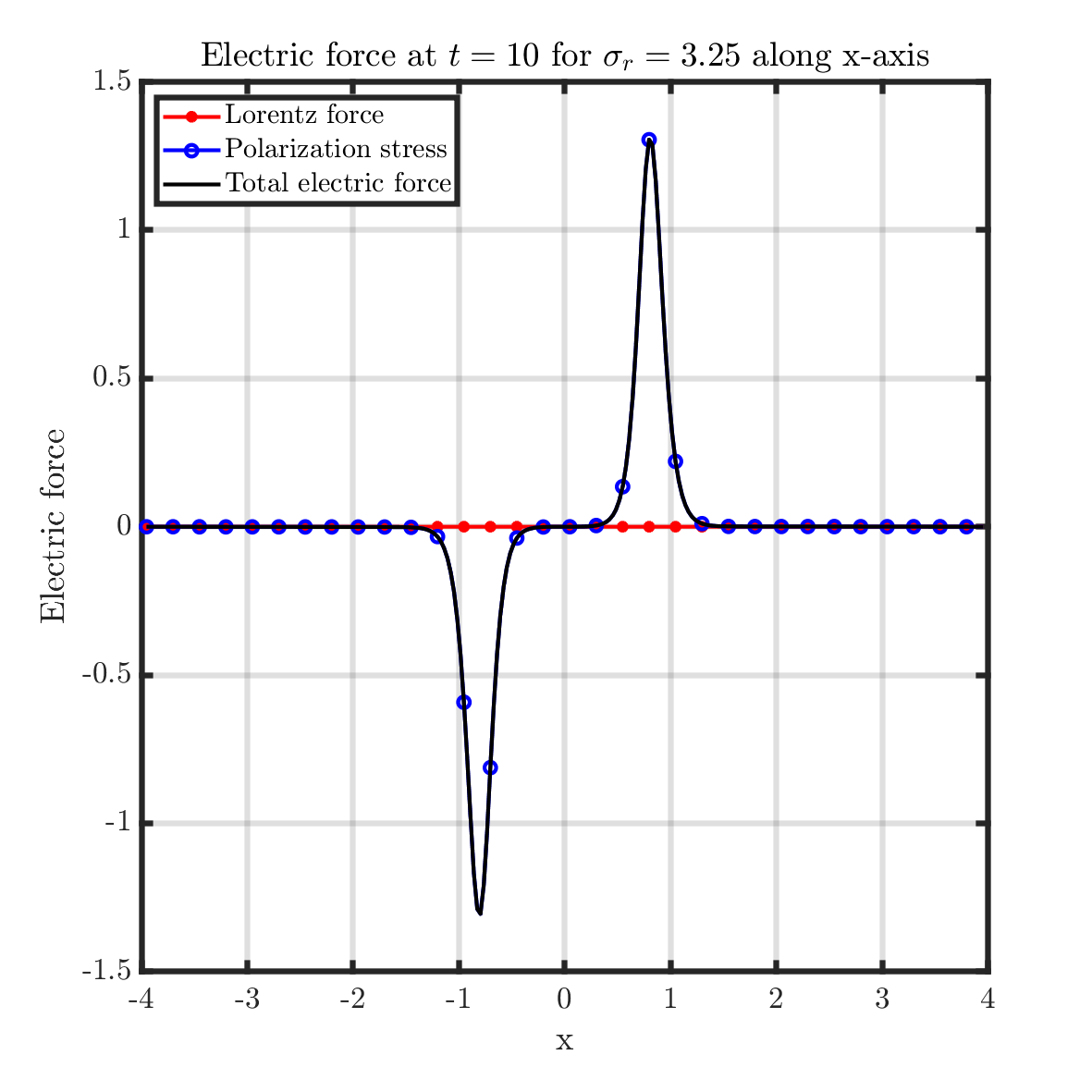}
		\includegraphics[width=0.325\textwidth]{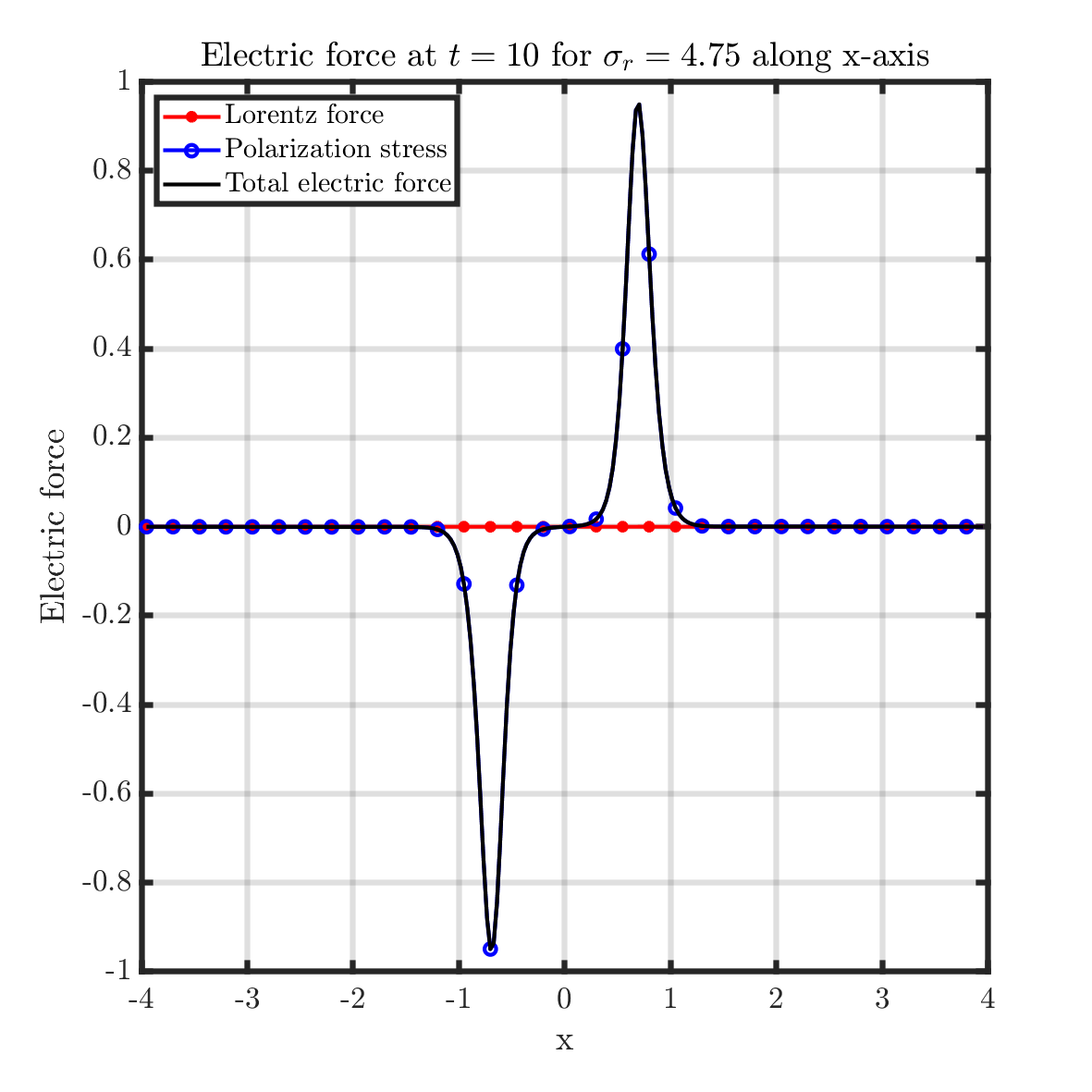}\\
		\includegraphics[width=0.325\textwidth]{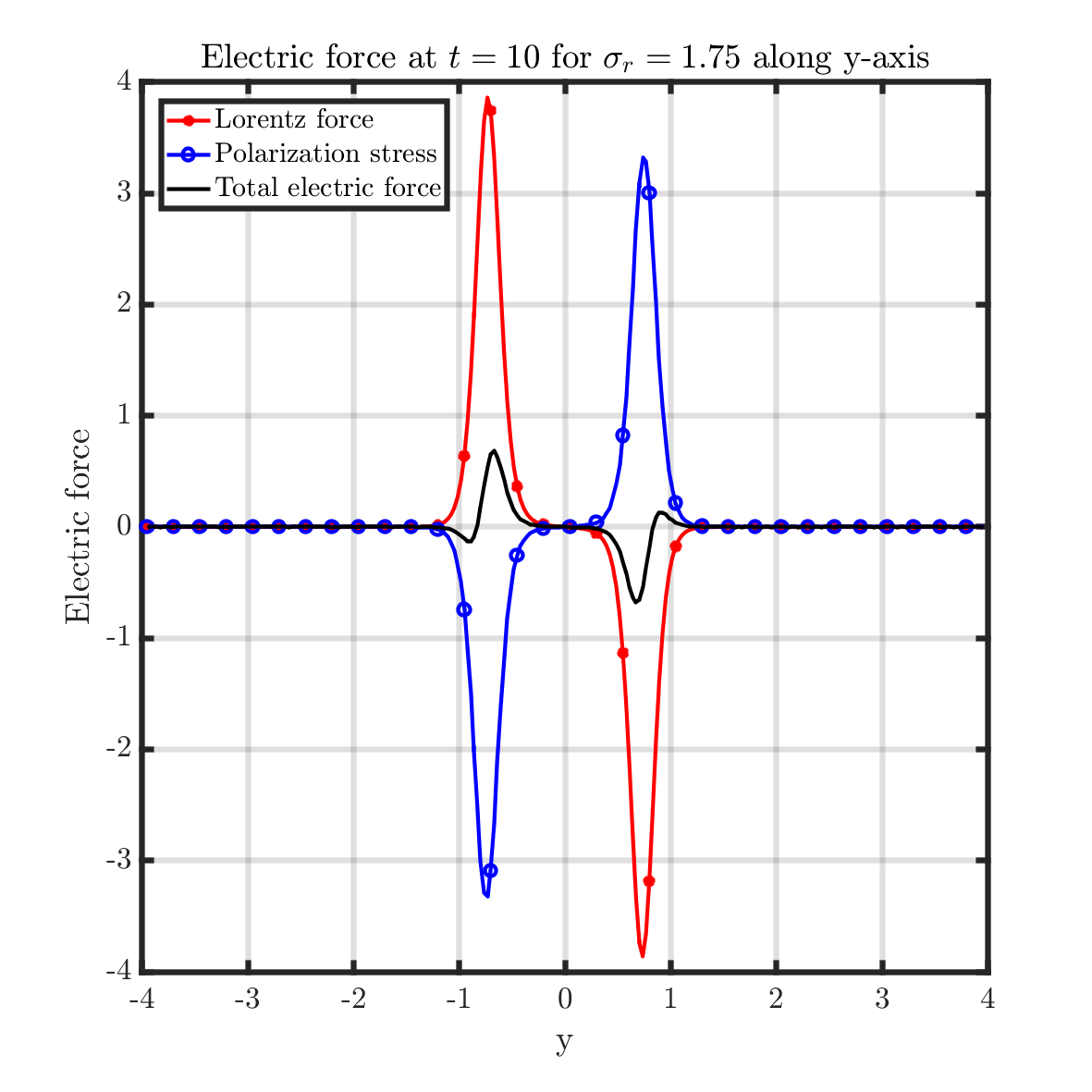}
		\includegraphics[width=0.325\textwidth]{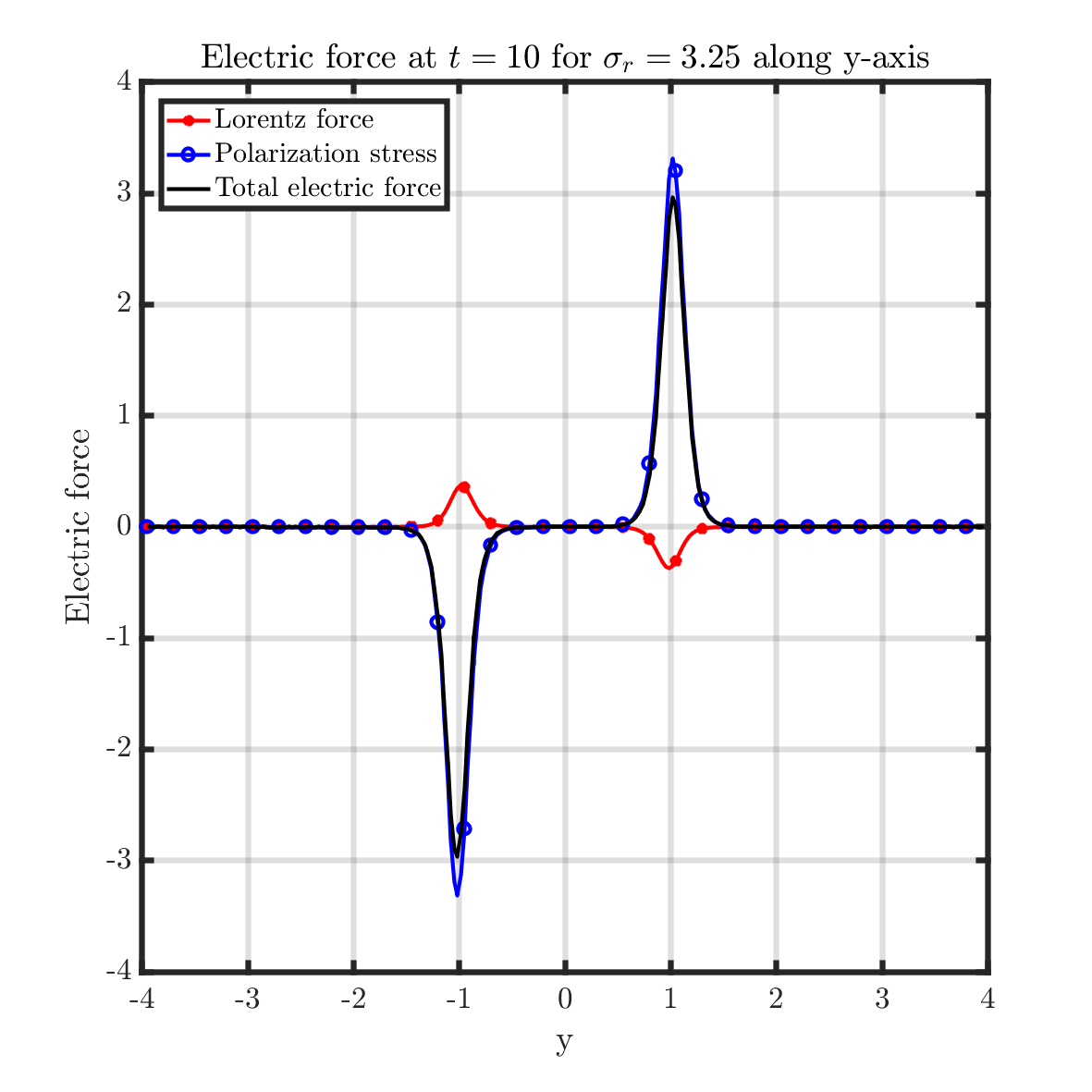}
		\includegraphics[width=0.325\textwidth]{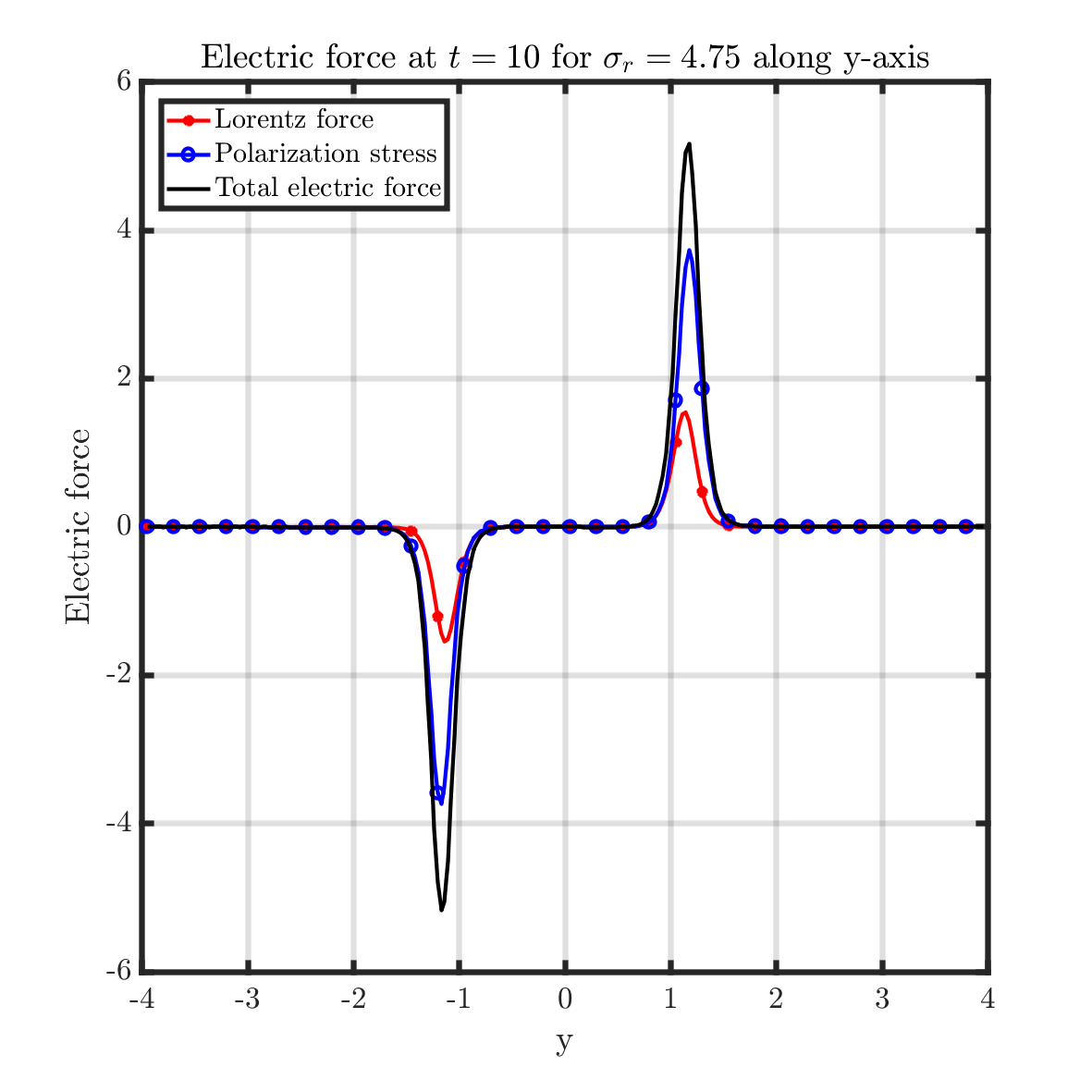} 
	\end{center} 
	\caption{The electric force along x-axis (top) and y-axis (bottom) for $\sigma_{r} = 1.75$ (left), $\sigma_{r} = 3.25$ (middle) 
		and $\sigma_{r} = 4.75$ (right) at $t = 10$ for example \ref{eqn: initial single drop} in section \ref{sec: comparison with sharp model}. 
		In each figure, the red solid line with star symbol shows the Loerentz force, 
		the blue solid line with circle shows the polarization stress and the black solid line shows the total electric force. 
		The rest parameters are chosen as $\epsilon_{r} = 3.5$, $Ca_{E} = 1$.}
	\label{fig: electric force for single drop}
\end{figure}

In Fig. \ref{fig: electric force for single drop}, we present a detailed distribution of electric force along the $x$-axis (top) and $y$-axis (bottom) for droplets with different values of $\sigma_r$: $\sigma_{r} = 1.75$ (left), $\sigma_{r} = 3.25$ (middle), and $\sigma_{r} = 4.75$ (right). In all cases, the Lorentz force is negligible since the net charge density is small and the total force is outward along the $x$ direction.
When $\sigma_r=1.75$, the Lorentz force is larger than the polarization force in the $y$-axis direction, resulting in the compression of the droplet by the electric field. Therefore, under the total electric force, the droplet appears oblate.
For $\sigma_r = 3.25$, the polarization force is larger in the $y$ direction. However, the total expansion force in the $y$ direction is larger than in the $x$ direction, and since the fluid is incompressible, the droplet is elongated in the $y$ direction with a prolate profile at equilibrium.
In the case of $\sigma_r = 4.75$, both the Lorentz force and the polarization force are pointing outward in the $y$ direction, resulting in a net force that is larger in the $y$ direction than in the $x$ direction. As a result, the droplet is elongated in the $y$ direction.
%Besides, we display some other physical information at equilibrium state in Figure 
%which are the electric field $\bf{E} = -\nabla\phi$,
%and the magnitude of electric force $\rho_e\nabla\phi$. 
%We can see that the positive ions accumulate at the top of the interface when $\sigma_{r} = 1.75$ 
%and $\sigma_{r} = 3.25$, whereas the negative ions on the contrary. 
%Nevertheless the positive ions accumulate at the bottom of the interface when $\sigma_{r} = 4.75$.  

%Figure  shows the magnitude and direction of 
%viscos force $\nabla\cdot\sigma_{\eta}$ (top), 
%the electric force $\nabla\cdot\sigma_{e}$ (middle) 
%and the surface force $\nabla\cdot\sigma_{\psi}$ (bottom) 
%with different conductivity ratios 
%$\sigma_{r}=1.75$ (left), $\sigma_{r}=3.25$ (middle), $\sigma_{r}=4.75$ (right).
%Results show that the surface force is almost equal at every direction 
%and the magnitude of the viscos force is small so that it may cause a small deformation effect. 
%Therefore the electric force dominates the deformation of drop. 

\subsection{Electro-coalescence}\label{sec: tow drops merge}
%\subsubsection{Capacitance effect for double drops}
In this section, we conduct a series of numerical experiments to explore the electro-coalescence. Electro-coalescence refers to the process of two or more suspended droplets or particles coming into contact and merging under the influence of an applied electric field. It is a phenomenon commonly observed in various electrokinetic systems and has significant implications in fields such as microfluidics, emulsion stability, and particle manipulation.

\begin{figure}
	\begin{center}
		\includegraphics[width=0.245\textwidth]{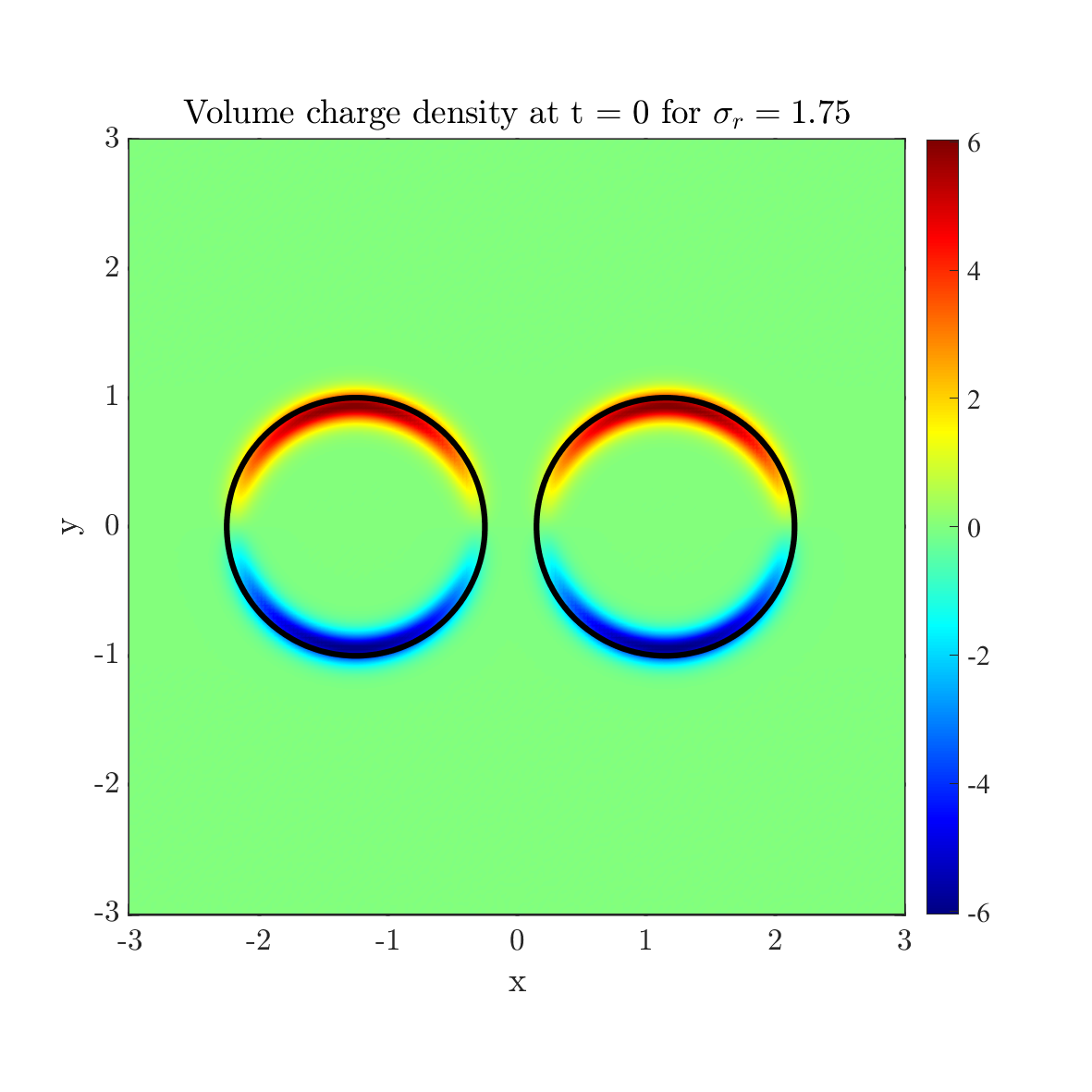}
		\includegraphics[width=0.245\textwidth]{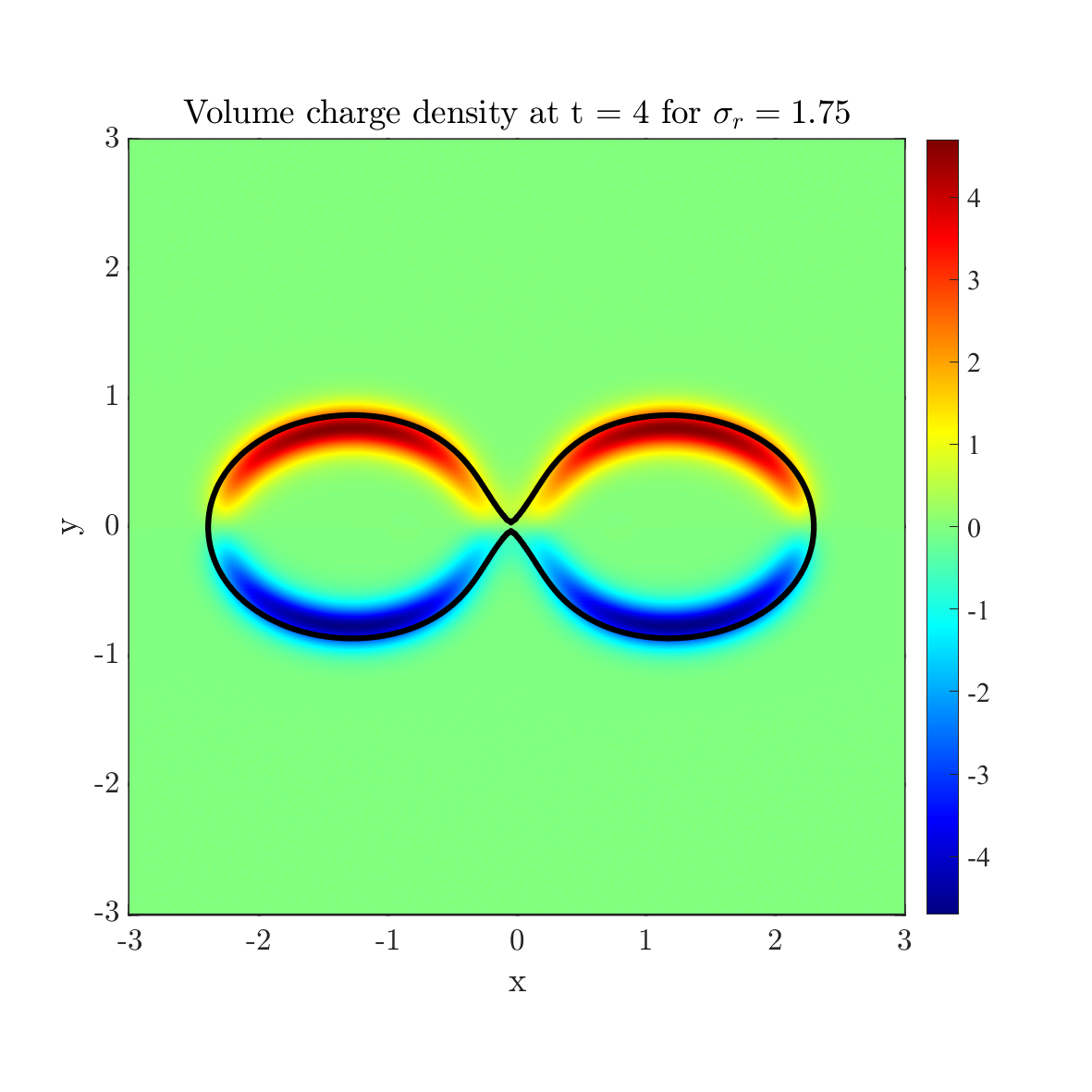}
		\includegraphics[width=0.245\textwidth]{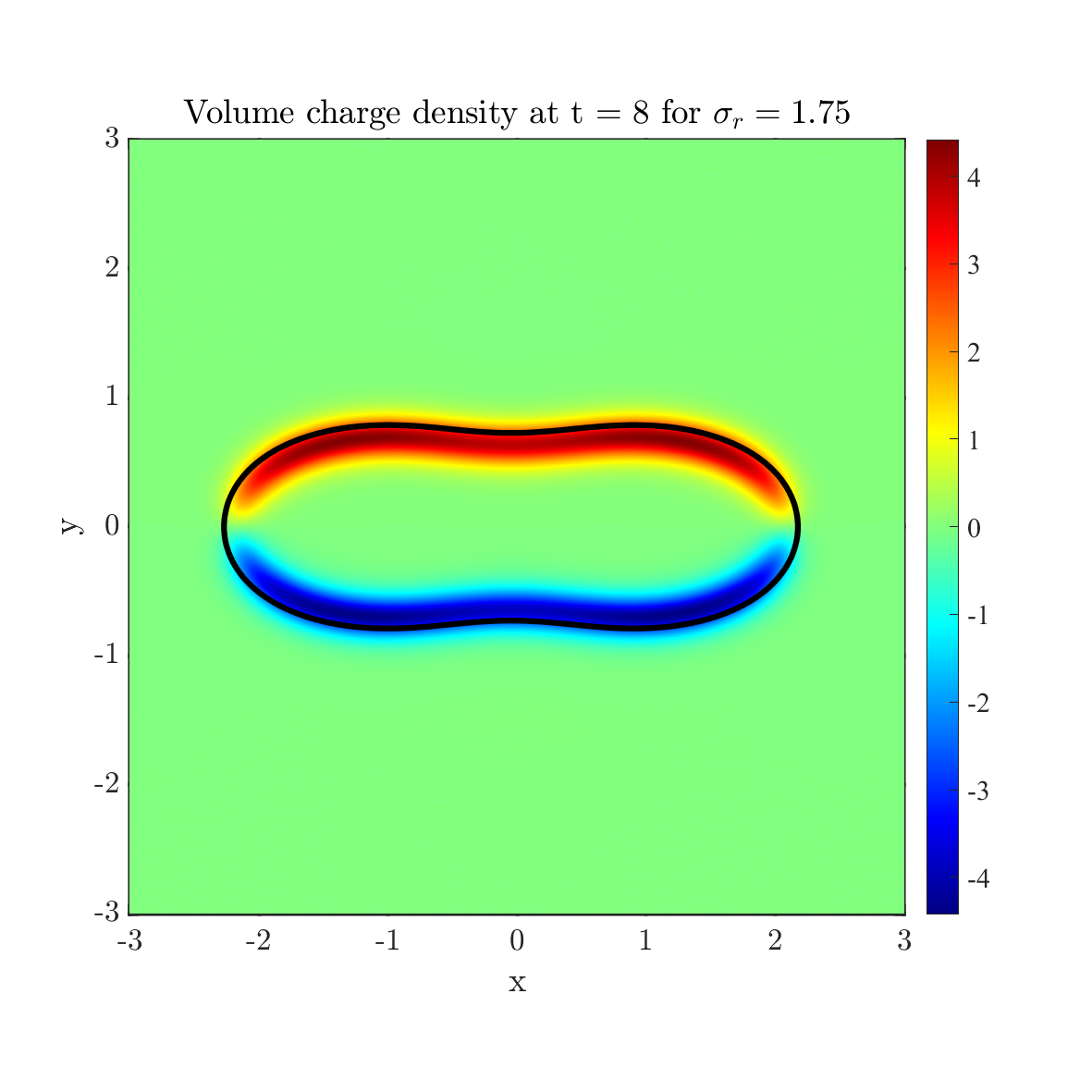}
		\includegraphics[width=0.245\textwidth]{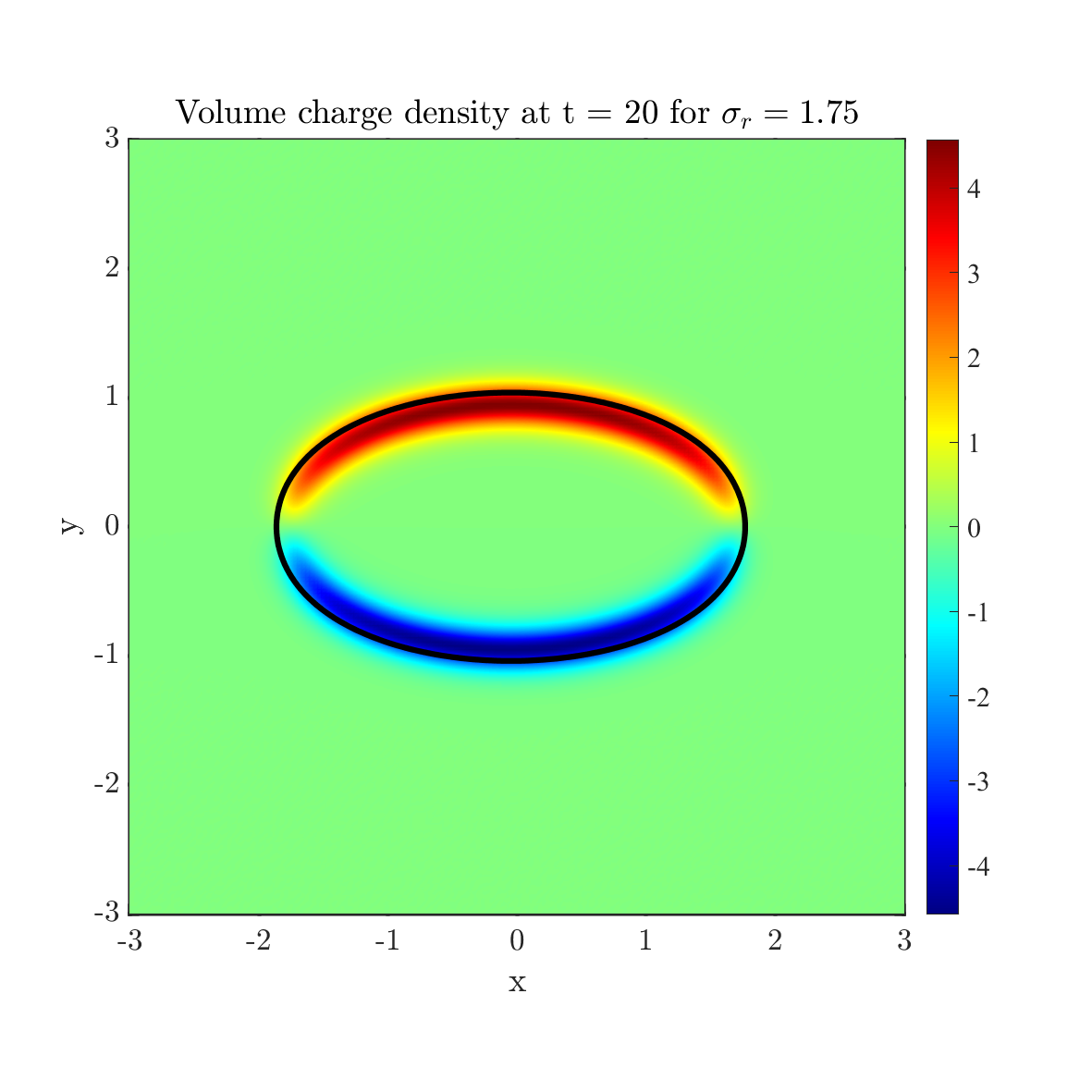}
		\includegraphics[width=0.245\textwidth]{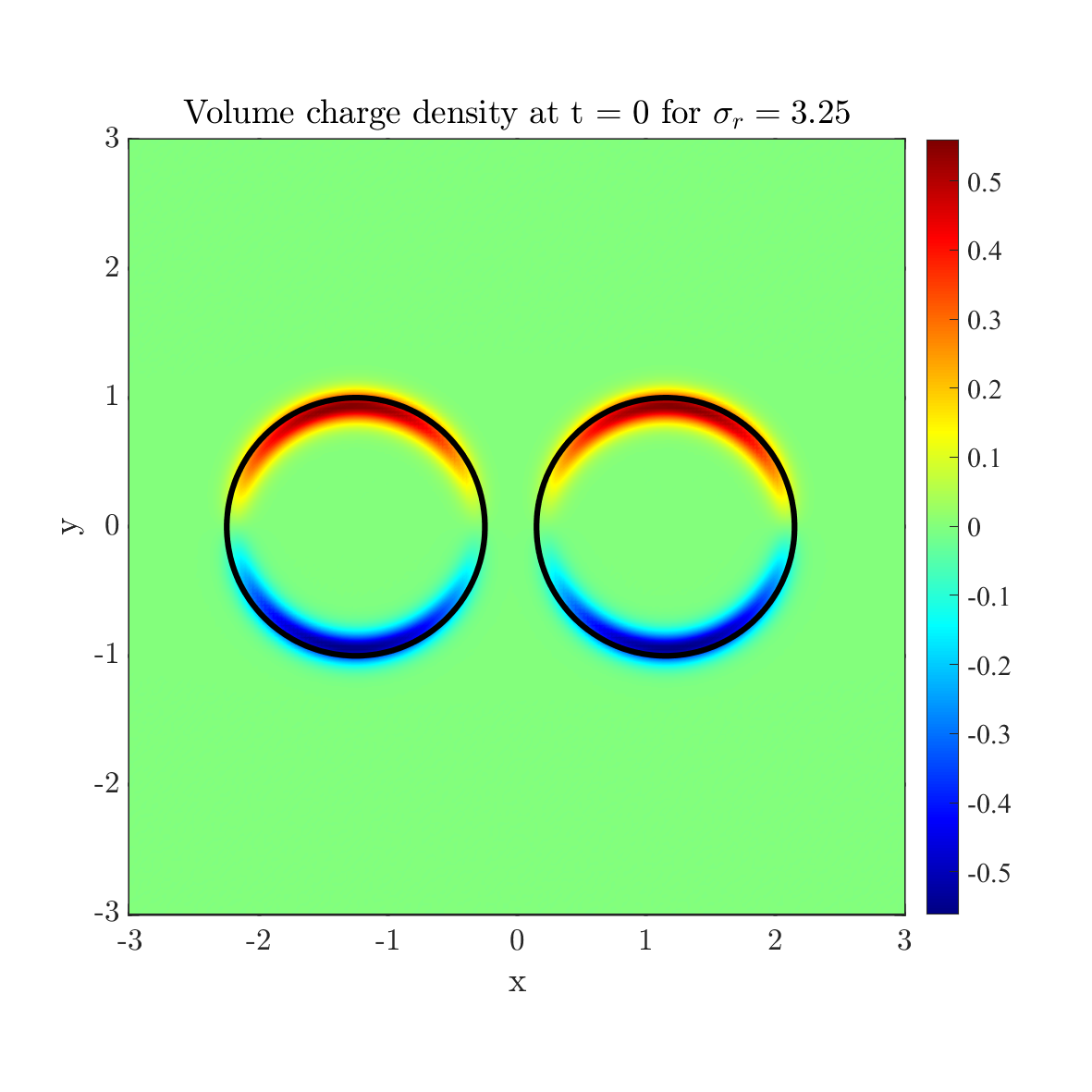}
		\includegraphics[width=0.245\textwidth]{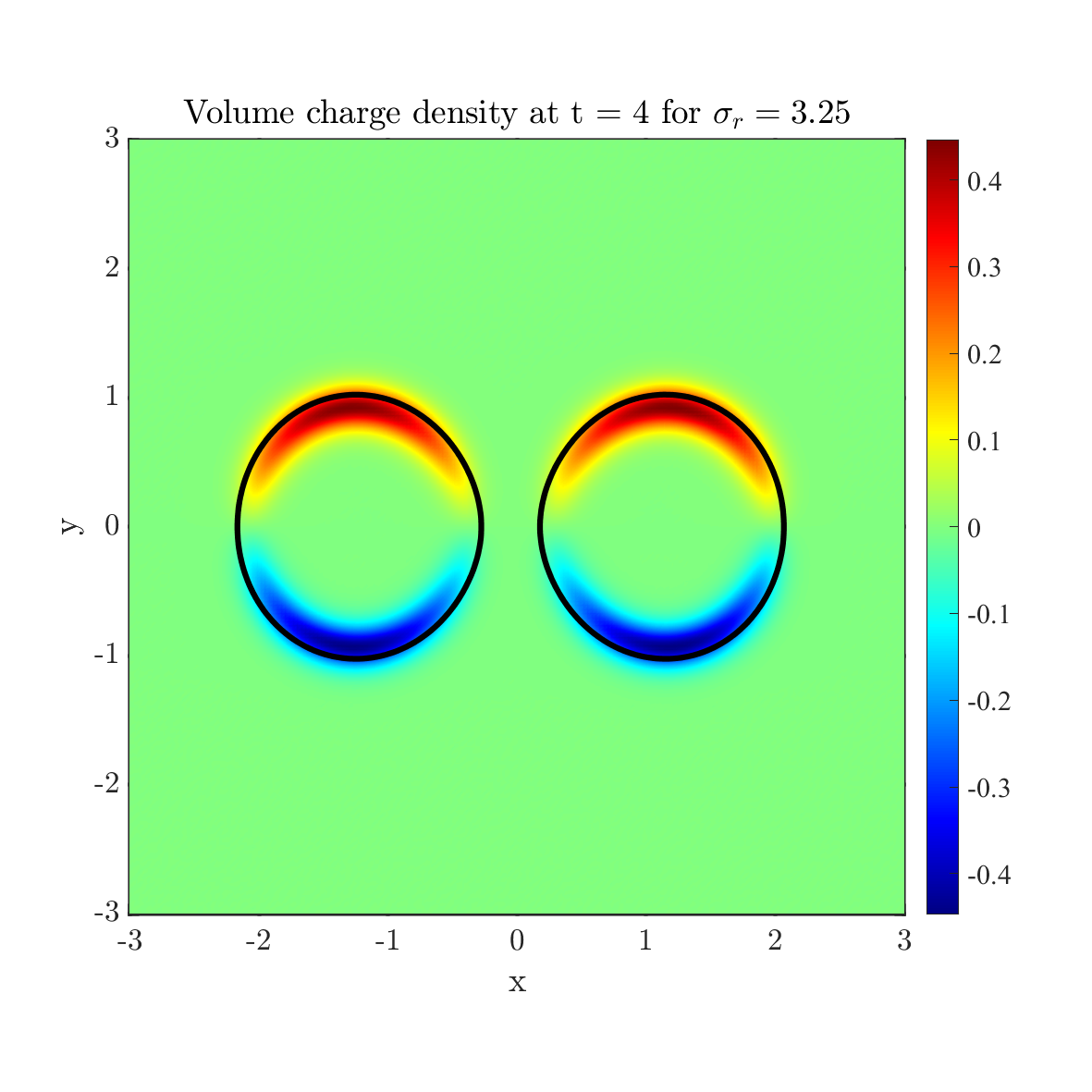}
		\includegraphics[width=0.245\textwidth]{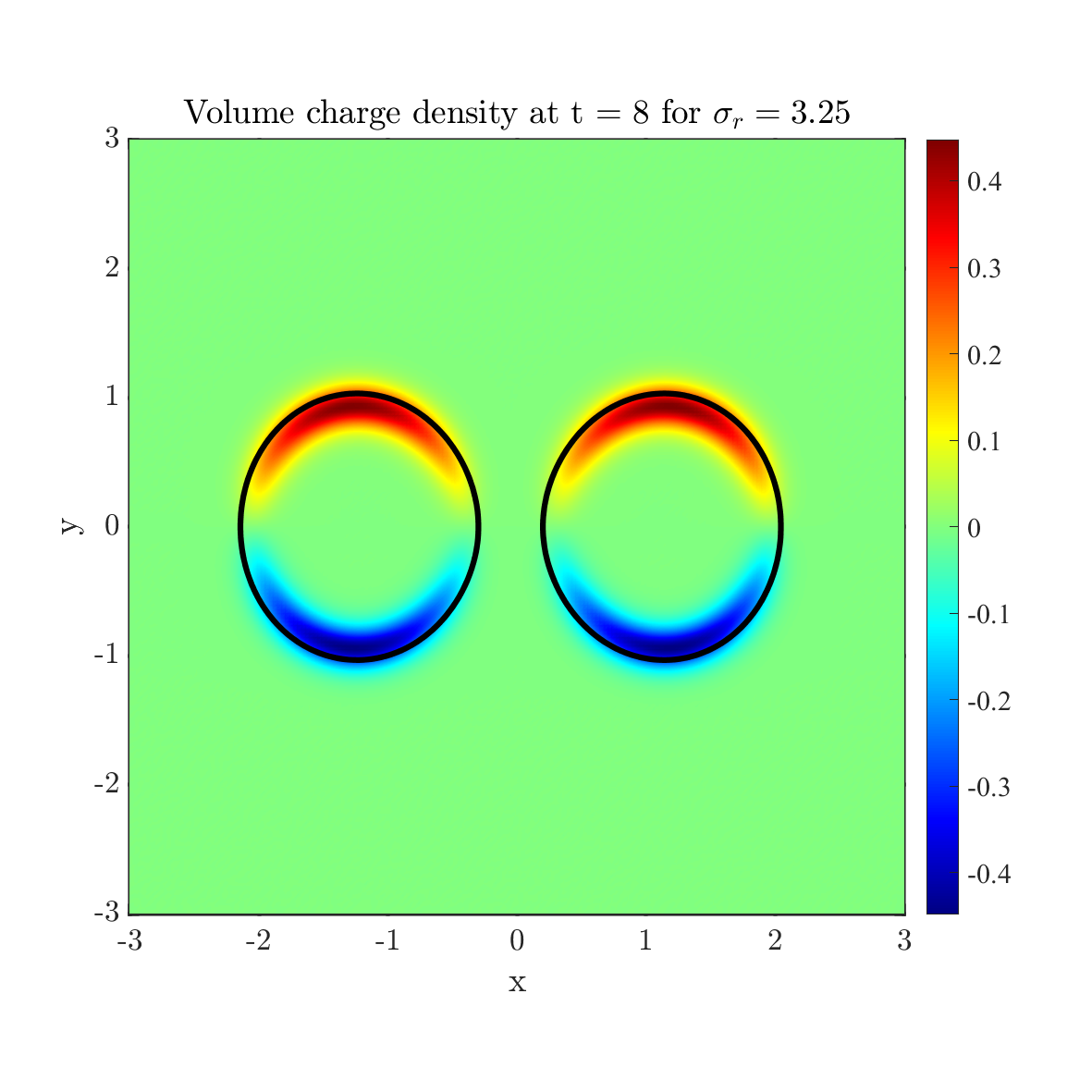}
		\includegraphics[width=0.245\textwidth]{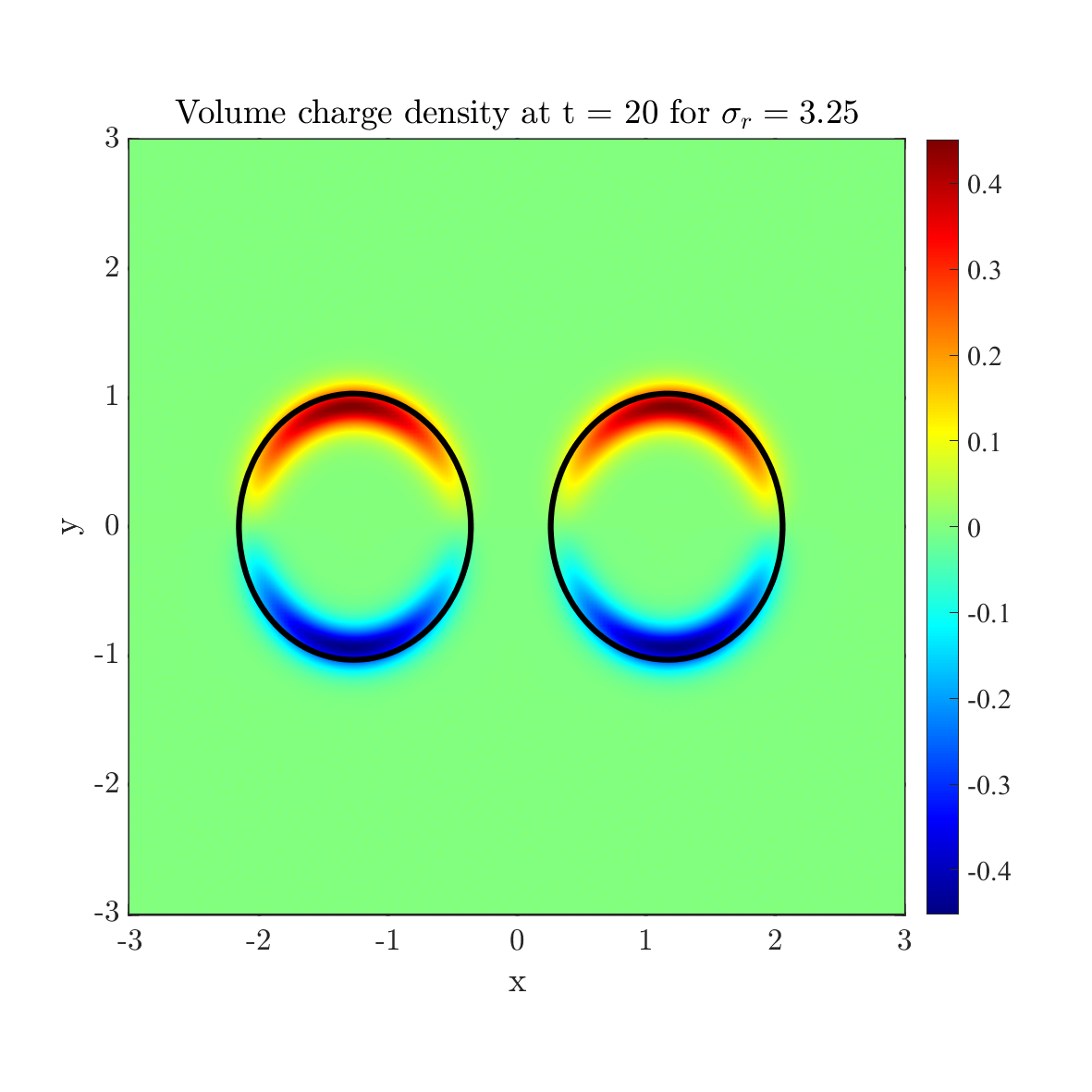}
		\includegraphics[width=0.245\textwidth]{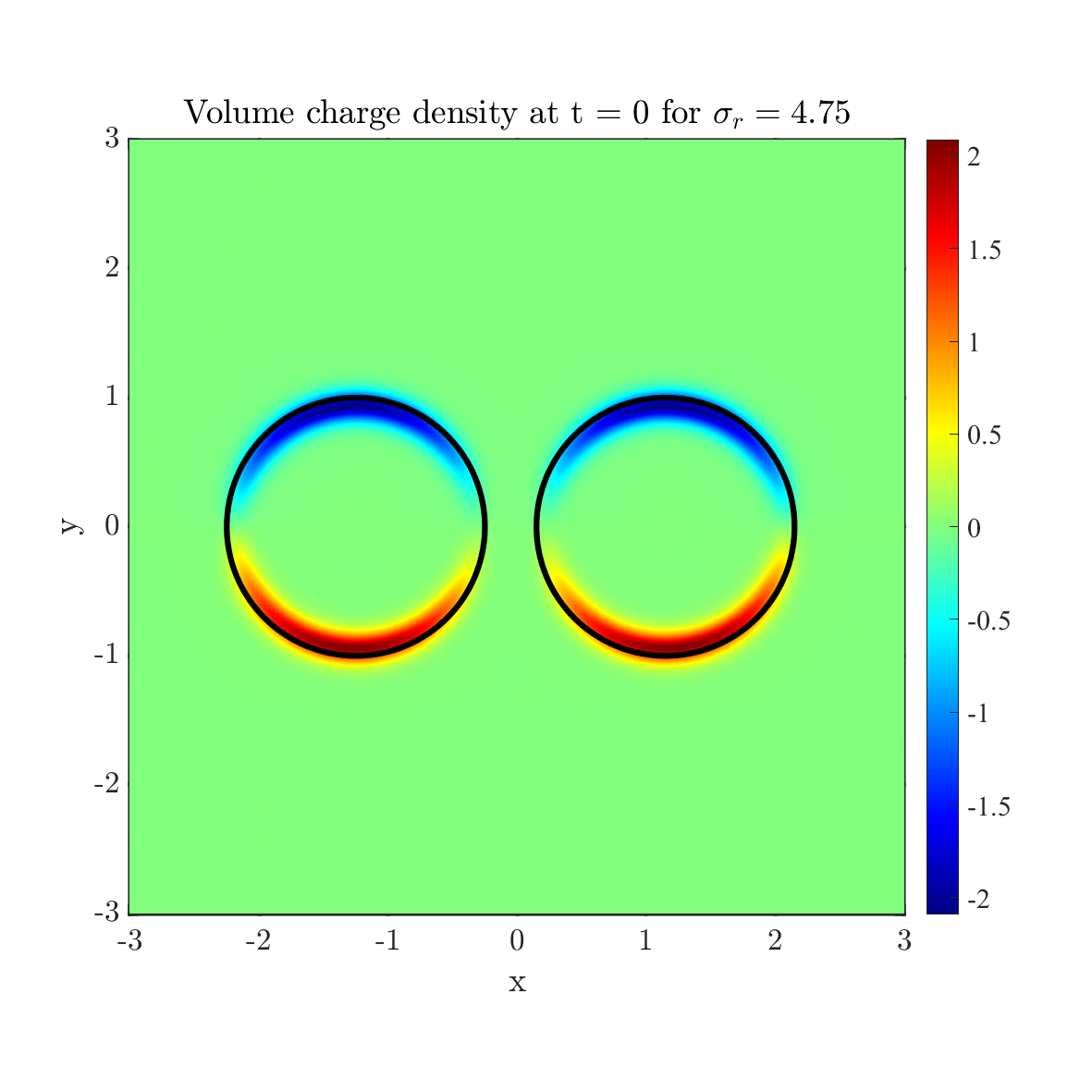}
		\includegraphics[width=0.245\textwidth]{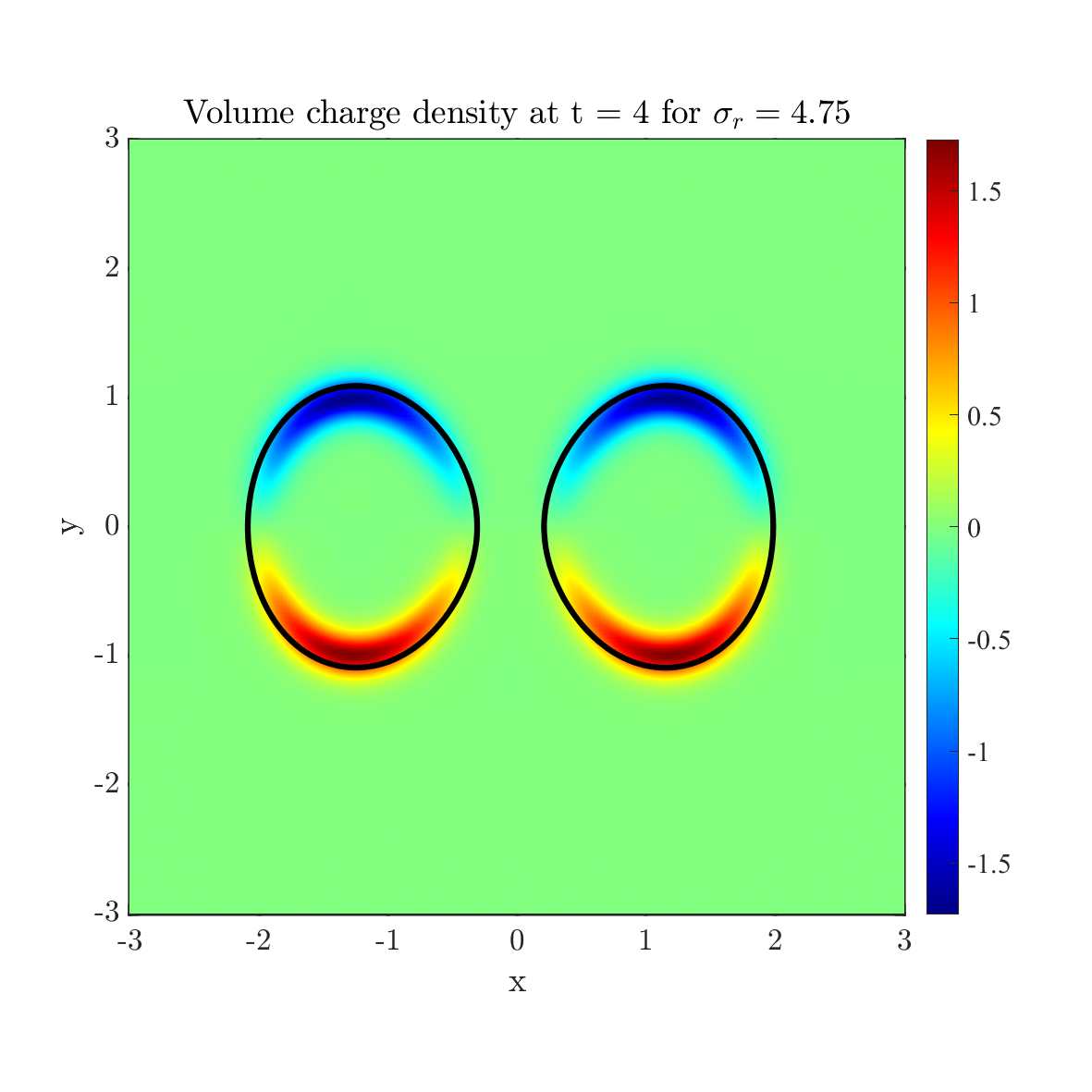}
		\includegraphics[width=0.245\textwidth]{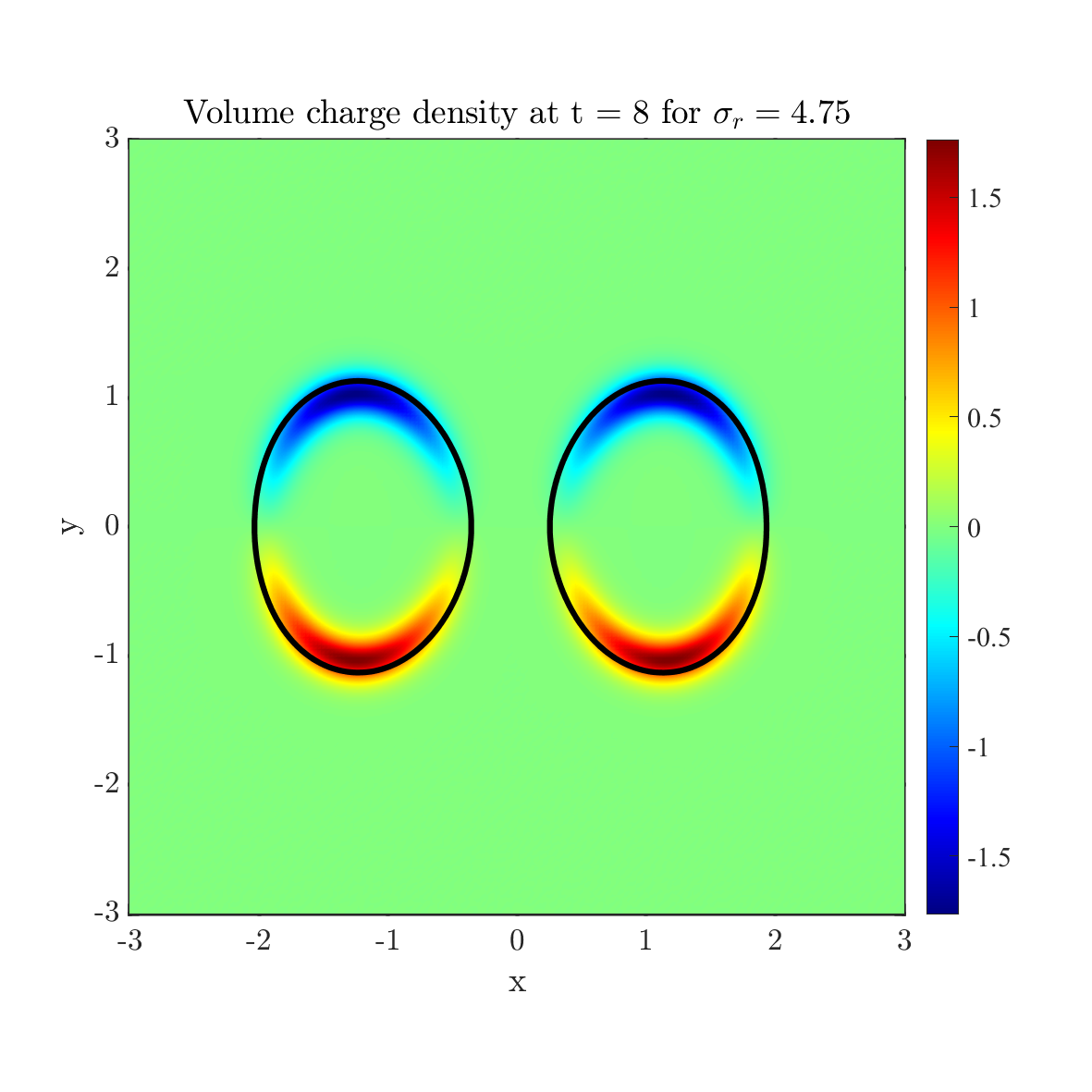}
		\includegraphics[width=0.245\textwidth]{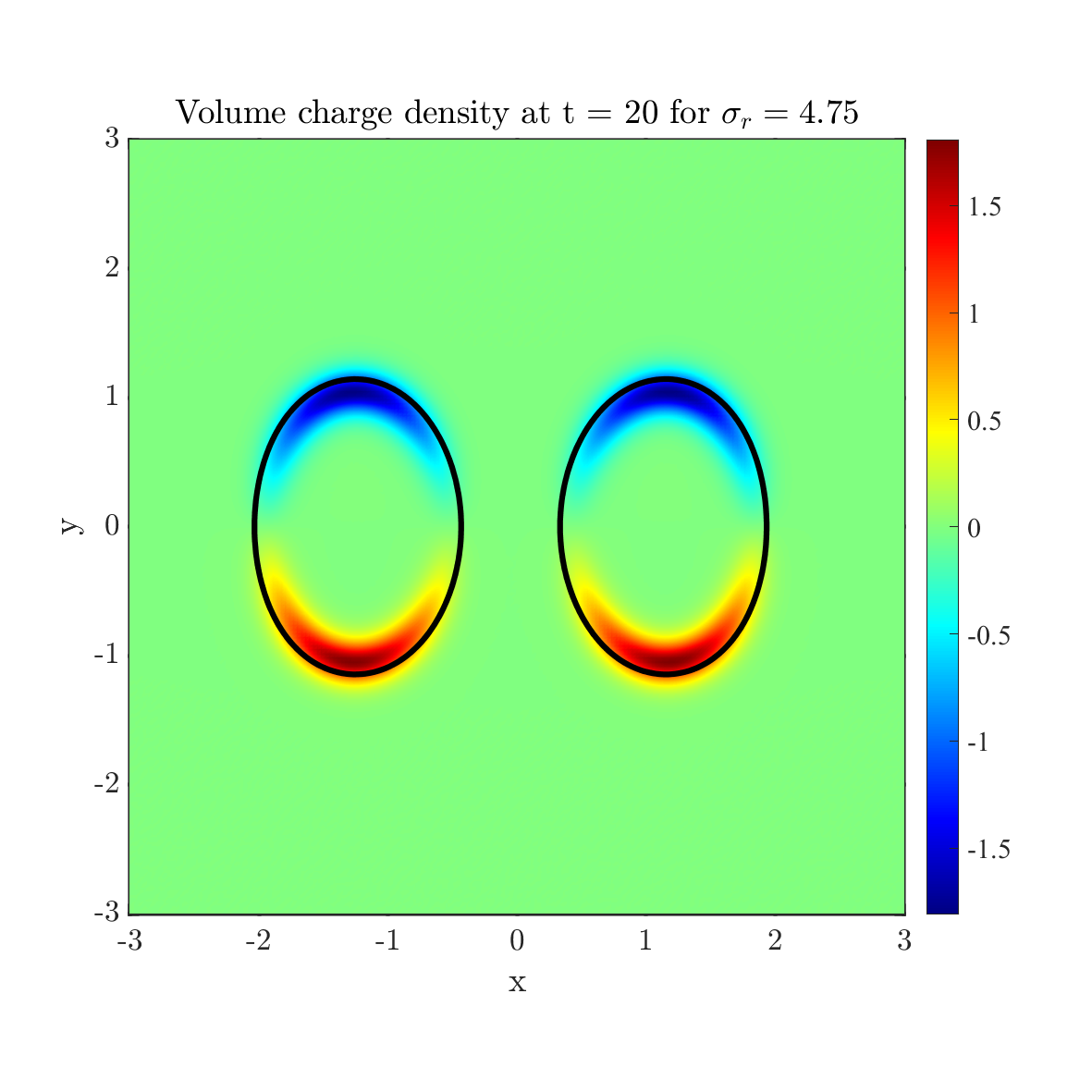}
	\end{center} 
	\caption{The merge effect for different conductivity ratios
		$\sigma_{r} = 1.75$ (top), $\sigma_{r} = 3.25$ (middle), $\sigma_{r} = 4.75$ (bottom) 
		at $t = 0$, $t = 4$, $t = 8$ and $t = 20$ from left to right, respectively 
		for the example \ref{eqn: initial two single drops 1} in section \ref{sec: tow drops merge}. 
		In each figure, the black solid line shows the zero level set ($\psi=0$) 
		to describe the location of droplet. 
		The rest parameters are chosen as $\epsilon_{r} = 3.5$, $Ca_{E} = 1$.}
	\label{fig: merge for two drops without cm}
\end{figure}

The computational domain and all the parameters are kept the same as the former section 
\ref{sec: comparison with sharp model}.  
The initial profile for double drops is shown as follows (see Fig. \ref{fig: merge for two drops without cm} first column)
\begin{equation}\label{eqn: initial two single drops 1}
	\psi\left(x,y,0\right) 
	= \tanh{\frac{1 - \sqrt{\left(x-1.35\right)^{2}+y^{2}}}{\sqrt{2}\delta}}
	+ \tanh{\frac{1 - \sqrt{\left(x+1.25\right)^{2}+y^{2}}}{\sqrt{2}\delta}}+1. 
\end{equation}
and each of the two droplets are equal the droplet in Eq. \eqref{eqn: initial single drop}. When $\sigma_r = 1.75$, 
due to the oblate deformation, two droplets merge together and the charge redistributed 
which is same as the previous session result at the equilibrium. While for the other two cases, 
since the deformations of droplets are prolates, the distance between two droplets is increased.
The distributions of the electric forces could be found in 
Appendix Fig. \ref{fig: Lorentz force in merge for two drops without cm}-\ref{fig: polarization stress merge for two drops without cm}.

%The results are shown in Fig. \ref{fig: merge for two drops without cm}.  

\begin{figure}
	\begin{center}
		\includegraphics[width=0.17\textwidth]{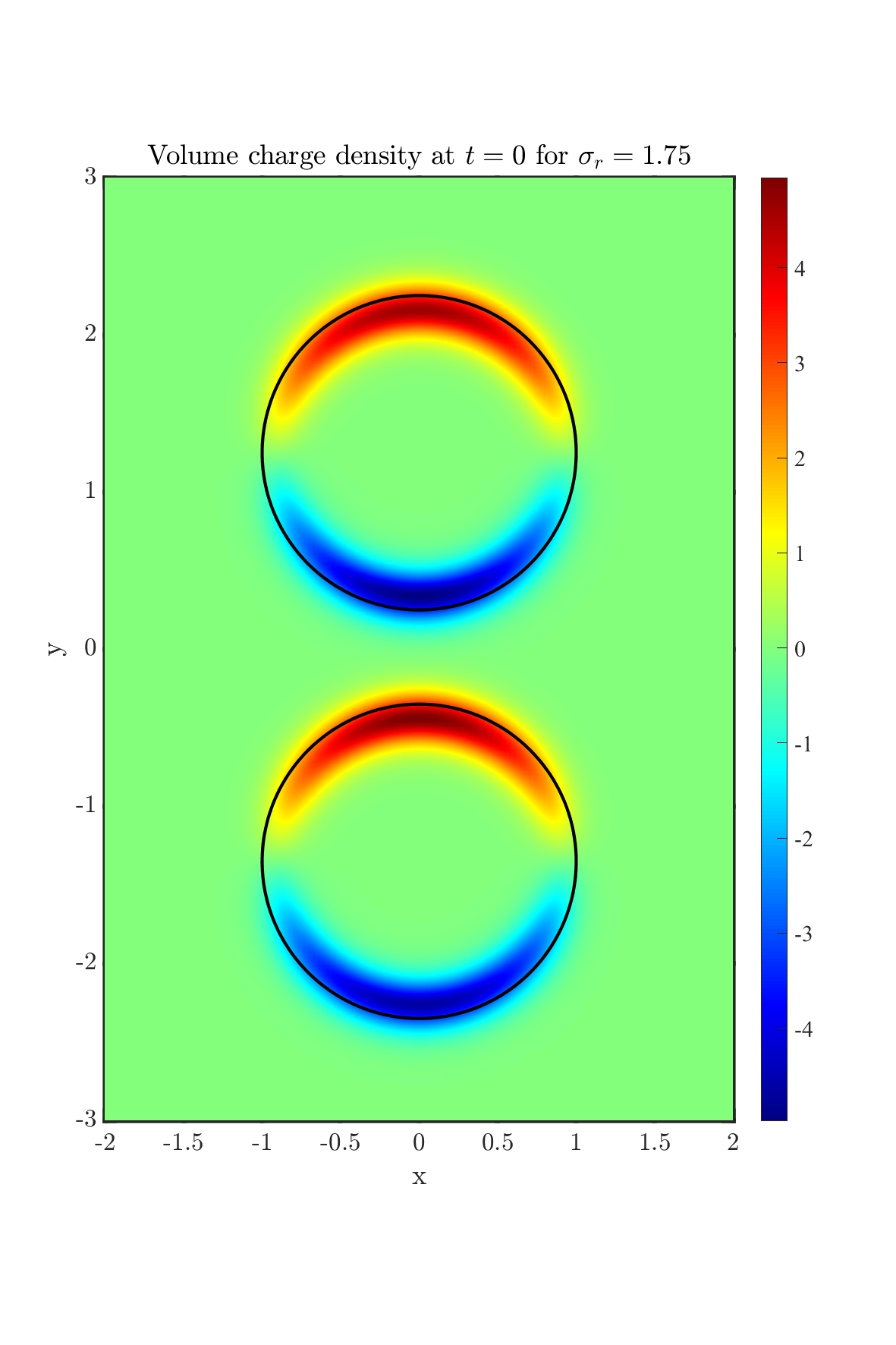}
		\includegraphics[width=0.17\textwidth]{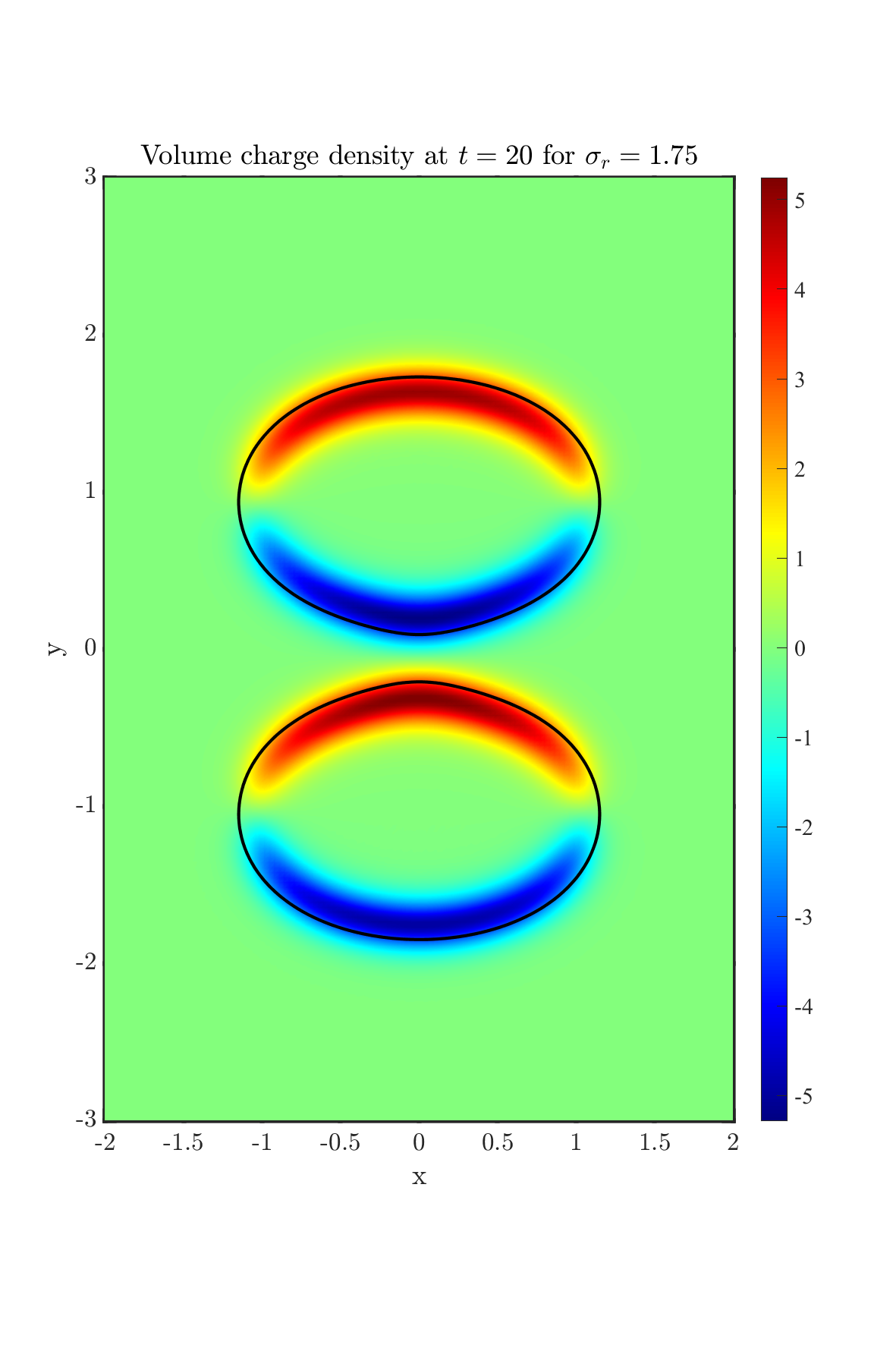}
		\includegraphics[width=0.17\textwidth]{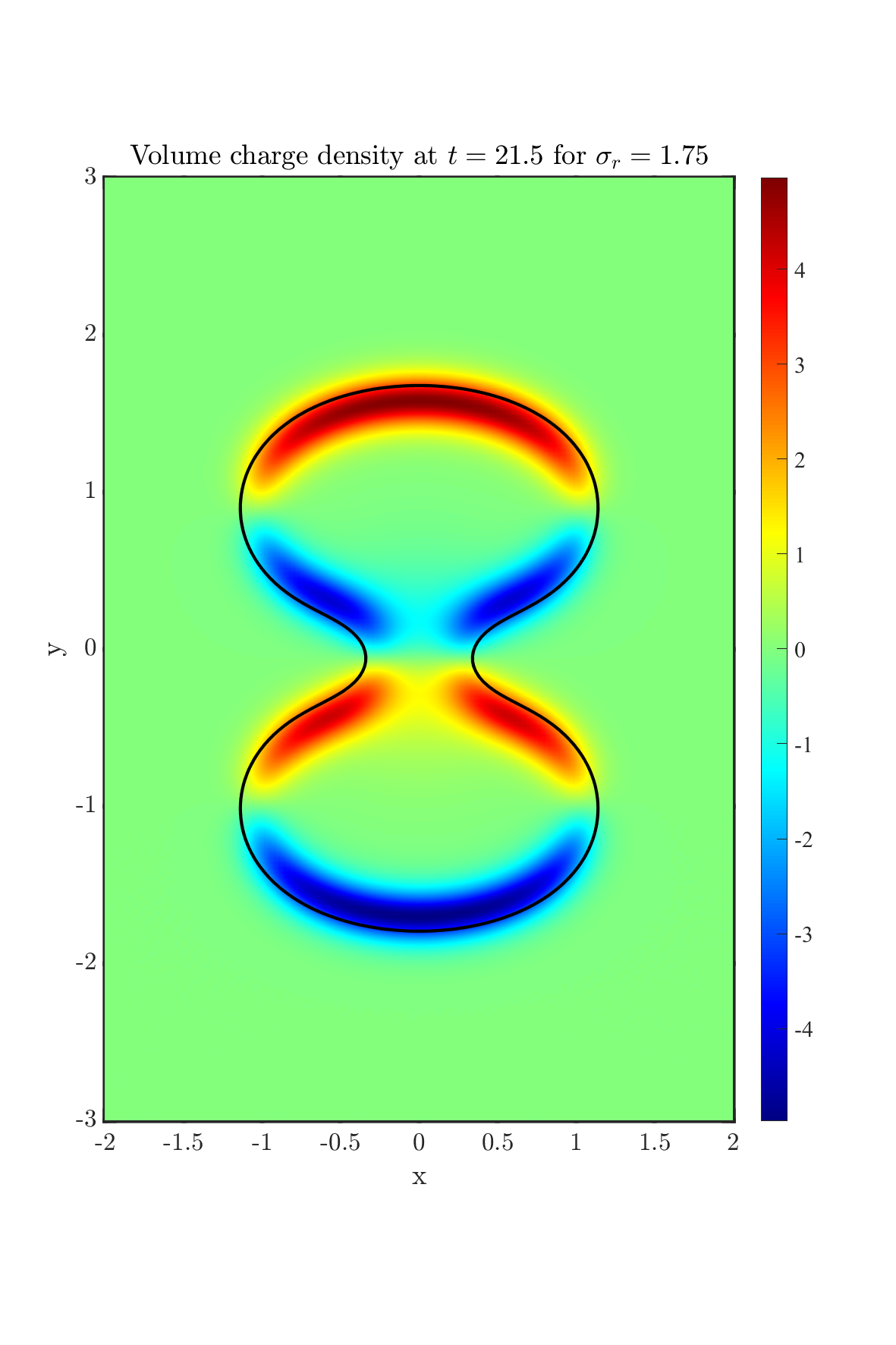}
		\includegraphics[width=0.17\textwidth]{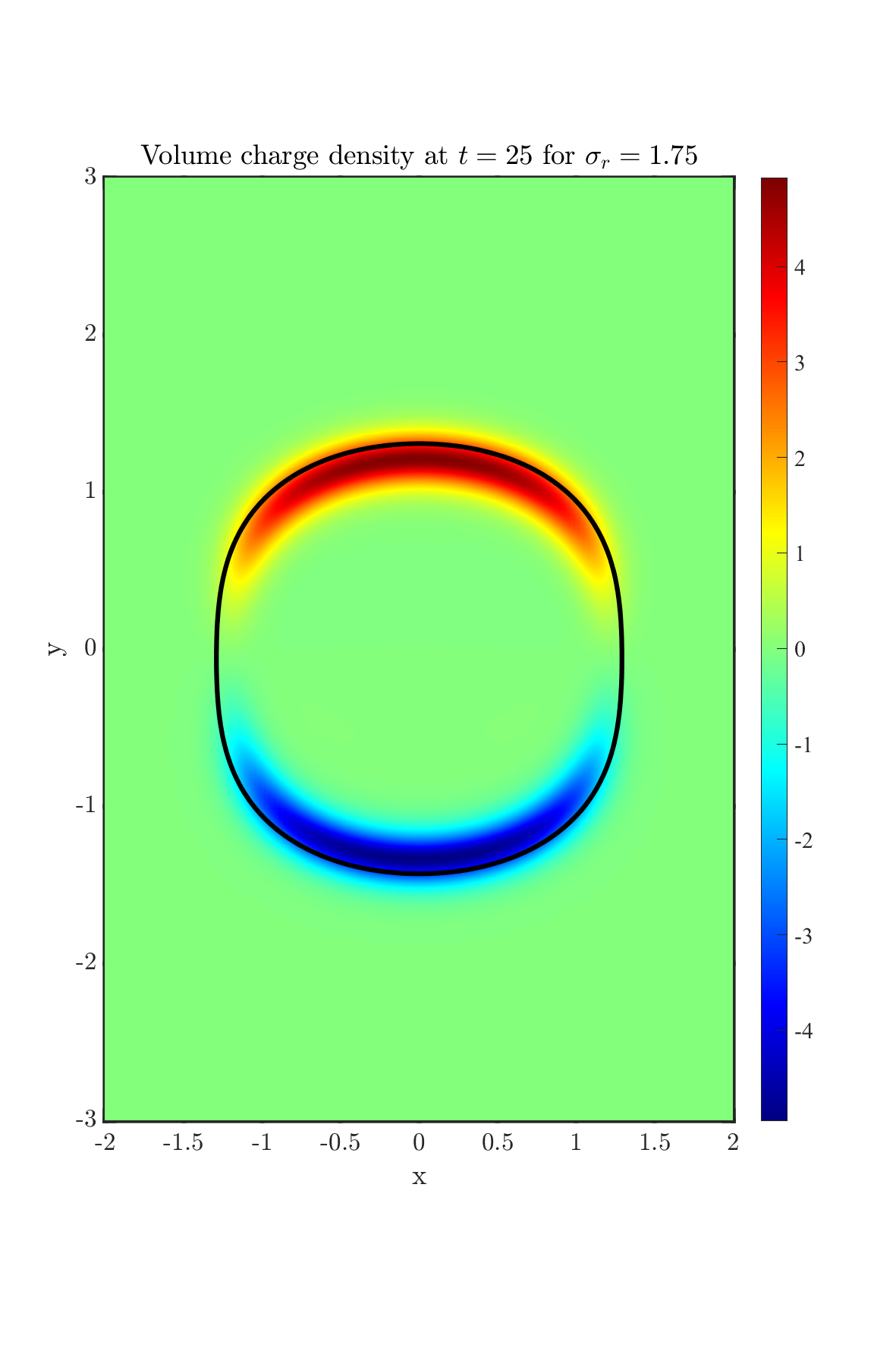}
		\includegraphics[width=0.17\textwidth]{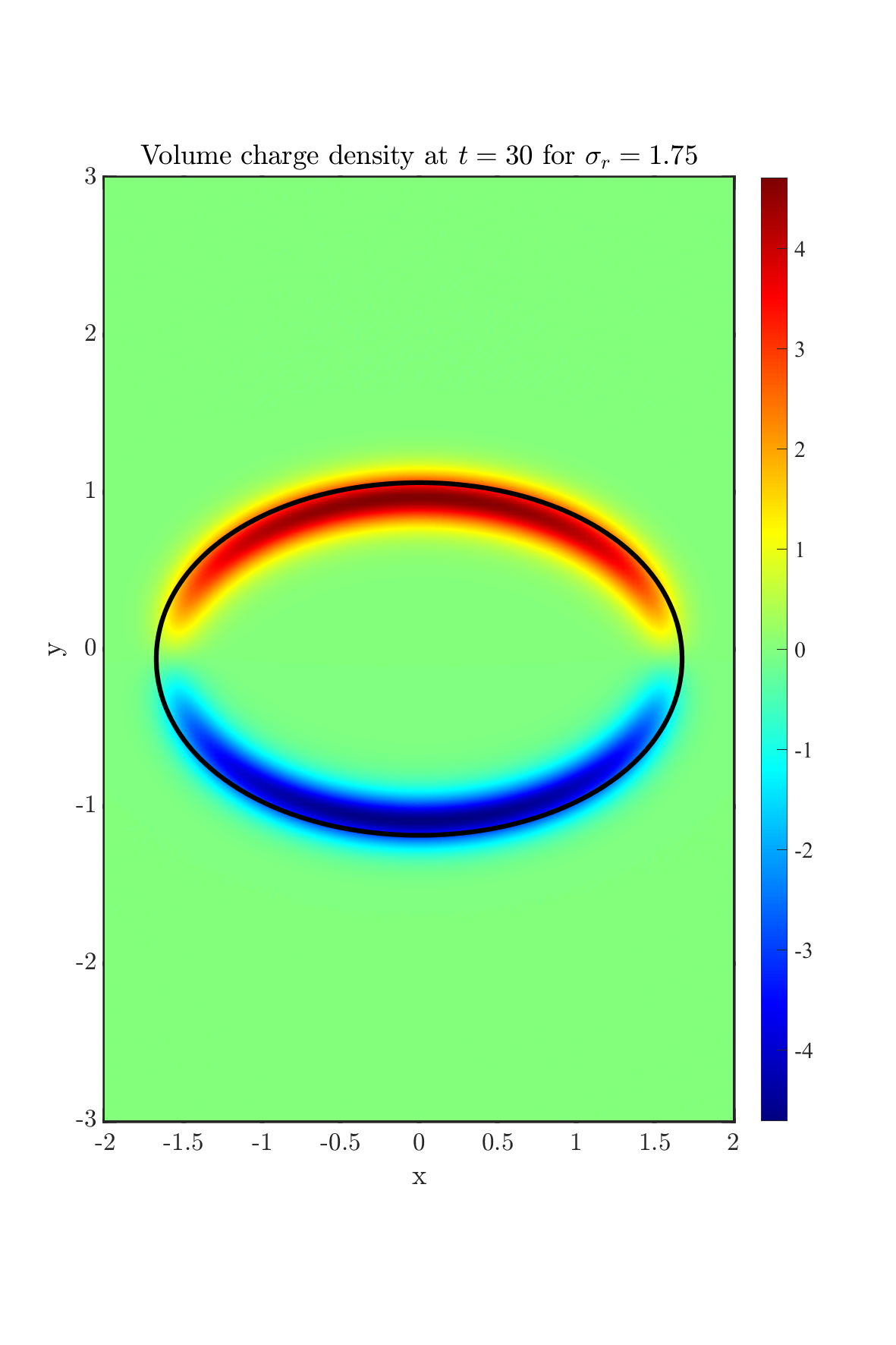}
		\includegraphics[width=0.17\textwidth]{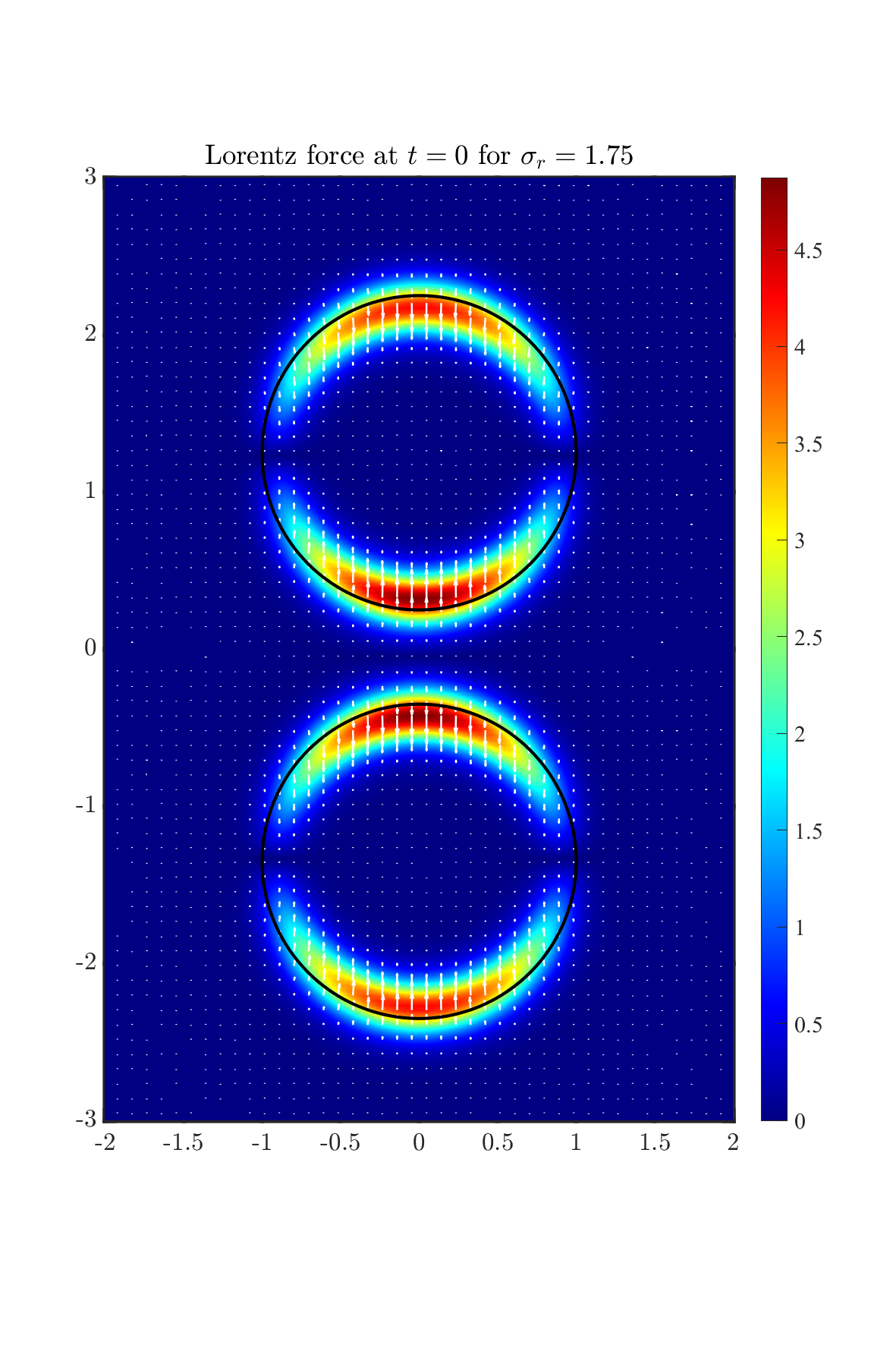}
		\includegraphics[width=0.17\textwidth]{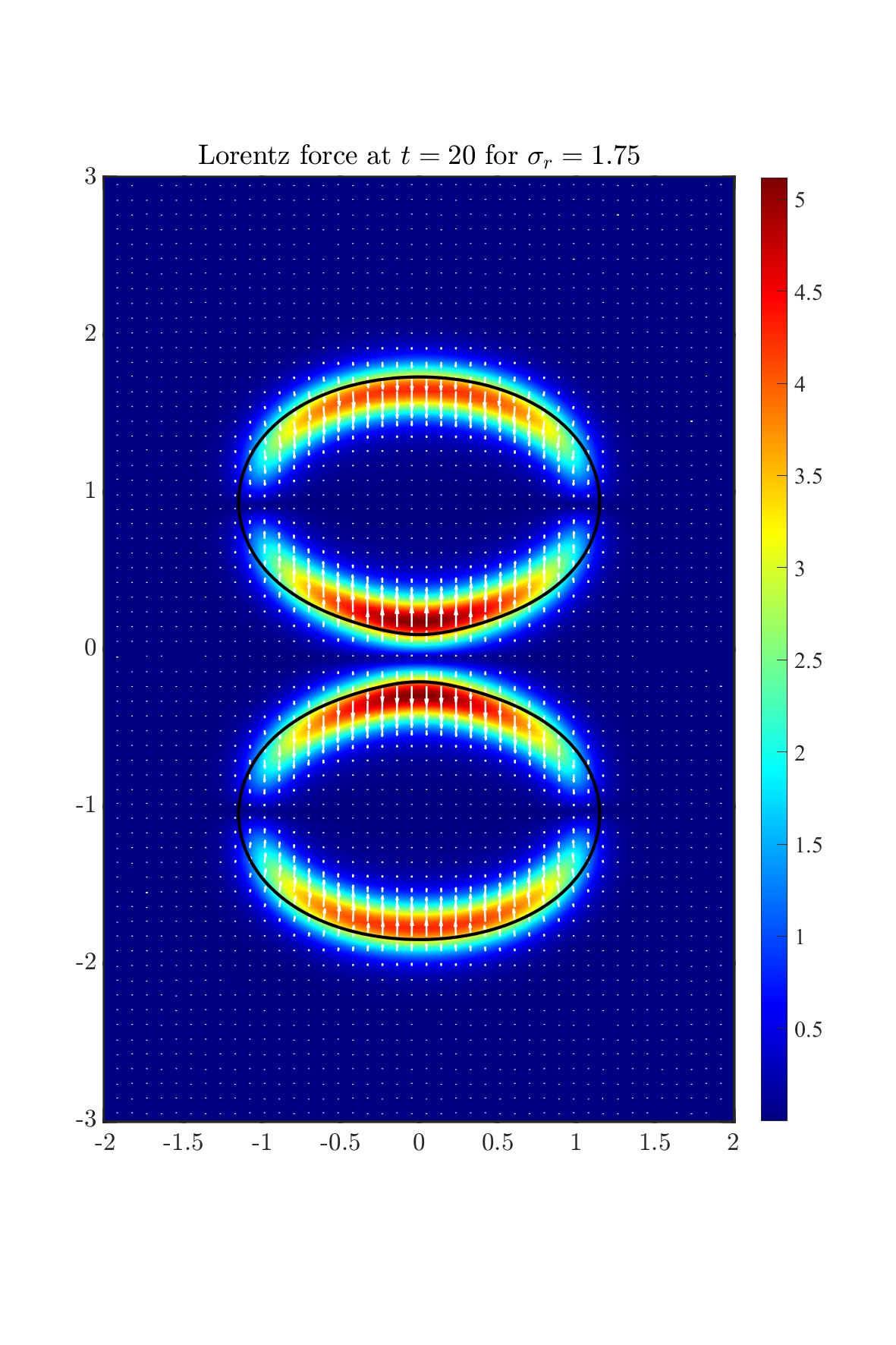}
		\includegraphics[width=0.17\textwidth]{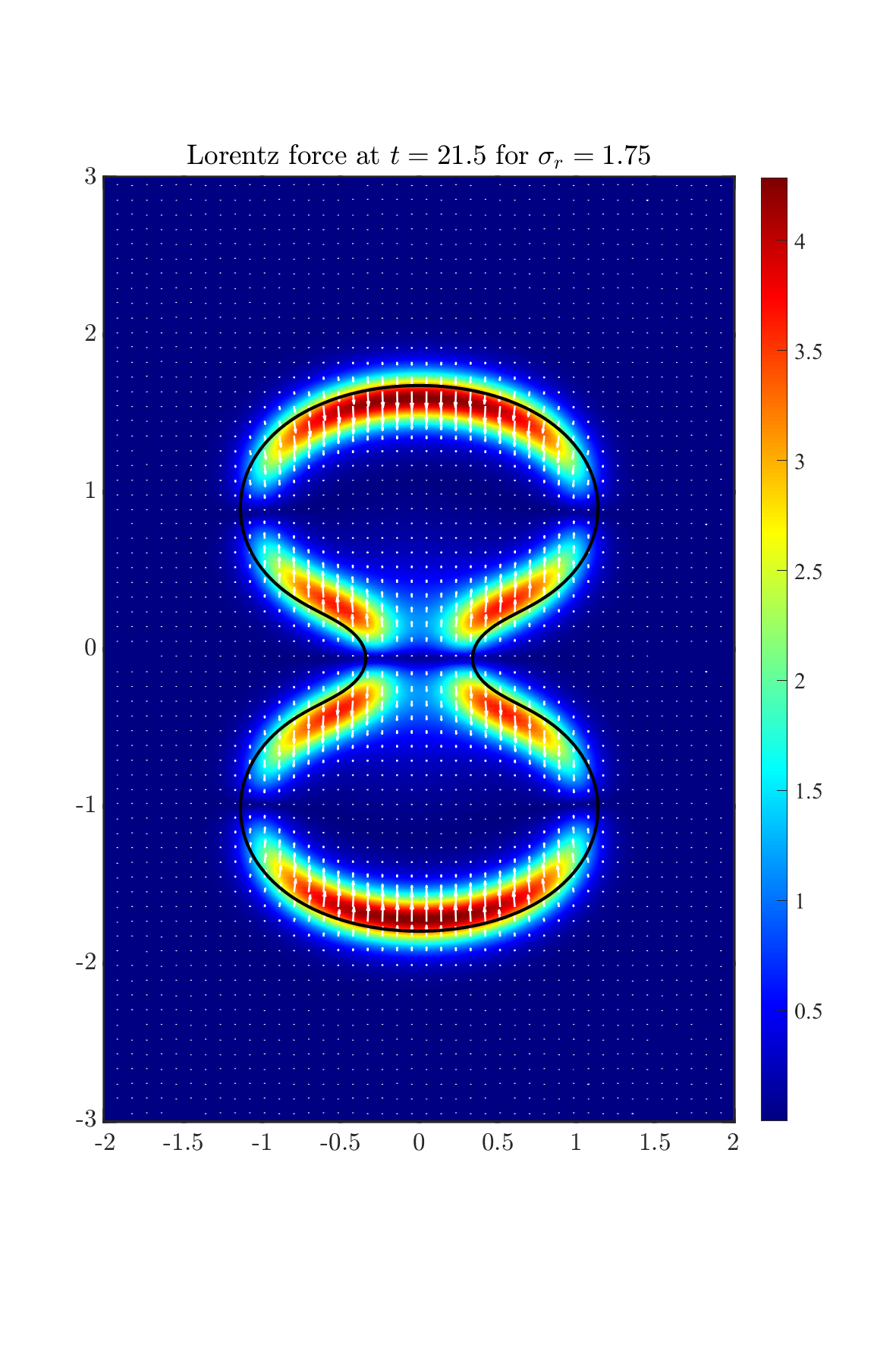}
		\includegraphics[width=0.17\textwidth]{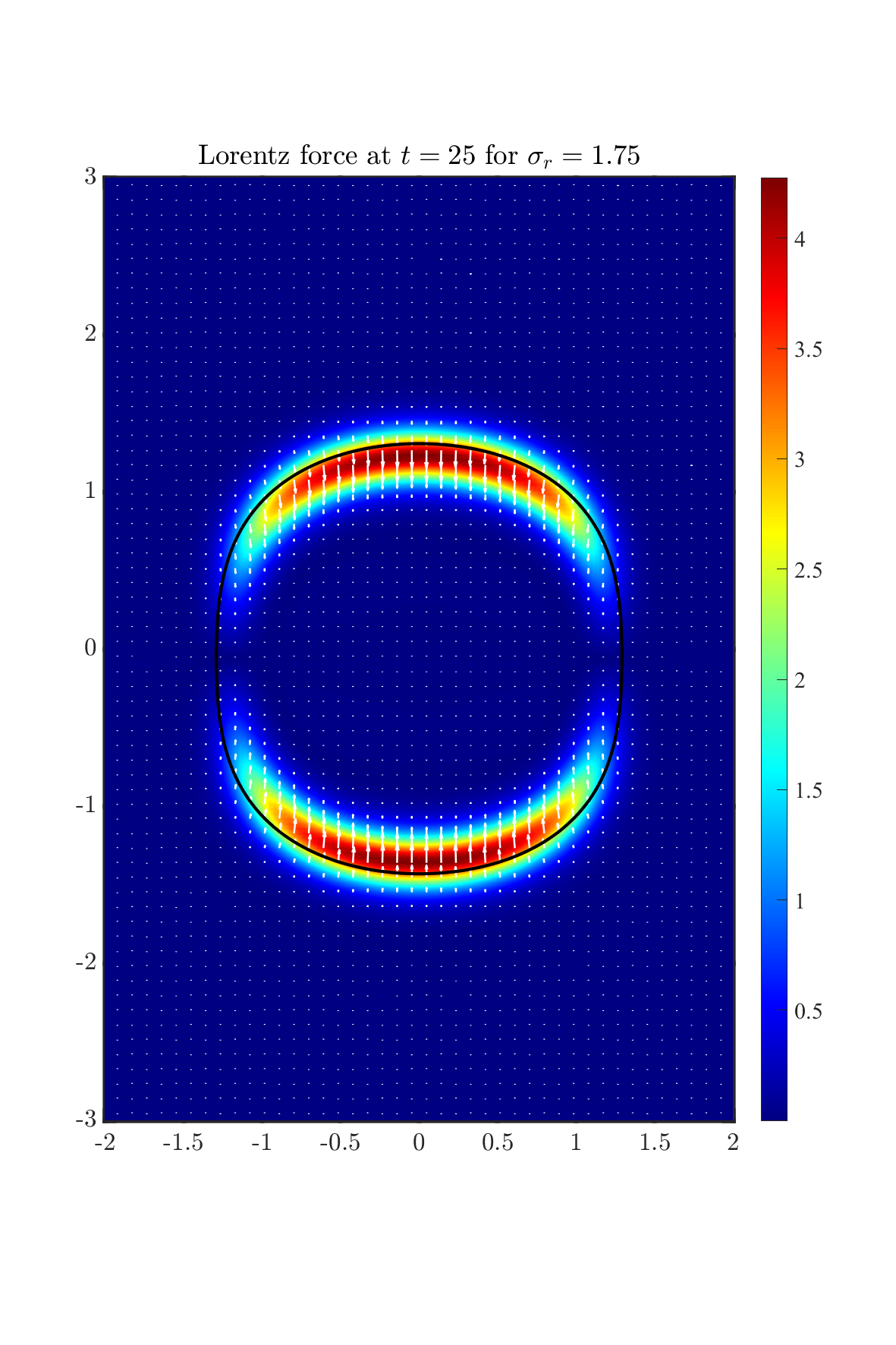}
		\includegraphics[width=0.17\textwidth]{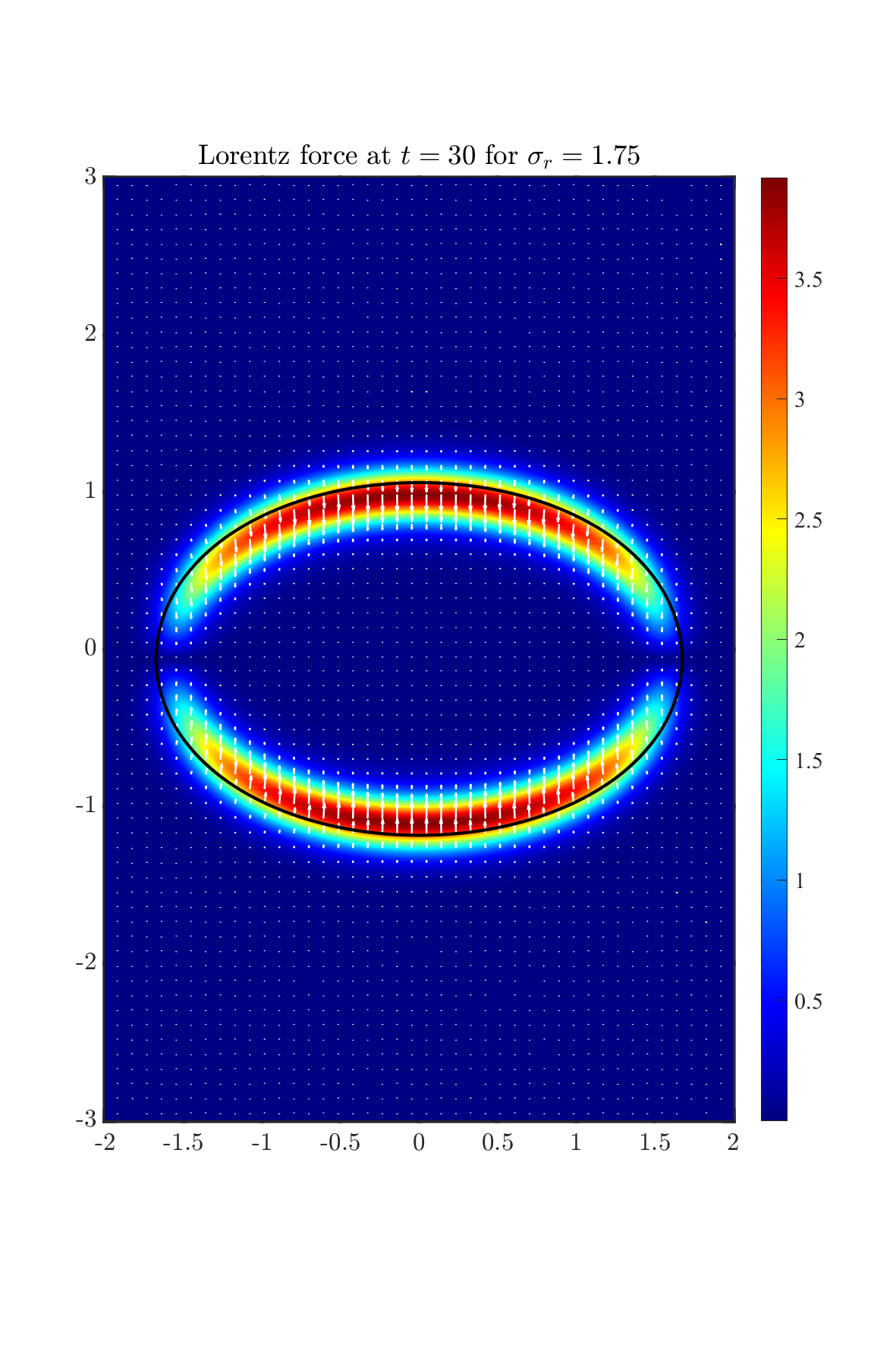}
		\includegraphics[width=0.17\textwidth]{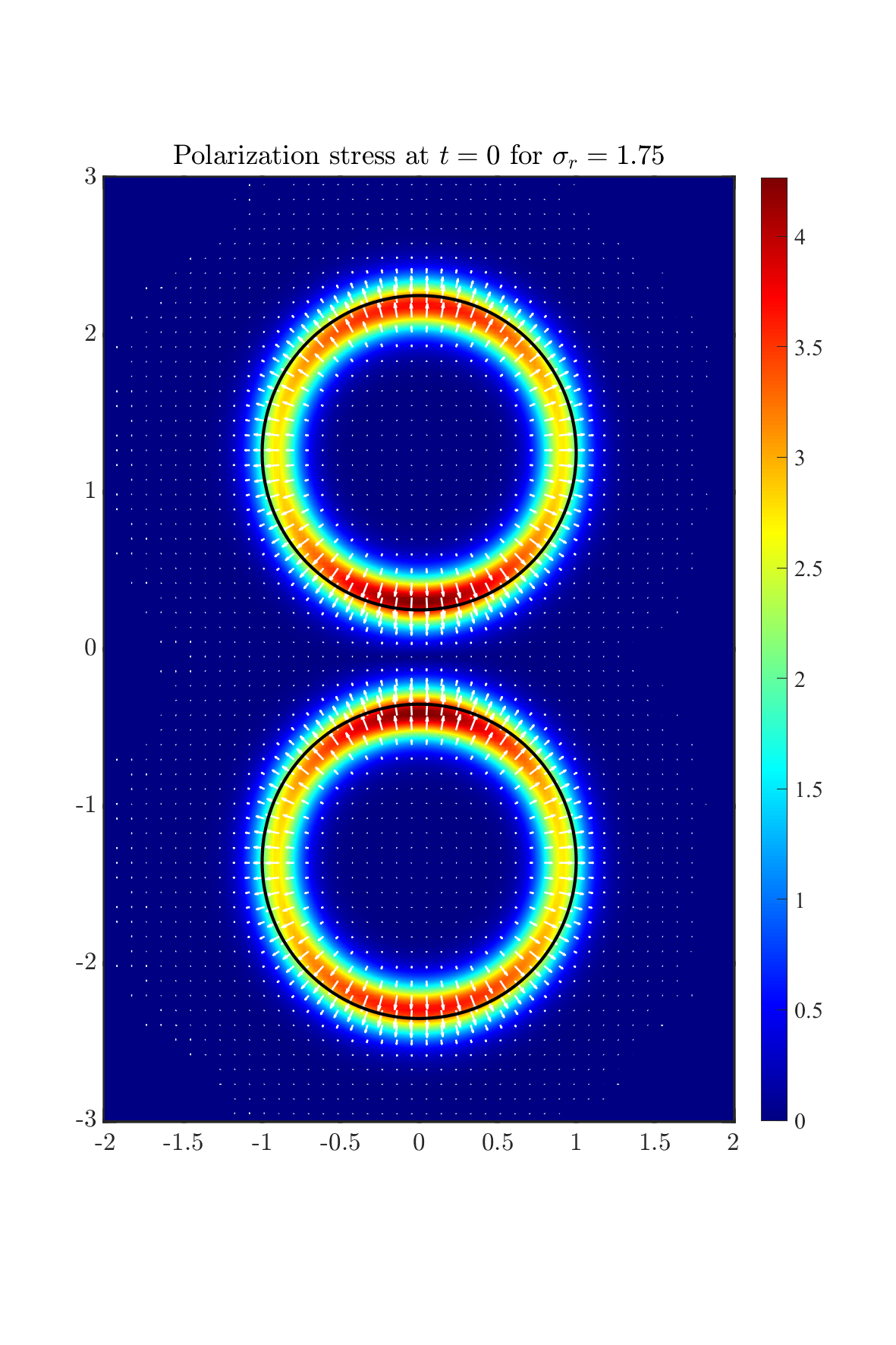}
		\includegraphics[width=0.17\textwidth]{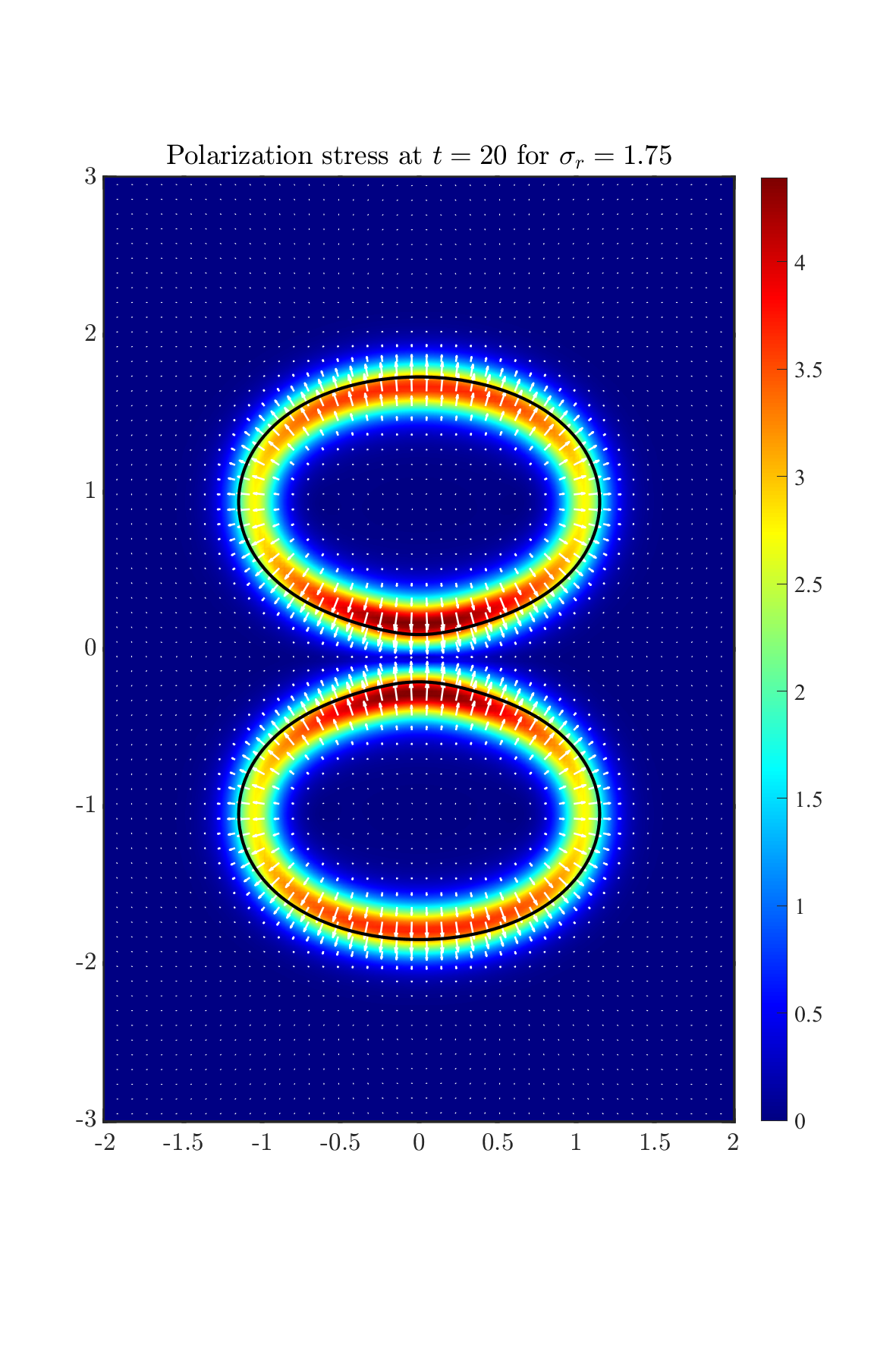}
		\includegraphics[width=0.17\textwidth]{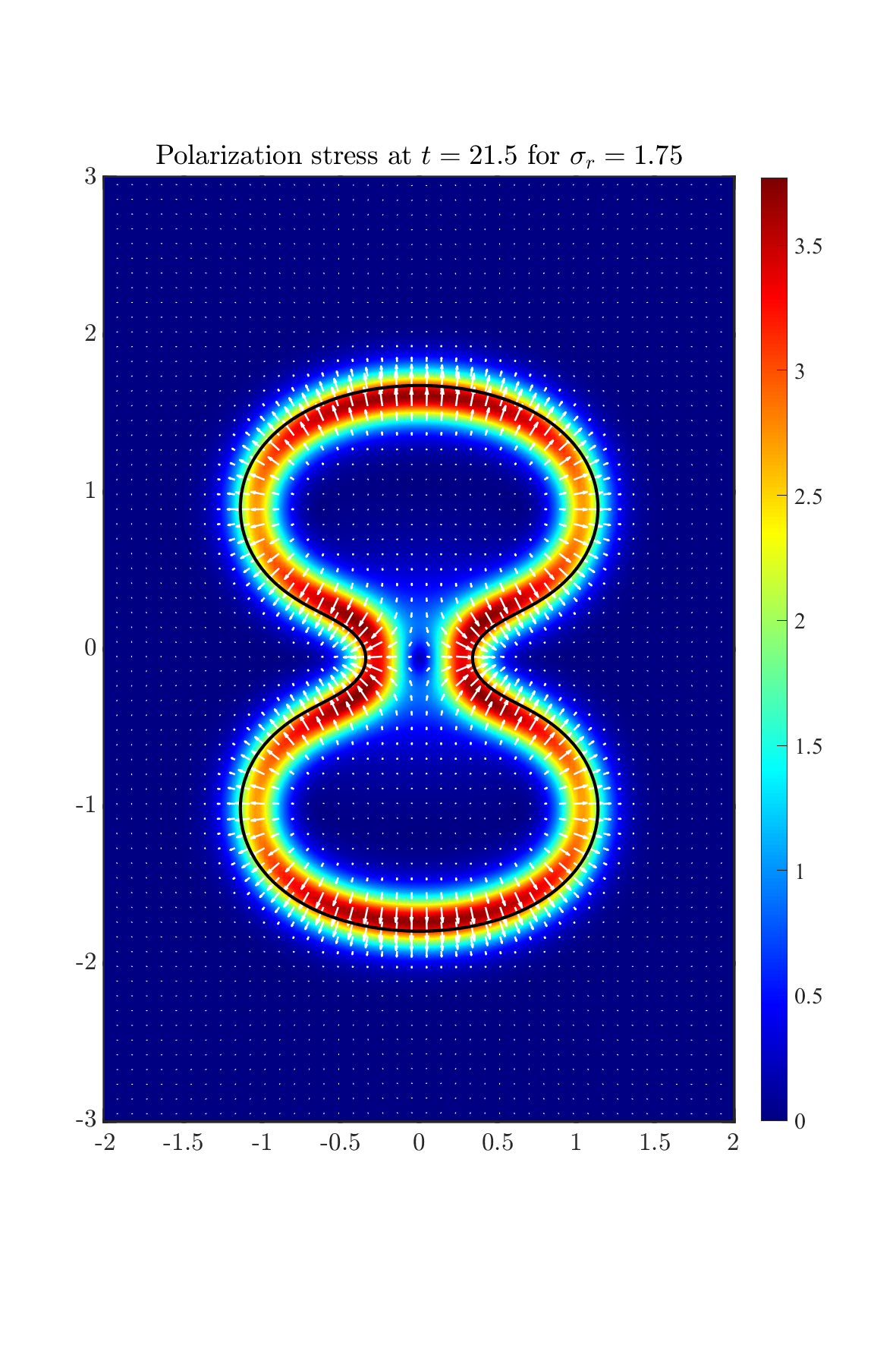}
		\includegraphics[width=0.17\textwidth]{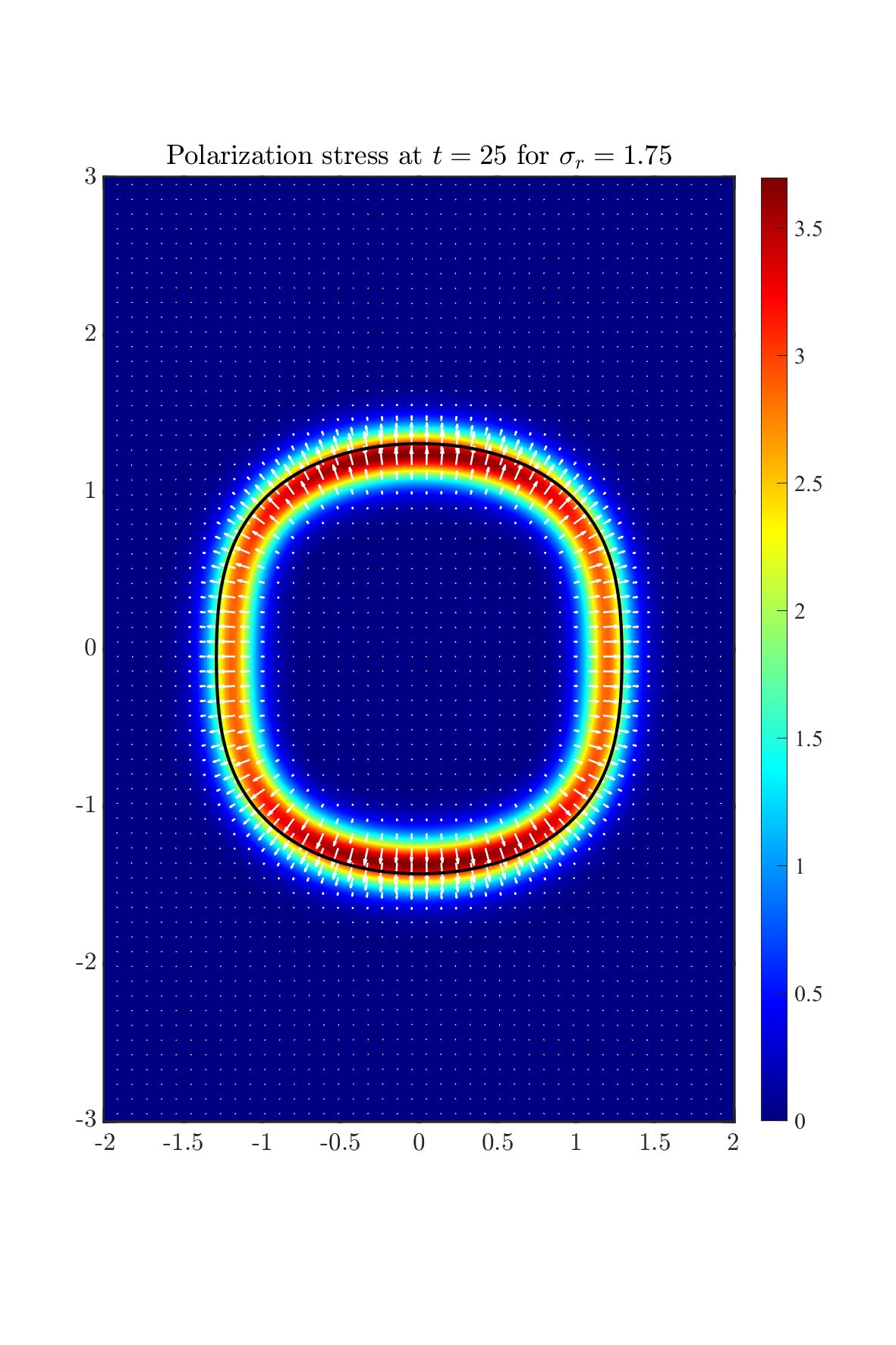}
		\includegraphics[width=0.17\textwidth]{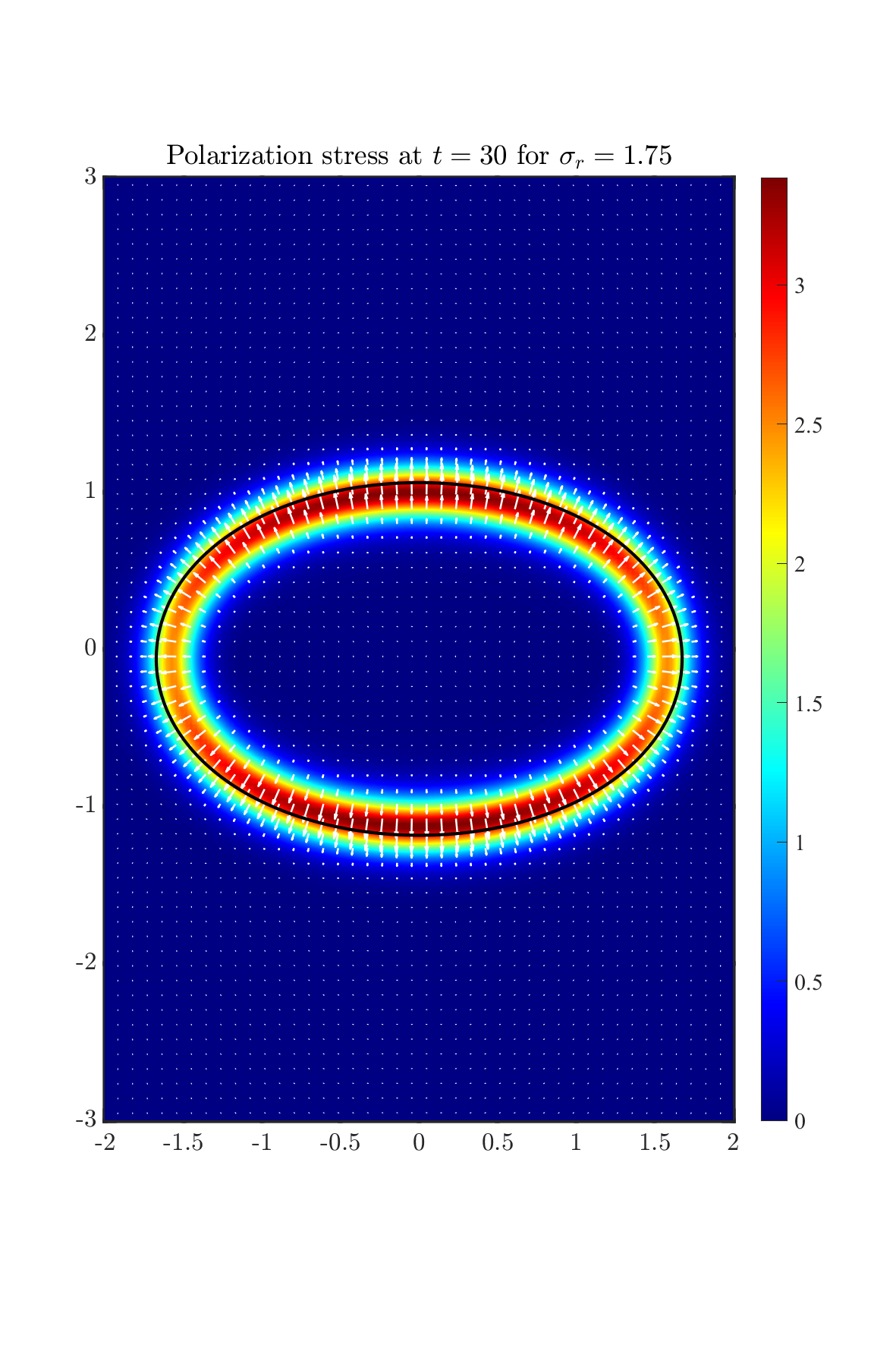}
	\end{center} 
	\caption{The merge effect for conductivity ratios
		$\sigma_{r} = 1.75$ at $t = 0$, $t = 20$, $t = 21.5$, $t = 25$ and $t = 30$ 
		from left to right, respectively 
		for the example \ref{eqn: initial two single drops 2} in section \ref{sec: tow drops merge}. 
		In each figure, the black solid line shows the zero level set ($\psi=0$) 
		to label the location of droplet. 
		The volume charge density (top), Lorentz force (middle) and polarization stress (bottom) are presented. 
		The rest parameters are chosen as $\epsilon_{r} = 3.5$, $Ca_{E} = 1$.}
	\label{fig: merge 1 for two drops without cm}
\end{figure}

\begin{figure}
	\begin{center}
		\includegraphics[width=0.32\textwidth]{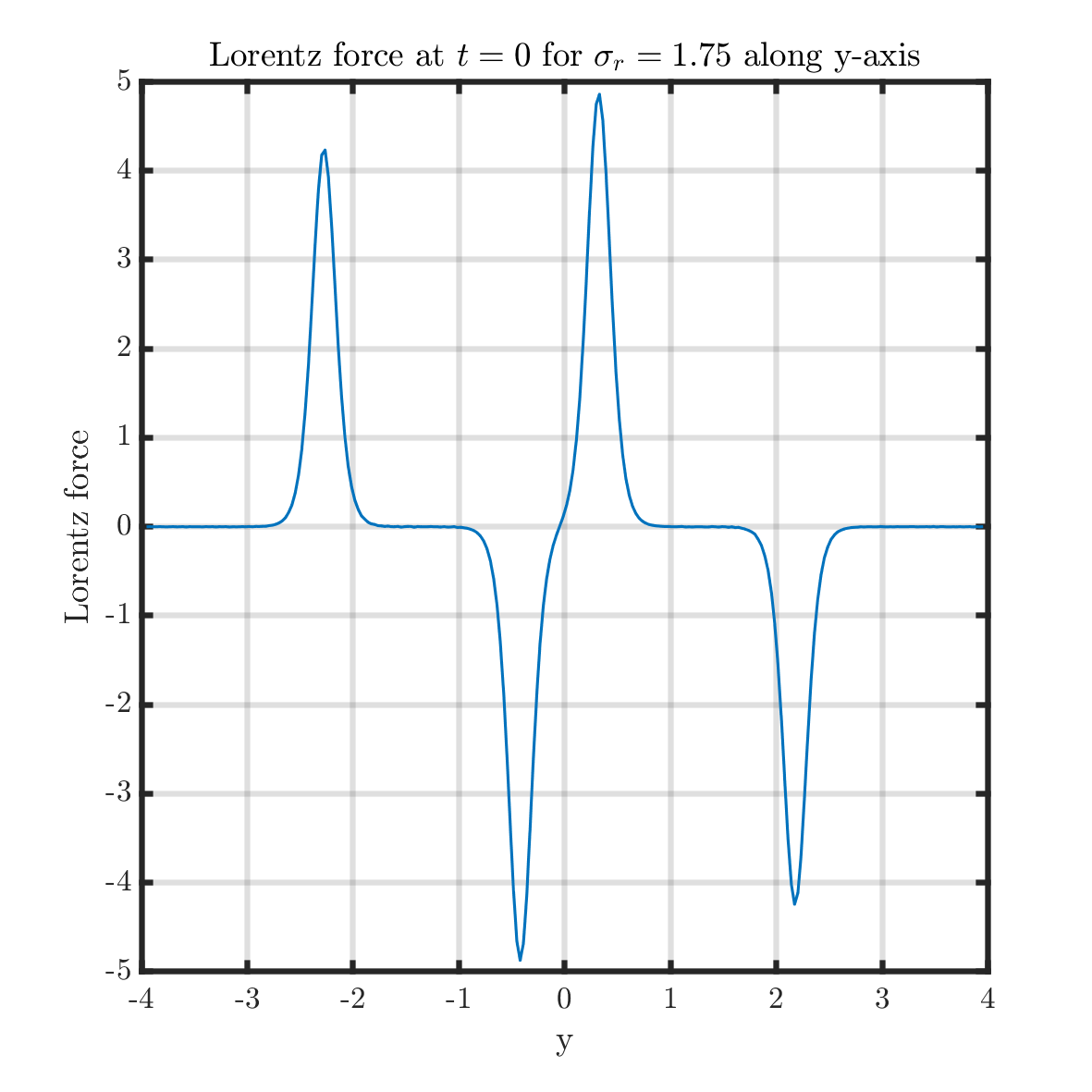}
		\includegraphics[width=0.32\textwidth]{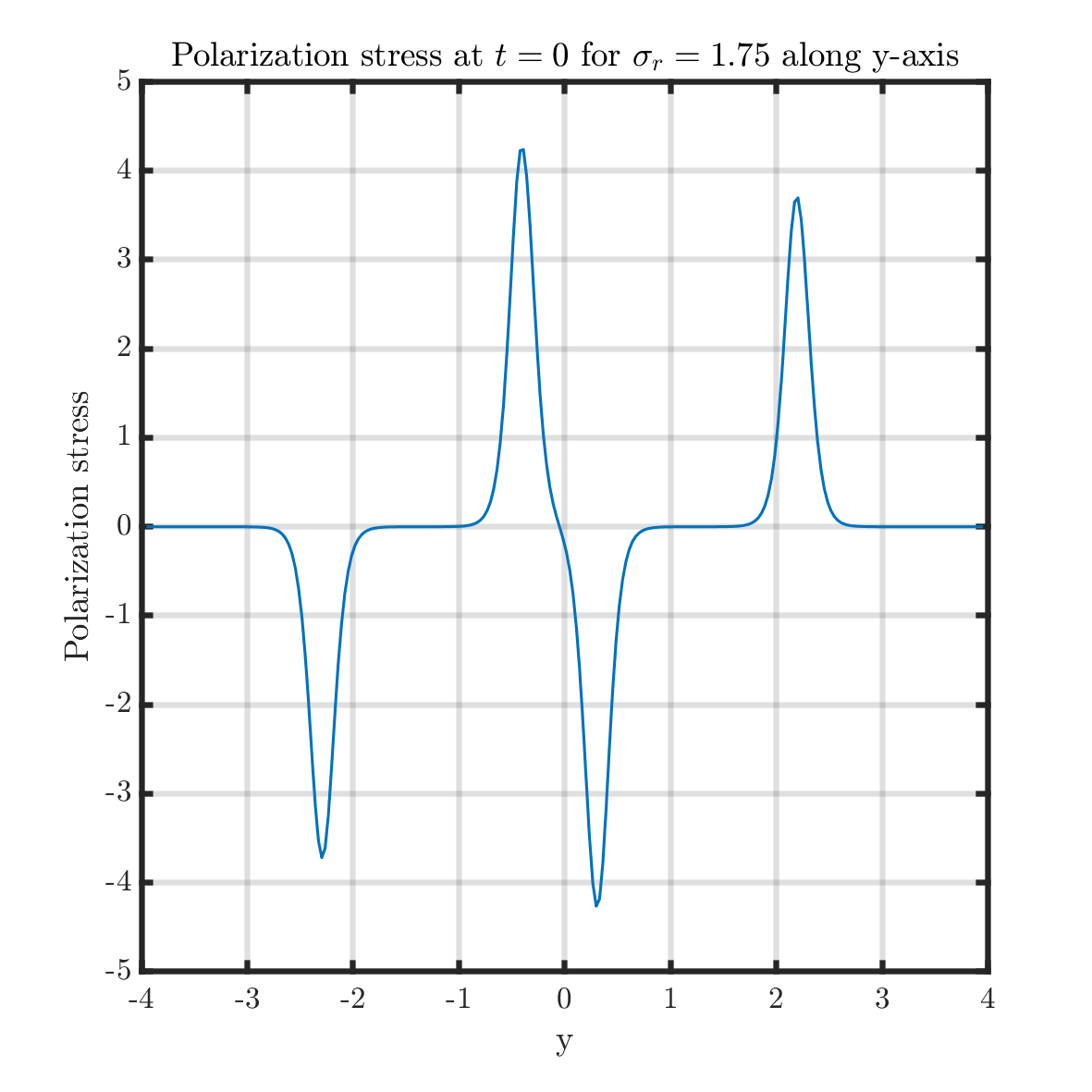}
		\includegraphics[width=0.32\textwidth]{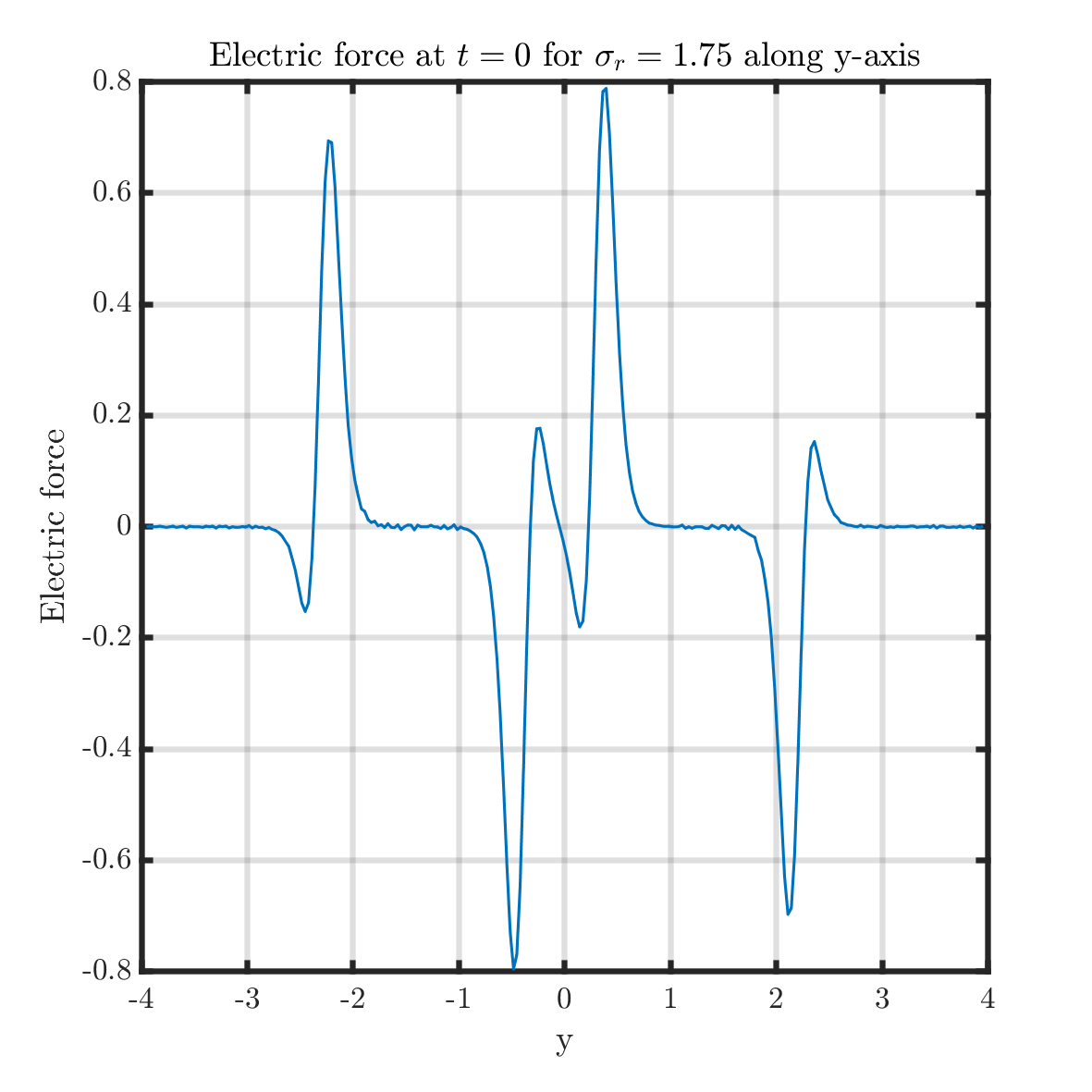}
		\includegraphics[width=0.32\textwidth]{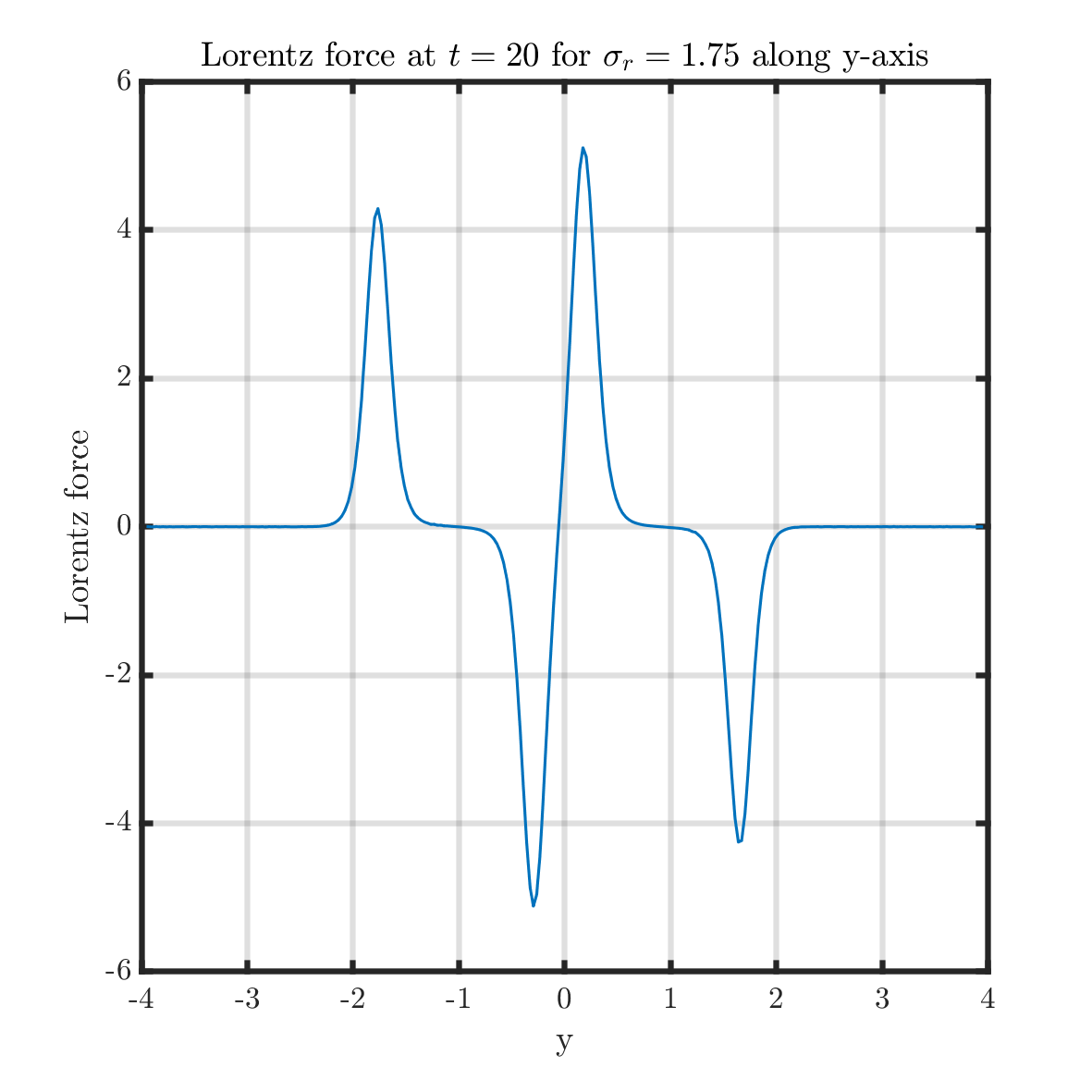}
		\includegraphics[width=0.32\textwidth]{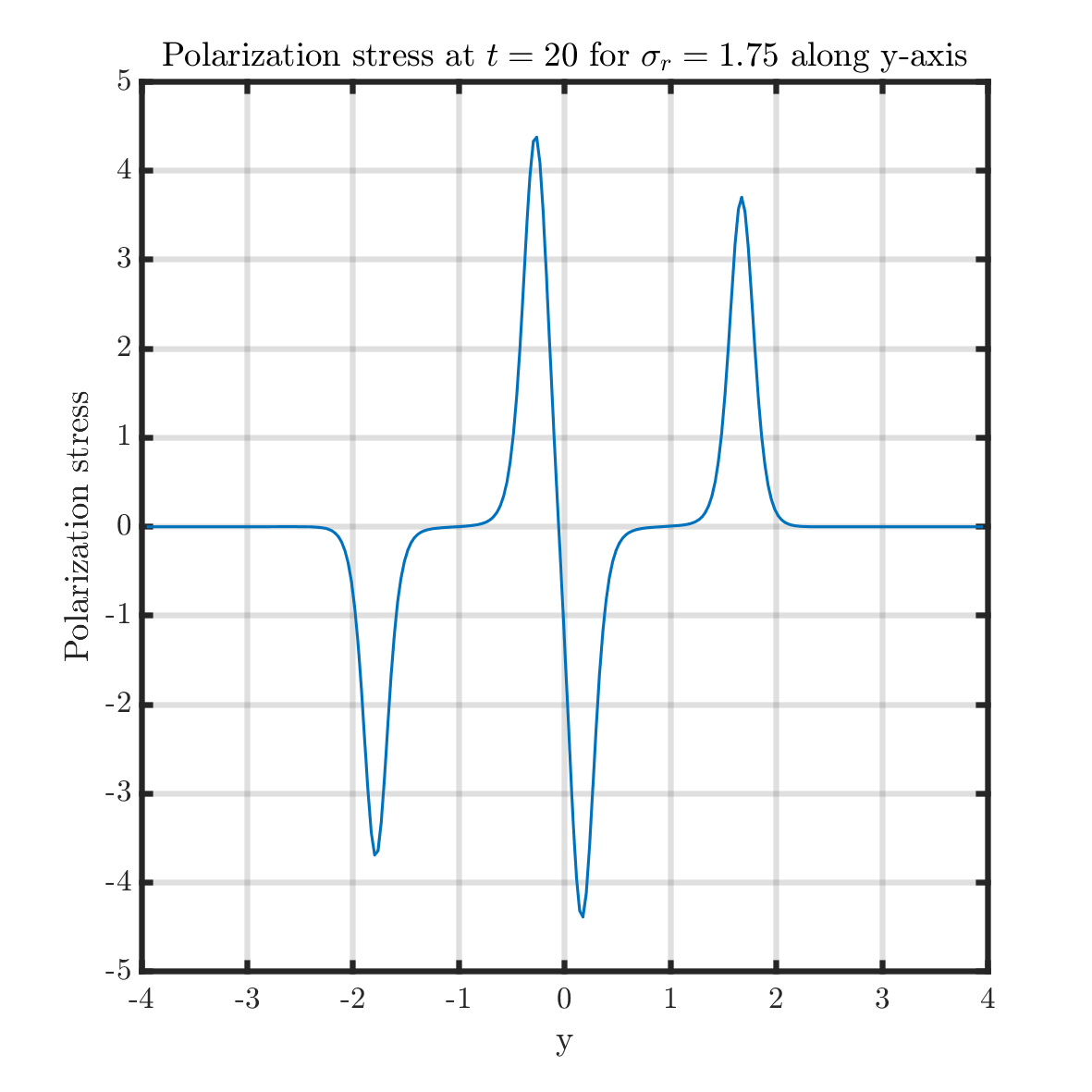}
		\includegraphics[width=0.32\textwidth]{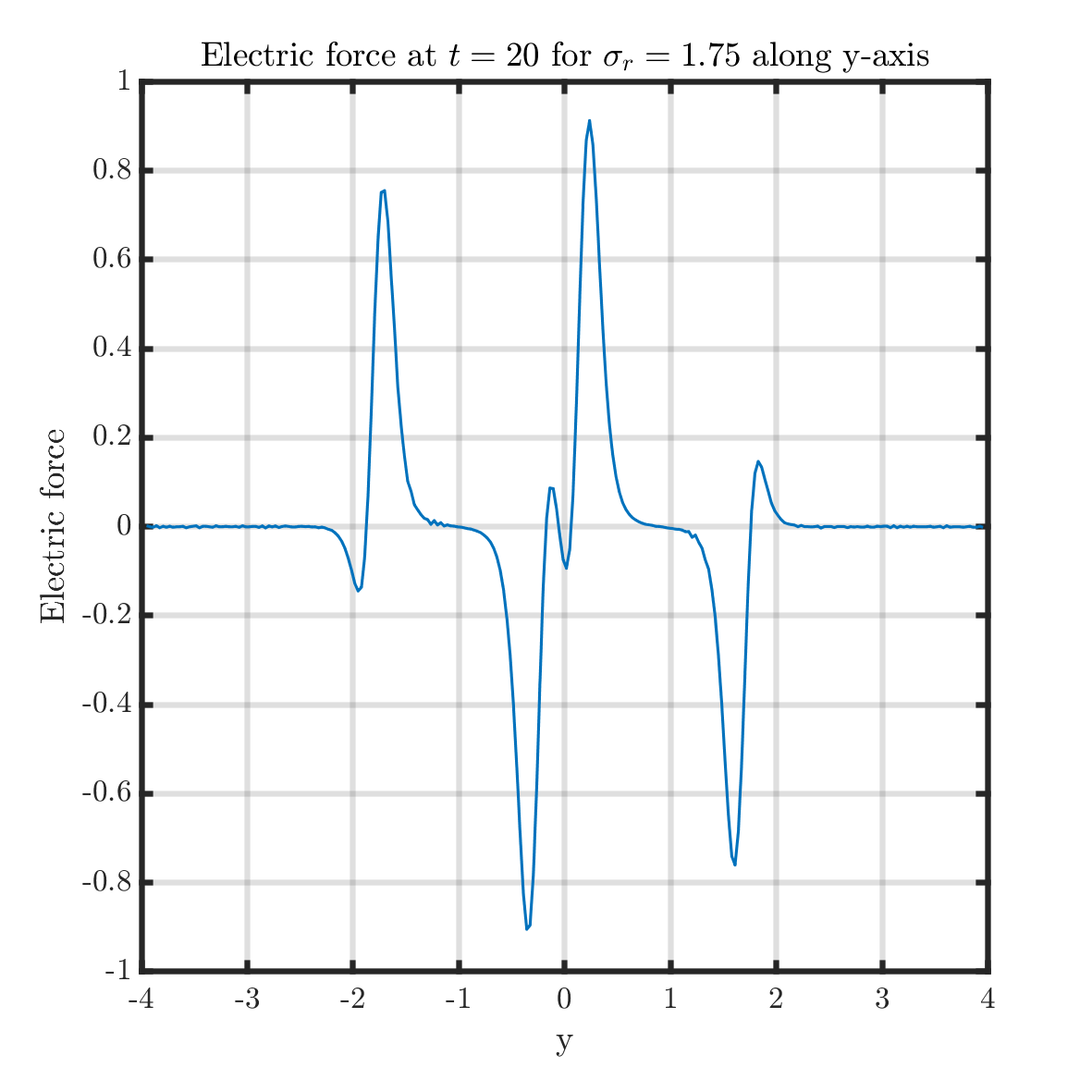}
	\end{center} 
	\caption{Electric force  distribution for 2 vertical droplets with $\sigma_r=1.75$ along 1D line $x=0$ 
		for the example \ref{eqn: initial two single drops 2} in section \ref{sec: tow drops merge}. 
		Top: $t = 0$; Bottom: $t = 20$. 
		When $y<0$, it is the force on the lower droplet; $y>0$, it is the force on the upper droplet. }
	\label{fig: merge 1 for two drops}
\end{figure}

In the next, we fix the conductivity ratio to be $\sigma_r =1.75$ and position two droplets vertically. 
The initial position of two droplets are set to be 
\begin{equation}\label{eqn: initial two single drops 2}
	\psi\left(x,y,0\right) 
	= \tanh{\frac{1 - \sqrt{x^{2}+\left(y-1.35\right)^{2}}}{\sqrt{2}\delta}}
	+ \tanh{\frac{1 - \sqrt{x^{2}+\left(y+1.25\right)^{2}}}{\sqrt{2}\delta}}+1. 
\end{equation} 
As we can see in Fig. \ref{fig: merge 1 for two drops without cm}, due to the oblate deformation, 
two droplets don't merge together first. The positive ions accumulate on the top of each drop 
and the negative ions accumulate on the bottom of each drop. 
However, under the influence of the electric field, 
the top bubble will migrate down, and the bottom bubble will migrate up and finally the two bubbles 
will touch each other and merge into one droplet. The big droplet deforms from prolate to oblate 
and the volume charge redistribute which is same as the previous session result at the equilibrium (top). 
The distributions of Lorentz force (middle) and polarization stress (bottom) are also presented. 
The direction of  Lorentz force is also from outside to inside, which is not affected by the location of ions, 
and the direction of polarization stress is from inside to outside. 1D  distributions of forces along the line $x=0$ 
are presented in Fig. \ref{fig: merge 1 for two drops}.  
In the region between two droplets, the total electric force $\nabla\cdot\bm{\sigma}_e$ is attraction force leading 
to two droplets approaching each other.  

\subsection{Capacitance effect}\label{sec: capacitance single drop}
\begin{figure}
	\begin{center}
		\includegraphics[width=1.0\textwidth]{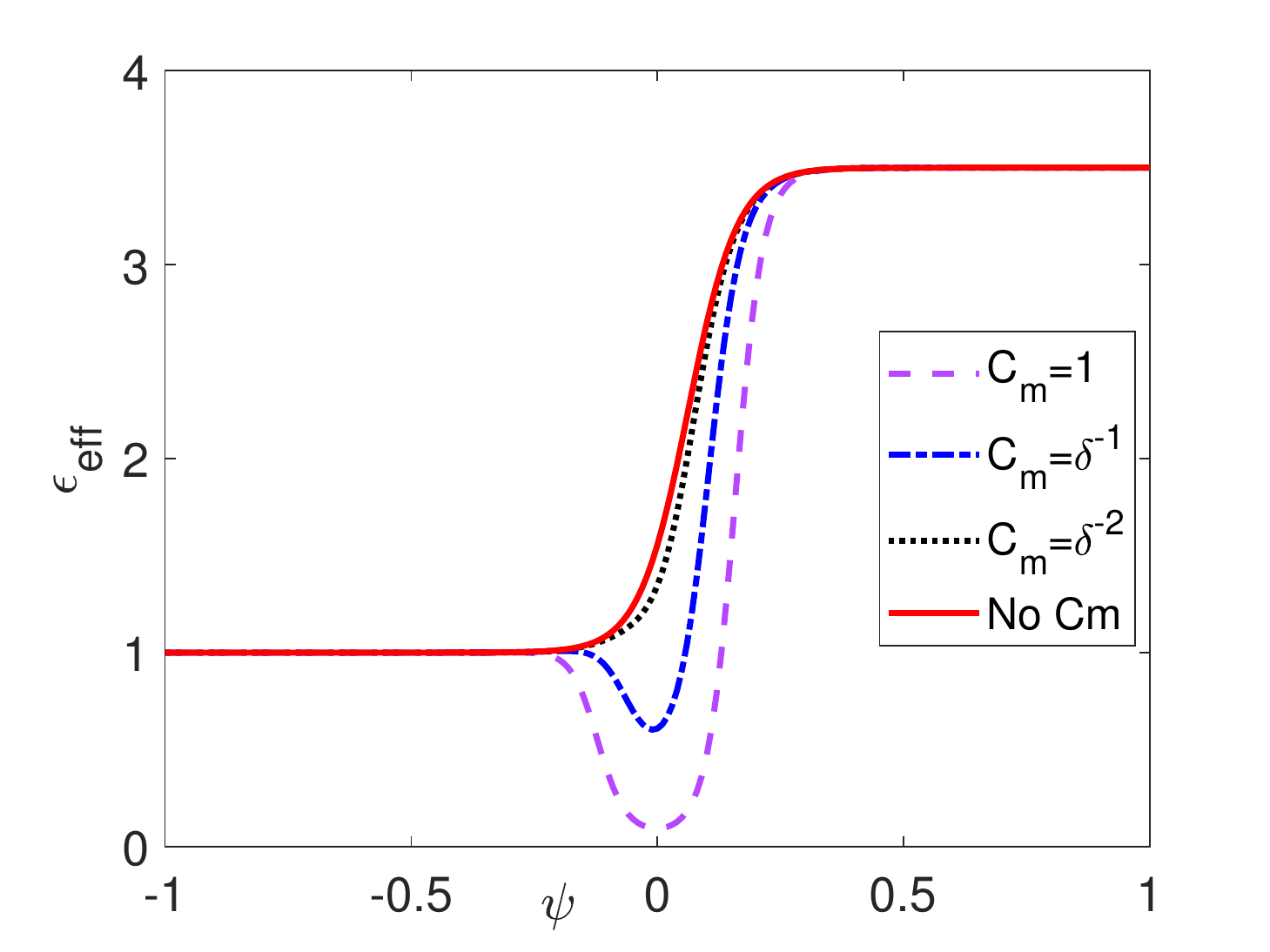}
	\end{center}
	\caption{Effective dielectric coefficient with respect label function when  $\epsilon_r=1.75$.  Red: without $C_m$; Purple: $C_m = \delta^{-2}$; Blue: $C_m = \delta^{-1}$; Black: $C_m = 1$.  }
	\label{fig: eps_eff}
\end{figure}

In this section, we focus on the capacitance effect acting on the drop. 
We consider  a single drop in the applied electric field with finite capacitance $C_m = 1,\delta^{-1}, \delta^{-2}$ and compare with the results without capacitance in Section \ref{sec: comparison with sharp model}. All the other parameters are set to be same as Section \ref{sec: comparison with sharp model}. 
When  the capacitance $C_m$  is set to be  $1,\delta^{-1}, \delta^{-2}$, it means the electric permittivity near the interface $\epsilon_m = \delta C_m$ is around 
$\delta, 1, \delta^{-1}$. The effective permittivity with different order of capacitance is shown in Fig. \ref{fig: eps_eff},  where the finite capacitance introduces a local perturbation near the interface. 

\begin{figure}
	\begin{center}
		\includegraphics[width=0.32\textwidth]{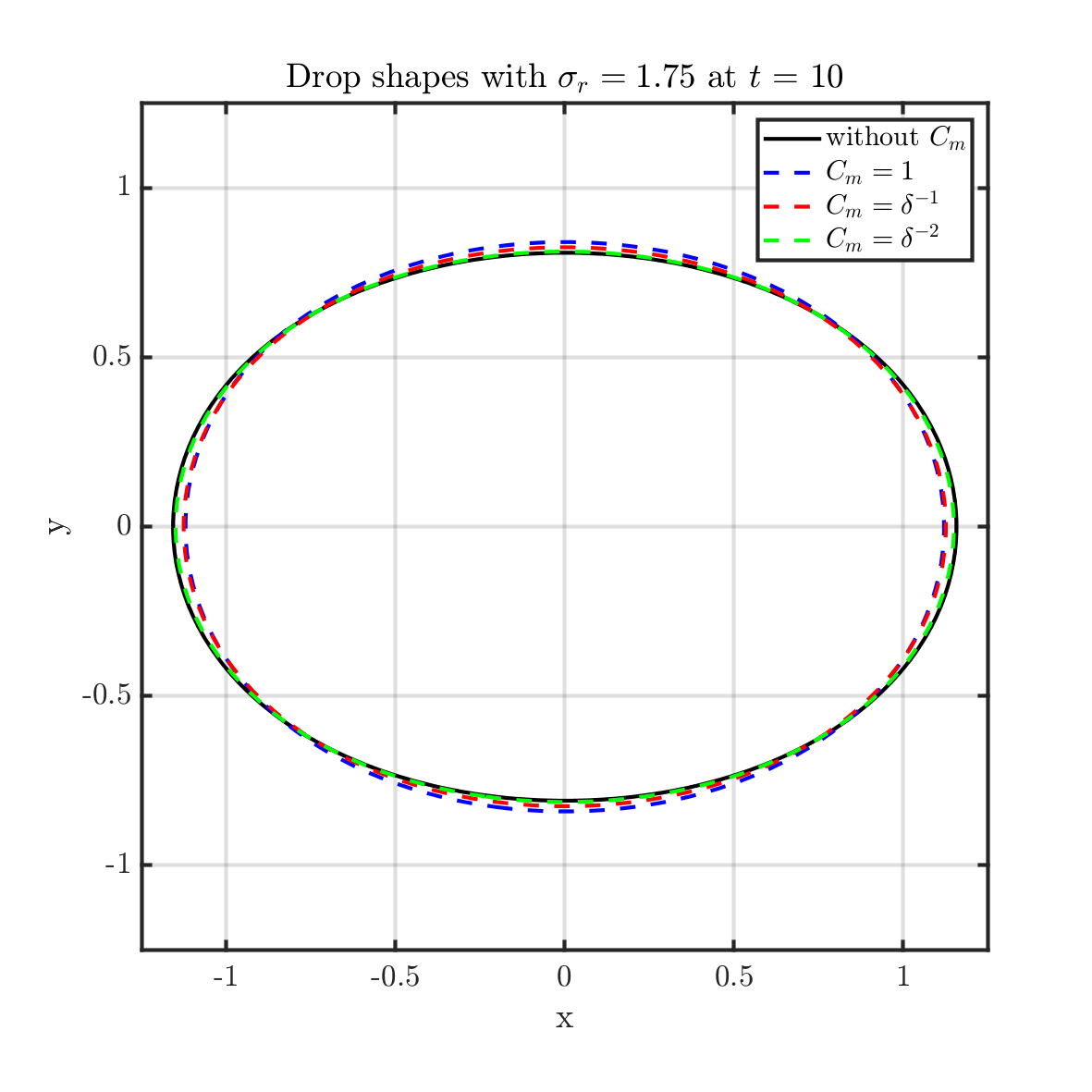}
		\includegraphics[width=0.32\textwidth]{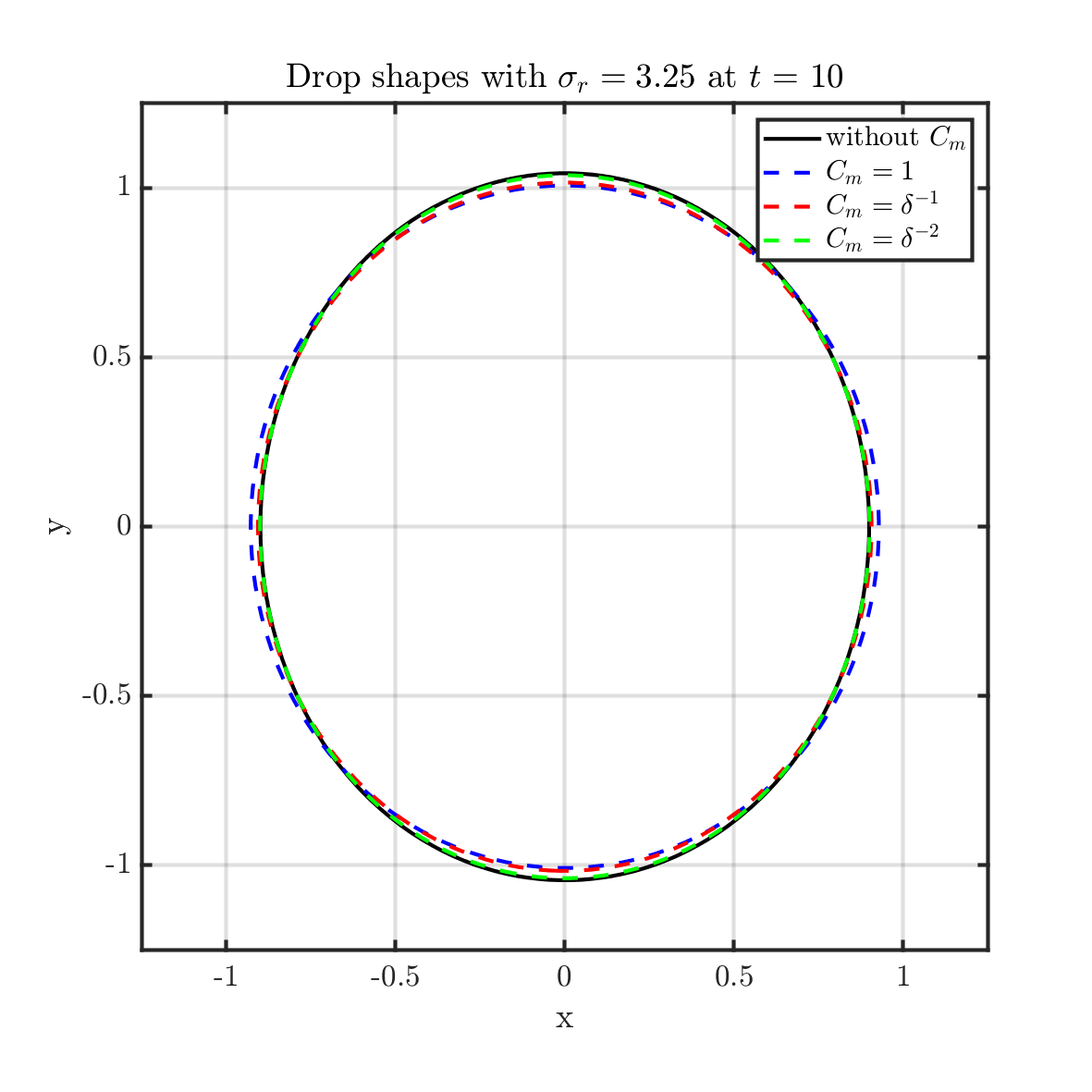}
		\includegraphics[width=0.32\textwidth]{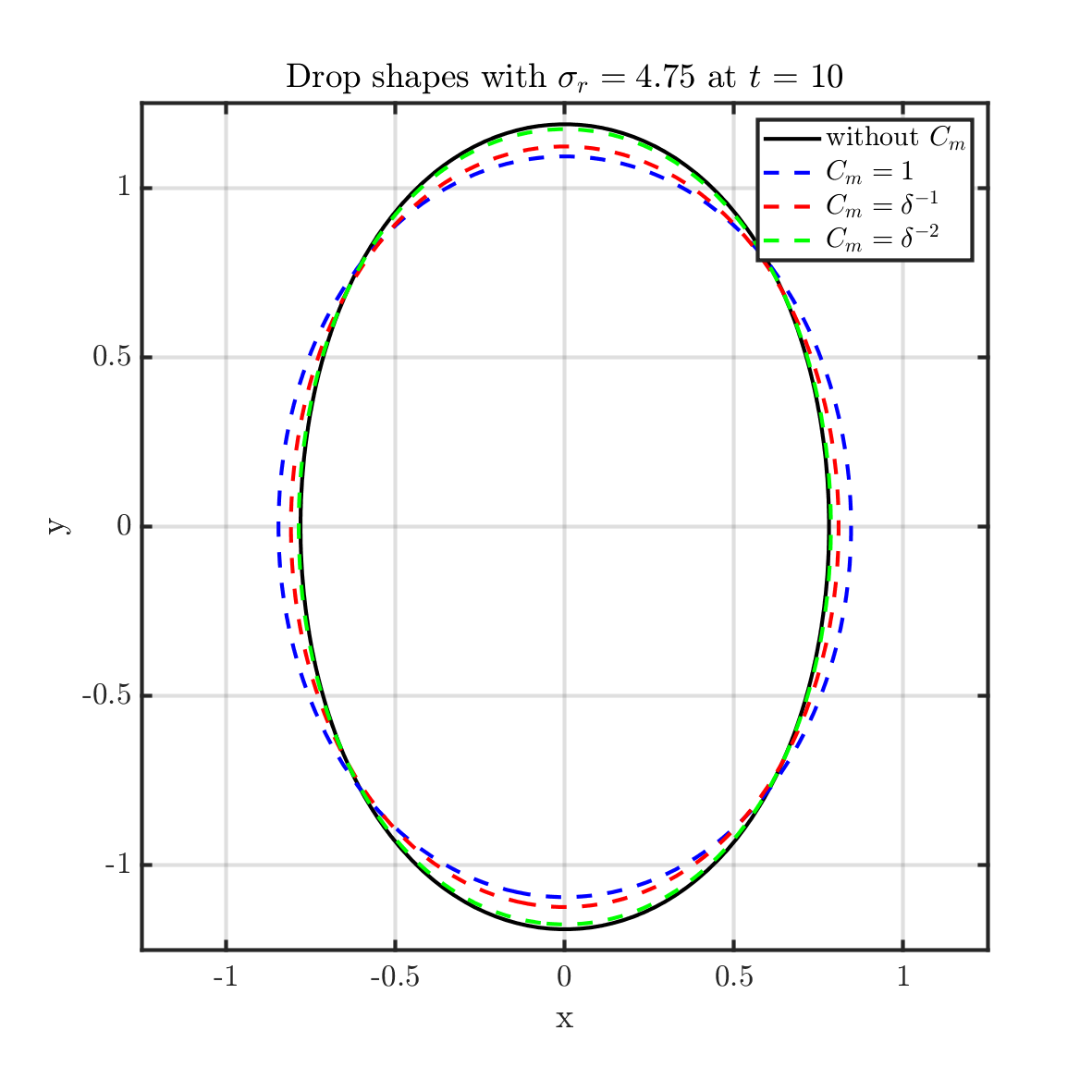}
	\end{center} 
	\caption{The behavior of drop shapes for different conductivity ratios
		$\sigma_{r} = 1.75$ (left), $\sigma_{r} = 3.25$ (middle), $\sigma_{r} = 4.75$ (right) 
		by considering different capacitances $C_{m}$ at $t = 10$. 
		In each figure, the solid line shows the zero level set ($\psi=0$), 
		where the black, blue, red and green lines show the drop shape with $C_{m} = 1$, 
		$C_{m} = \delta^{-1}$ and $C_{m} = \delta^{-2}$, respectively. 
		The rest parameters are chosen as $\epsilon_{r} = 3.5$, $Ca_{E} = 1$.
		The change of the drop shape indicates that a bigger capacitance will cause a more intense 
		drop change.}
	\label{fig: single drop with cm}
\end{figure}

\begin{figure}
	\begin{center}
		\includegraphics[width=0.32\textwidth]{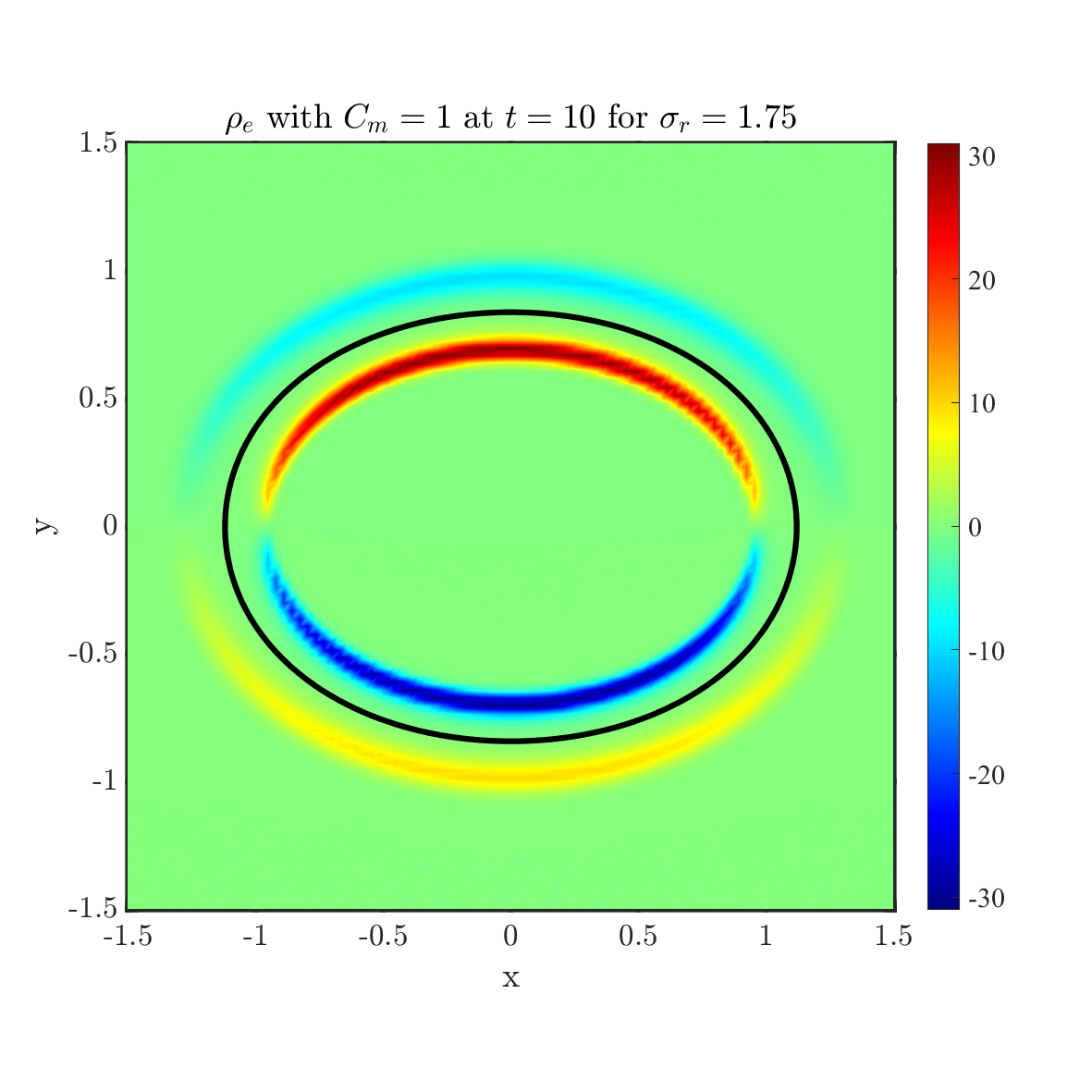}
		\includegraphics[width=0.32\textwidth]{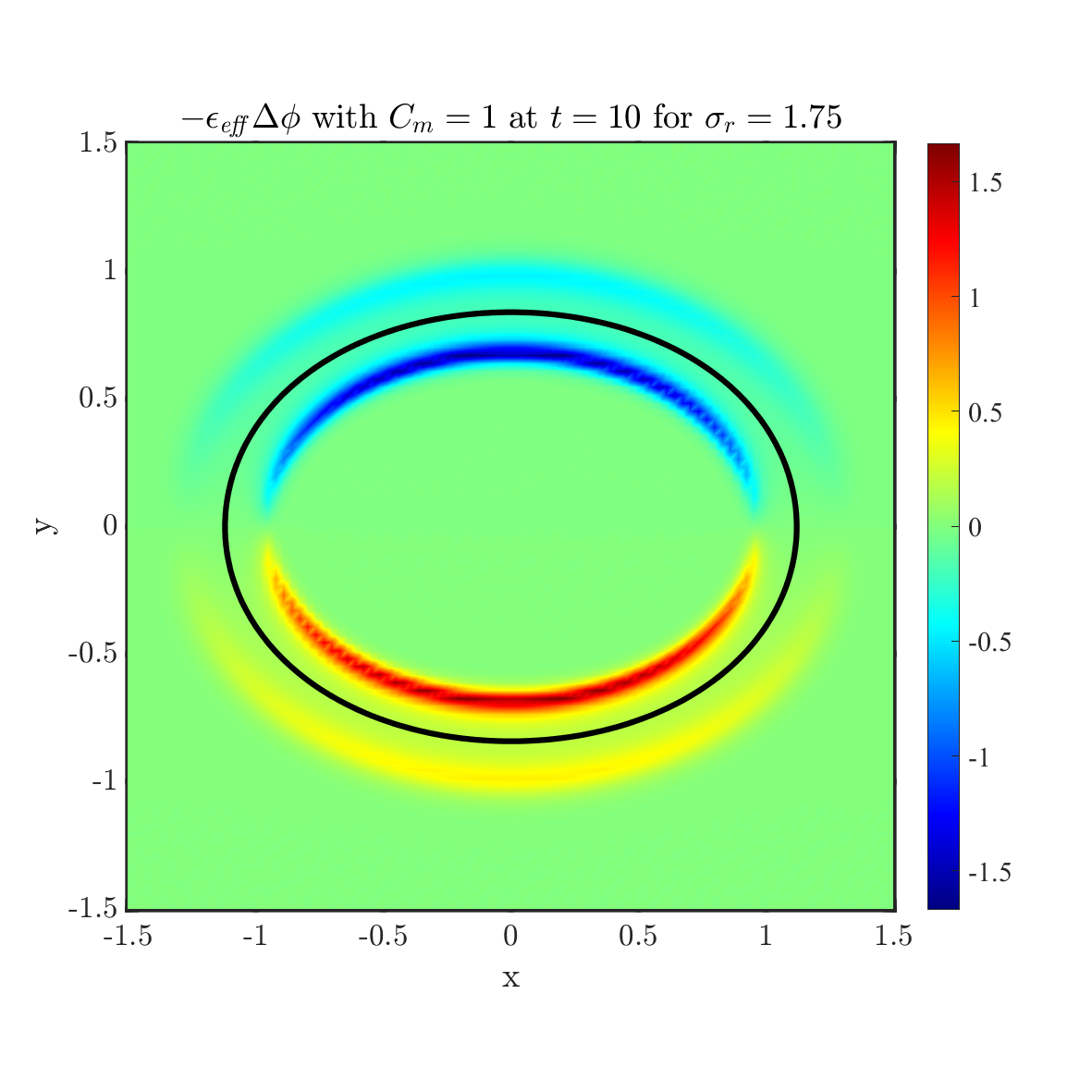}
		\includegraphics[width=0.32\textwidth]{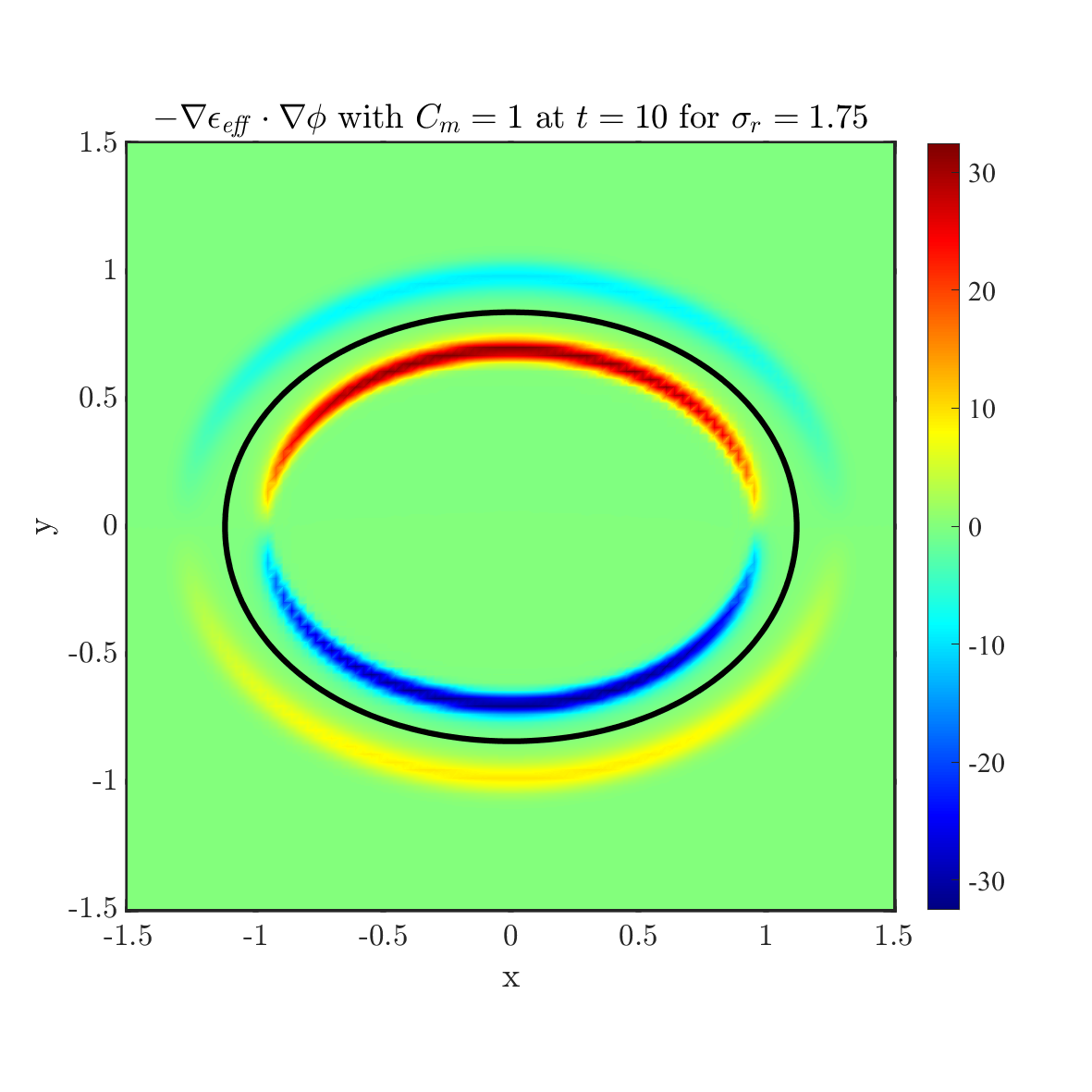}
		\includegraphics[width=0.32\textwidth]{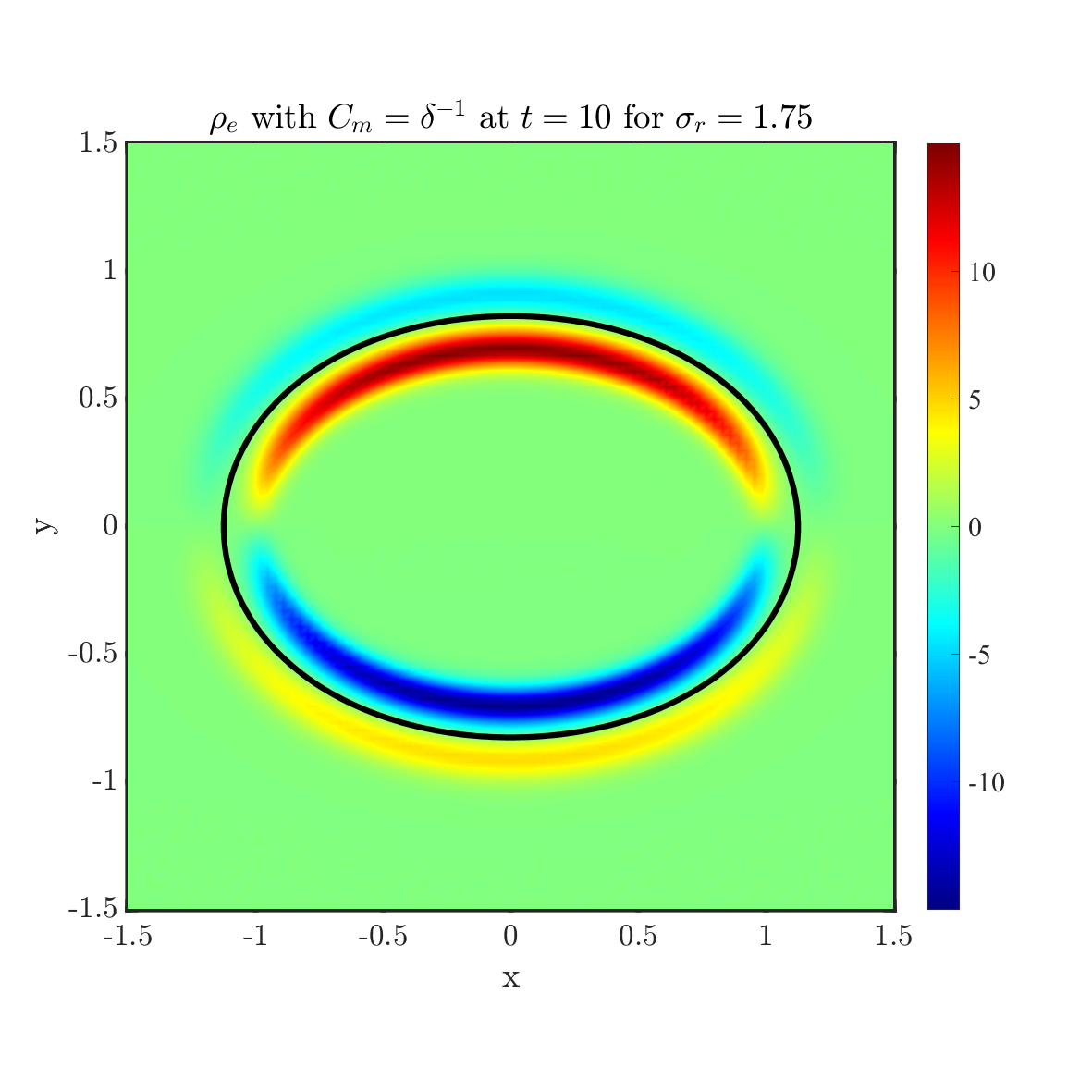}
		\includegraphics[width=0.32\textwidth]{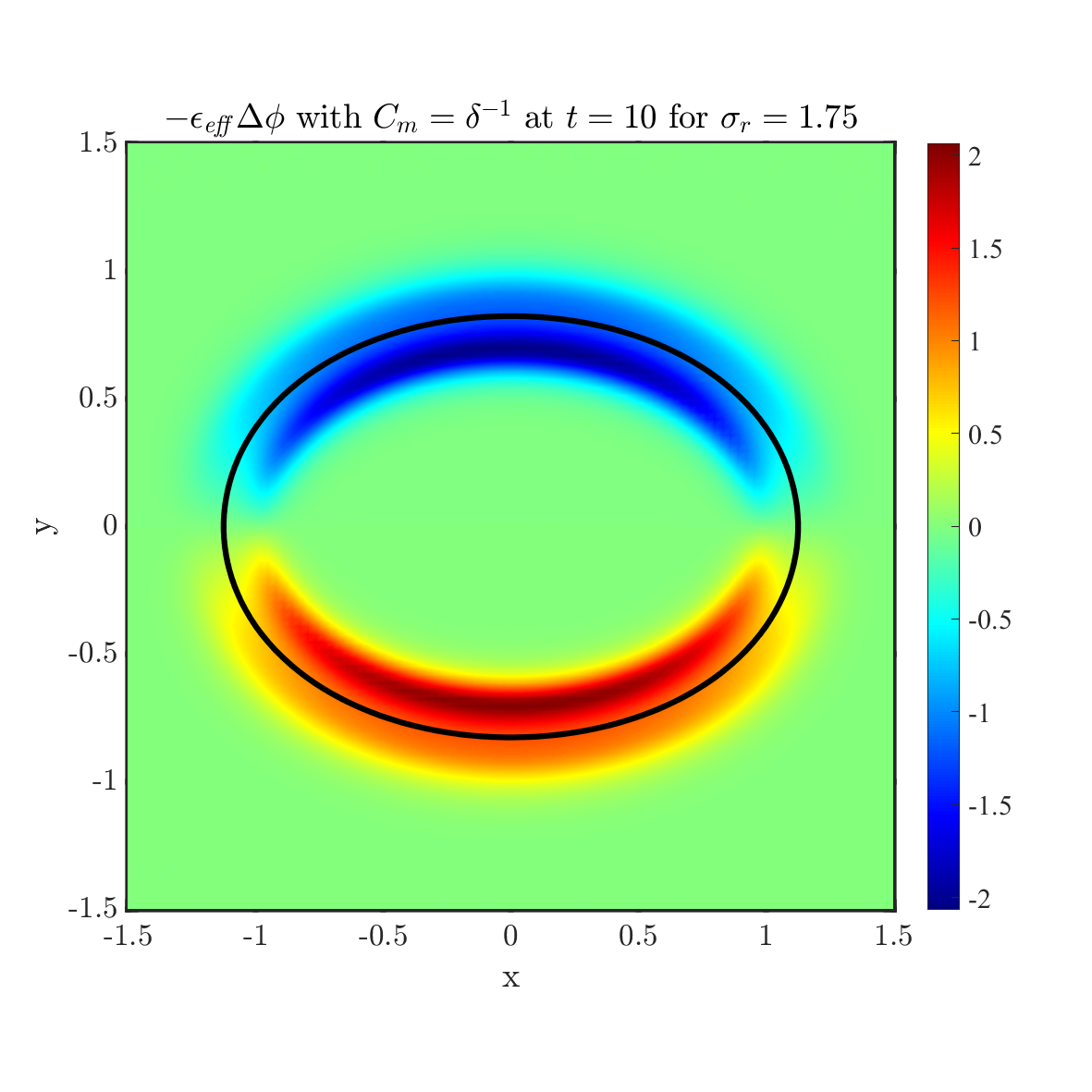}
		\includegraphics[width=0.32\textwidth]{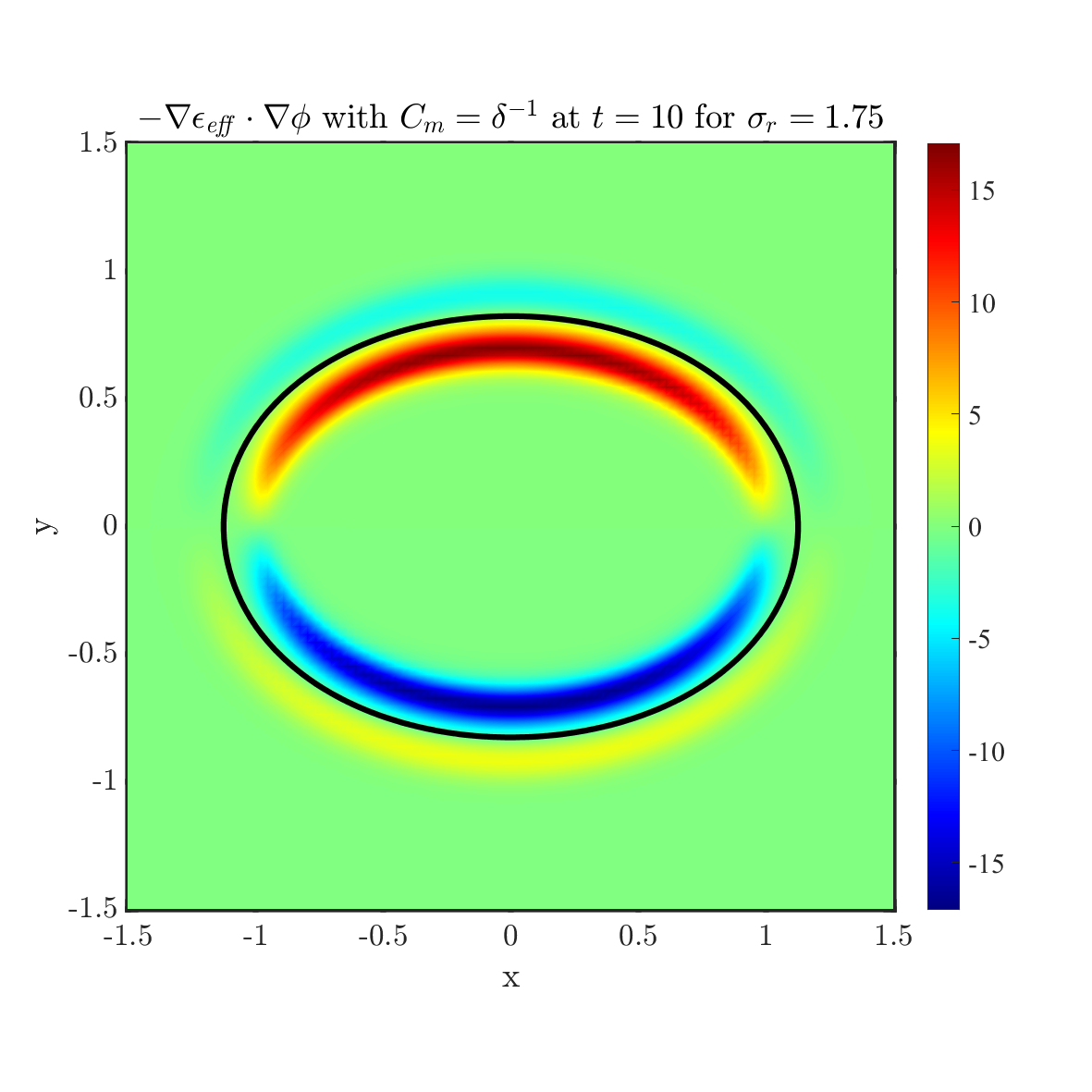}
		\includegraphics[width=0.32\textwidth]{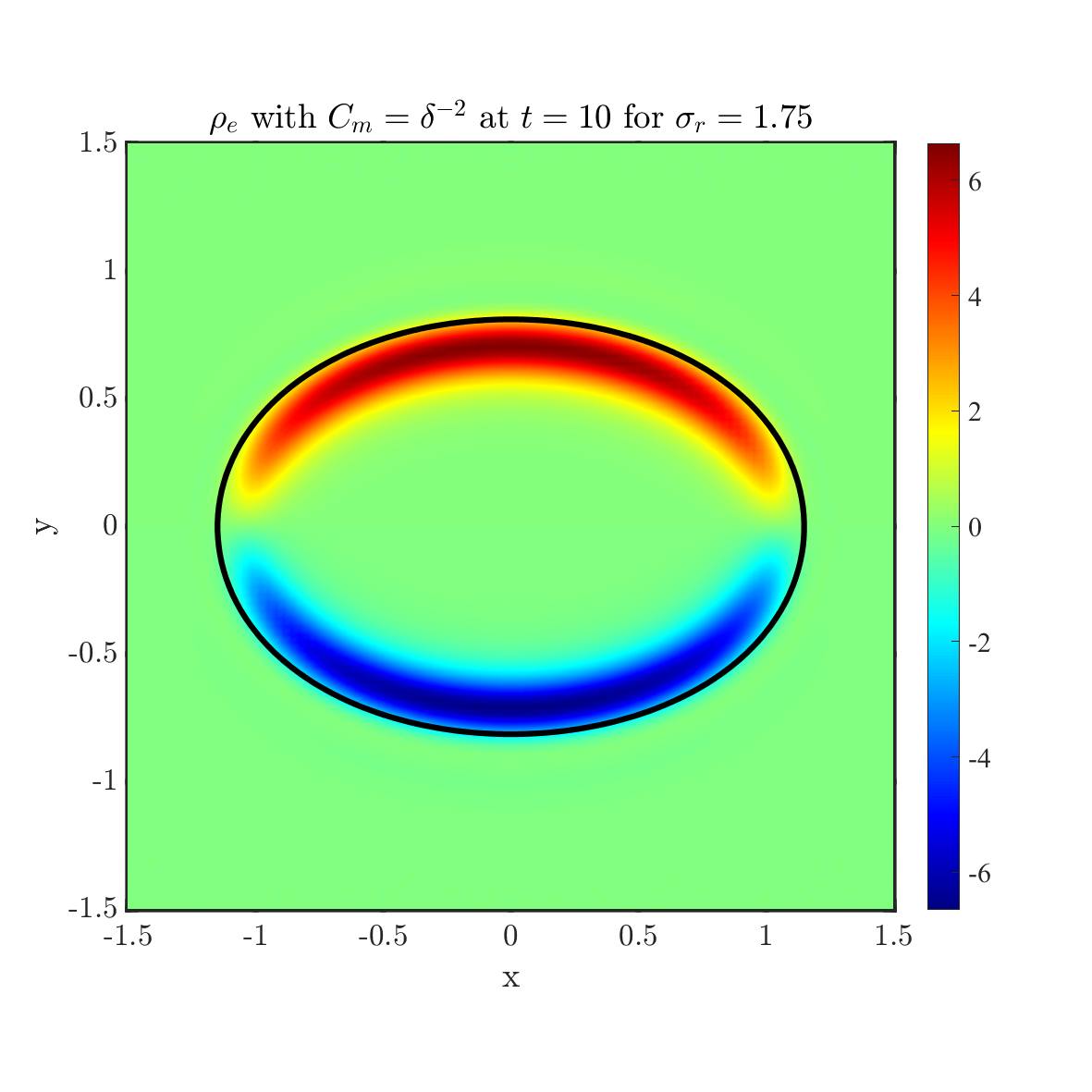}
		\includegraphics[width=0.32\textwidth]{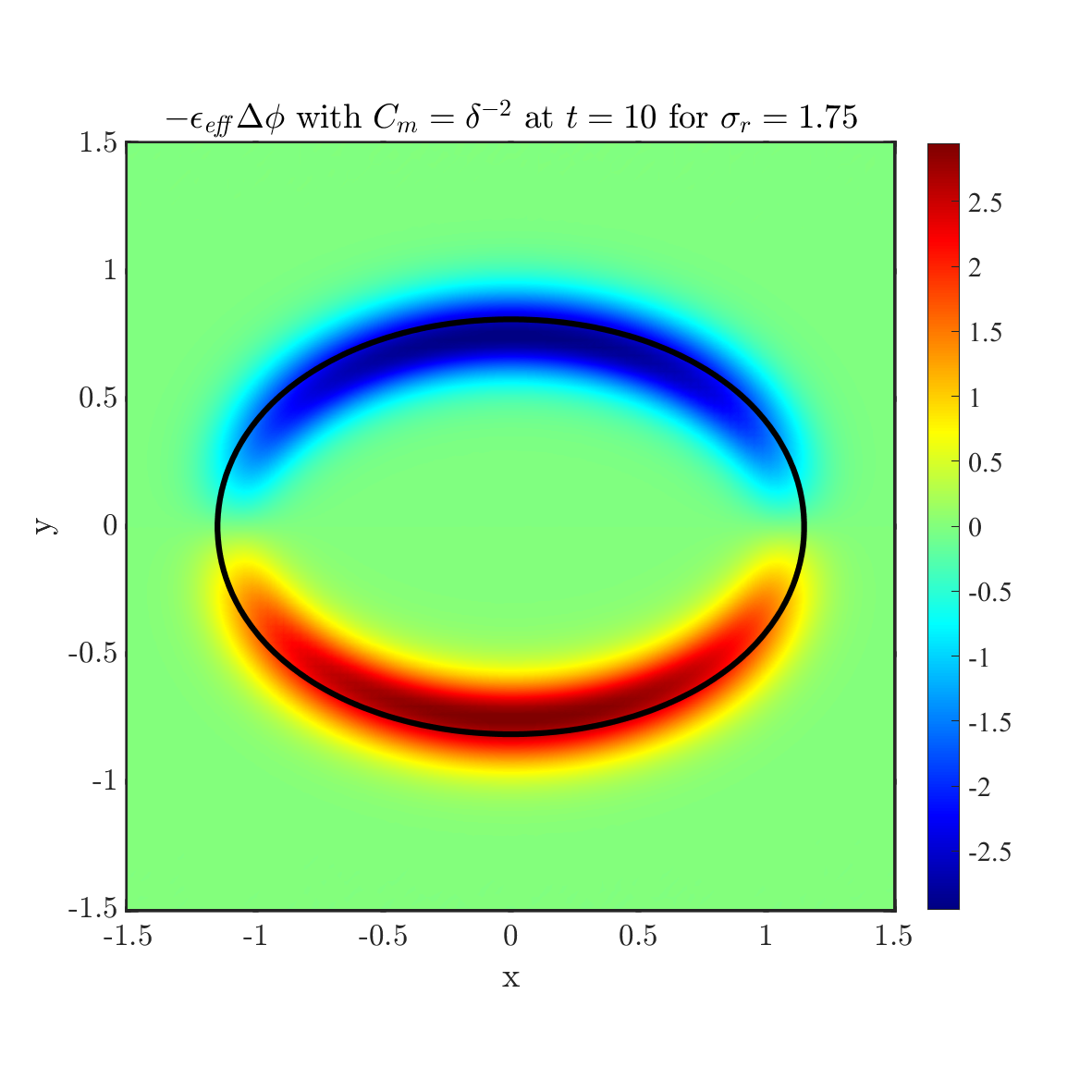}
		\includegraphics[width=0.32\textwidth]{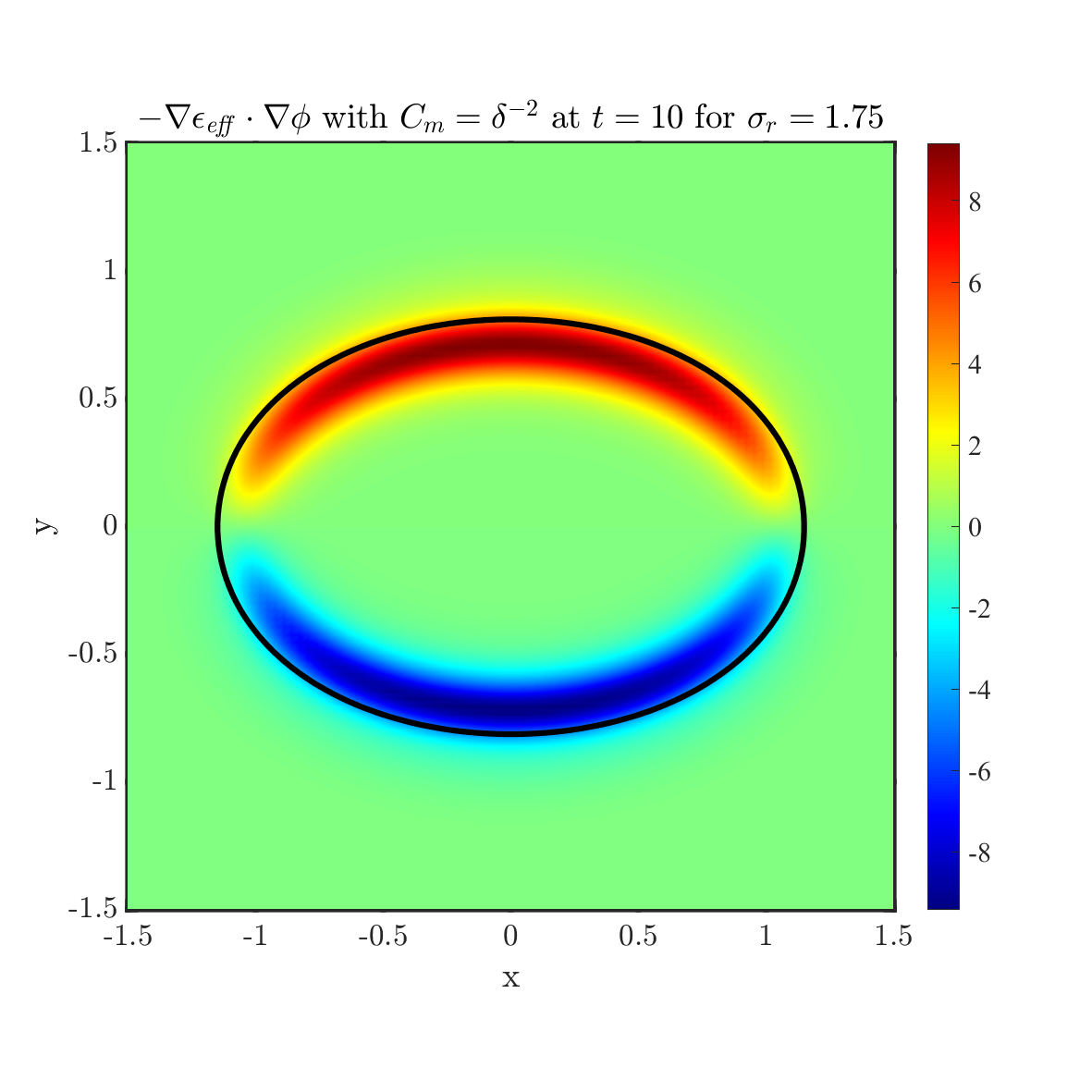}
		\includegraphics[width=0.32\textwidth]{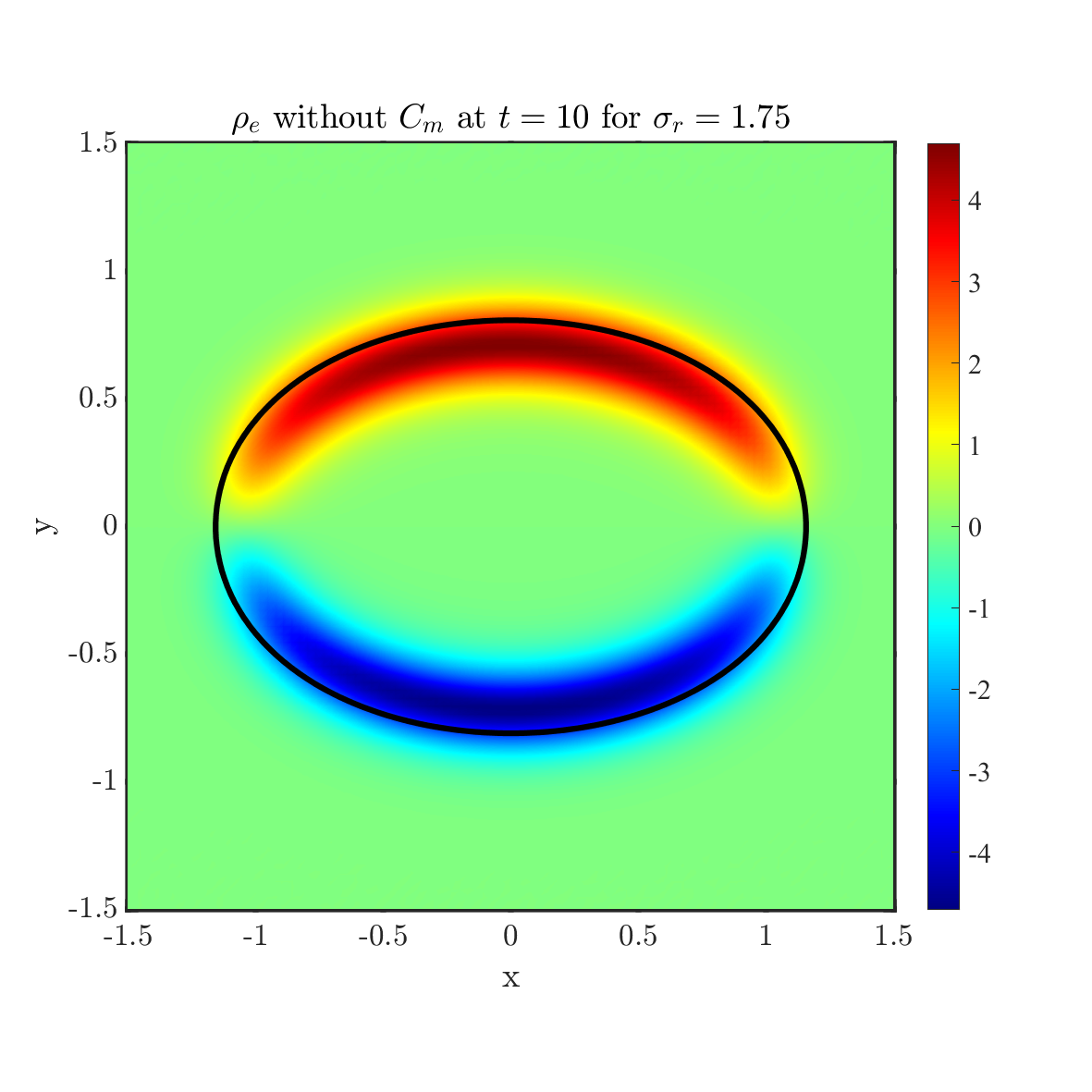}
		\includegraphics[width=0.32\textwidth]{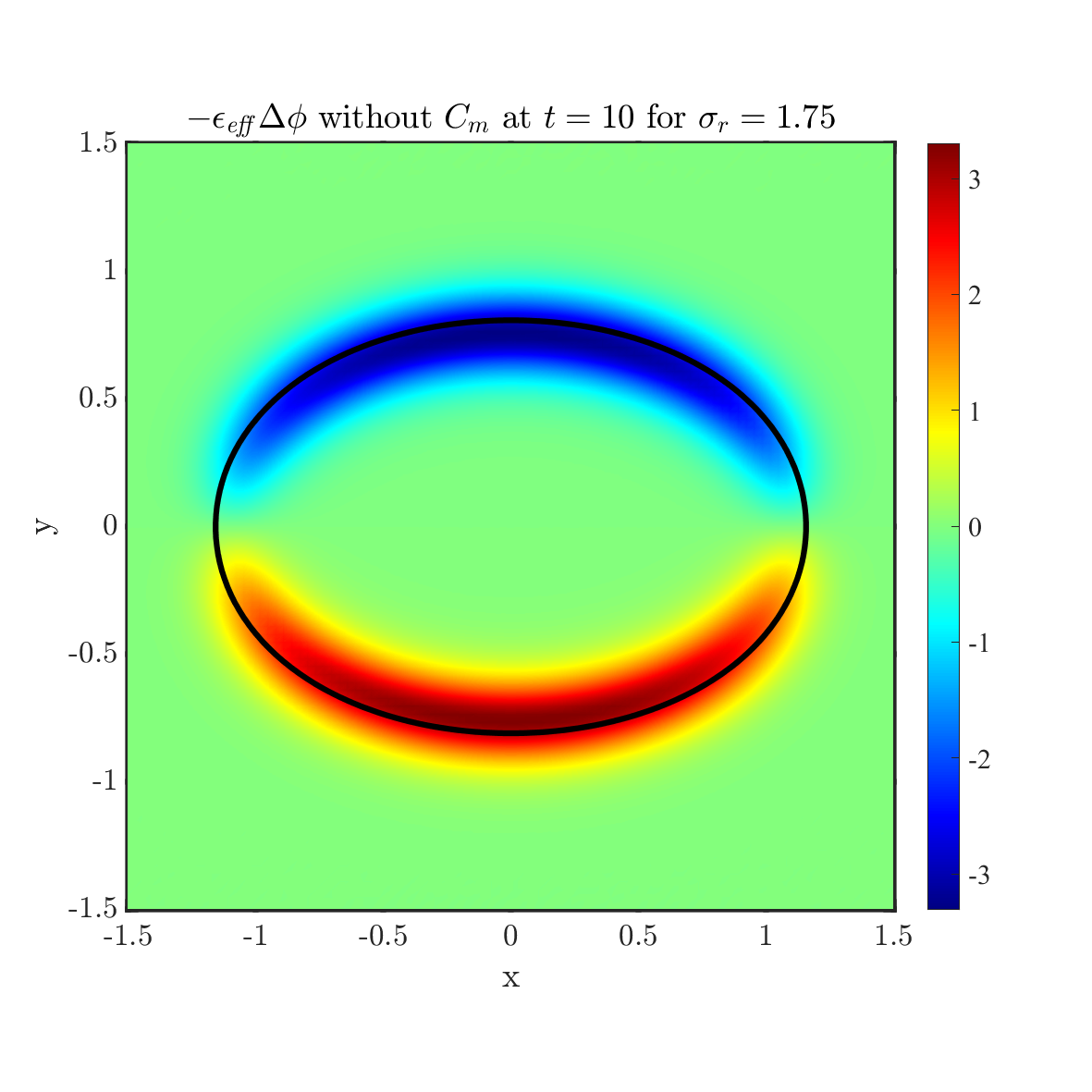}
		\includegraphics[width=0.32\textwidth]{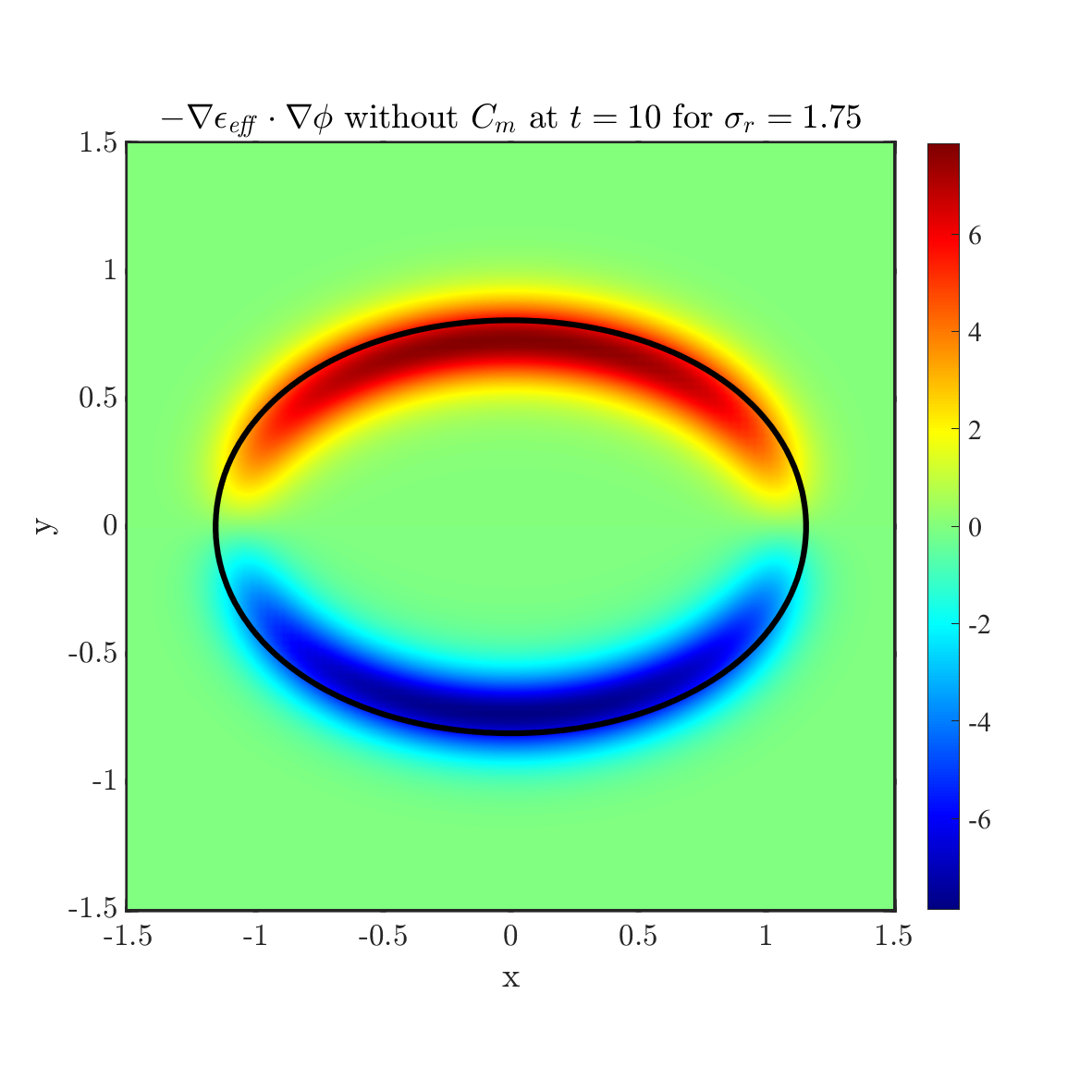}
	\end{center} 
	\caption{The net charge distribution for conductivity ratio $\sigma_{r} = 1.75$ 
		by considering different capacitances $C_{m} = 1$, $C_{m} = \delta^{-1}$, $C_{m} = \delta^{-2}$ and no $C_{m}$ 
		from top to bottom 	at time $t = 10$. In each figure, the solid line shows the zero level set ($\psi=0$).
		The rest parameters are chosen as $\epsilon_{r} = 3.5$, $Ca_{E} = 1$.}
	\label{fig: net charge for single drop with 3 cm sigma175}
\end{figure}

The shape behavior with different conductivity ratios 
$\sigma_{r} = 1.75$ (left), $\sigma_{r} = 3.25$ (middle) and $\sigma_{r} = 4.75$ (right) 
with different capacitance at the final time $t=10$ is presented in 
Figure \ref{fig: single drop with cm}.  The black solid lines denotes the interfaces of the droplets without capacitance for reference. 
Whereas the colorful dash lines are the interfaces of droplets with different capacitance $C_{m} = 1$ (blue), $C_{m} = \delta^{-1}$ (red) and $C_{m} = \delta^{-2}$ (green). 
It shows that the deformation decreases as the capacitance decrease in all three cases.  The distribution of the net charge
is presented  in Fig. \ref{fig: net charge for single drop with 3 cm sigma175} and Fig. \ref{fig: net charge for single drop with 3 cm sigma35}-\ref{fig: net charge for single drop with 3 cm sigma475} in Appendix. It shows that the distribution of net charge is affected near the interface when the capacitance is considered. 
With capacitance,  the total net charge $\rho_e = -\nabla\cdot(\epsilon_{\it eff}\nabla\phi) = -\epsilon_{\it eff}\Delta\phi-\nabla\epsilon_{\it eff}\cdot\nabla\phi$ is formed by the variation of electric field and the variation of electric permittivity. The results illustrate that the main difference is induced by the variation of the electric permittivity (third column). The derivative $\frac{\partial\epsilon_{\it eff}}{\partial \psi}$  changes sign across the interface and leads to accumulation of counter-ions at the outer interface. 

\section{Discussion on time scales}\label{sec:time scale}
In Eq.\eqref{eqn: charge}, if we introduce the dimensionless ratios
% we focus on comparing the phase field leaky dielectric model we proposed with the net charge model 
%by adding the net charge equation shown as \eqref{eqn: charge} via a series of numerical experiments, 
%which means we use a correction method for the electric potential to carry the numerical validations. 
%%The first correction method computes the net charge by considering \eqref{eqn: charge}. 
%First of all, we dimensionlize the net charge equation by using the following dimensionless parameters: 
\begin{equation}
	t_{E2M} = \frac{\tilde{t}_{E}}{\tilde{t}}, \quad t_{E2D} = \frac{\tilde{t}_{E}}{\tilde{t}_{D}}, 
\end{equation}
it could be written as 
\begin{equation}\label{eqn: dimensionless charge}
	t_{E2M}\left(\frac{\partial \rho_e}{\partial t} 
	+ \nabla \cdot \left(\bm{u}\rho_e\right) \right)
	= \zeta^{2}\nabla\cdot\left(\sigma_{c}\nabla\phi\right)
	+t_{E2D}\nabla\cdot\left( D\nabla \rho_e\right). 
\end{equation}
As we mentioned previously, as electric relaxation time is  fast enough, i.e. $t_{E2M}<<1$ and $t_{E2D}<<1$, the leaky dielectric model \eqref{eq:leakydielectricsystem} is  achieved.   In this section,  we would like to compare the leaky dielectric model with the net charge model \eqref{eqn: dimensionless charge} under different time scale ratios. For the convenience of discussion, we consider $t_{E2M} = t_{E2D}$ in the following. 
So, we will not mention $t_{E2D}$ no longer. The setup is same as in Section \ref{sec: comparison with sharp model}.

\begin{figure}
\begin{center}
\includegraphics[width=0.32\textwidth]{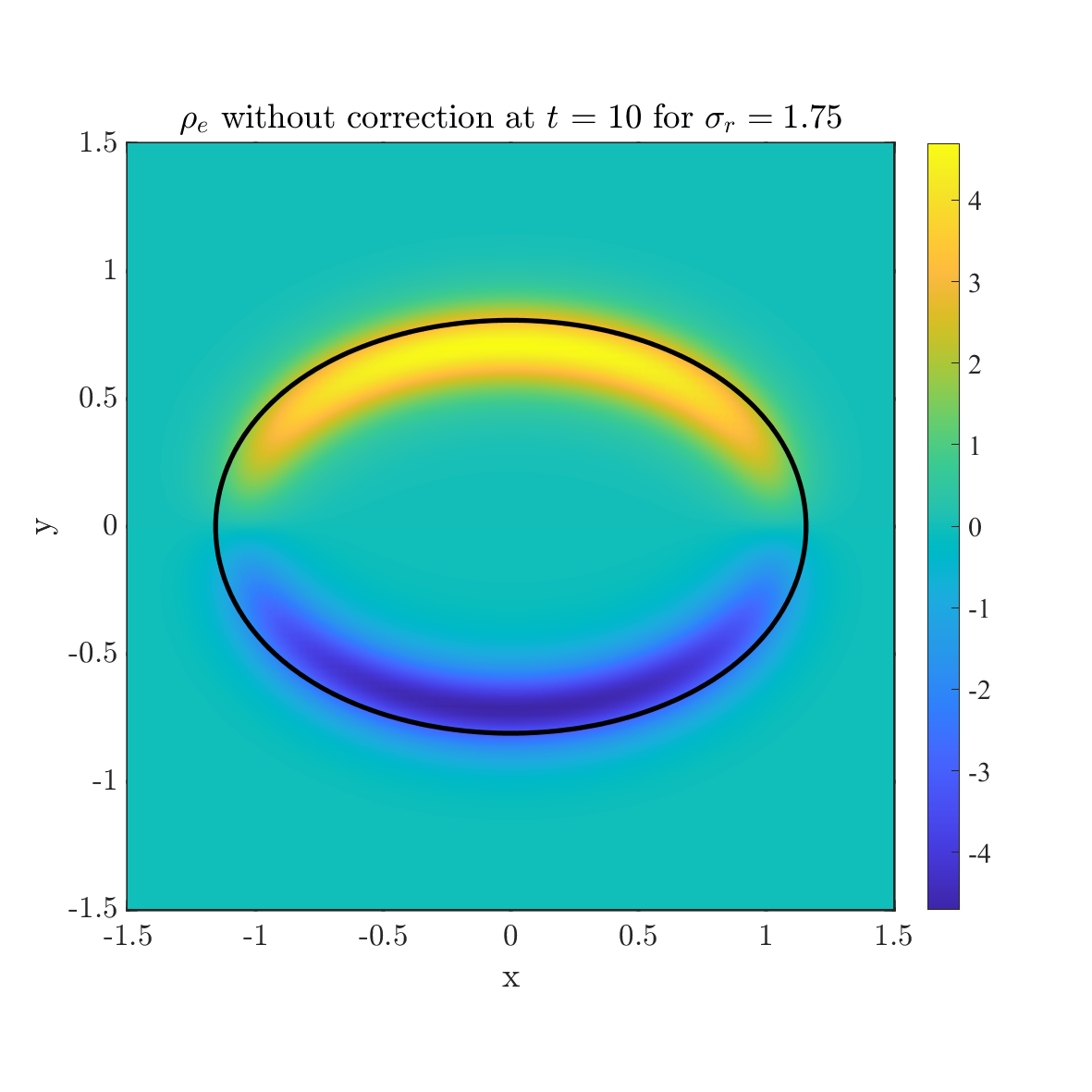}
\includegraphics[width=0.32\textwidth]{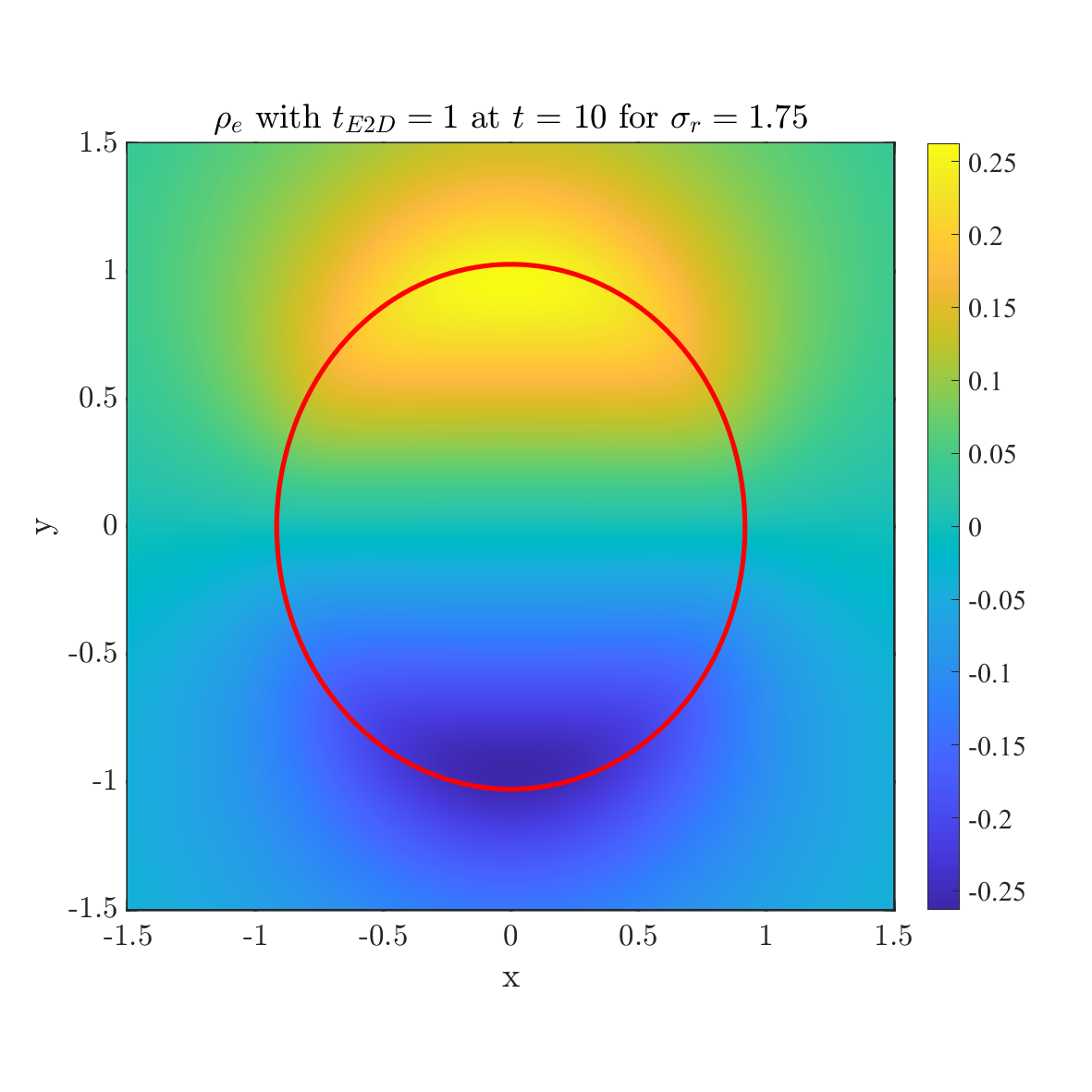}
\includegraphics[width=0.32\textwidth]{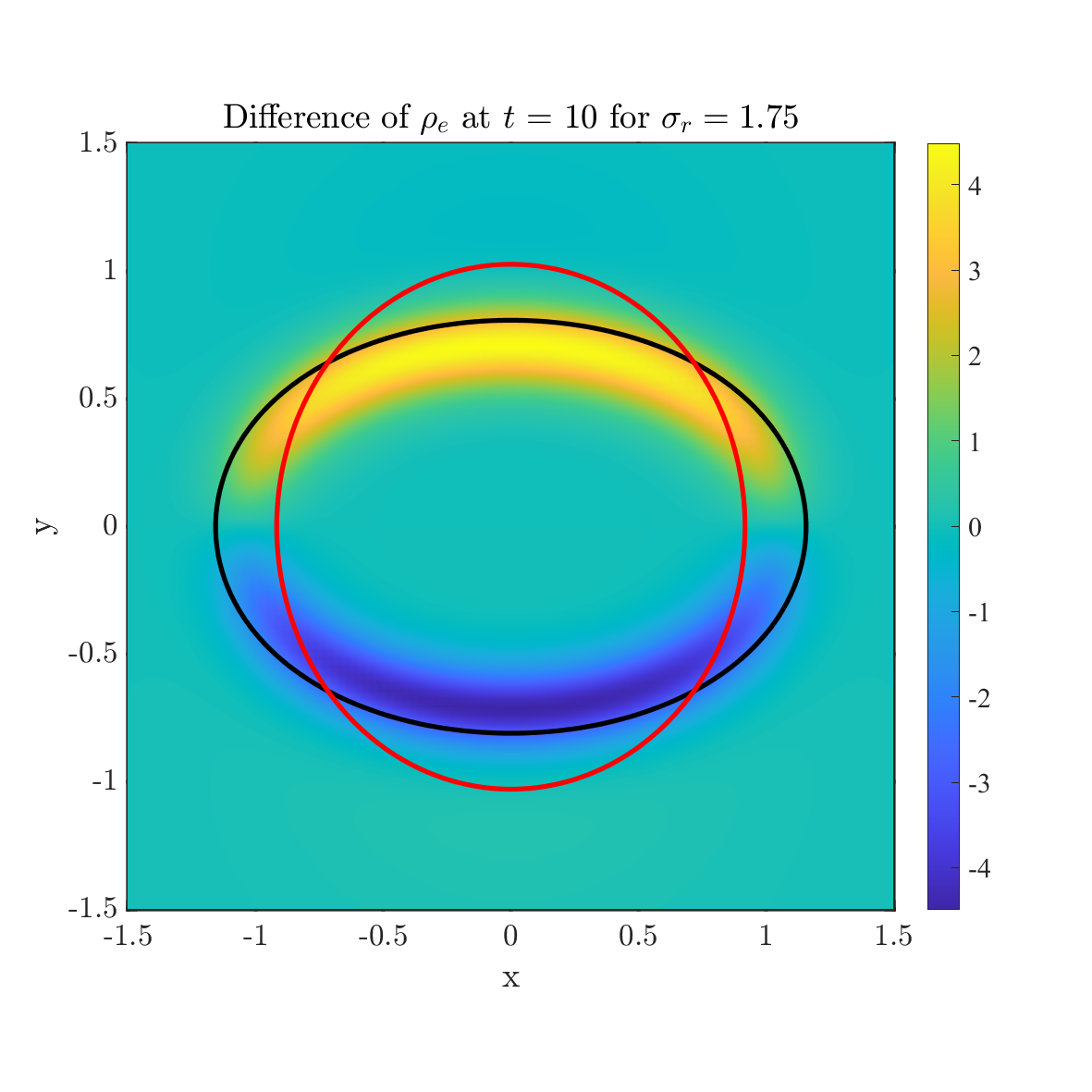}
\includegraphics[width=0.32\textwidth]{no_correction_175_no_Cm.png}
\includegraphics[width=0.32\textwidth]{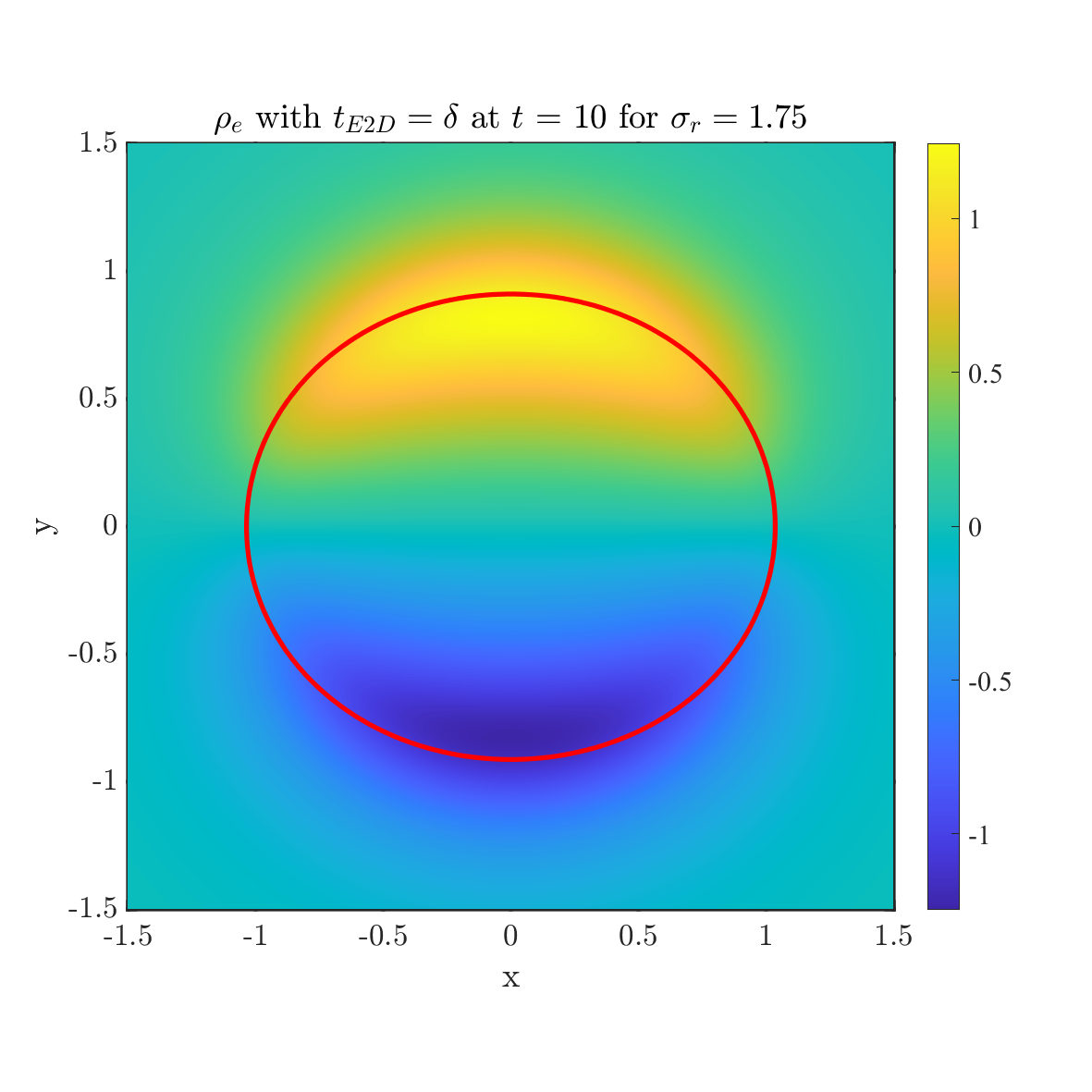}
\includegraphics[width=0.32\textwidth]{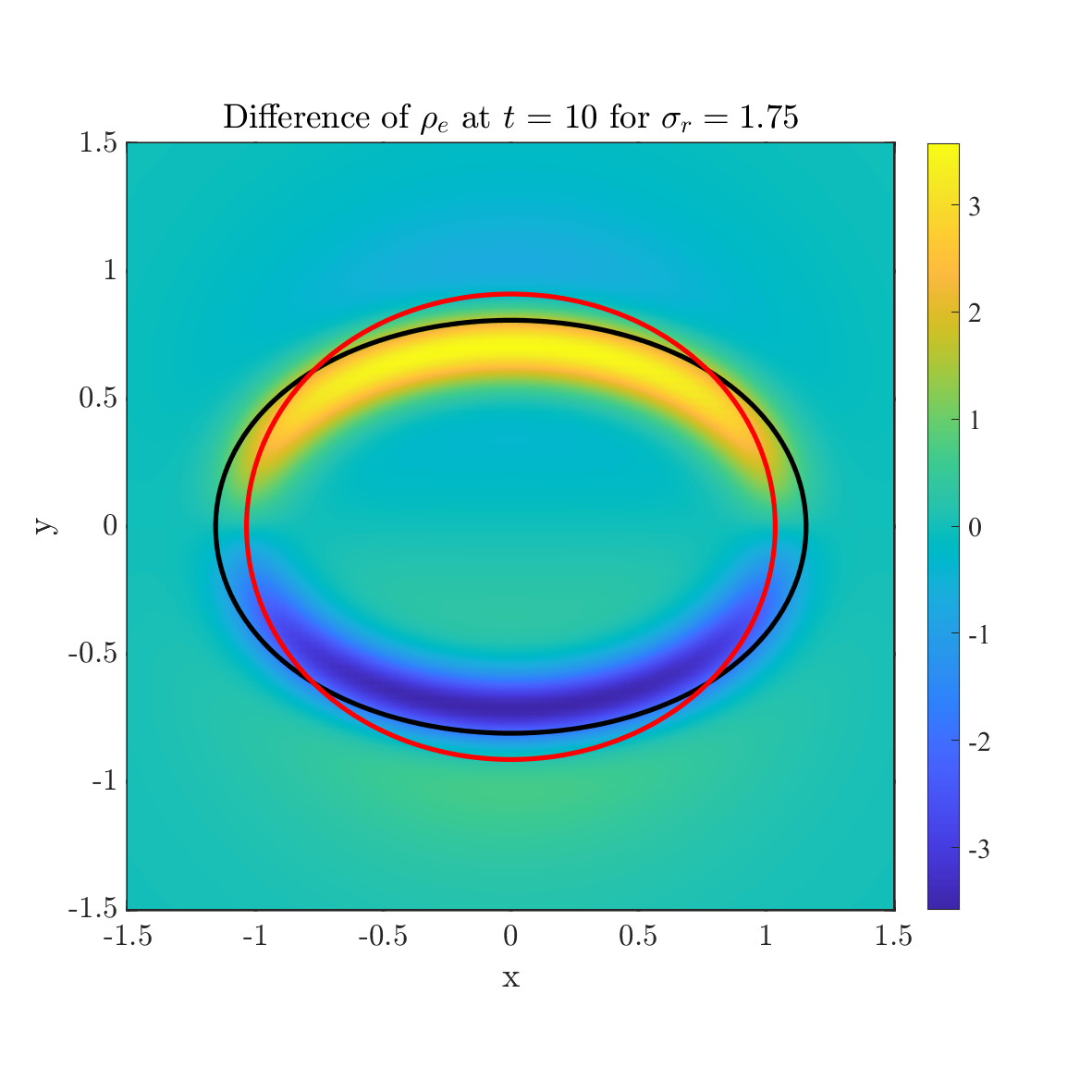}
\includegraphics[width=0.32\textwidth]{no_correction_175_no_Cm.png}
\includegraphics[width=0.32\textwidth]{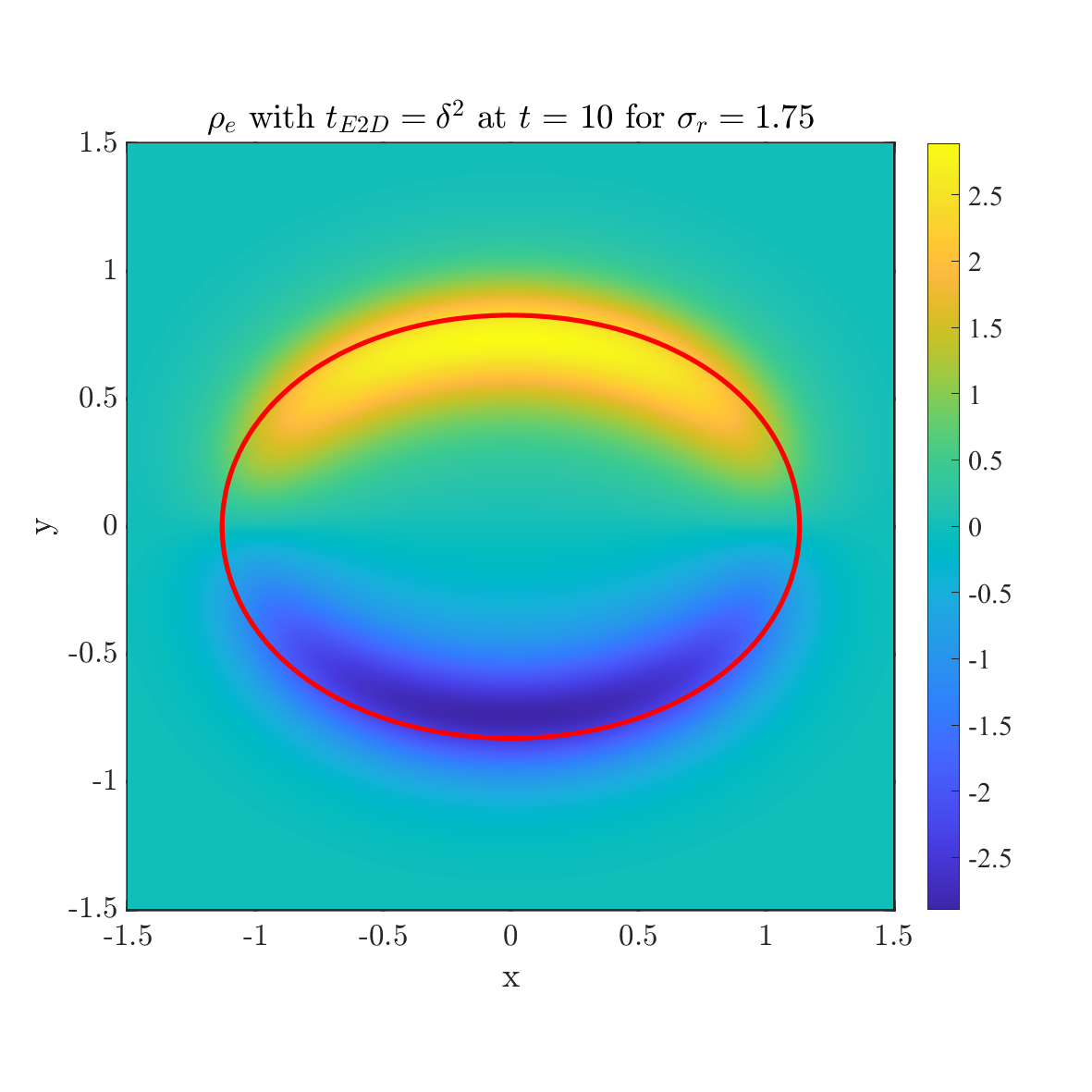}
\includegraphics[width=0.32\textwidth]{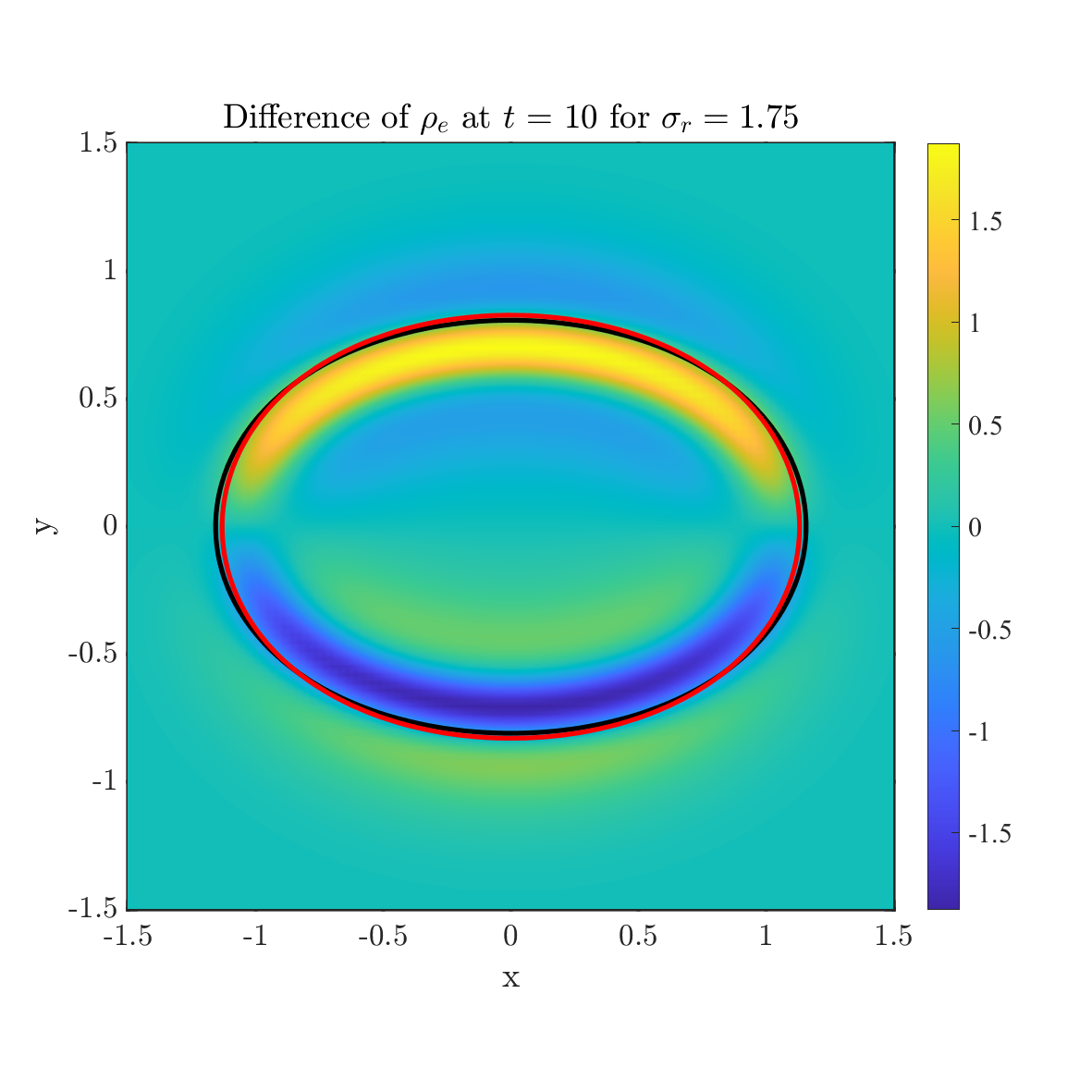}
\end{center}
\caption{The comparison between leaky dielectric model (left) and net charge model (middle) at $t = 10$. 
The conductivity ratio is $\sigma_{r} = 1.75$. 
And the difference between them is shown in the right column. 
relaxation time $t_{E2M} = t_{E2D} = 1$ (top), $t_{E2M} = t_{E2D} =\delta$ (middle) and $t_{E2M} = t_{E2D} = \delta^{2}$ (bottom) are considered here. 
In each figure, the solid line shows the zero level set ($\psi=0$) where the black line shows the drop shape without correction 
and the red line shows the drop shape with correction. 
The rest parameters are chosen as $\epsilon_{r} = 3.5$, $Ca_{E} = 1$.}
\label{fig: correction2_175}
\end{figure}

Firstly, we compare the phase field leaky dielectric model with a net charge model, both as approximations of the original PNP-NS-CH model, where  the capacitance on the diffuse interface is not considered.
In Figs.   \ref{fig: correction2_175}, \ref{fig: correction2_325} and \ref{fig: correction2_475}, 
the influence of different time scale $t_{E2M} = 1, ~\delta,~ \delta^{2}$  on droplets profiles and charge densities are shown 
for three conductivity ratio $\sigma_{r} = 1.75,~3.25,~4.75$, respectively. 
In each figure, the left column is the charge density of the leaky dielectric model with black lines for the  interfaces, 
the second column is the charge density of the net charge model with red lines for the interfaces and last columns is the difference between two models.  

\begin{figure} 
\begin{center}
\includegraphics[width=0.3\textwidth]{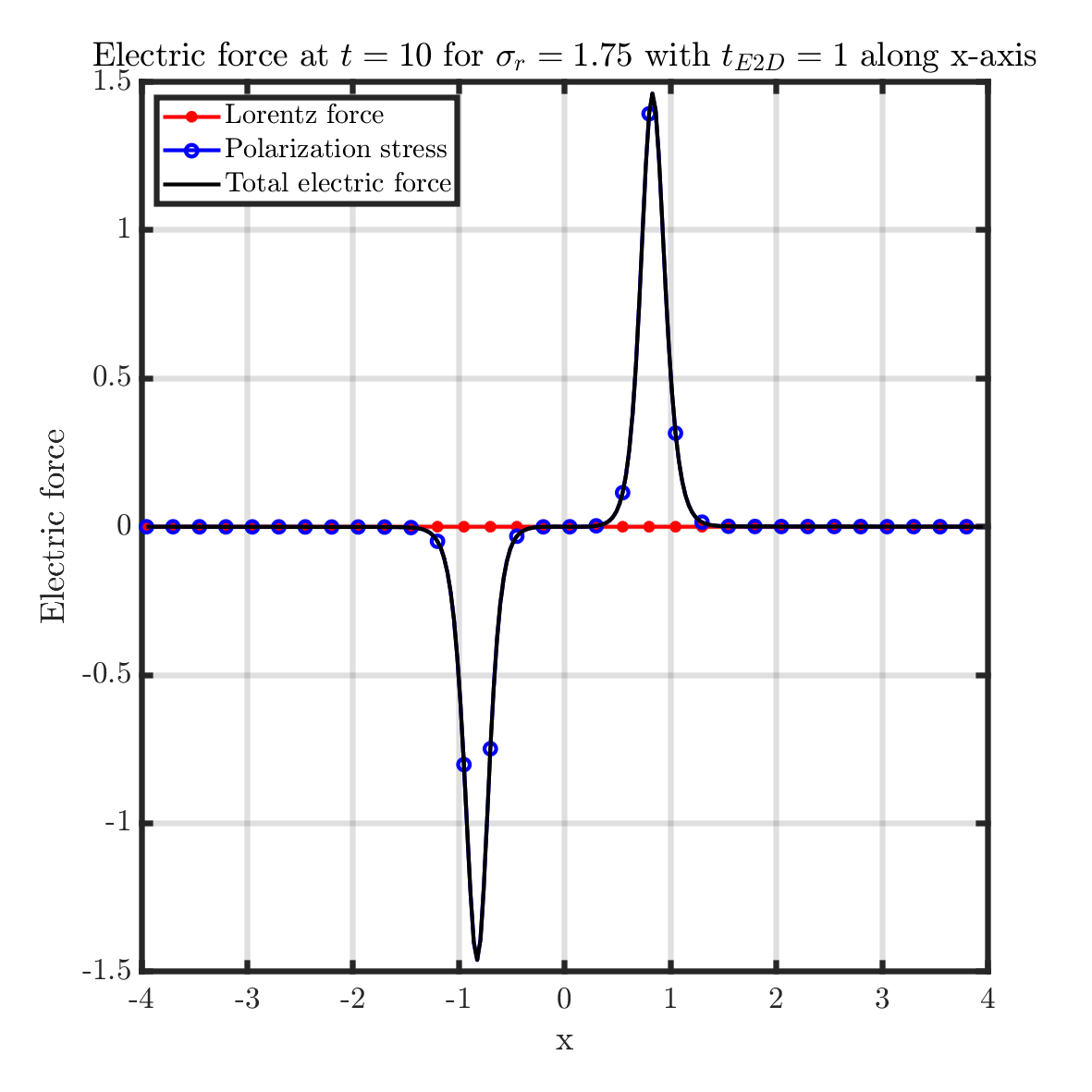}
\includegraphics[width=0.3\textwidth]{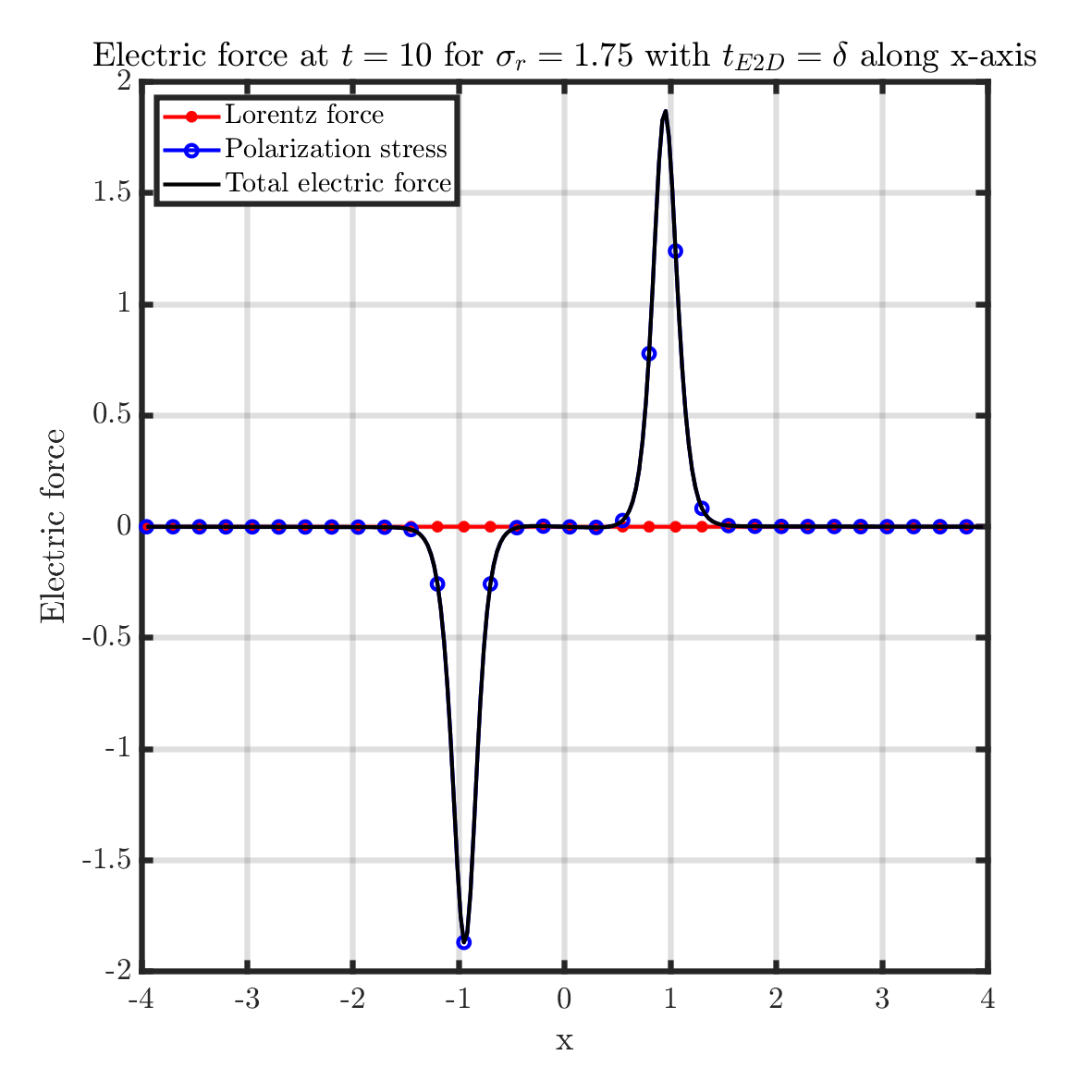}
\includegraphics[width=0.3\textwidth]{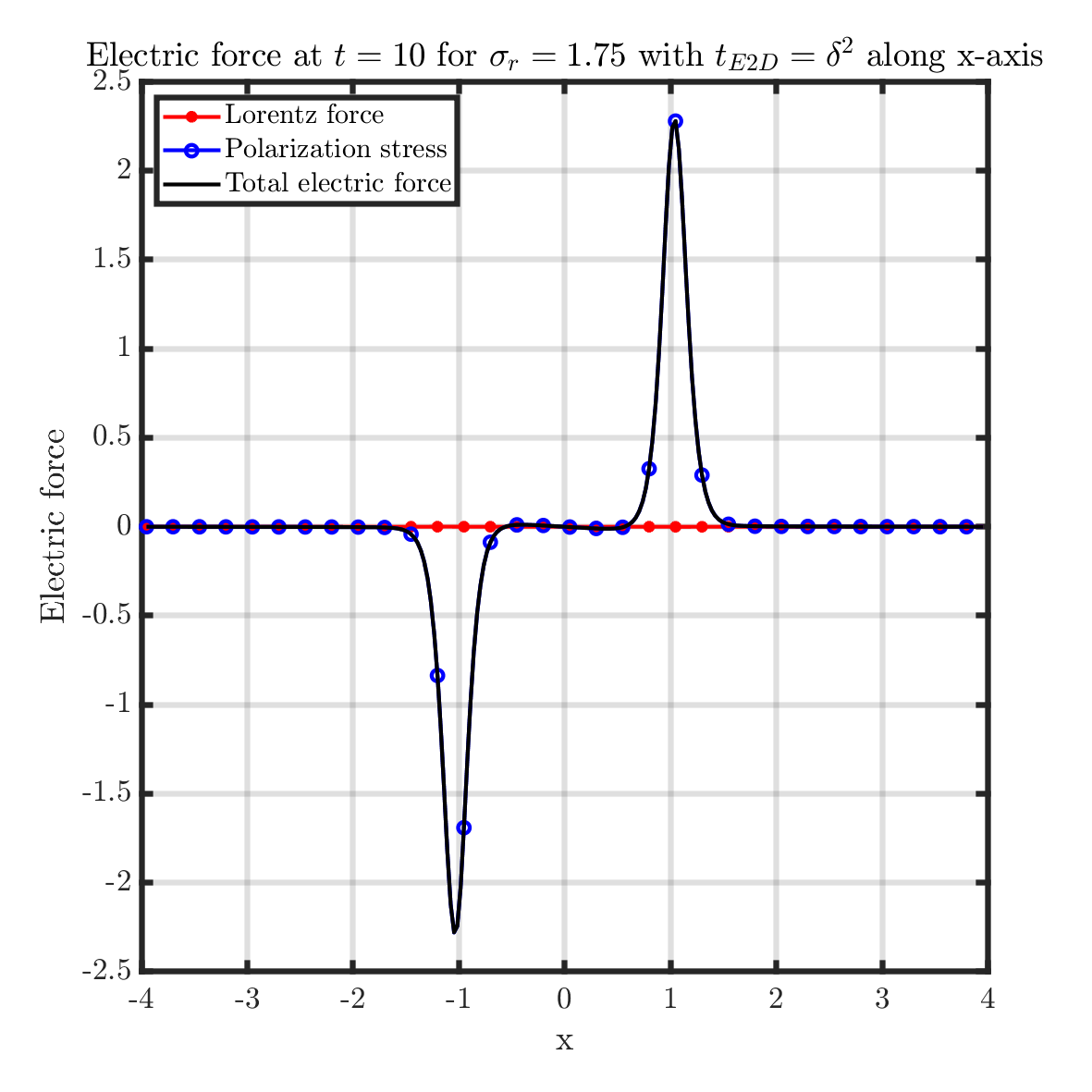} 
\includegraphics[width=0.3\textwidth]{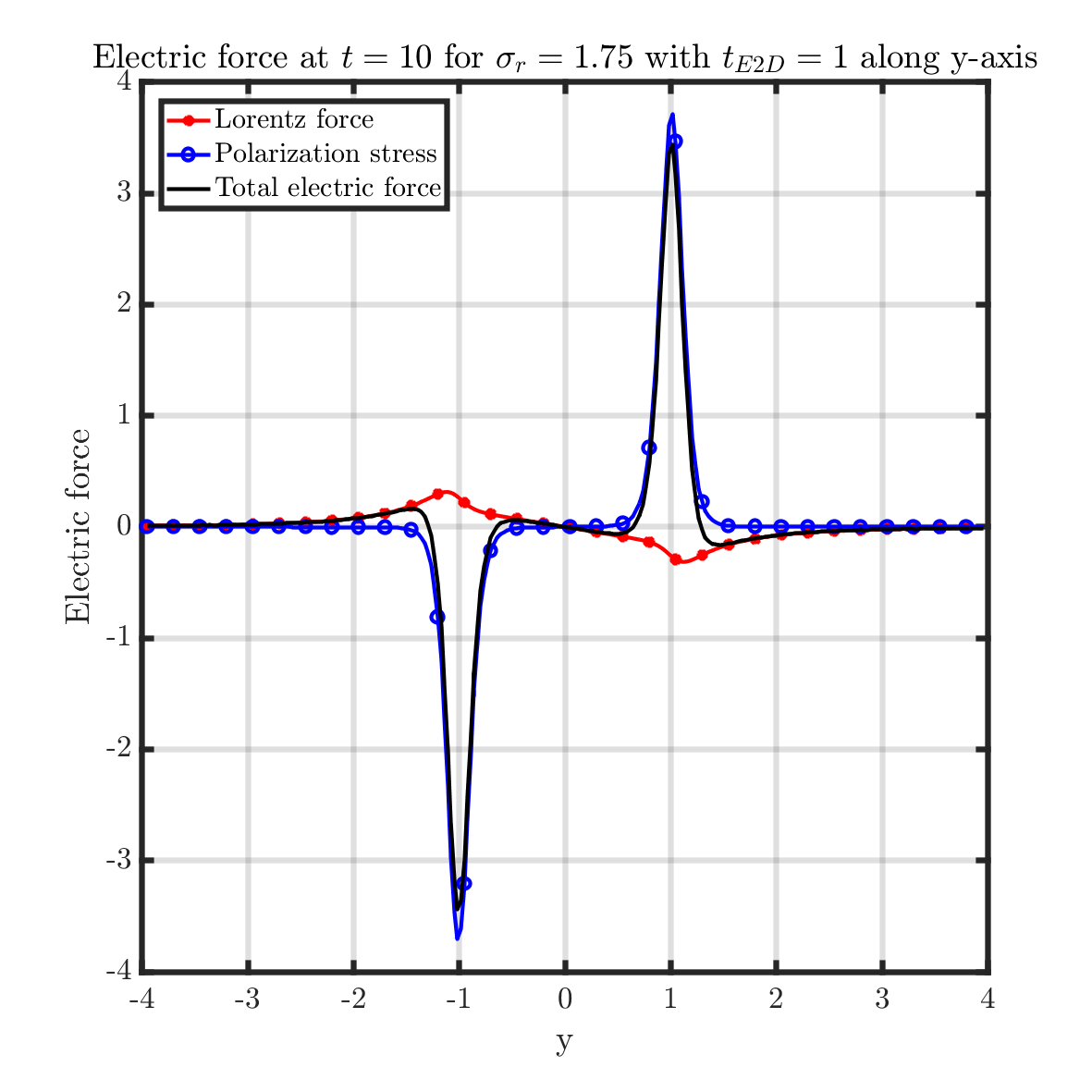}
\includegraphics[width=0.3\textwidth]{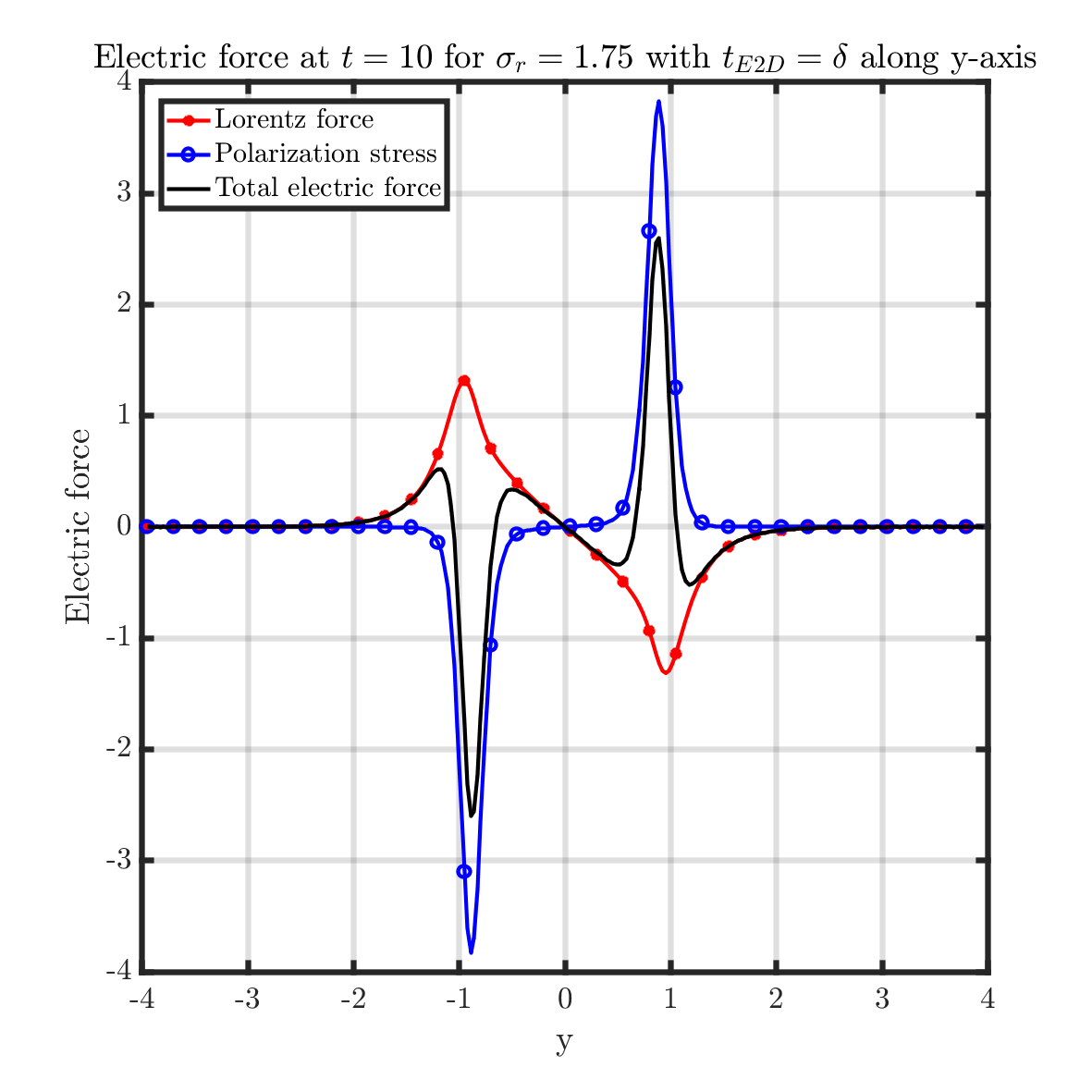}
\includegraphics[width=0.3\textwidth]{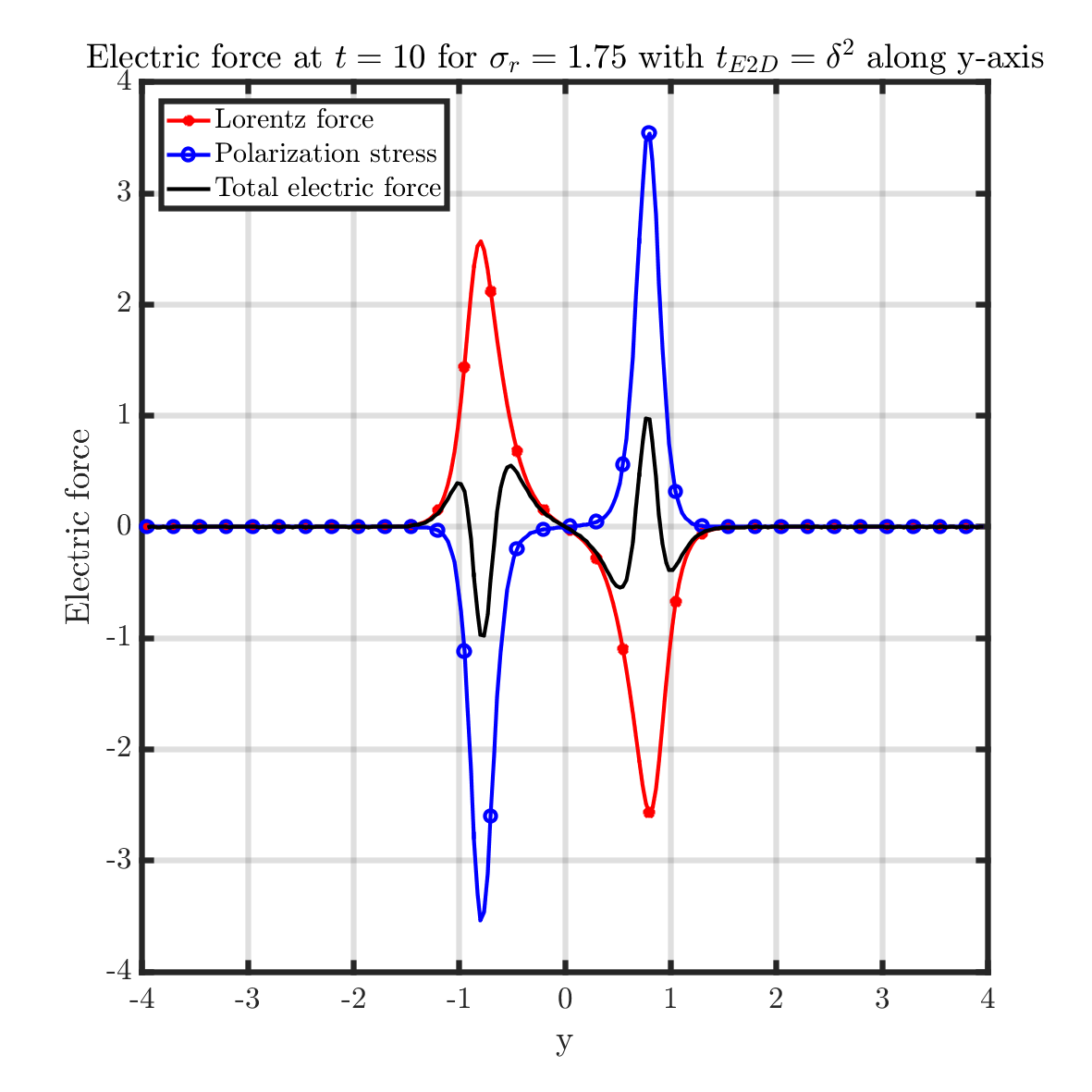} 
\end{center}
\caption{The electric force along x-axis (top) and y-axis (bottom) for $t_{E2D} = 1$ (left), $t_{E2D} = \delta$ (middle) and $t_{E2D} = \delta^{2}$ (right) about conductivity ratio $\sigma_{r} = 1.75$ at $t = 10$. 
In each figure, the red solid line with star symbol shows the Loerentz force, the blue solid line with circle shows the polarization stress and the black solid line shows the total electric force. 
The rest parameters are chosen as $\epsilon_{r} = 3.5$, $Ca_{E} = 1$.}
\label{fig: electric force correction2_175}
\end{figure}

For all three $\sigma_r$ cases, it is confirmed that the leaky dielectric model provides a reasonable approximation of the net charge model when the electric relaxation is fast, i.e., $t_{E2M} = \delta^{2}$. However, as the time ratio increases, the difference between the two models becomes larger at equilibrium due to the diffusion effect. Particularly, for the case of $\sigma_r = 1.75$, the droplet is prolate in the net charge model. The electric forces along the x-axis and y-axis are illustrated in Fig. \ref{fig: electric force correction2_175} for $t_{E2M} = 1$ (left), $t_{E2M} = \delta$ (middle), and $t_{E2M} = \delta^{2}$ (right). As before, the Lorentz force is negligible, and the polarization force induces expansion in the x-axis direction outward. However, in the y-axis direction, the compressed Lorentz force is weakened, and the polarization force dominates in all three cases, when the diffusion is considered, compared with the leaky dielectric model. The larger the $t_{E2M}$, the smaller the $F_L$. When the time ratio is $t_{E2M} = 1$, the expansion along the y-axis is larger than that along the x-axis, and the drop shape is prolate, which is completely different from the equilibrium in the leaky dielectric case. For small $t_{E2M}$, the equilibrium profile is oblate, which is the same as in the leaky-dielectric model, but for different reasons. In the leaky-dielectric case, it is a result of compression in the y-direction and expansion in the x-direction. When the full dynamics are considered, it is the outcome of competition between the expansion in two directions.   

%We can see that the net charge has a wider diffusion range which meets our expectation for that a larger $t_{E2M}$ causes faster diffusion of the net charge. 
%As for deformation, when we take a large $t_{E2M}$,  the direction of deformation for $\sigma_{r} = 1.75$ is changed from oblate to prolate. 
%The deformation is almostly the same for $\sigma_{r} = 3.25$ no matter which $t_{E2M}$ we take. 
%For $\sigma_{r} = 4.75$, different values of $t_{E2M}$ will not cause the direction change of deformation, 
%but there is a small deformation with small $t_{E2M}$. 

Figs \ref{fig: correction2_175_with_cm}-\ref{fig: correction2_475_with_cm} perform the behavior of drop shapes 
and distributation of the net charge for conductivity ratio $\sigma_{r} = 1.75$, $\sigma_{r} = 3.25$ and $\sigma_{r} = 4.75$ 
by considering there is no capacitance on the interface (left) and there is capacitance on the interface (middle) about the net charge model. 
And the difference between them is shown on the right column. 
Comparing with the leaky dirlectric model, the net charge model causes a lower charge accumulation on the interface due to the effect of diffusion. 
The capacitance effect which leads to a smaller capacitance causes a smaller deformation is also obtained for the net charge model. 
This result is consistent with the conclusion in section \ref{sec: capacitance single drop}. 

Figs \ref{fig: correction2_175_Cm_1_tE}-\ref{fig: correction2_175_Cm_2_tE} present the effect of relaxation time for the net charge model 
where capacitances $C_{m} = 1$ and $C_{m} = \delta^{2}$ are added to the interface, respectively. 
We choose $t_{E2D} = \delta^{2}$ as the reference and the relaxation time effect is consistent with the results shown in \ref{fig: correction2_175}. 

\section{Conclusions}\label{sec: conclusions}
In this paper, a mathematical model is developed to describe the deformation of droplets under the influence of electric field 
with different electric permittivity and conductivity, using the energy variation method. 
Specifically, the capacitance of the two-phase interface is considered using the harmonic average and the phase field function. 
The diffusive-interface leaky dielectric model is obtained when the electric relaxation time is sufficiently fast. 
After conducting careful asymptotic analysis, the sharp interface limit of this model is found to be consistent with the existing sharp interface model.

To demonstrate the effectiveness of the proposed model, a variety of numerical experiments are carried out, 
including the convergence test, comparison with previous sharp interface models, and deformations with topology change. 
The equilibrium profile of leaky dielectric droplets under a static electric field is mainly determined by the competition 
between Lorentz force and polarization stress. 
The former is induced by the accumulation of net charge, while the latter is induced by the variation of electric permittivity near the interface. 
The distribution of net charge is determined by the ratio between permittivity and conductivity $\frac{\epsilon_r}{\sigma_r}$. 
If it is greater than one, positive charges are accumulated near the side with higher potential and droplet is compressed to be oblate. 
While if it smaller than one, positive charges are accumulated near the low potential side and the droplets is elongated into prolate shape.   

Furthermore, the effect of capacitance on the interface is studied, 
revealing that the presence of counter-ions on the opposite side of the interfaces decreases the deformability of droplets. 
Finally, the impact of time scales on the deformation is discussed when the dynamics of net charge density $\rho_e$ is considered. 
The results confirm that the leaky dielectric model is a reasonable approximation when the electric time scale is fast enough. 
However, when the electric relaxation time and macro time scale are comparable, 
the diffusion of free charges leads to different droplet deformations due to the decrease in the Lorentz force with dispersed distribution of net charge.

Here, we have discussed the difference between the leaky dielectric model and the net charge model. In the next step, we plan to use an efficient numerical algorithm to directly compare the net charge model with the leaky dielectric model. Additionally, we note that in the leaky dielectric model, the electric potential is continuous with different slope across the interface. However, for a vesicle, the potential is discontinuous due to the capacitance and resistance effects of the membrane, and the total current is conserved across the interface. Building on the similar ideas presented in our previous work \cite{Qin2022111334}, we plan to extend our model to study vesicle electrohydrodynamics \cite{hu2016vesicle}. 

Overall, this paper presents a comprehensive analysis of the phase field leaky dielectric model for electrohydrodynamics in droplet systems. Through theoretical derivations and numerical experiments, we provide valuable insights into the behavior and characteristics of these systems, advancing our understanding in this field.

%
%\section{Citations and references}
%All papers included in the References section must be cited in the article, and vice versa. Citations should be included as, for example ``It has been shown \citep{Rogallo81} that...'' (using the {\verb}\citep}} command, part of the natbib package) ``recent work by \citet{Dennis85}...'' (using {\verb}\citet}}).
%The natbib package can be used to generate citation variations, as shown below.\\
%\verb#\citet[pp. 2-4]{Hwang70}#:\\
%\citet[pp. 2-4]{Hwang70} \\
%\verb#\citep[p. 6]{Worster92}#:\\
%\citep[p. 6]{Worster92}\\
%\verb#\citep[see][]{Koch83, Lee71, Linton92}#:\\
%\citep[see][]{Koch83, Lee71, Linton92}\\
%\verb#\citep[see][p. 18]{Martin80}#:\\
%\citep[see][p. 18]{Martin80}\\
%\verb#\citep{Brownell04,Brownell07,Ursell50,Wijngaarden68,Miller91}#:\\
%\citep{Brownell04,Brownell07,Ursell50,Wijngaarden68,Miller91}\\
%\citep{Briukhanovetal1967}\\
%\cite{Bouguet01}\\
%\citep{JosephSaut1990}\\
%
%The References section can either be built from individual \verb#\bibitem# commands, or can be built using BibTex. The BibTex files used to generate the references in this document can be found in the JFM {\LaTeX} template files folder provided on the website \href{https://www.cambridge.org/core/journals/journal-of-fluid-mechanics/information/author-instructions/preparing-your-materials}{here}.
%
%Where there are up to ten authors, all authors' names should be given in the reference list. Where there are more than ten authors, only the first name should appear, followed by {\it {et al.}}

%\backsection[Supplementary data]{\label{SupMat}Supplementary material and movies are available at \\https://doi.org/10.1017/jfm.2019...}

%\backsection[Acknowledgements]{Acknowledgements may be included at the end of the paper, before the References section or any appendices. Several anonymous individuals are thanked for contributions to these instructions.}

\section*{Acknowledgment}
This work is partly  supported by the National Natural Science Foundation 	of China (No. 12071190, 12201369, 12231004) and Natural Sciences and Engineering Research Council of Canada (NSERC).
%\backsection[Declaration of interests]{  The authors report no conflict of interest.}

%\backsection[Data availability statement]{The data that support the findings of this study are openly available in [repository name] at http://doi.org/[doi], reference number [reference number]. See JFM's \href{https://www.cambridge.org/core/journals/journal-of-fluid-mechanics/information/journal-policies/research-transparency}{research transparency policy} for more information}

%\section[Author ORCIDs]{Authors may include the ORCID identifers as follows.  Y. Qin, https://orcid.org/0000-0002-0027-8470; S. Xu, https://orcid.org/0000-0002-8207-7313.}
%
%\backsection[Author contributions]{Zheyu Qin: Model derivation; Simulations; Manuscript draft and revision. Huaxiong Huang: Project design; Manuscript revision. Zilong Song: Model derivation.   Shixin Xu: Model derivation; Project design and coordination; Manuscript draft and revision.}

\newpage
\appendix

\section{}\label{appA}

\begin{figure}
\begin{center}
\includegraphics[width=0.4\textwidth]{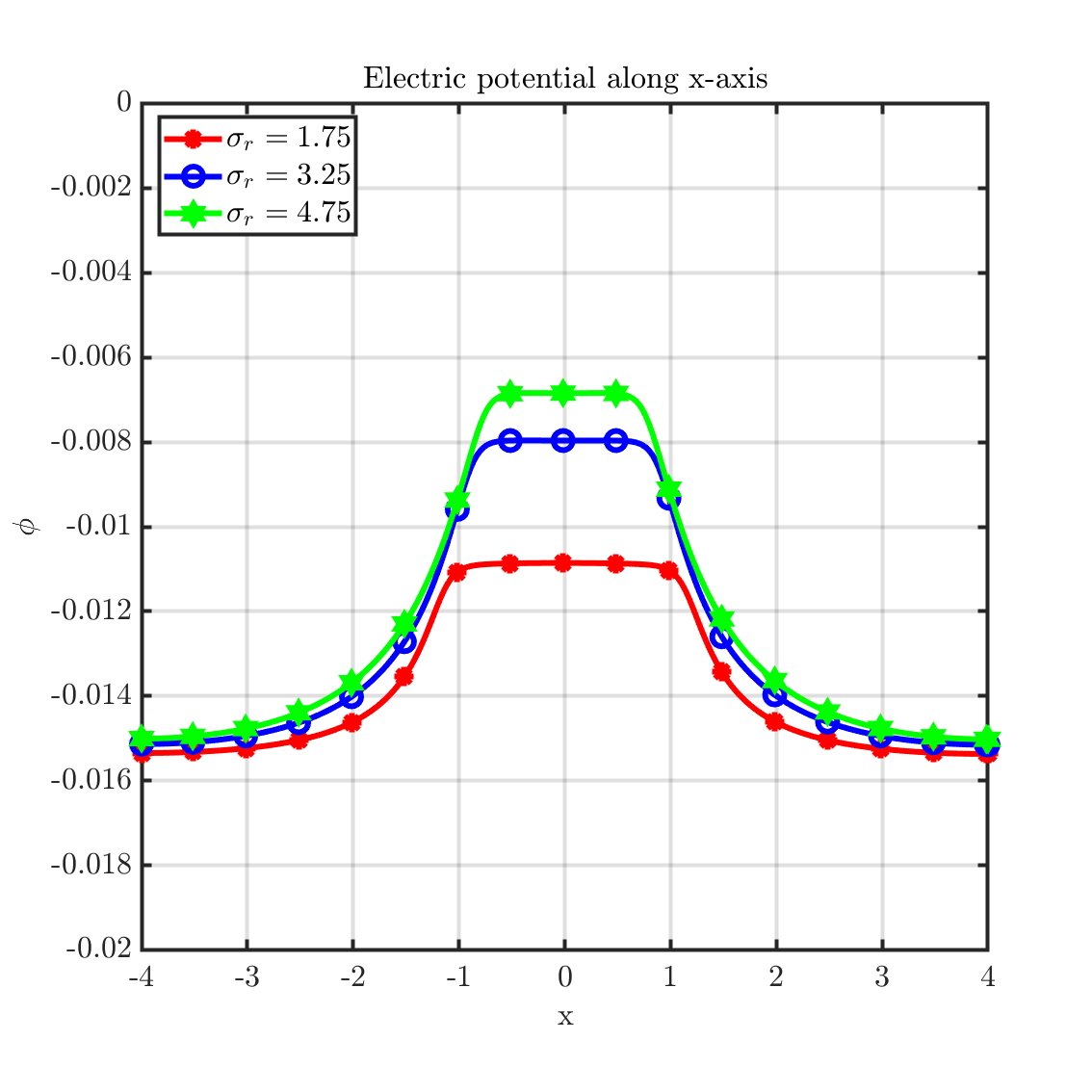}
\includegraphics[width=0.4\textwidth]{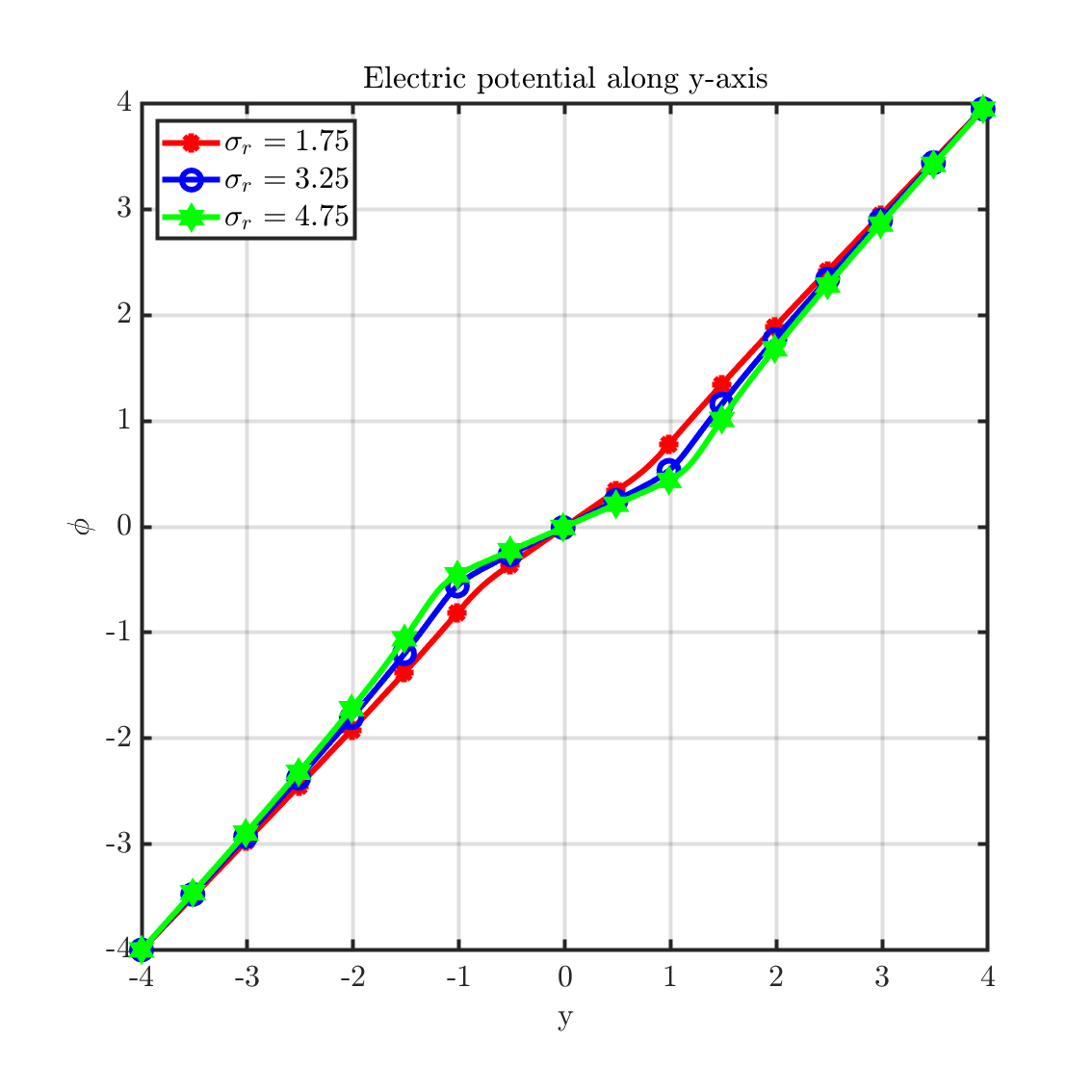}
\includegraphics[width=0.3\textwidth]{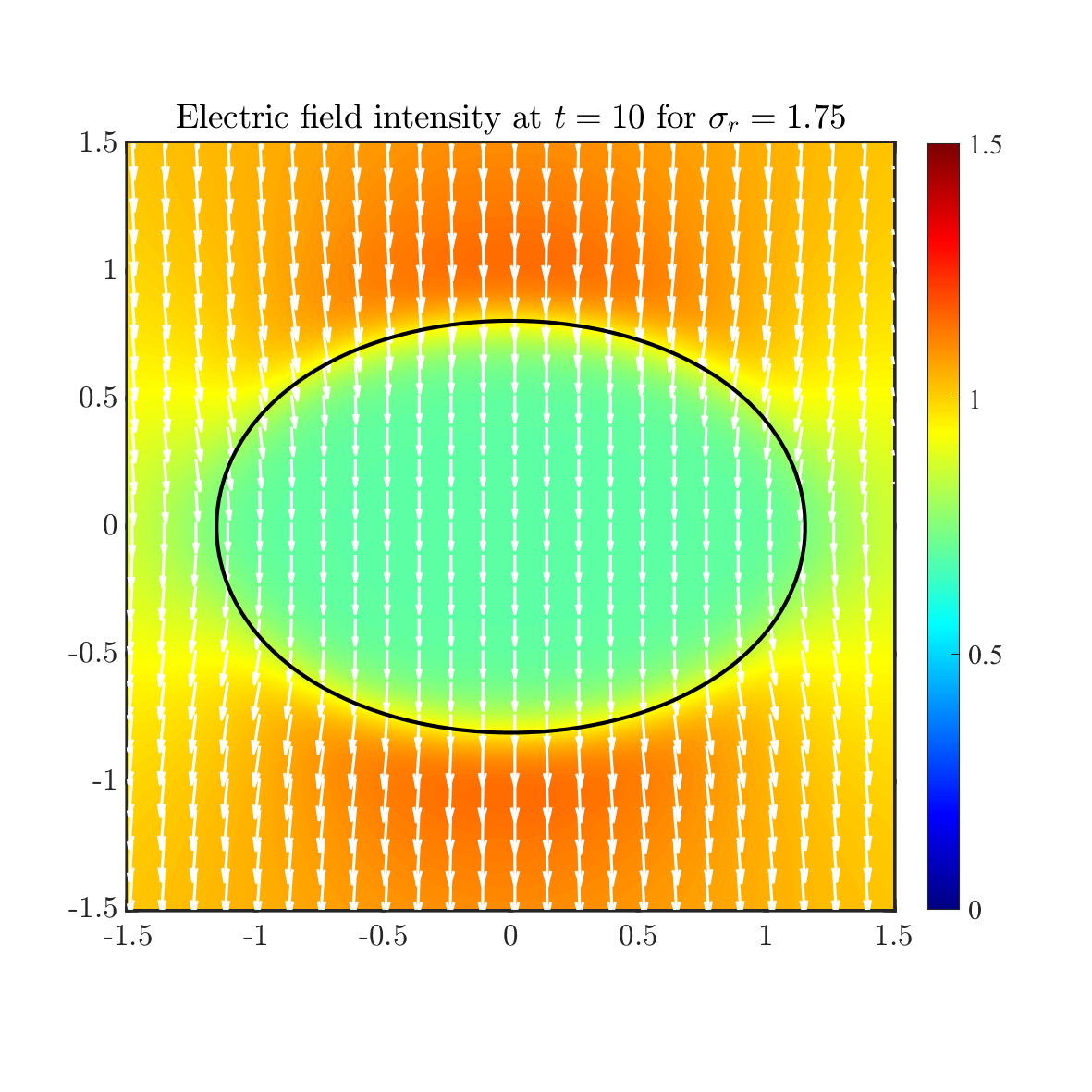}
\includegraphics[width=0.3\textwidth]{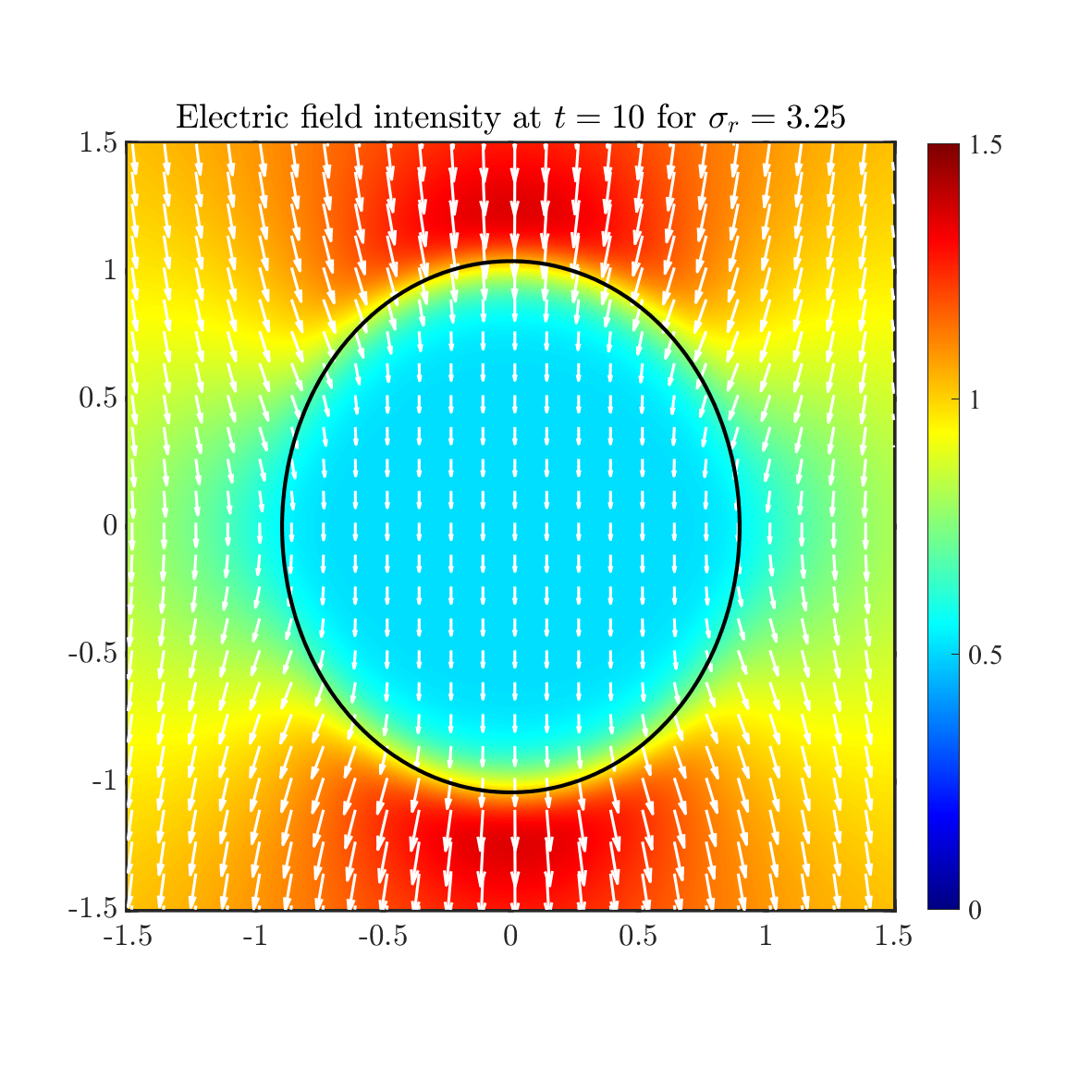}
\includegraphics[width=0.3\textwidth]{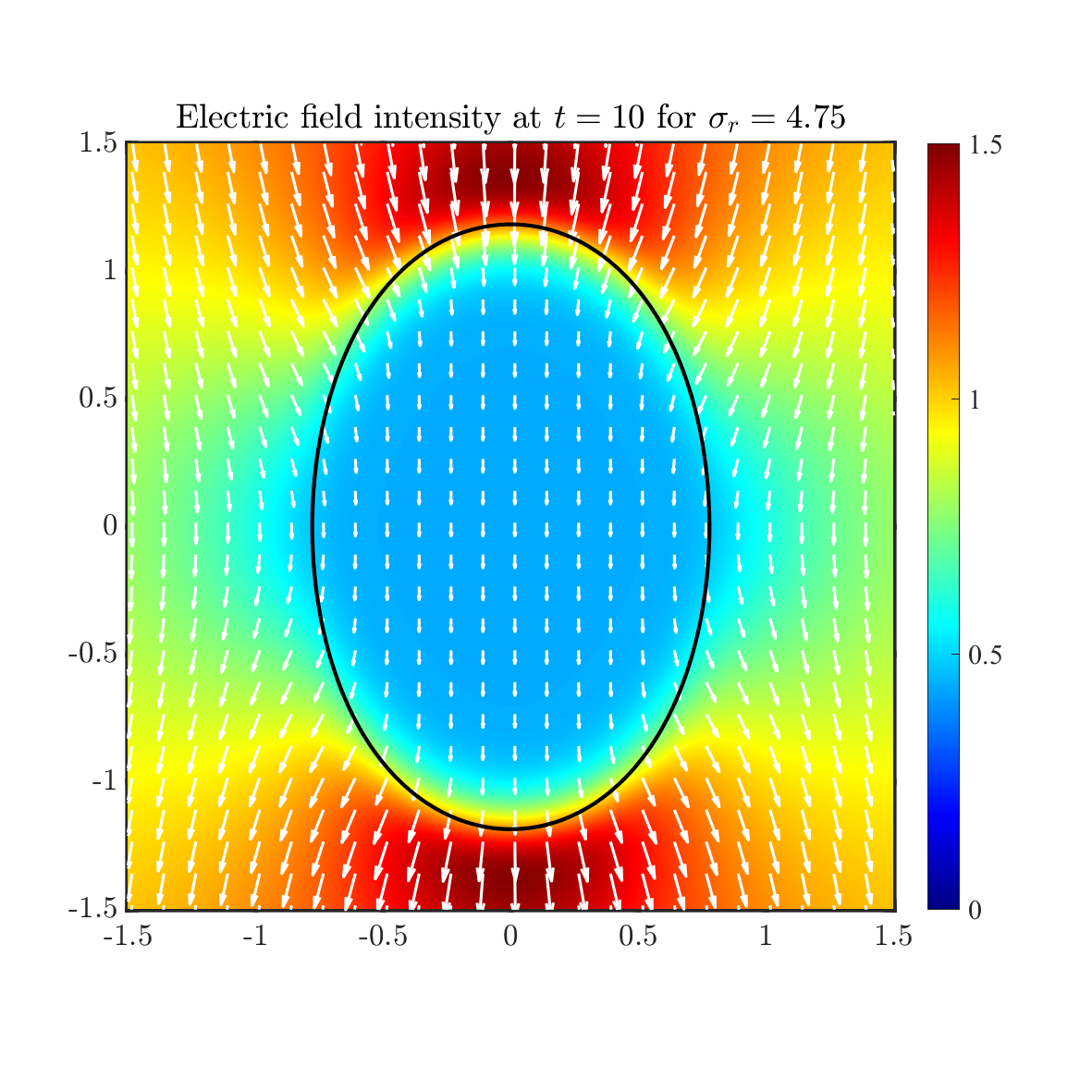}
\end{center} 
\caption{Top: electric potential distribution along $y=0$ (left) and $x=0$ (right); Bottom: electric field distribution in space with different $\sigma_r$. }
\label{fig: electric field intensity}
\end{figure}

\begin{figure}
\begin{center}
\includegraphics[width=0.245\textwidth]{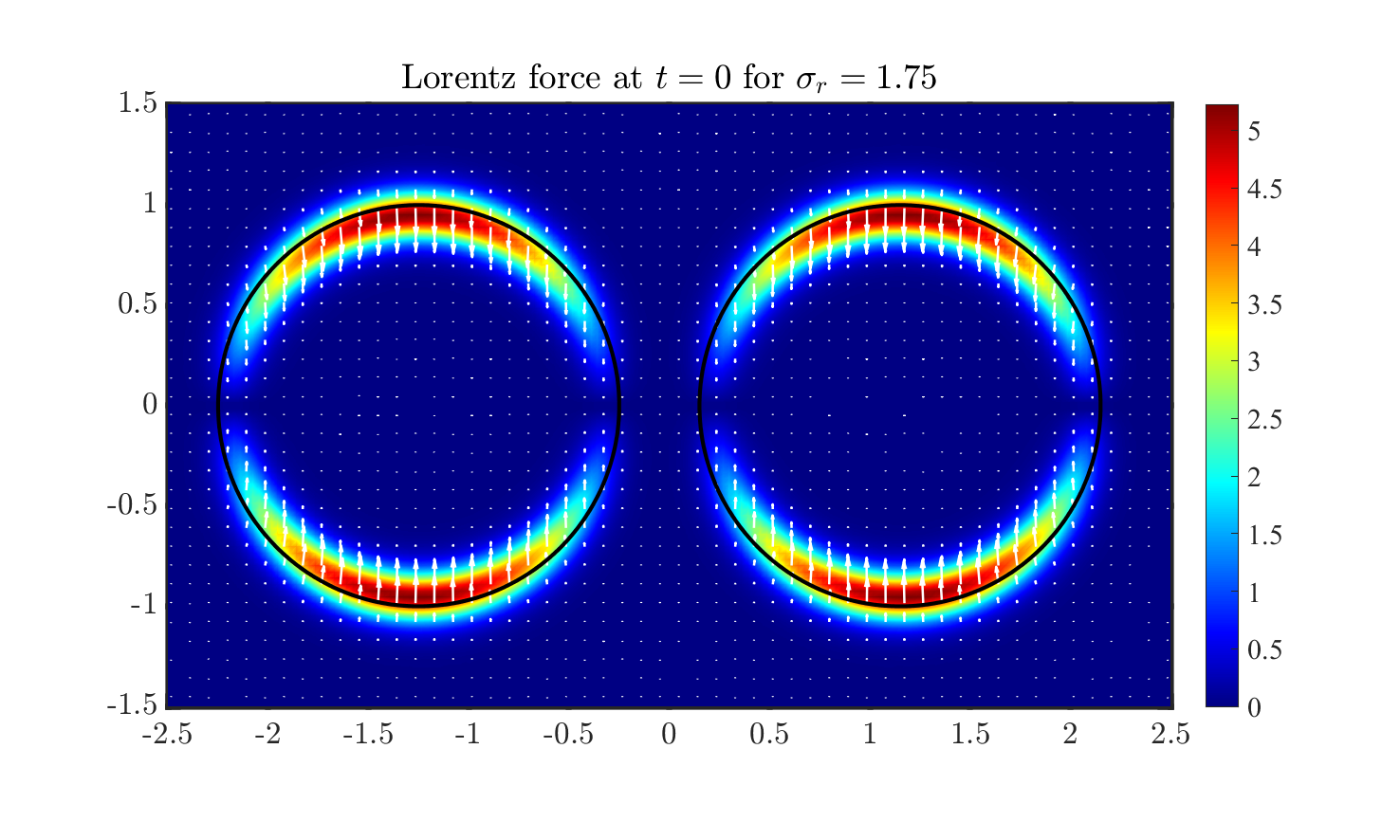}
\includegraphics[width=0.245\textwidth]{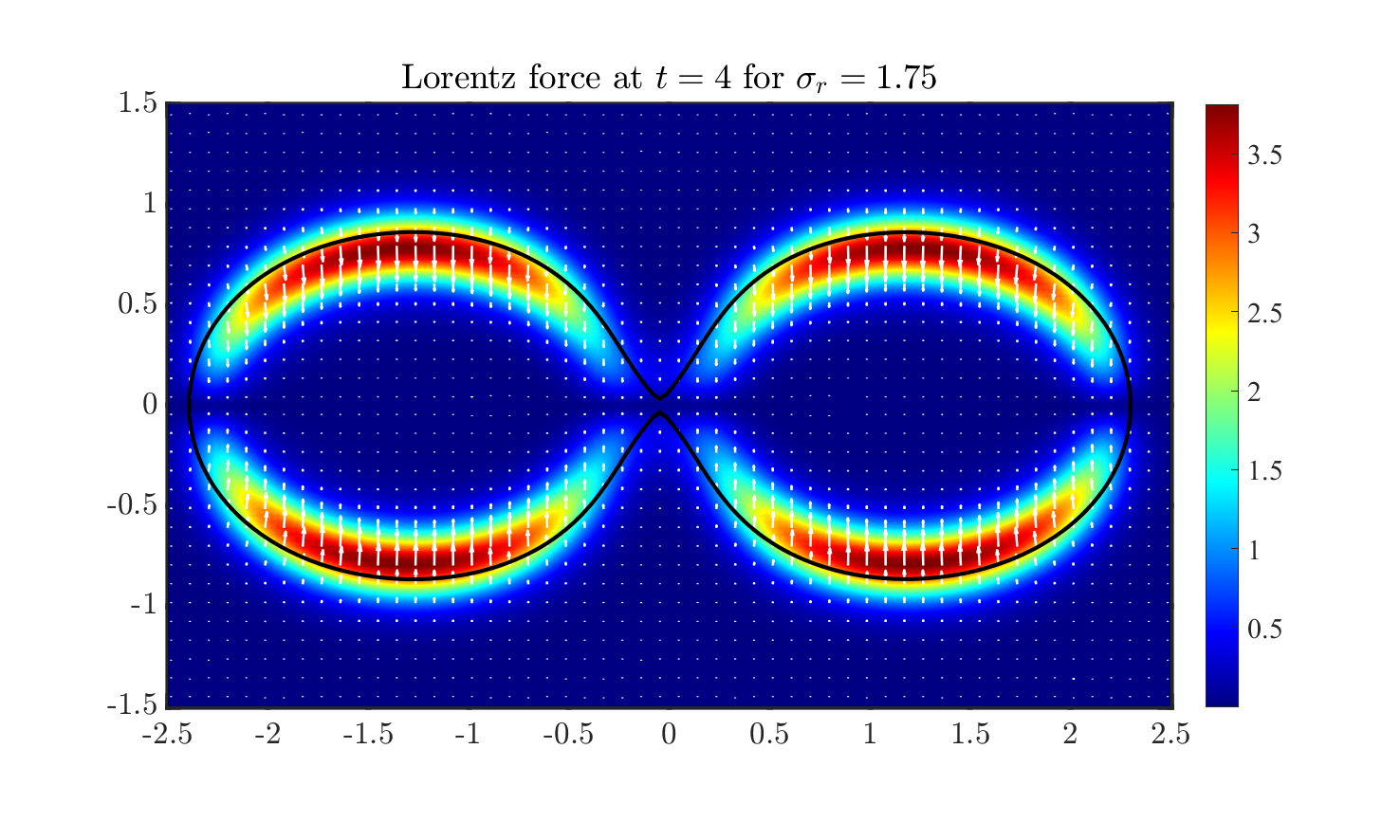}
\includegraphics[width=0.245\textwidth]{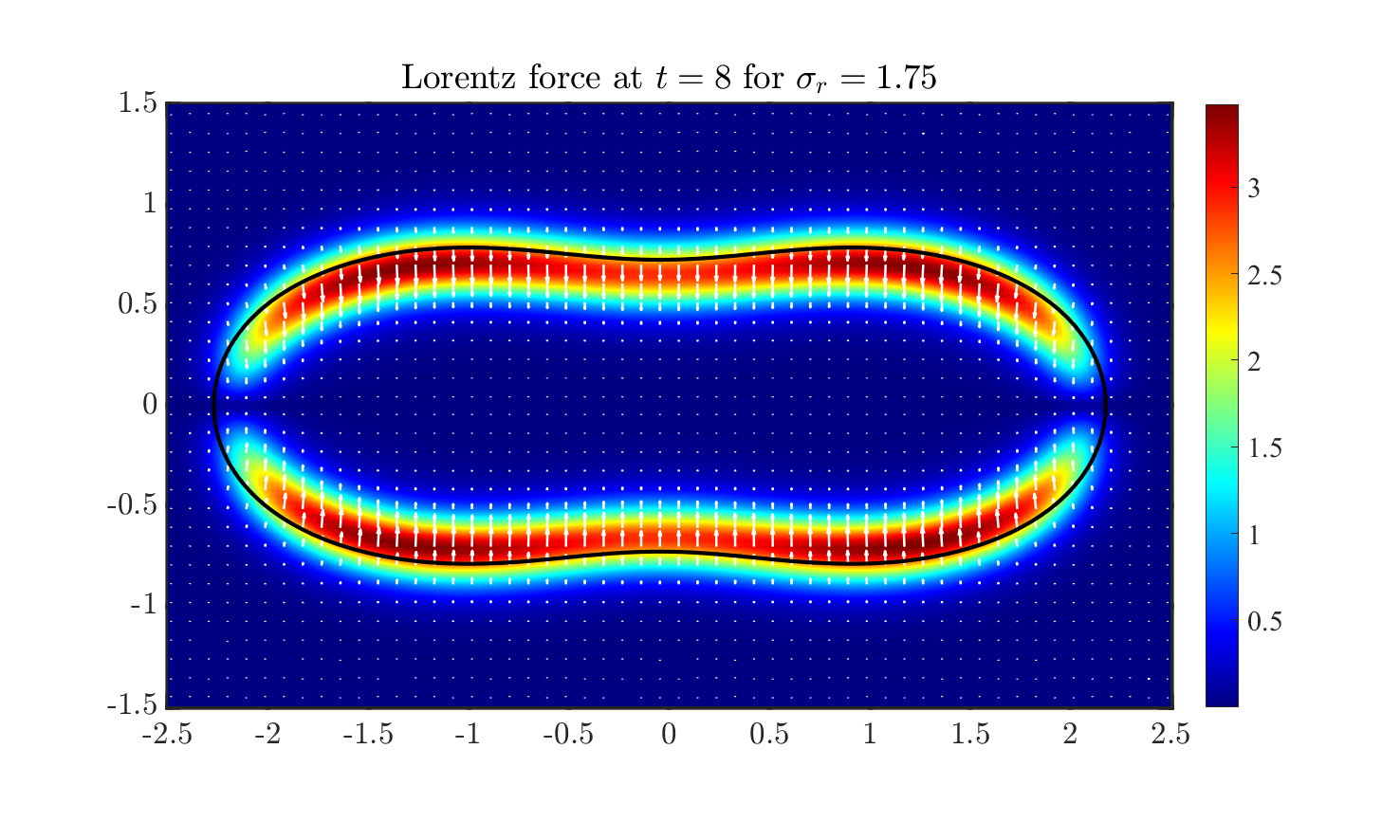}
\includegraphics[width=0.245\textwidth]{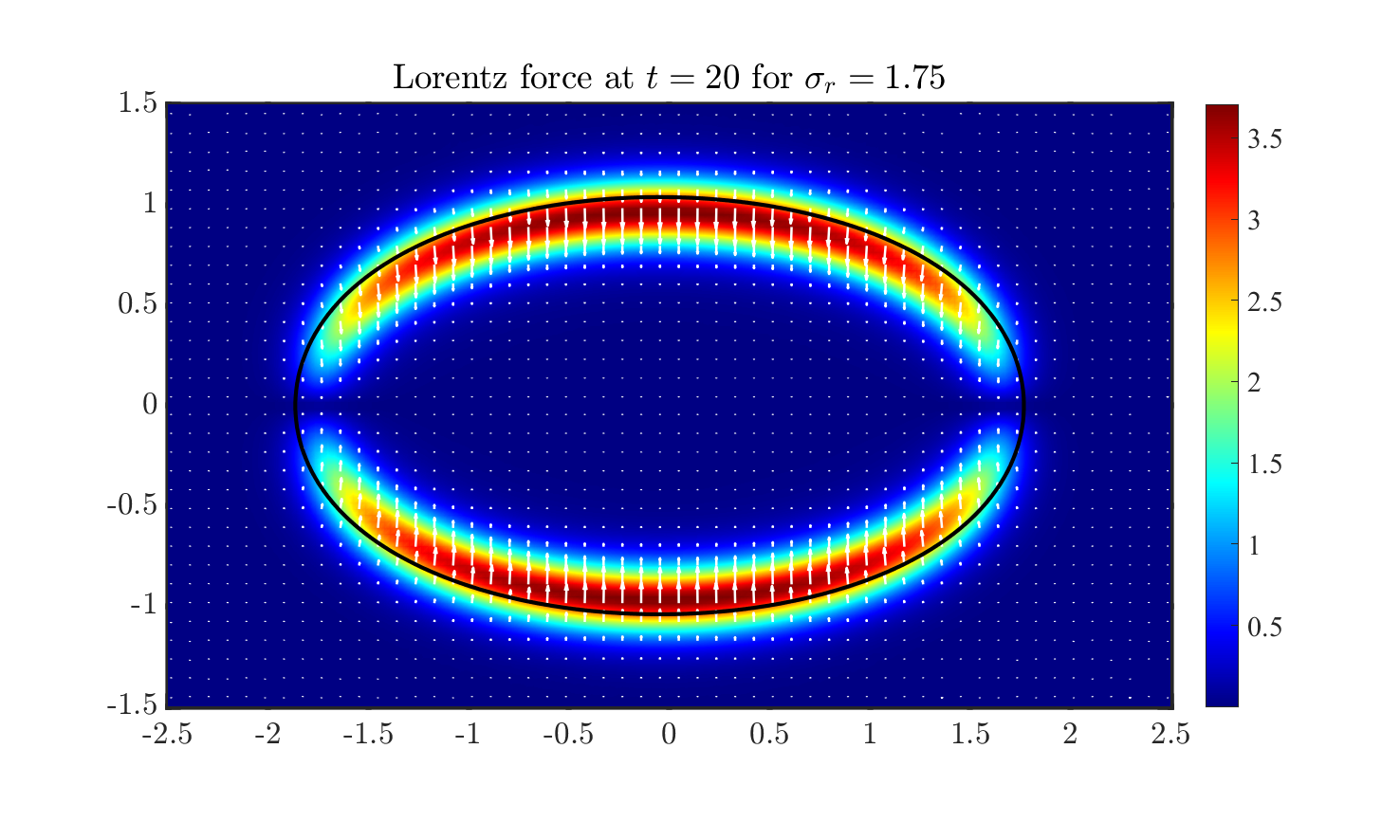}\\
\includegraphics[width=0.245\textwidth]{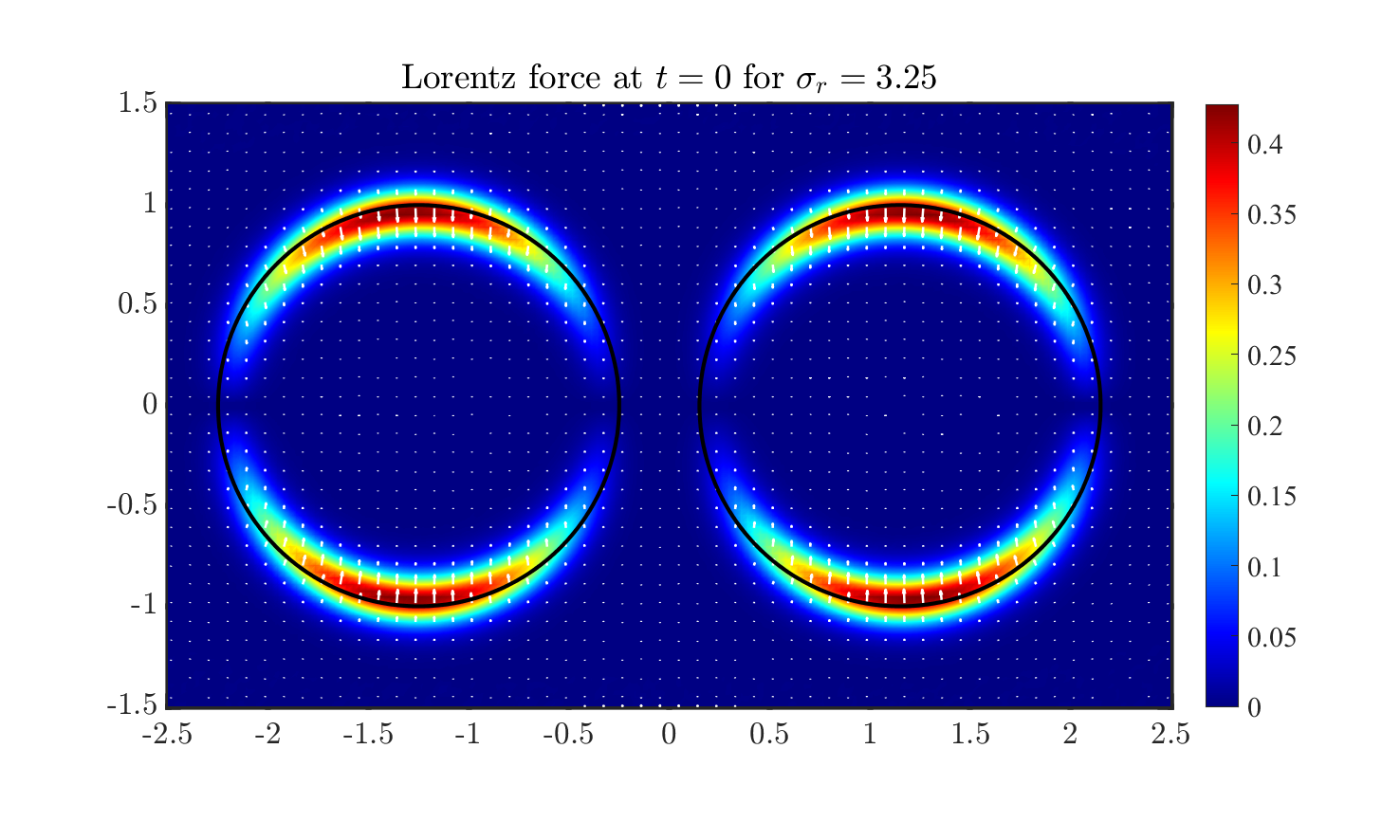}
\includegraphics[width=0.245\textwidth]{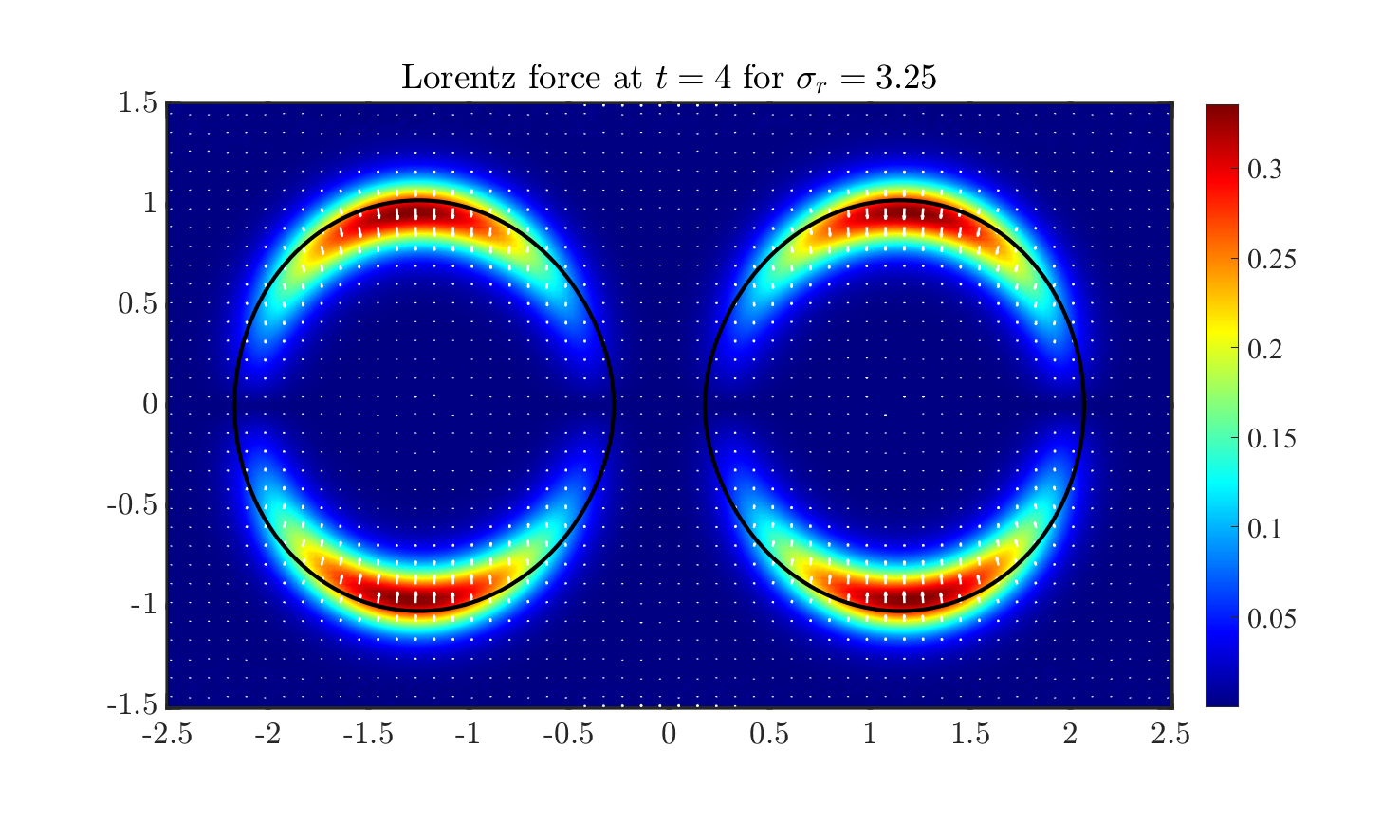}
\includegraphics[width=0.245\textwidth]{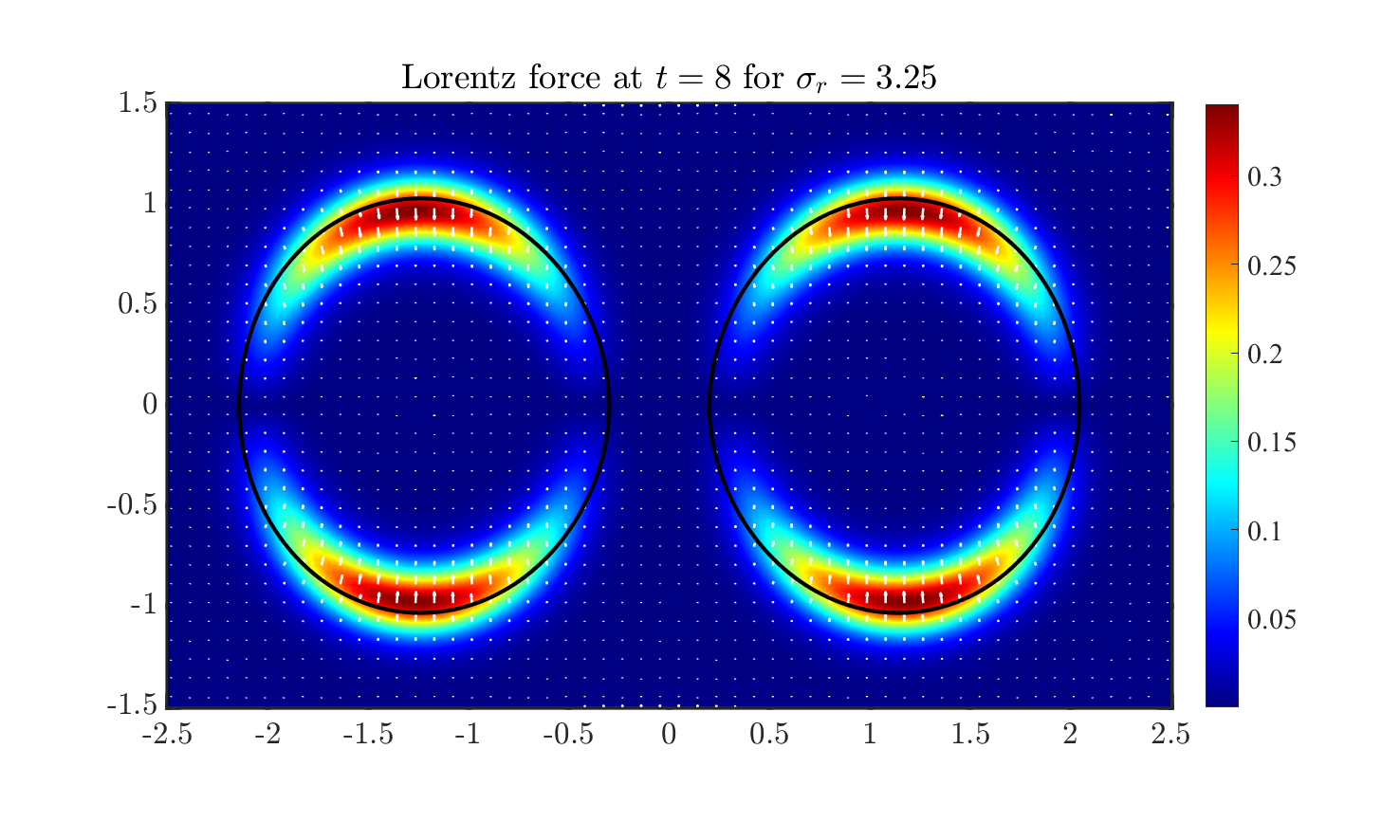}
\includegraphics[width=0.245\textwidth]{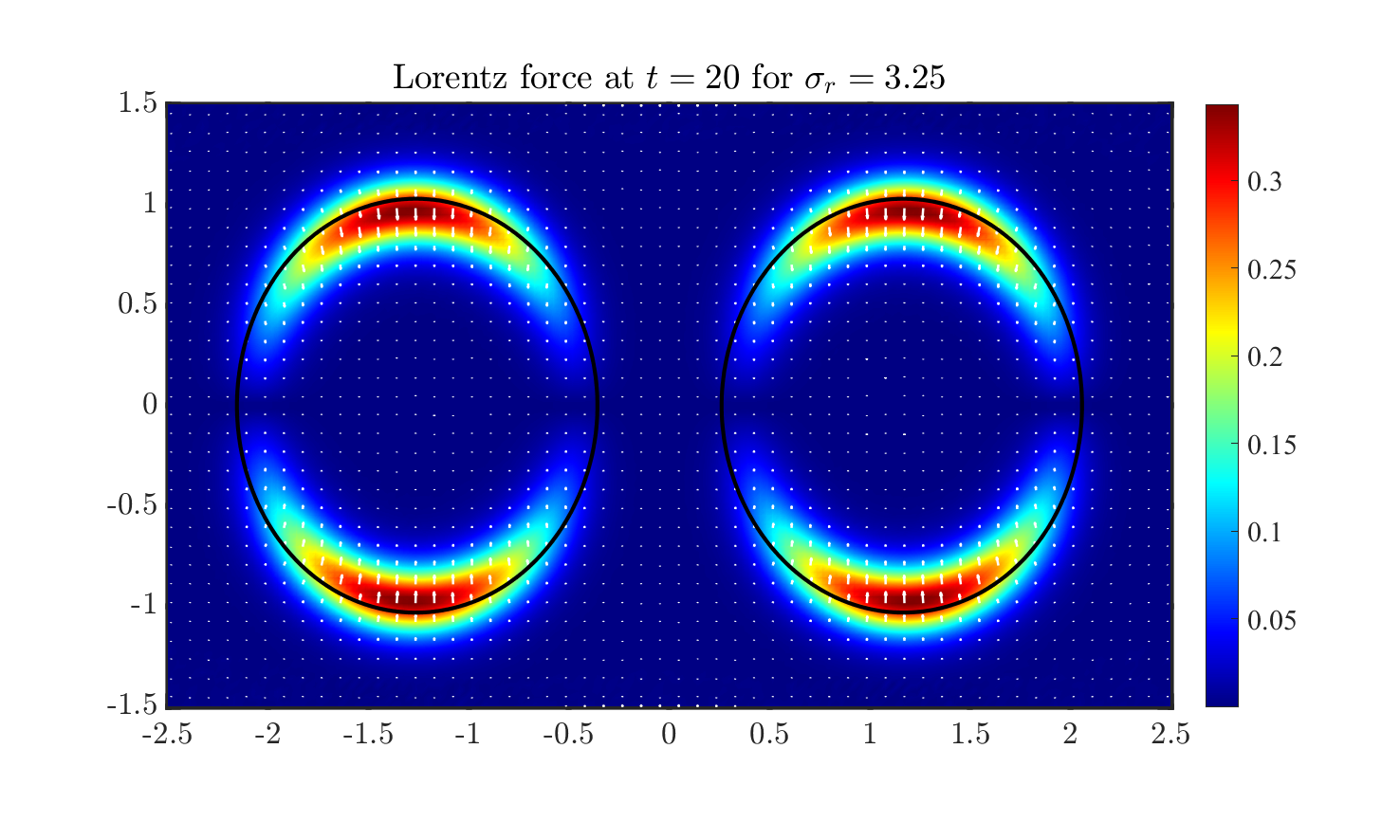}\\
\includegraphics[width=0.245\textwidth]{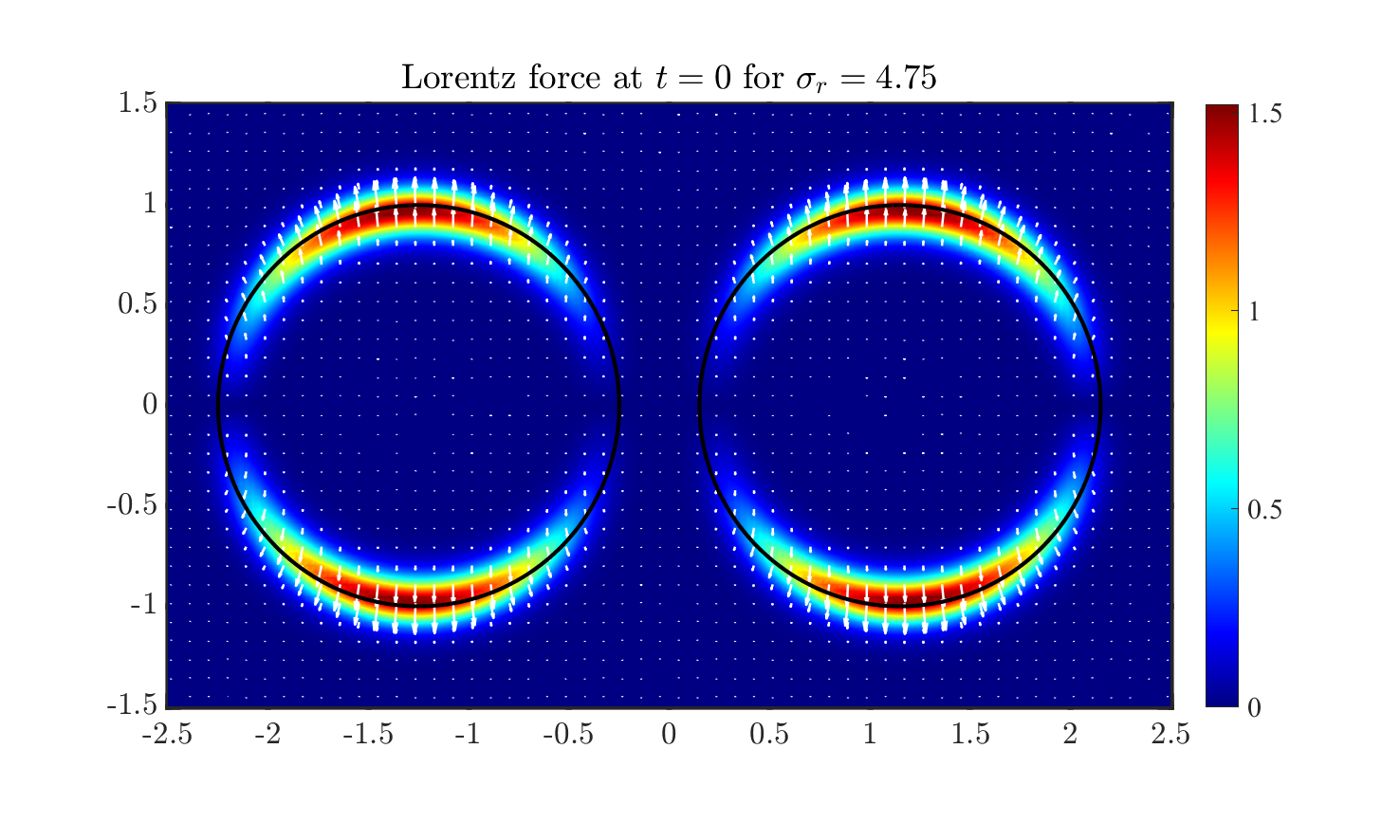}
\includegraphics[width=0.245\textwidth]{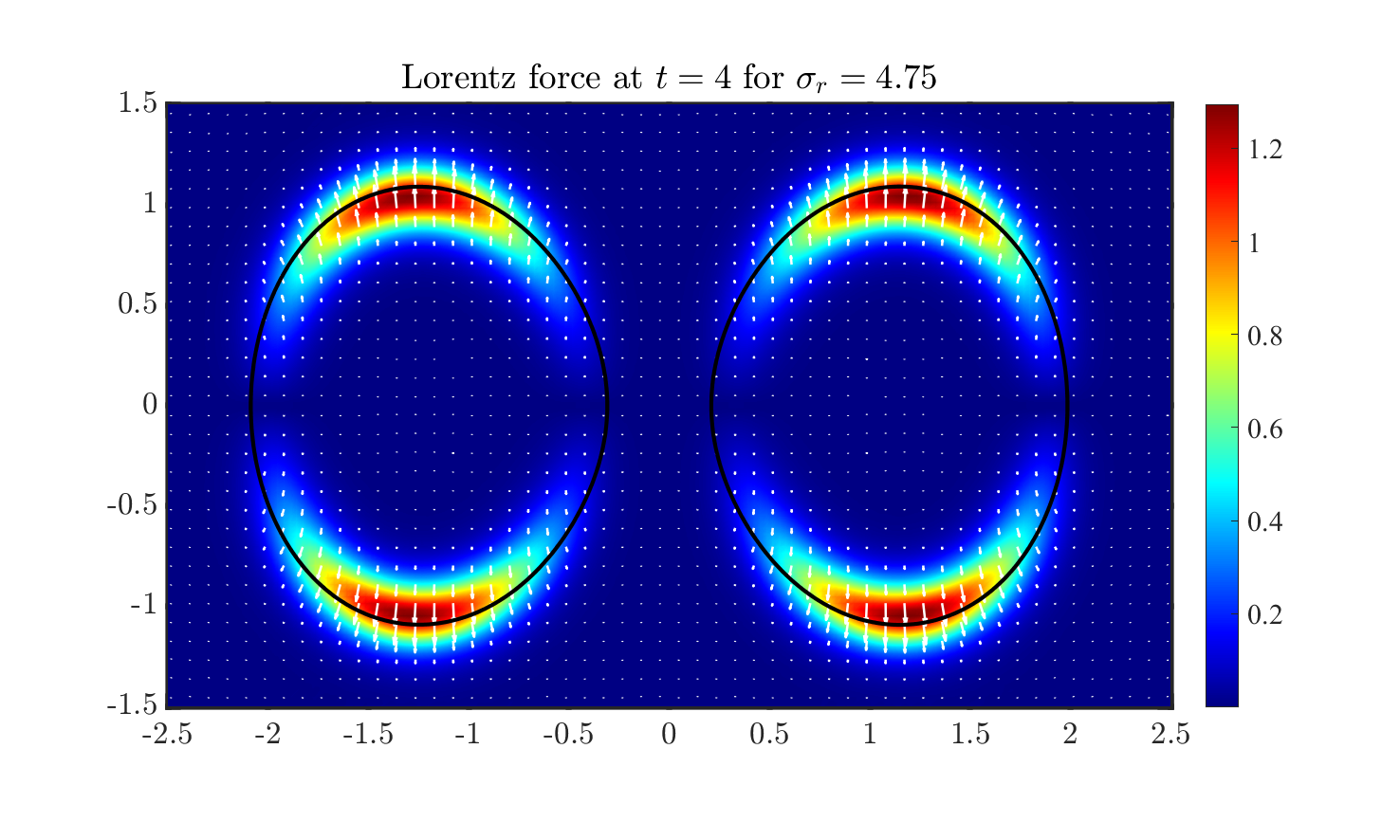}
\includegraphics[width=0.245\textwidth]{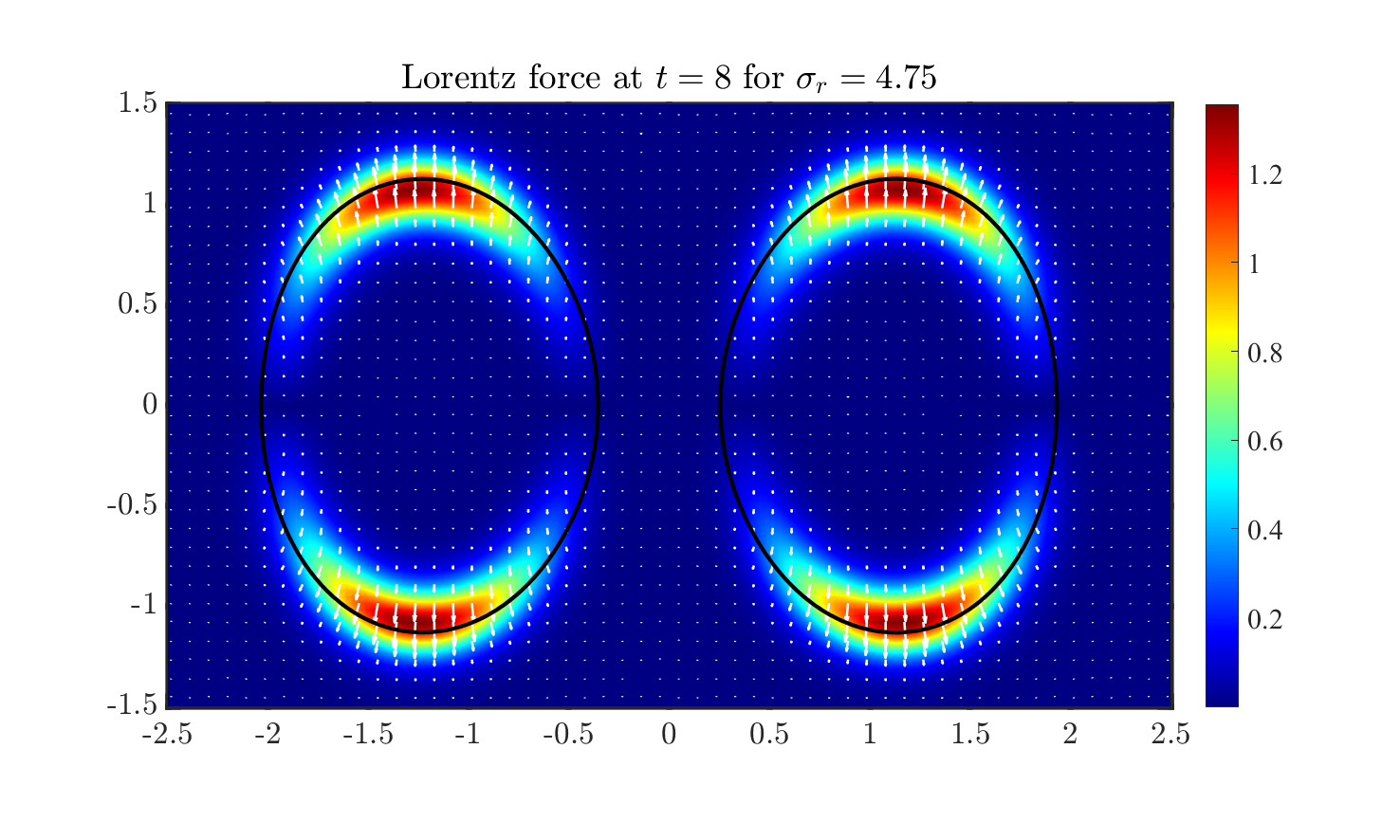}
\includegraphics[width=0.245\textwidth]{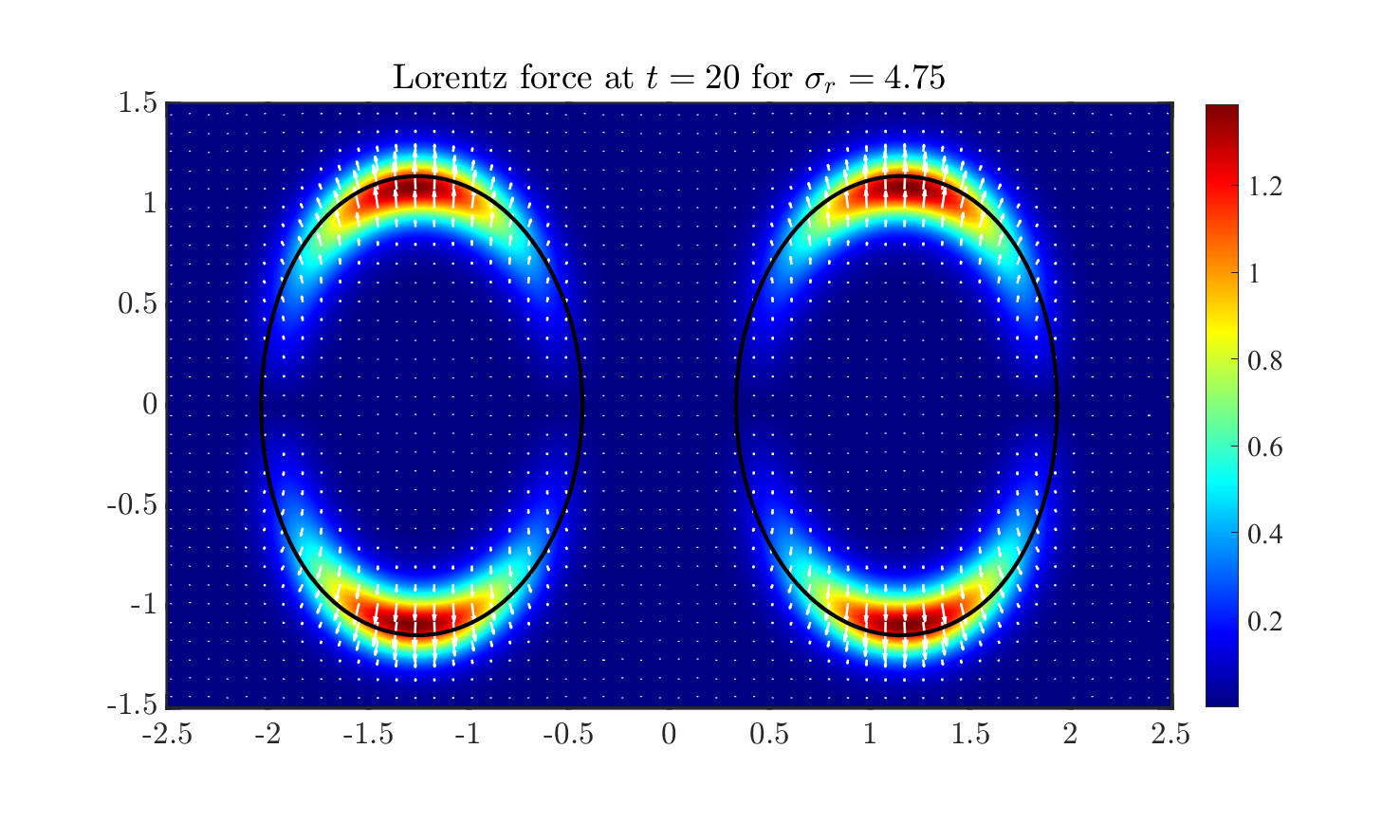}
\end{center} 
\caption{The Lorentz force in merge effect for different conductivity ratios
$\sigma_{r} = 1.75$ (top), $\sigma_{r} = 3.25$ (middle), $\sigma_{r} = 4.75$ (bottom) 
at $t = 0$, $t = 4$, $t = 8$ and $t = 20$ from left to right, respectively. 
In each figure, the solid line shows the zero level set ($\psi=0$). 
The rest parameters are chosen as $\epsilon_{r} = 3.5$, $Ca_{E} = 1$.}
\label{fig: Lorentz force in merge for two drops without cm}
\end{figure}

\begin{figure}
\begin{center}
\includegraphics[width=0.245\textwidth]{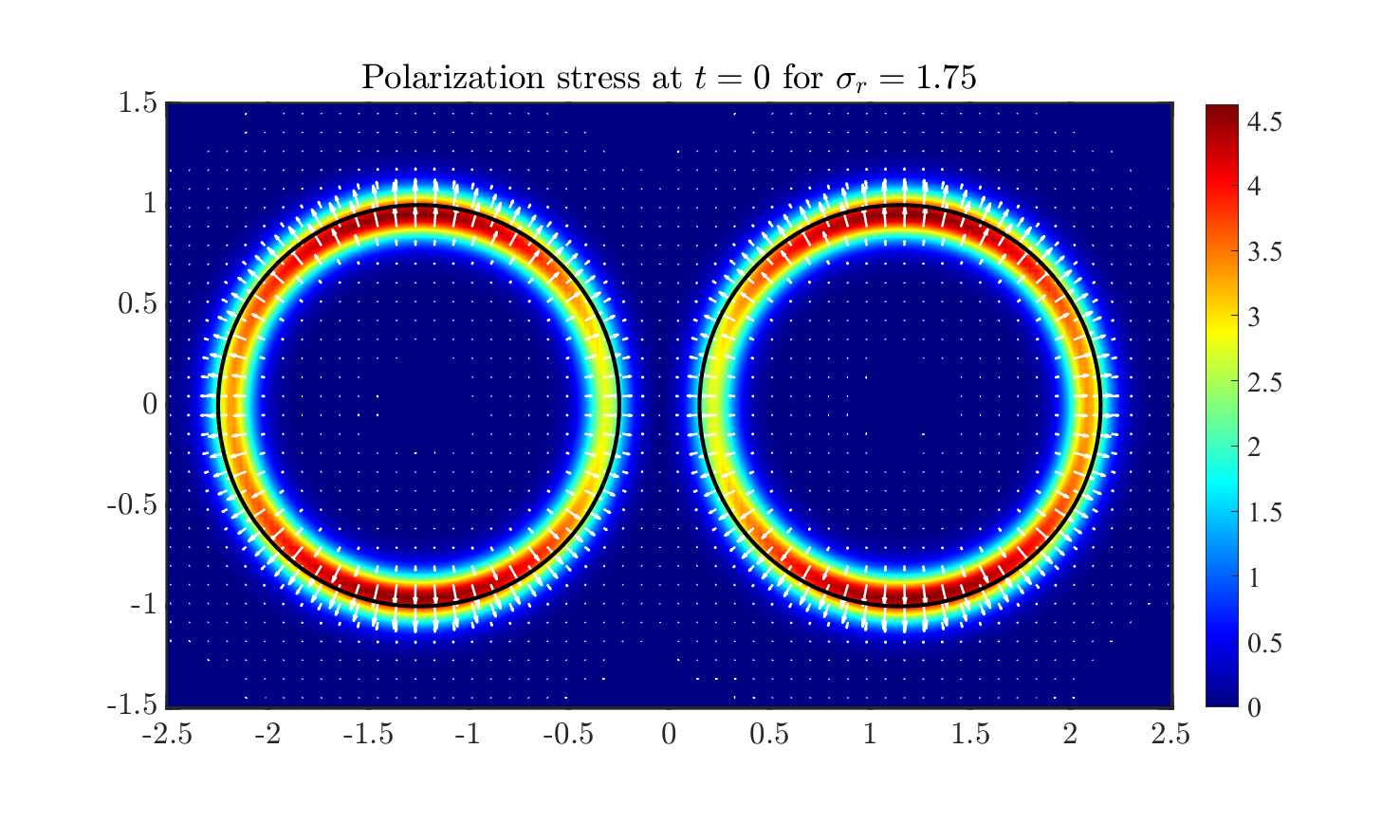}
\includegraphics[width=0.245\textwidth]{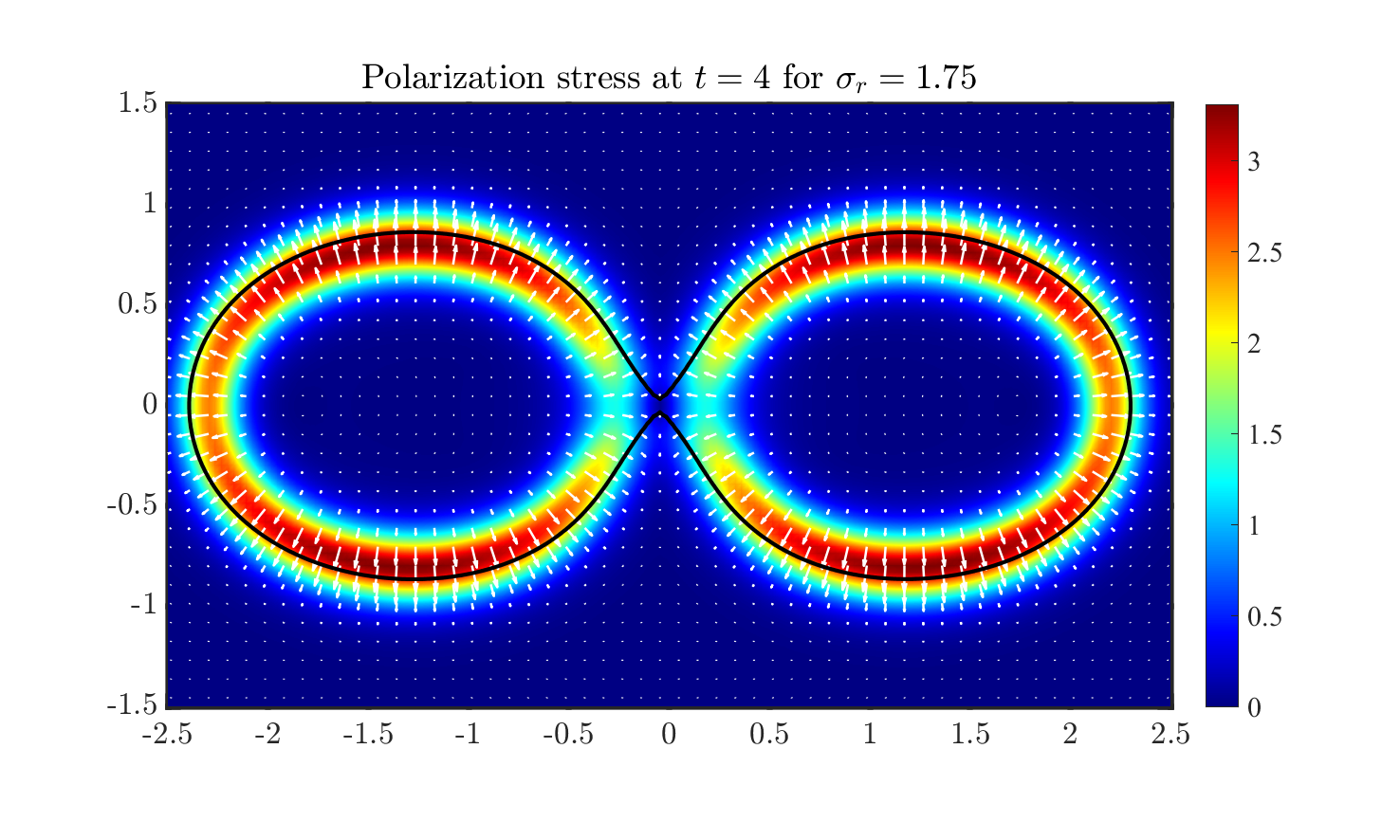}
\includegraphics[width=0.245\textwidth]{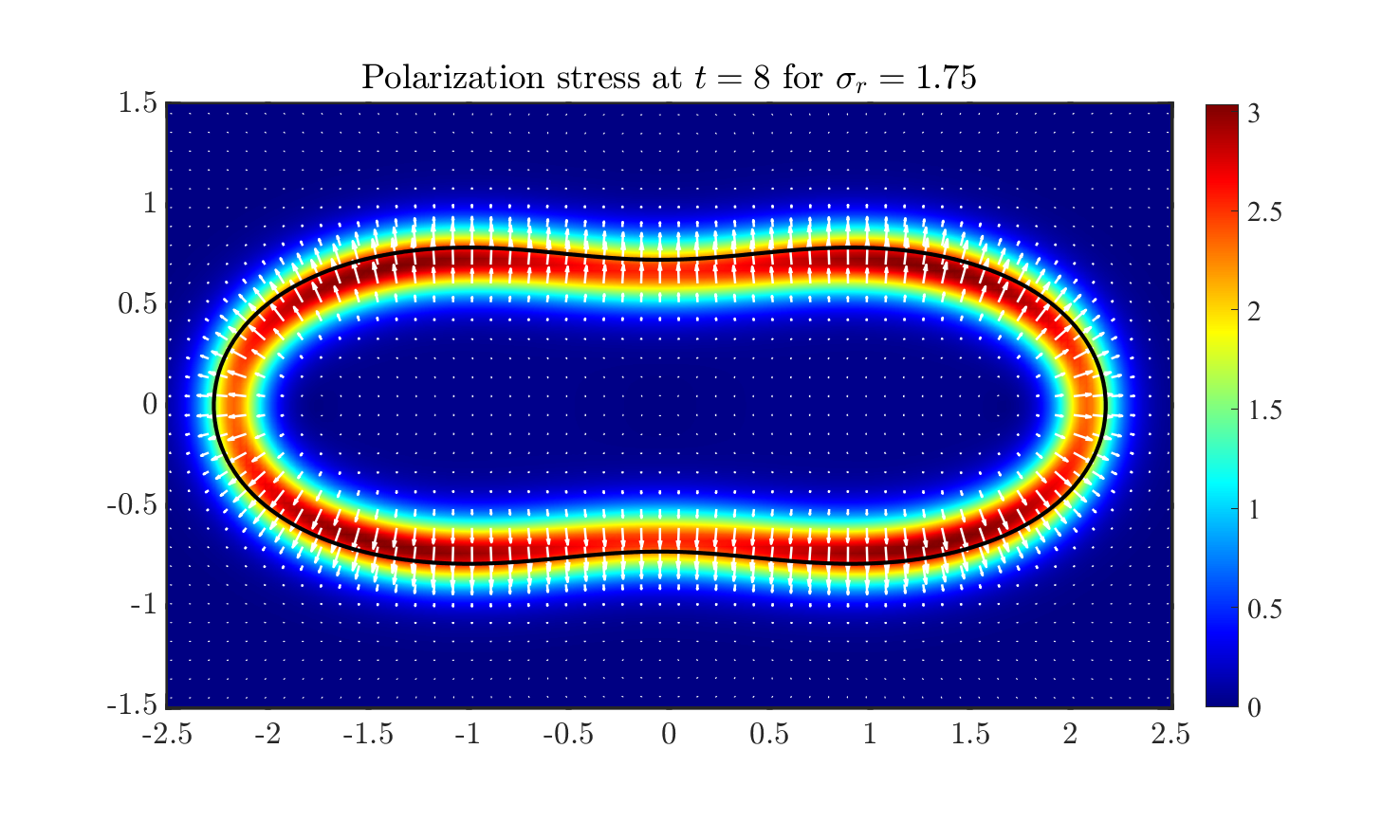}
\includegraphics[width=0.245\textwidth]{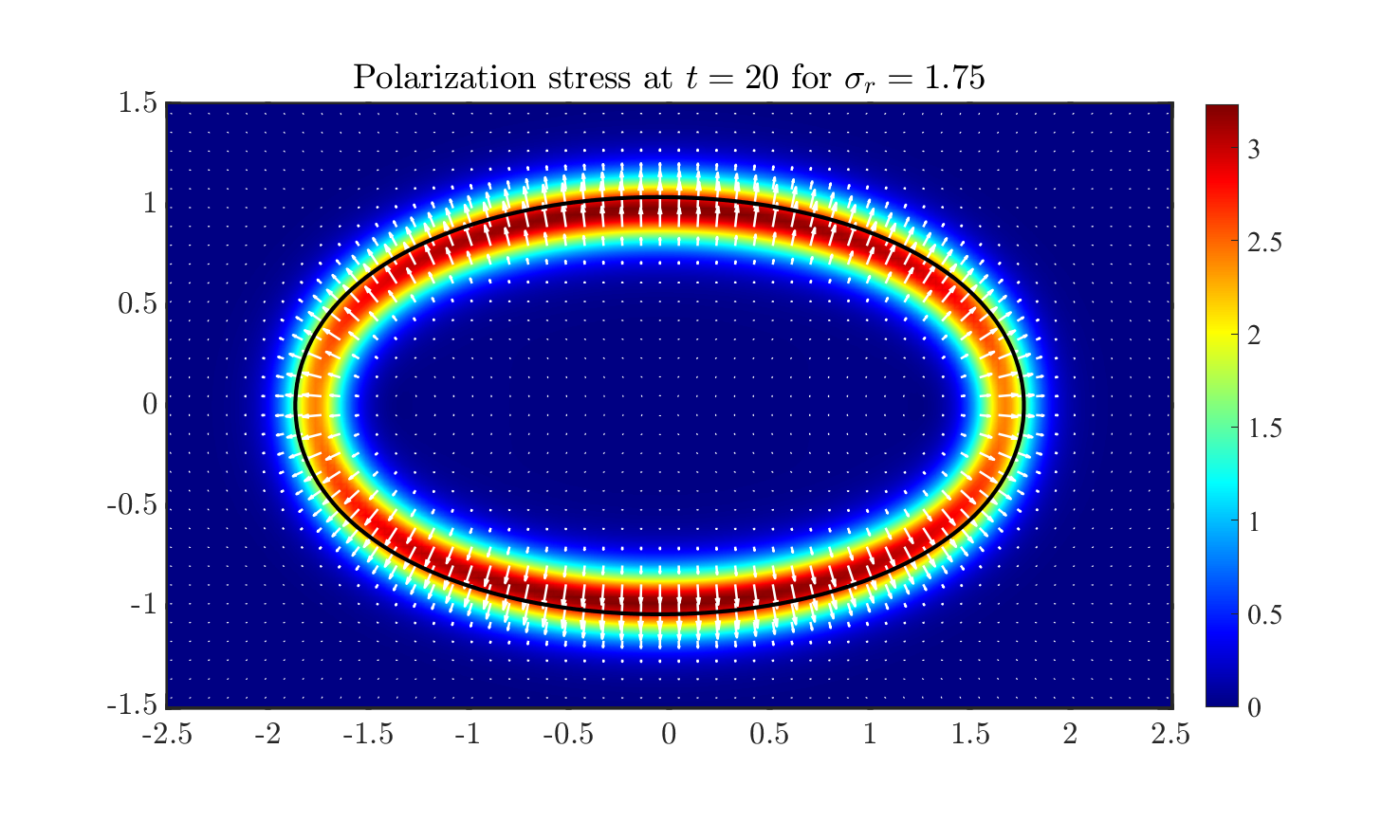}\\
\includegraphics[width=0.245\textwidth]{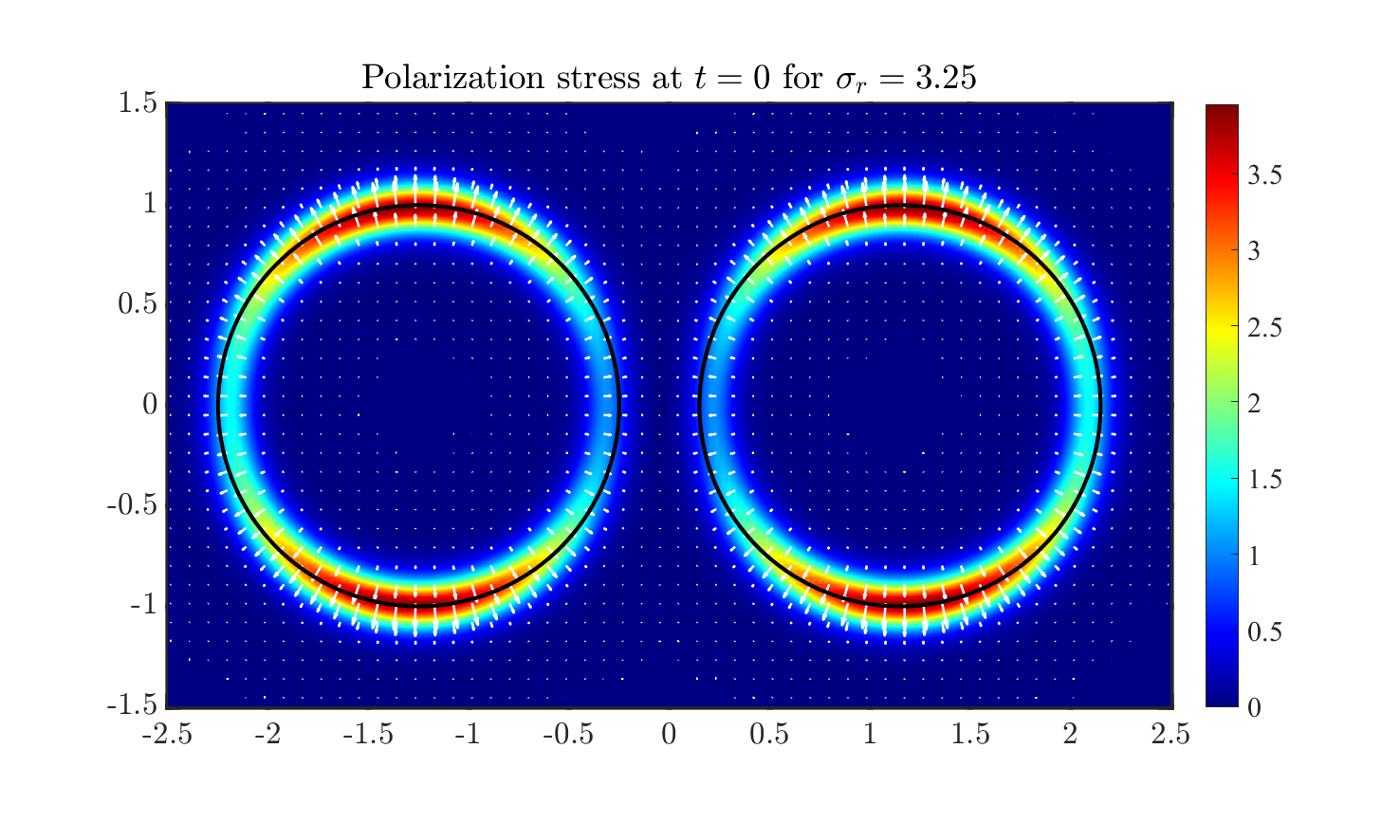}
\includegraphics[width=0.245\textwidth]{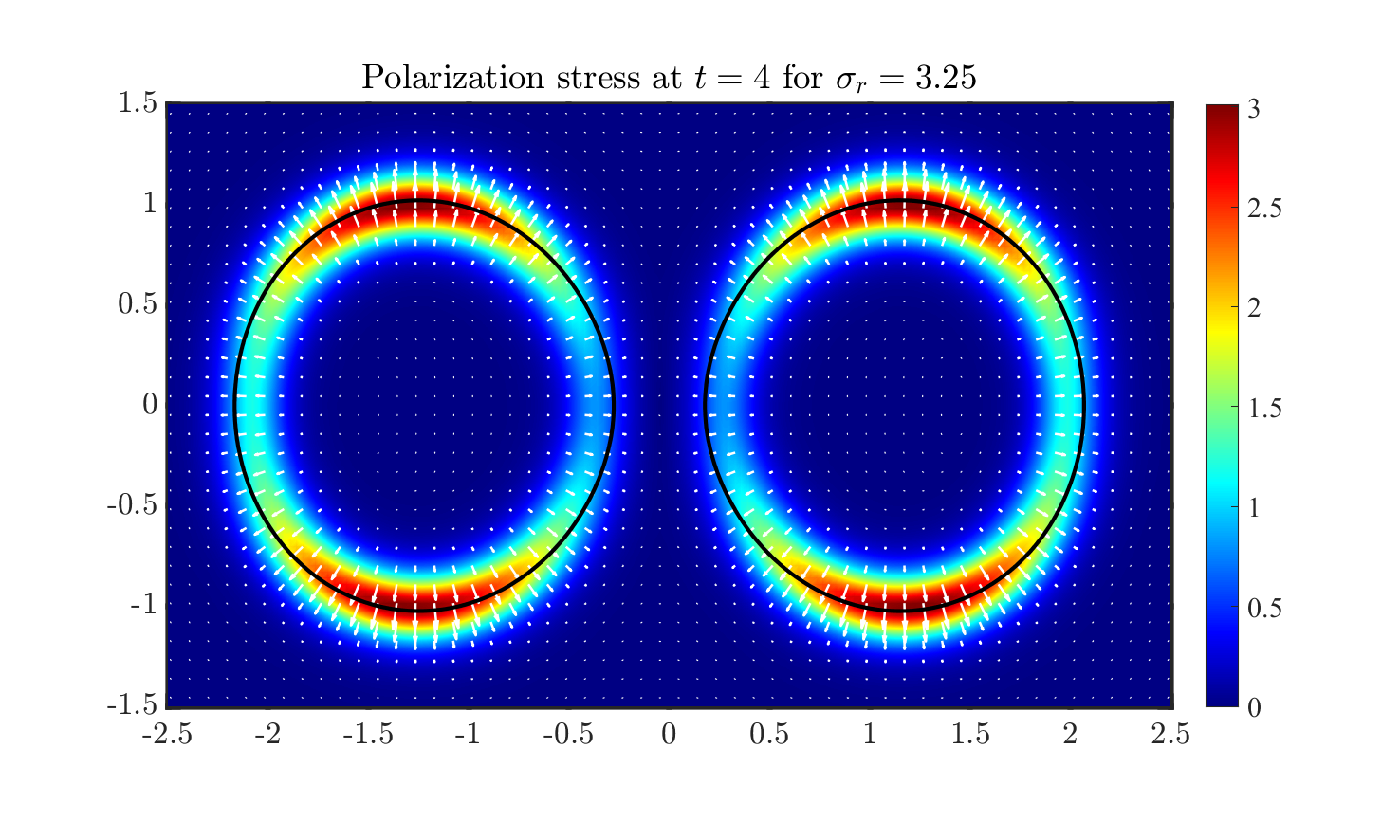}
\includegraphics[width=0.245\textwidth]{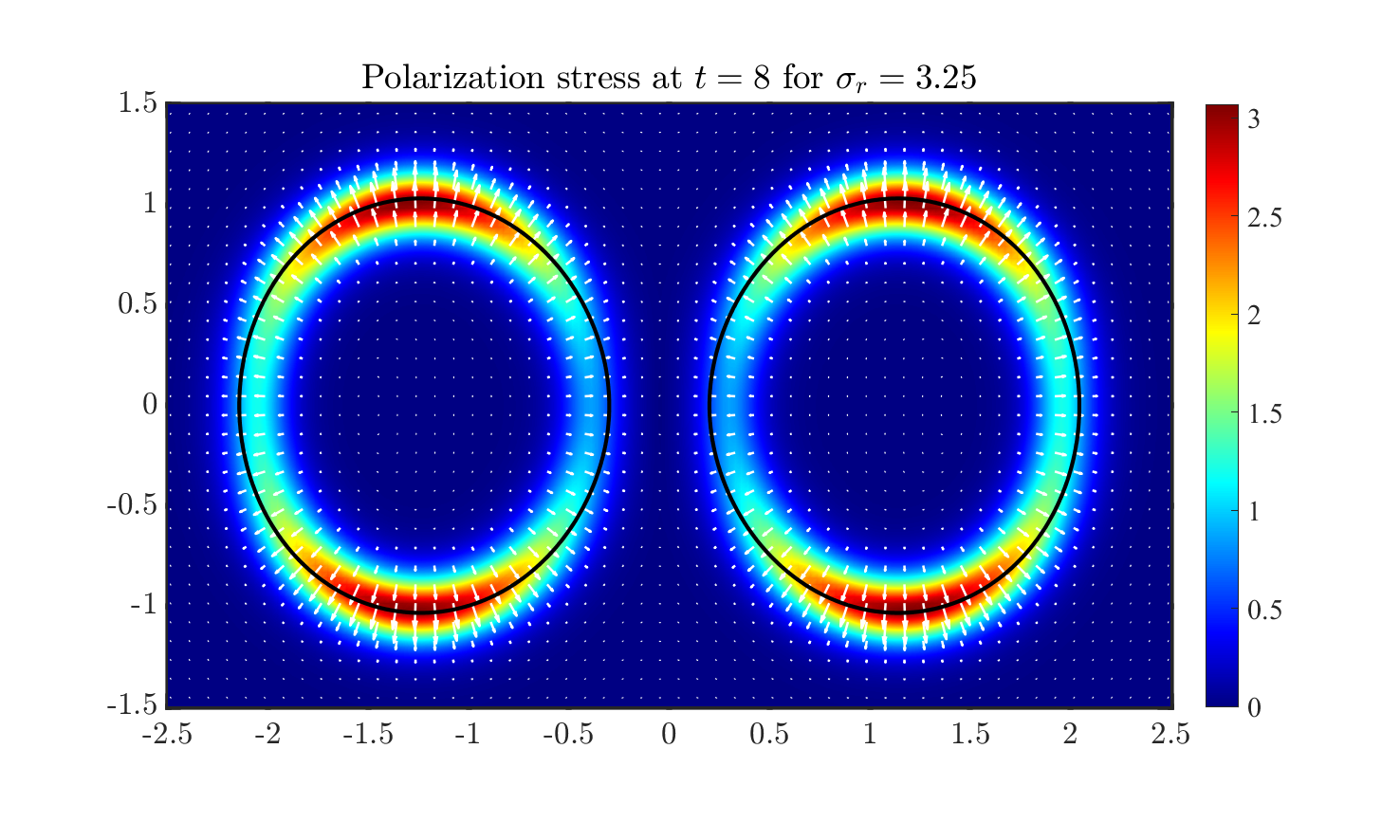}
\includegraphics[width=0.245\textwidth]{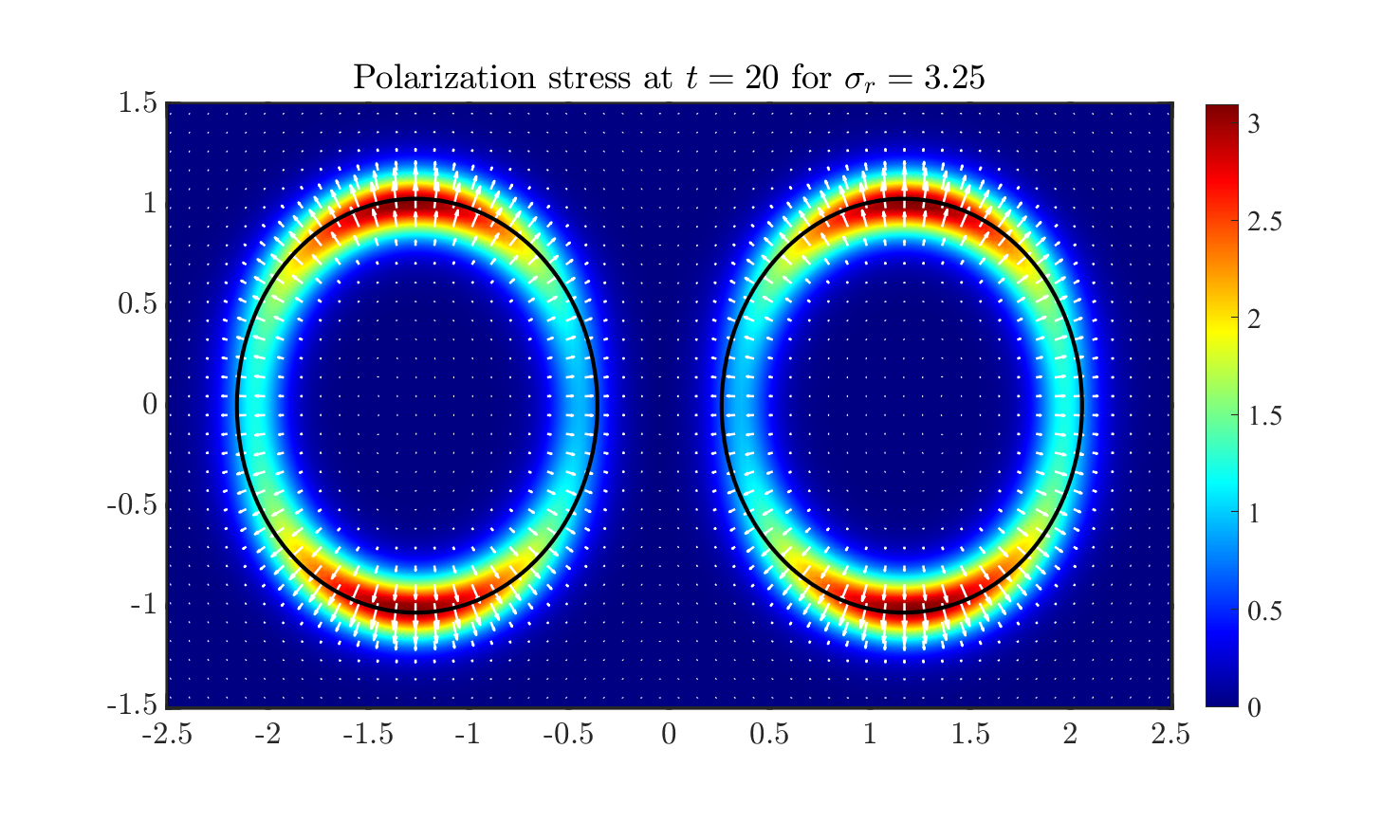}\\
\includegraphics[width=0.245\textwidth]{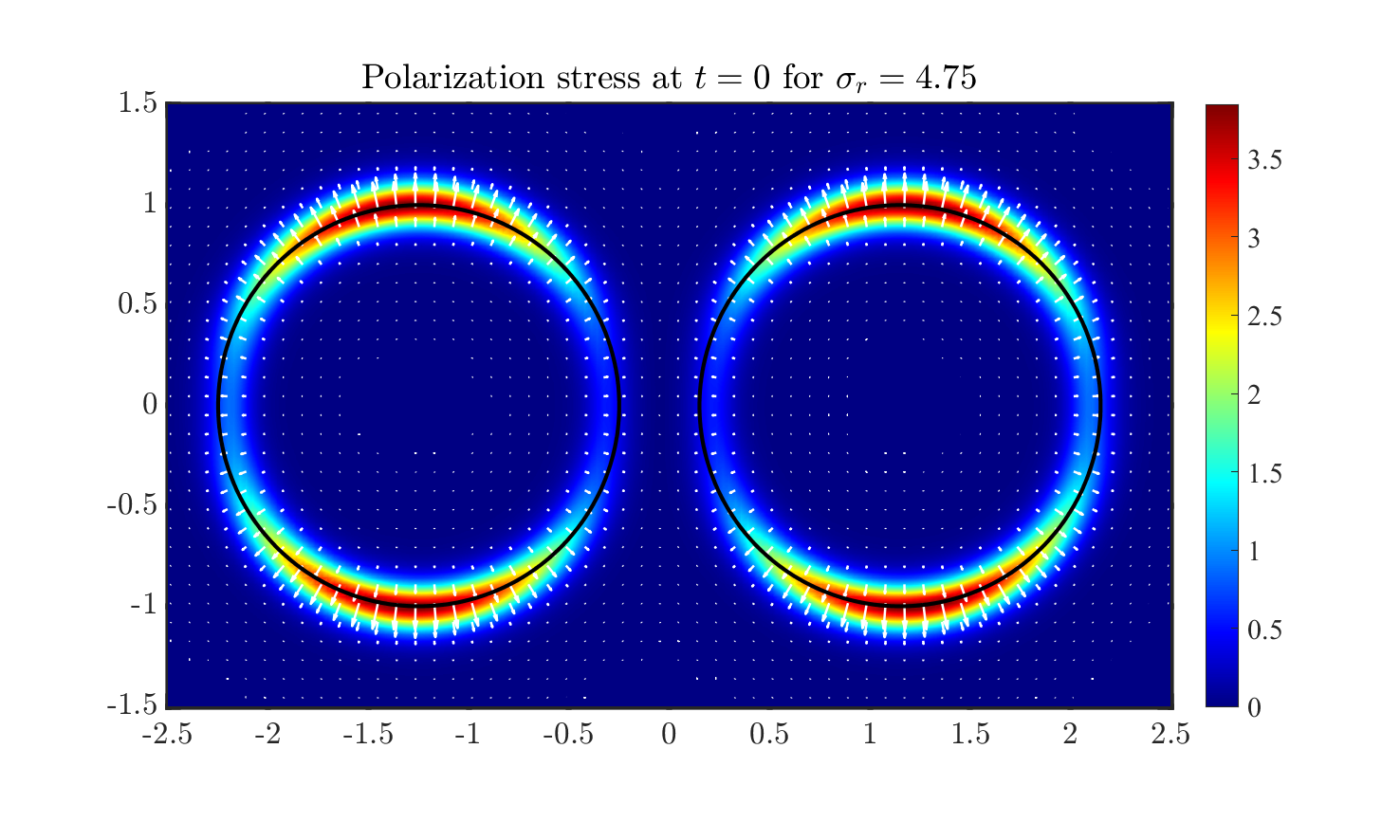}
\includegraphics[width=0.245\textwidth]{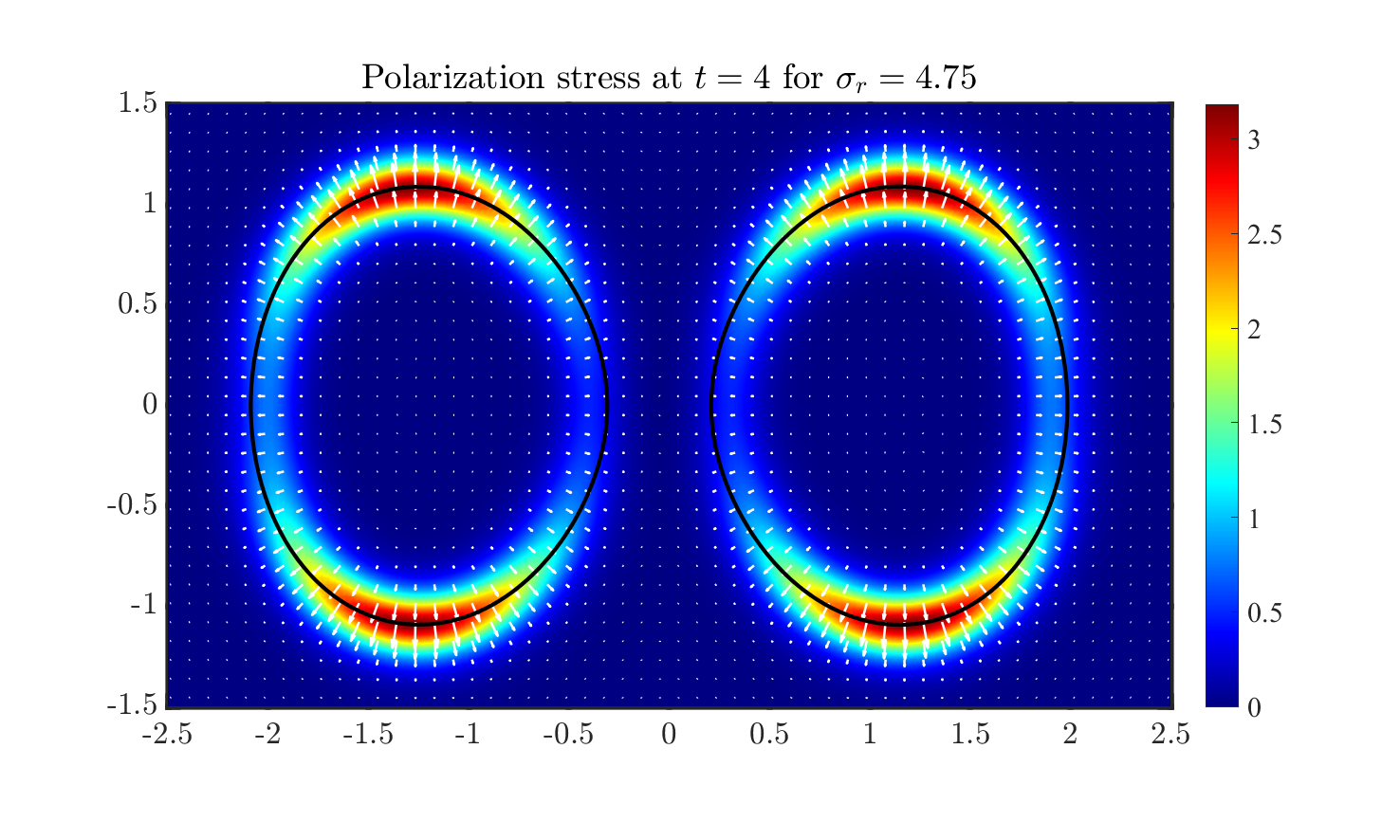}
\includegraphics[width=0.245\textwidth]{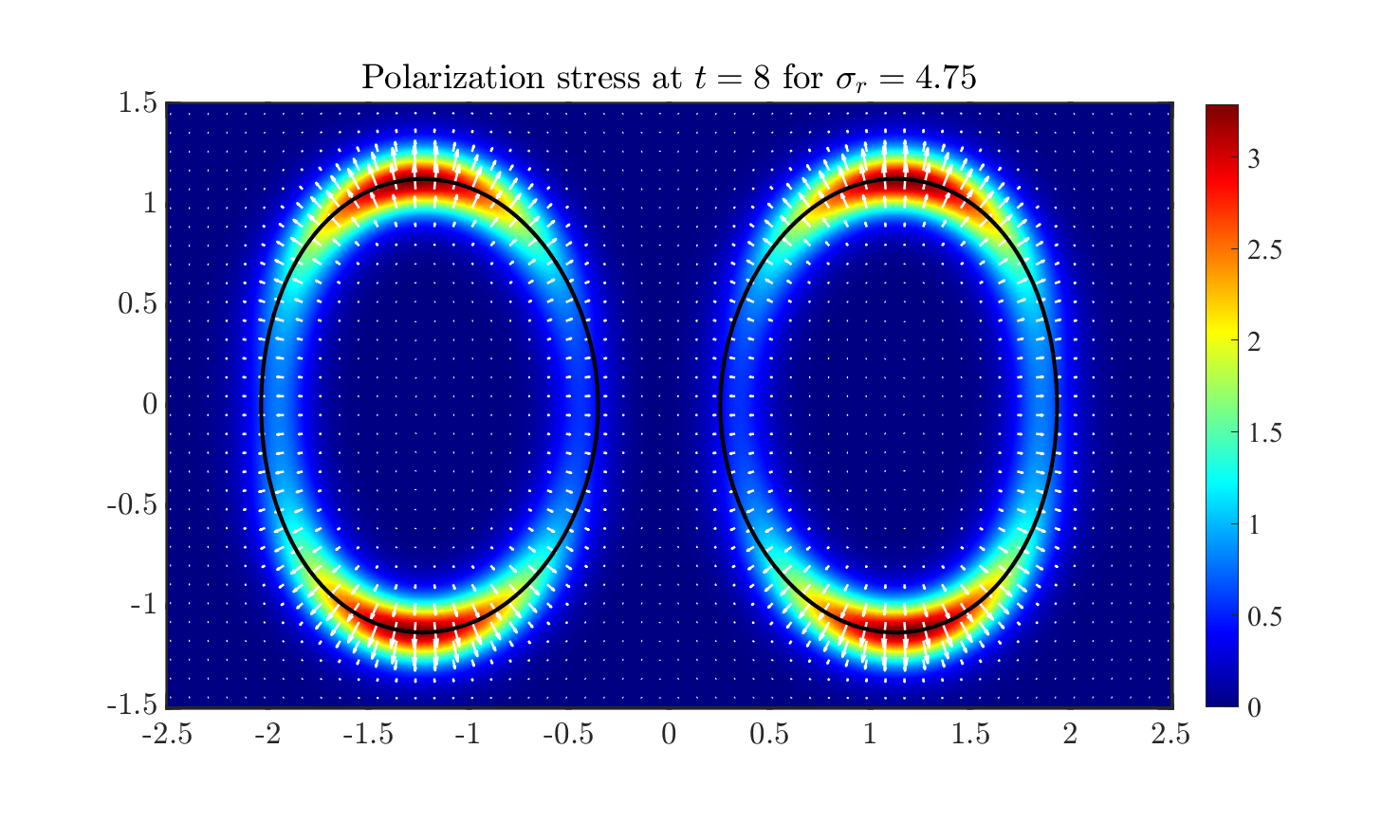}
\includegraphics[width=0.245\textwidth]{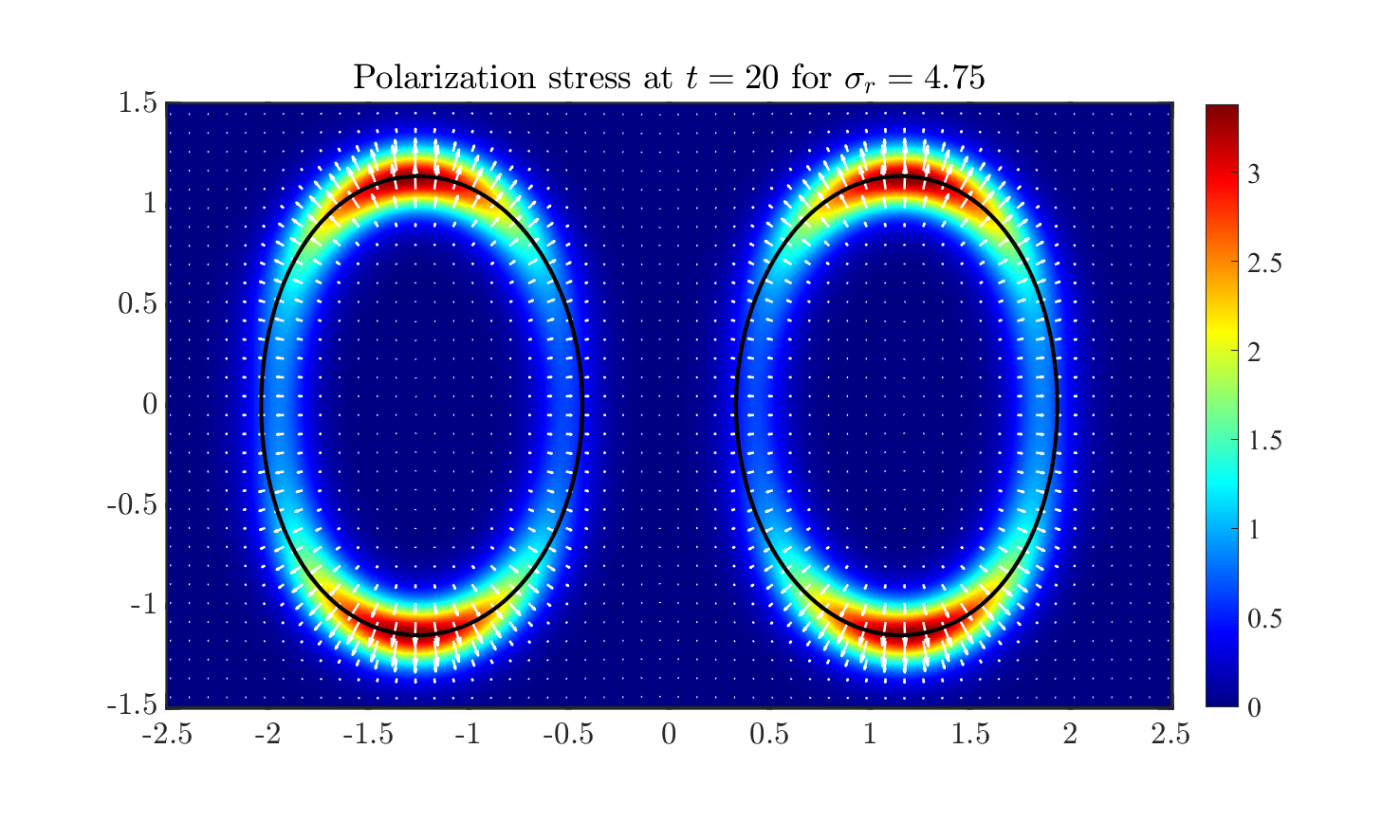}
\end{center} 
\caption{The polarization stress in merge effect for different conductivity ratios
$\sigma_{r} = 1.75$ (top), $\sigma_{r} = 3.25$ (middle), $\sigma_{r} = 4.75$ (bottom) 
at $t = 0$, $t = 4$, $t = 8$ and $t = 20$ from left to right, respectively. 
In each figure, the solid line shows the zero level set ($\psi=0$). 
The rest parameters are chosen as $\epsilon_{r} = 3.5$, $Ca_{E} = 1$.}
\label{fig: polarization stress merge for two drops without cm}
\end{figure}

\begin{figure}
\begin{center}
\includegraphics[width=0.32\textwidth]{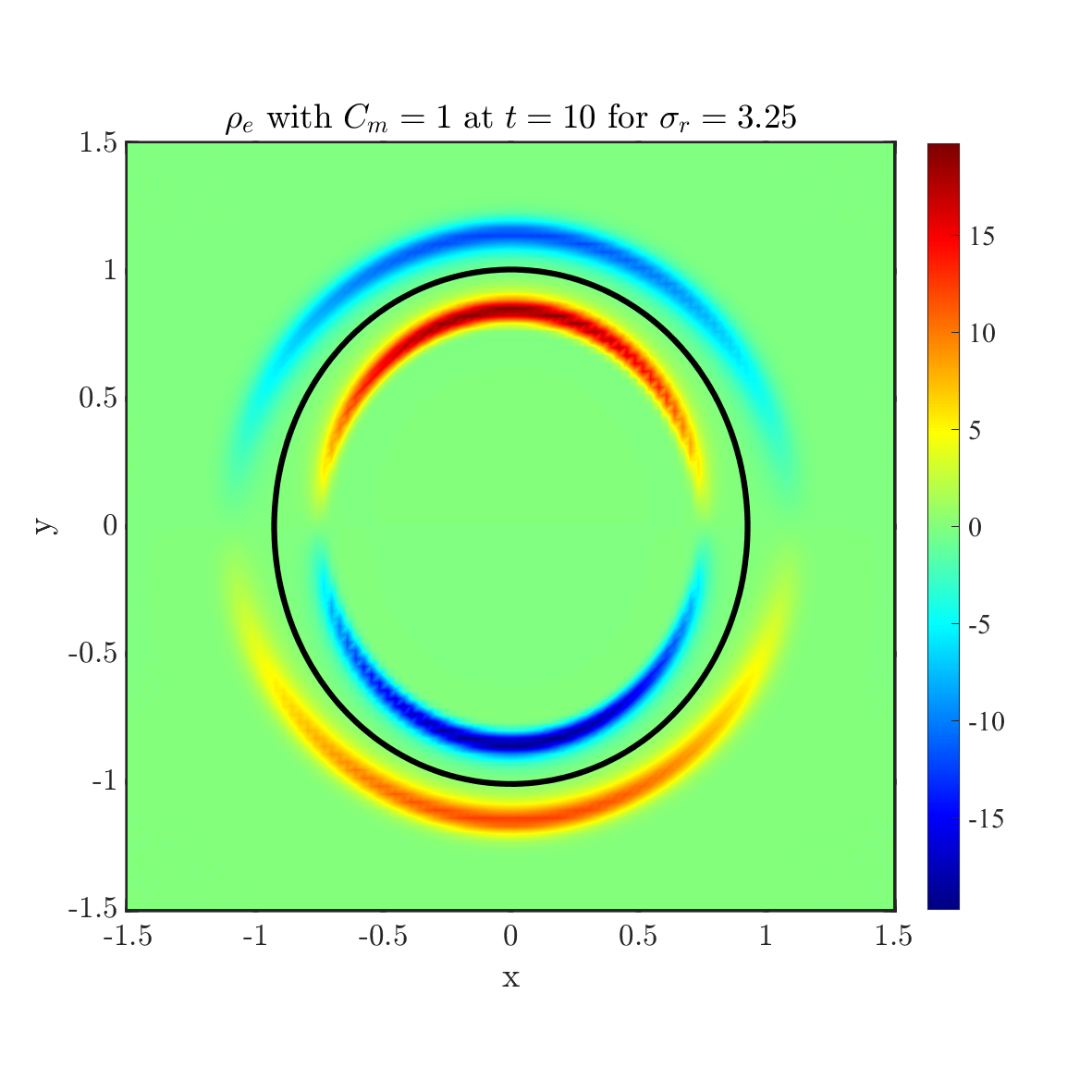}
\includegraphics[width=0.32\textwidth]{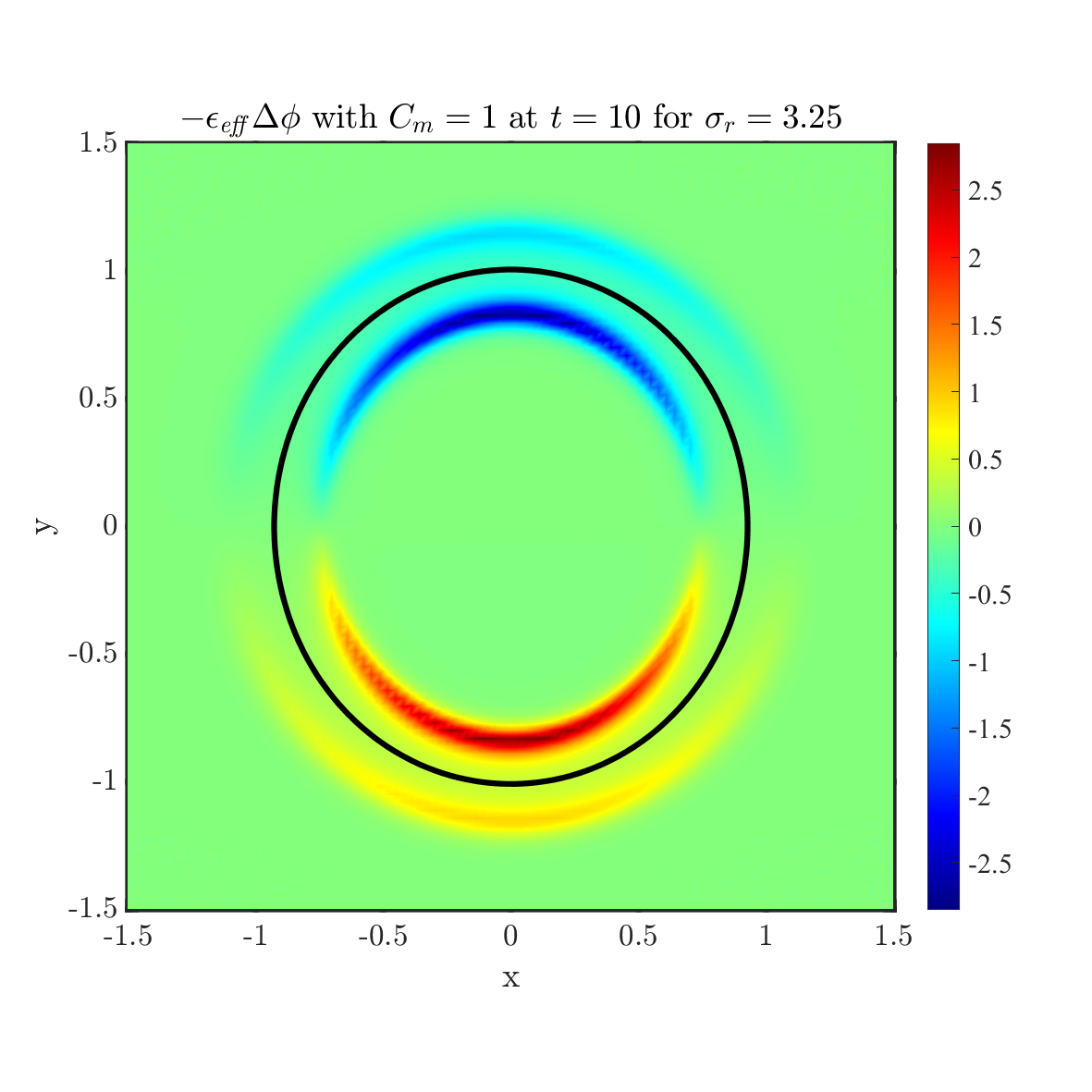}
\includegraphics[width=0.32\textwidth]{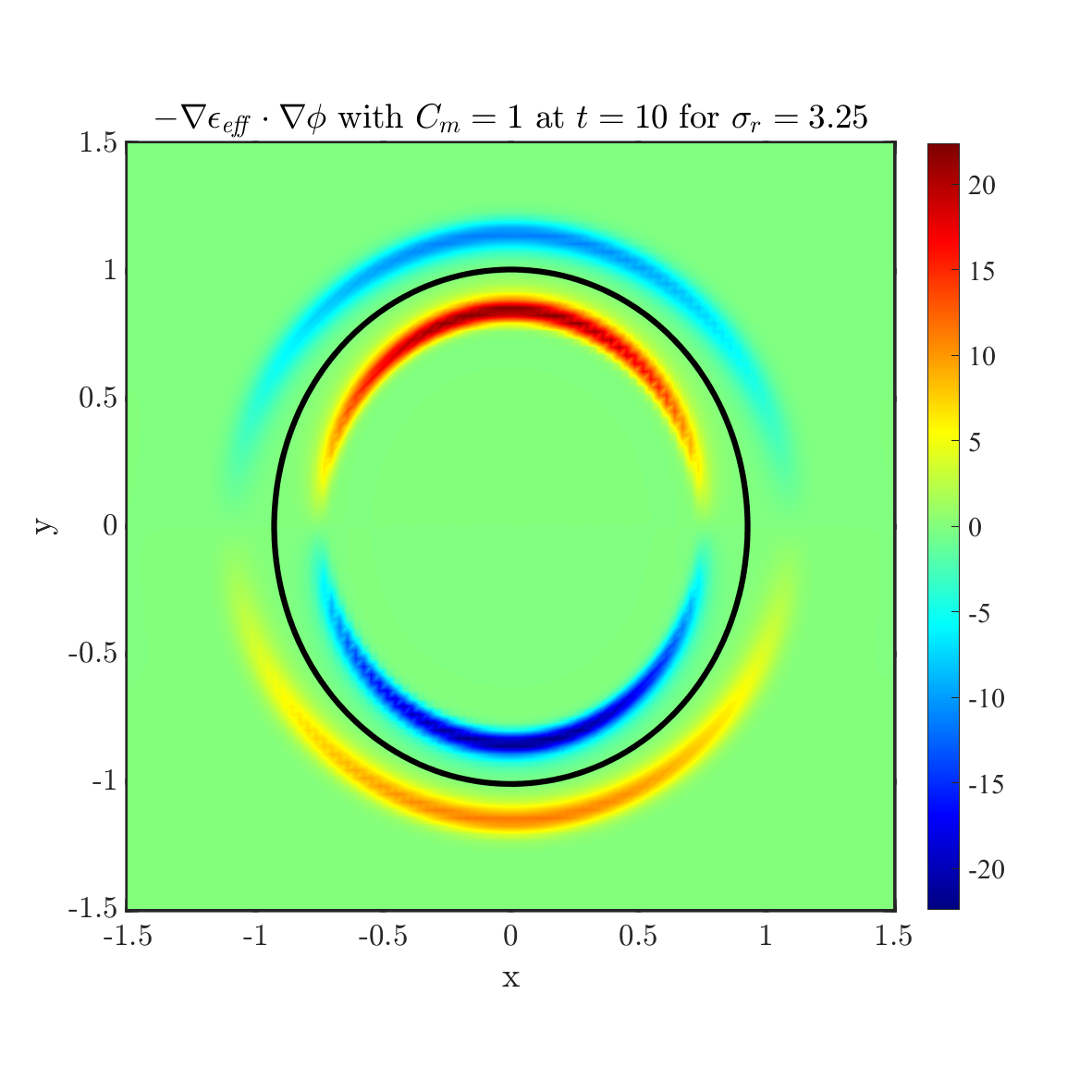}
\includegraphics[width=0.32\textwidth]{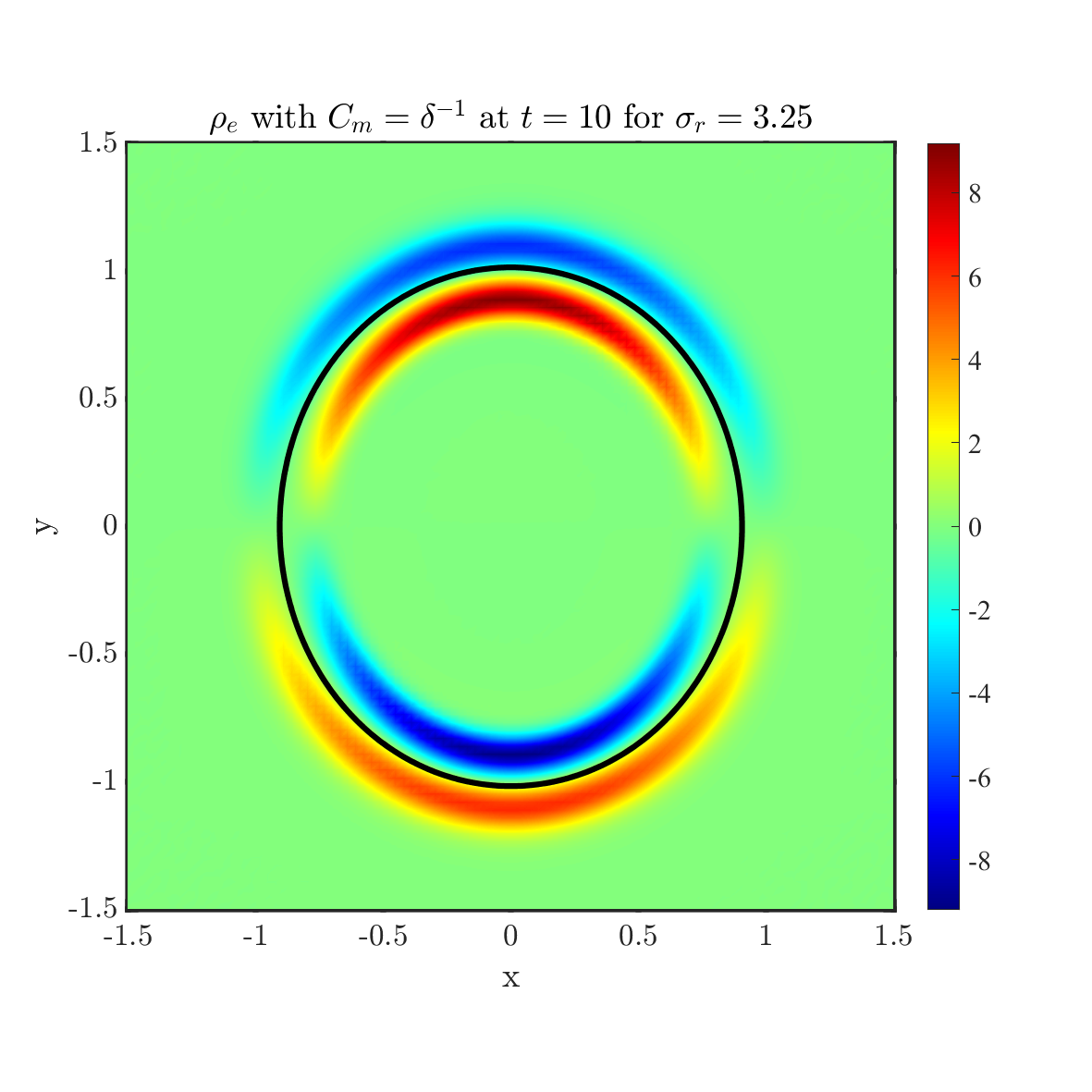}
\includegraphics[width=0.32\textwidth]{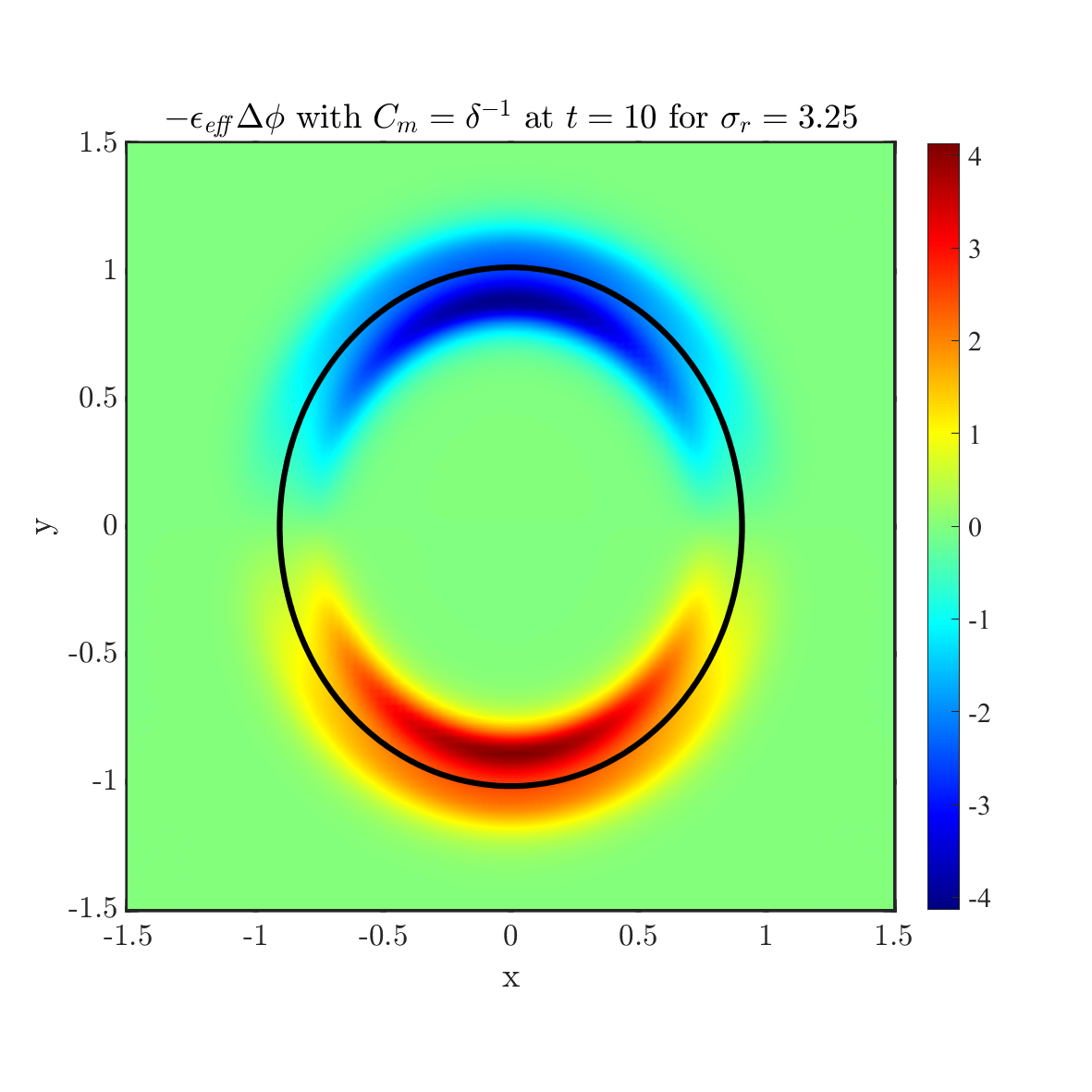}
\includegraphics[width=0.32\textwidth]{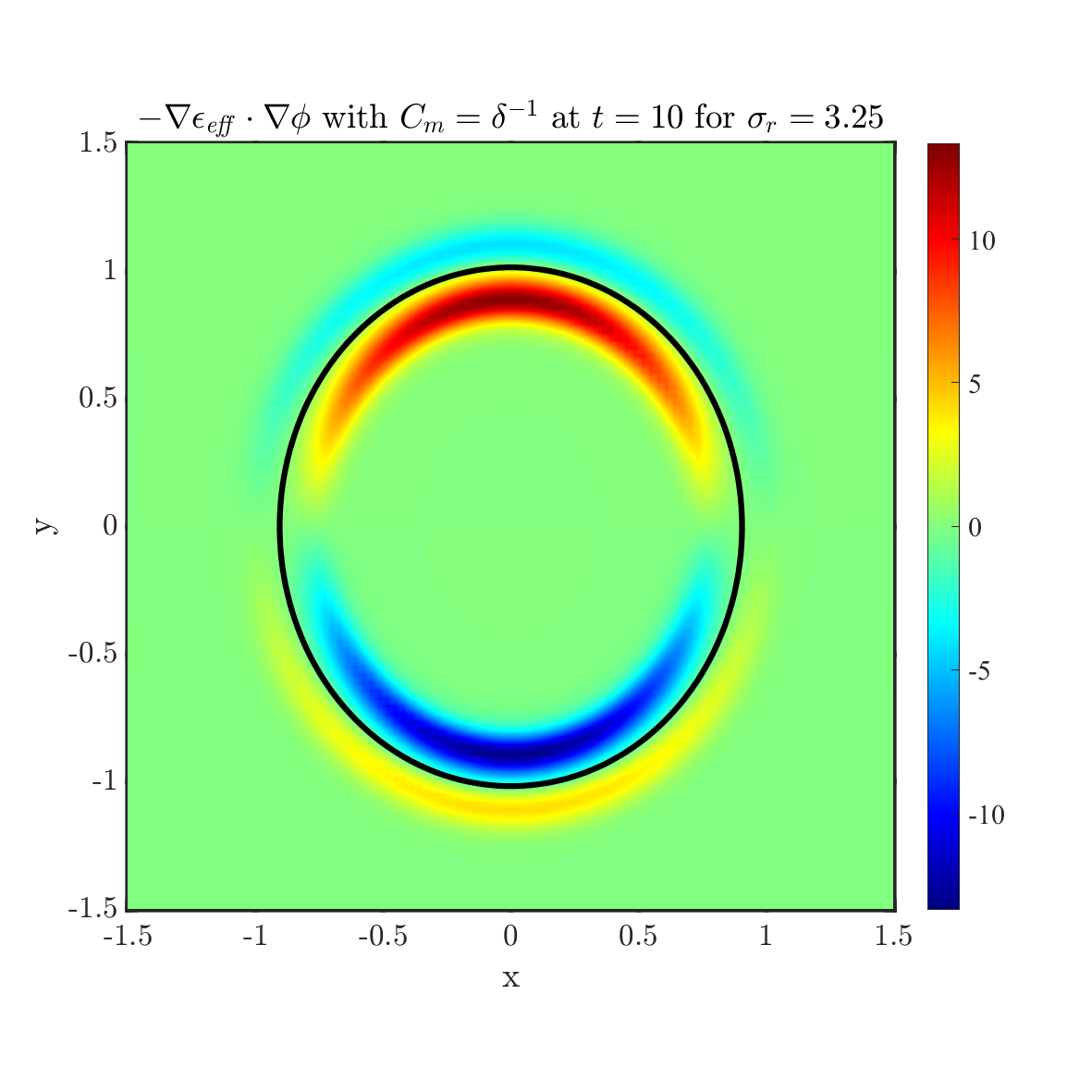}
\includegraphics[width=0.32\textwidth]{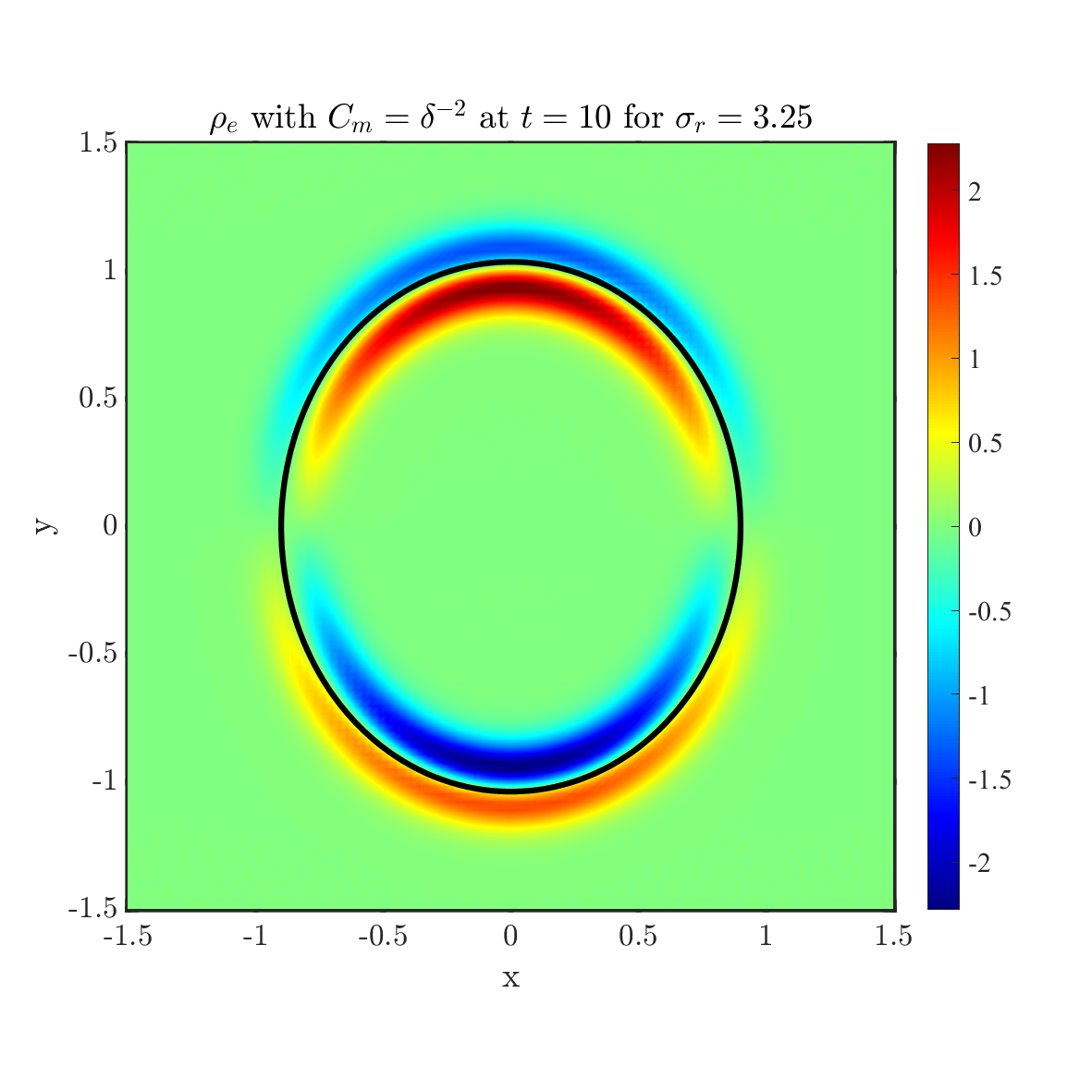}
\includegraphics[width=0.32\textwidth]{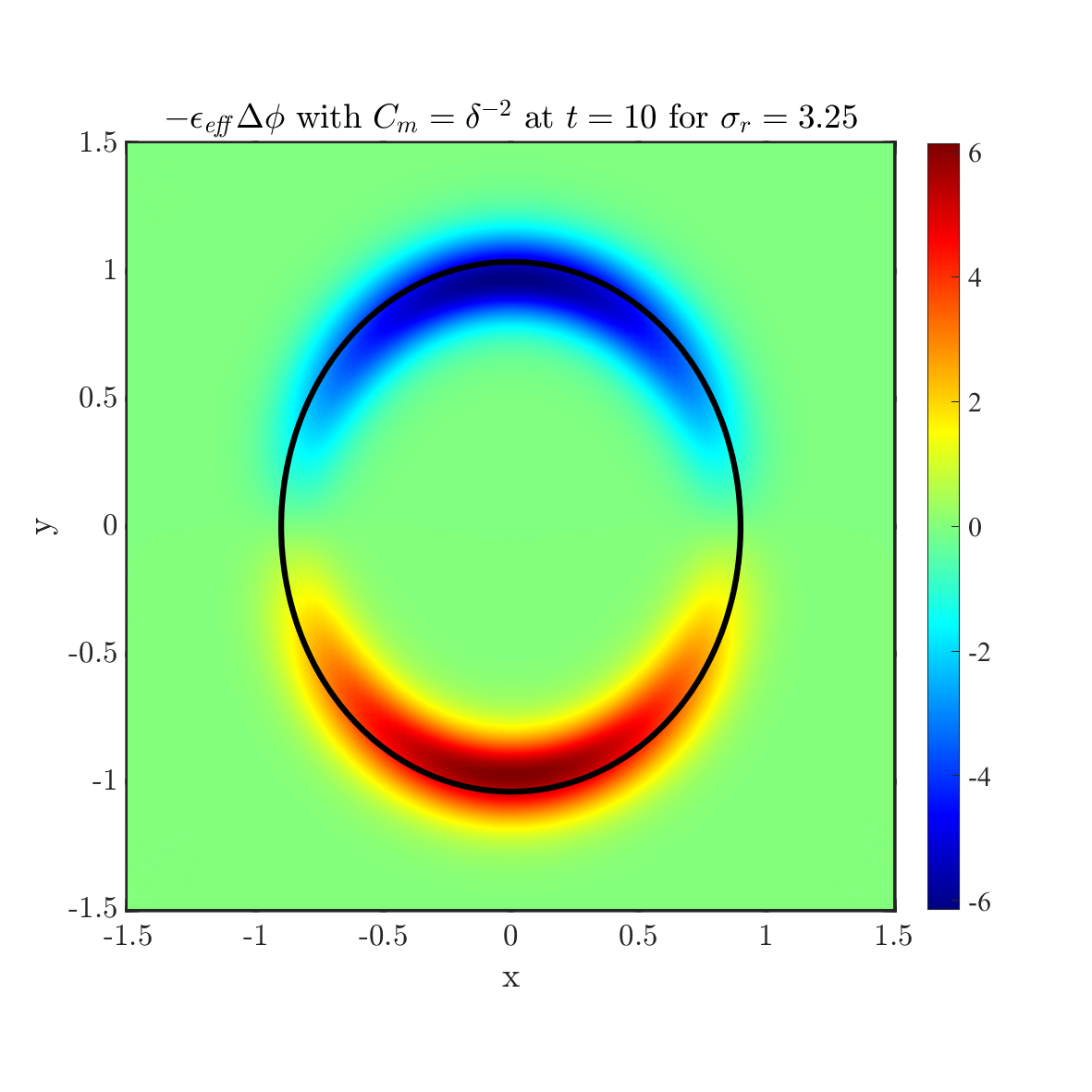}
\includegraphics[width=0.32\textwidth]{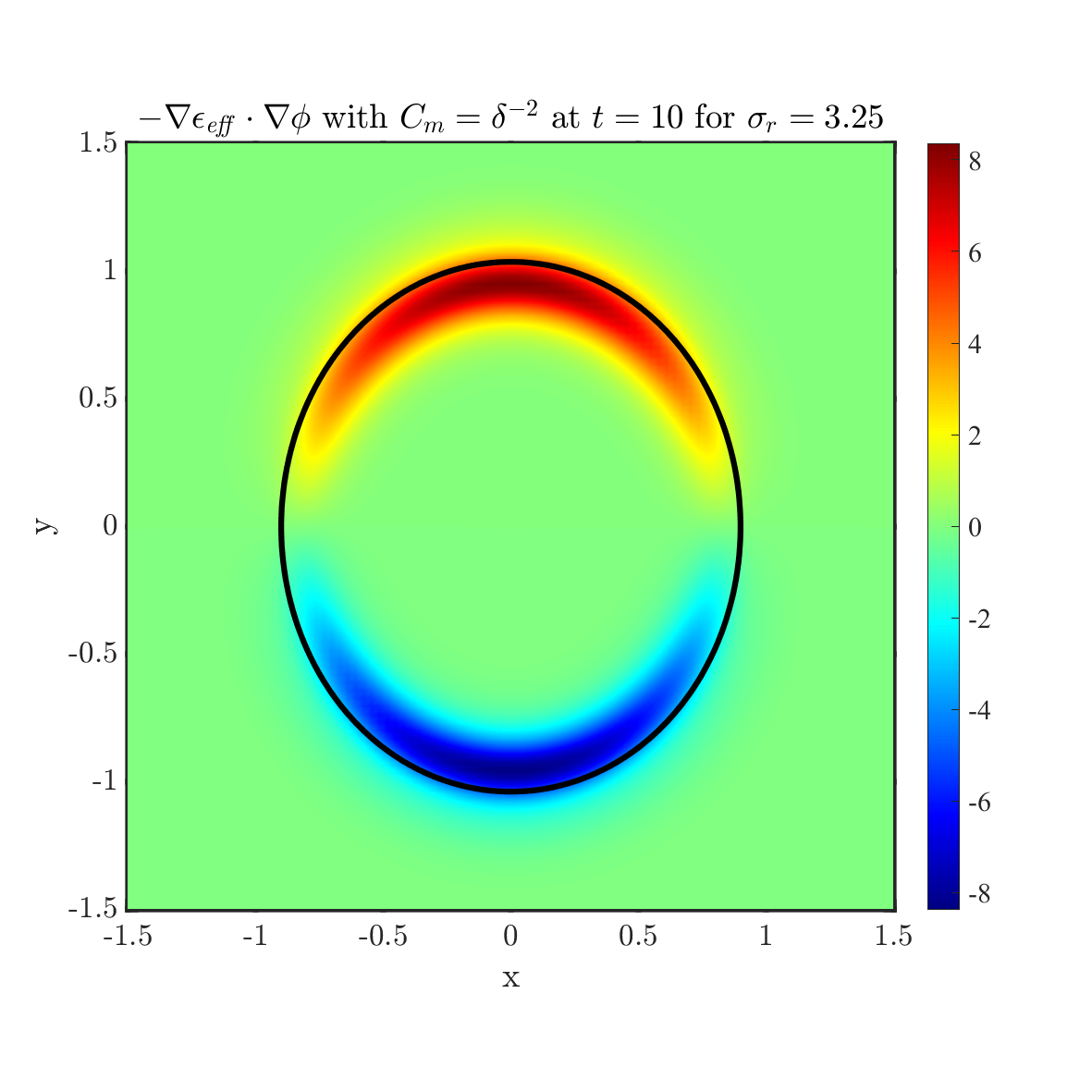}
\includegraphics[width=0.32\textwidth]{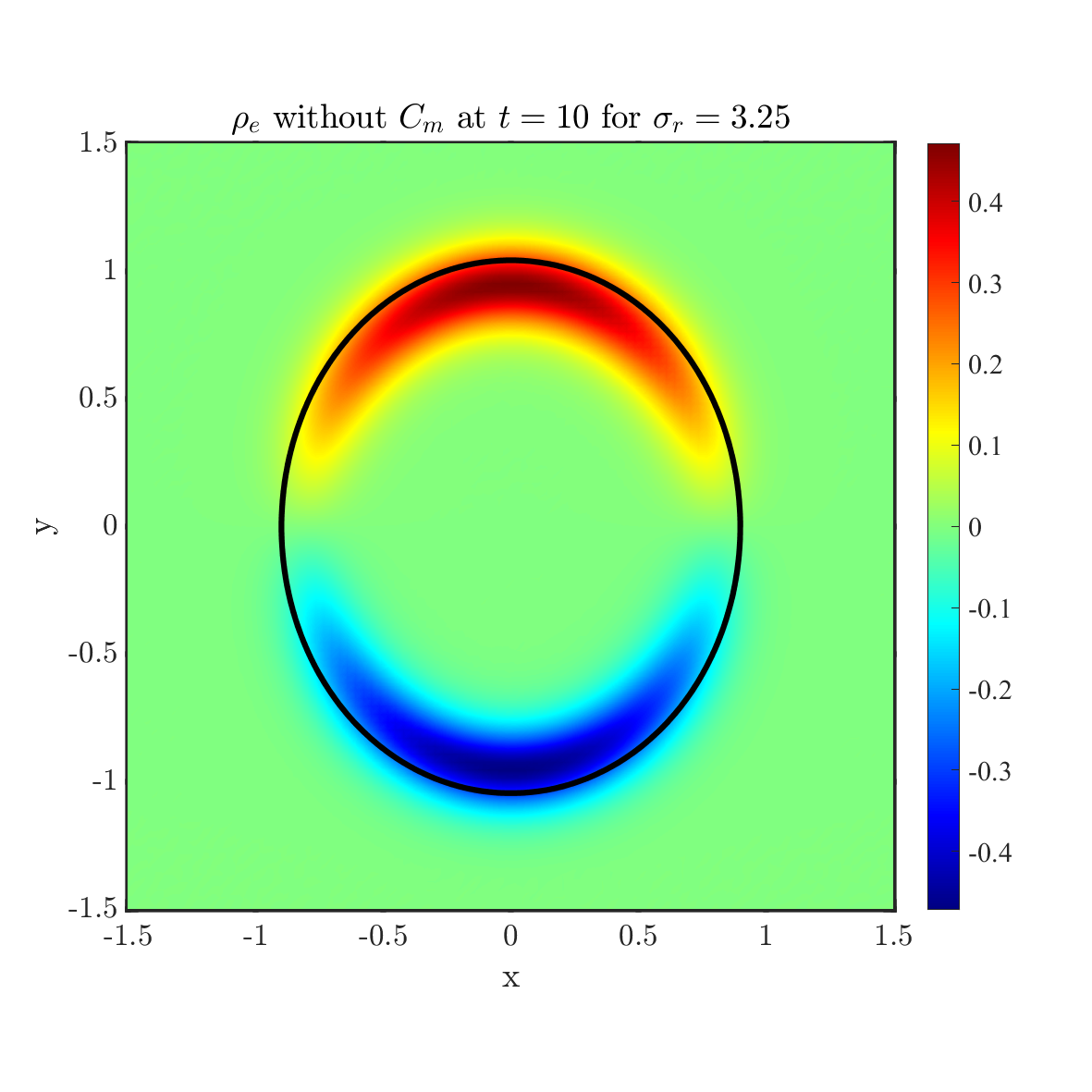}
\includegraphics[width=0.32\textwidth]{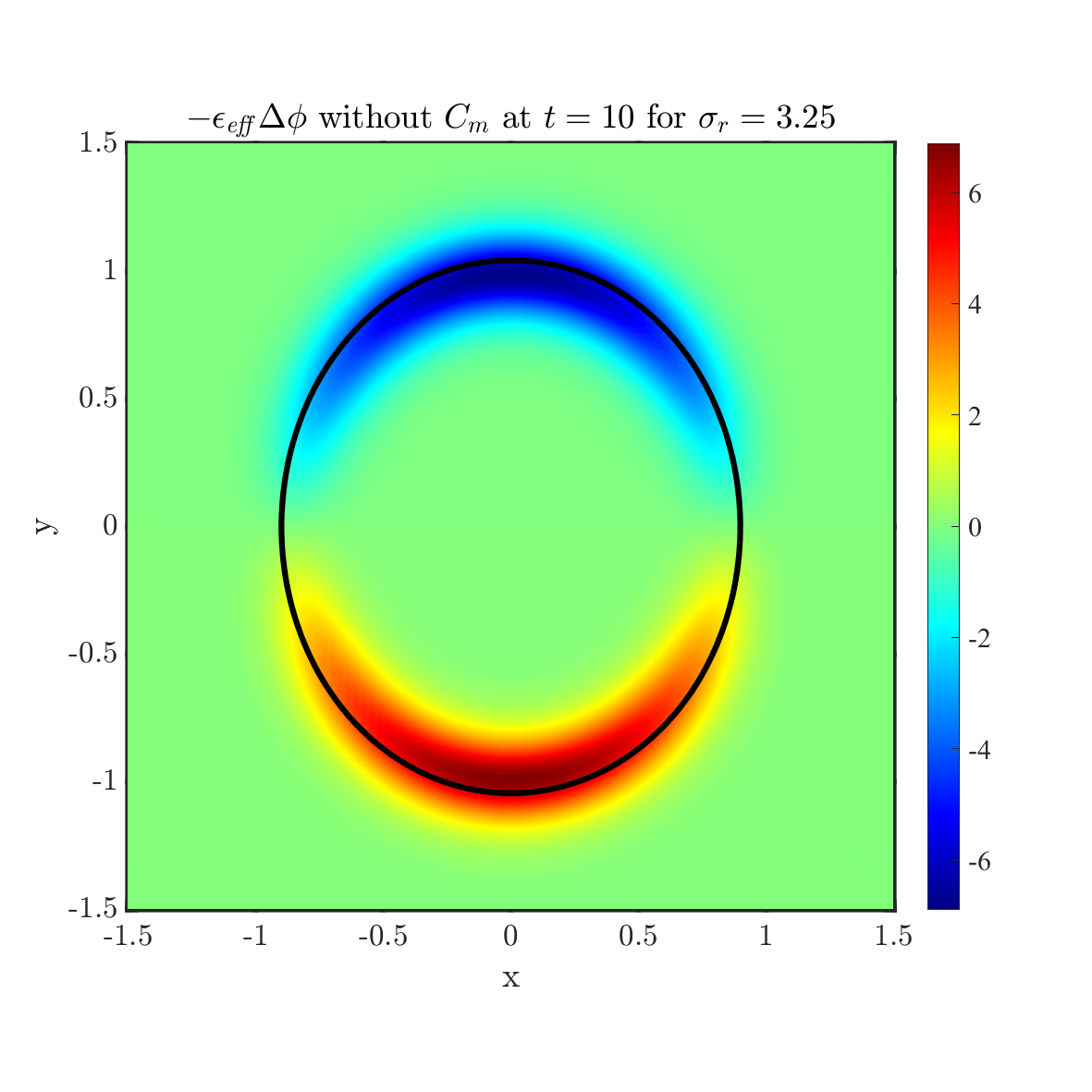}
\includegraphics[width=0.32\textwidth]{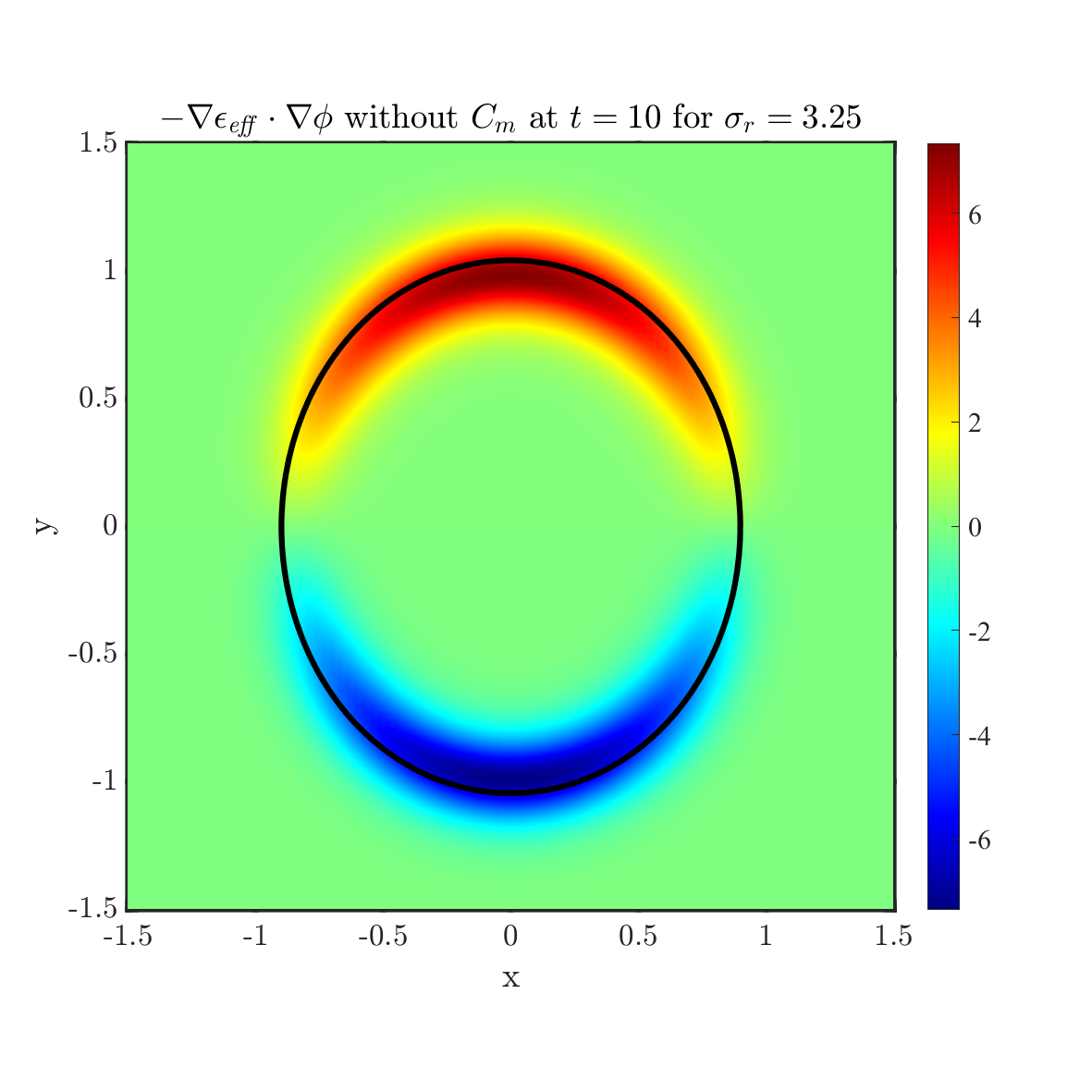}
\end{center} 
\caption{The net charge distribution for conductivity ratio $\sigma_{r} = 3.25$ 
by considering different capacitances $C_{m} = 1$, $C_{m} = \delta^{-1}$, $C_{m} = \delta^{-2}$ and no $C_{m}$ 
from top to bottom 	at time $t = 10$. In each figure, the solid line shows the zero level set ($\psi=0$).
The rest parameters are chosen as $\epsilon_{r} = 3.5$, $Ca_{E} = 1$.}
\label{fig: net charge for single drop with 3 cm sigma35}
\end{figure}

\begin{figure}
\begin{center}
\includegraphics[width=0.32\textwidth]{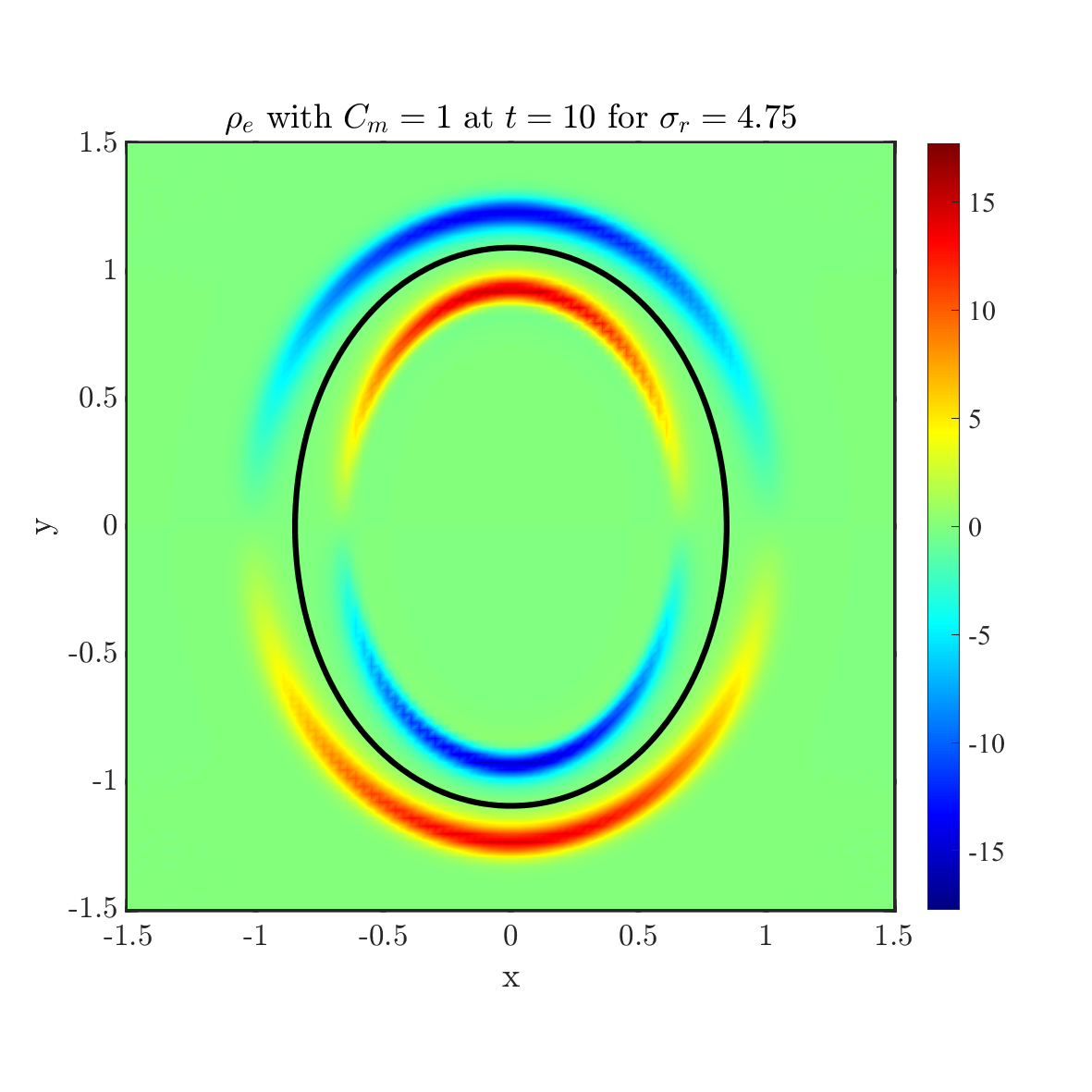}
\includegraphics[width=0.32\textwidth]{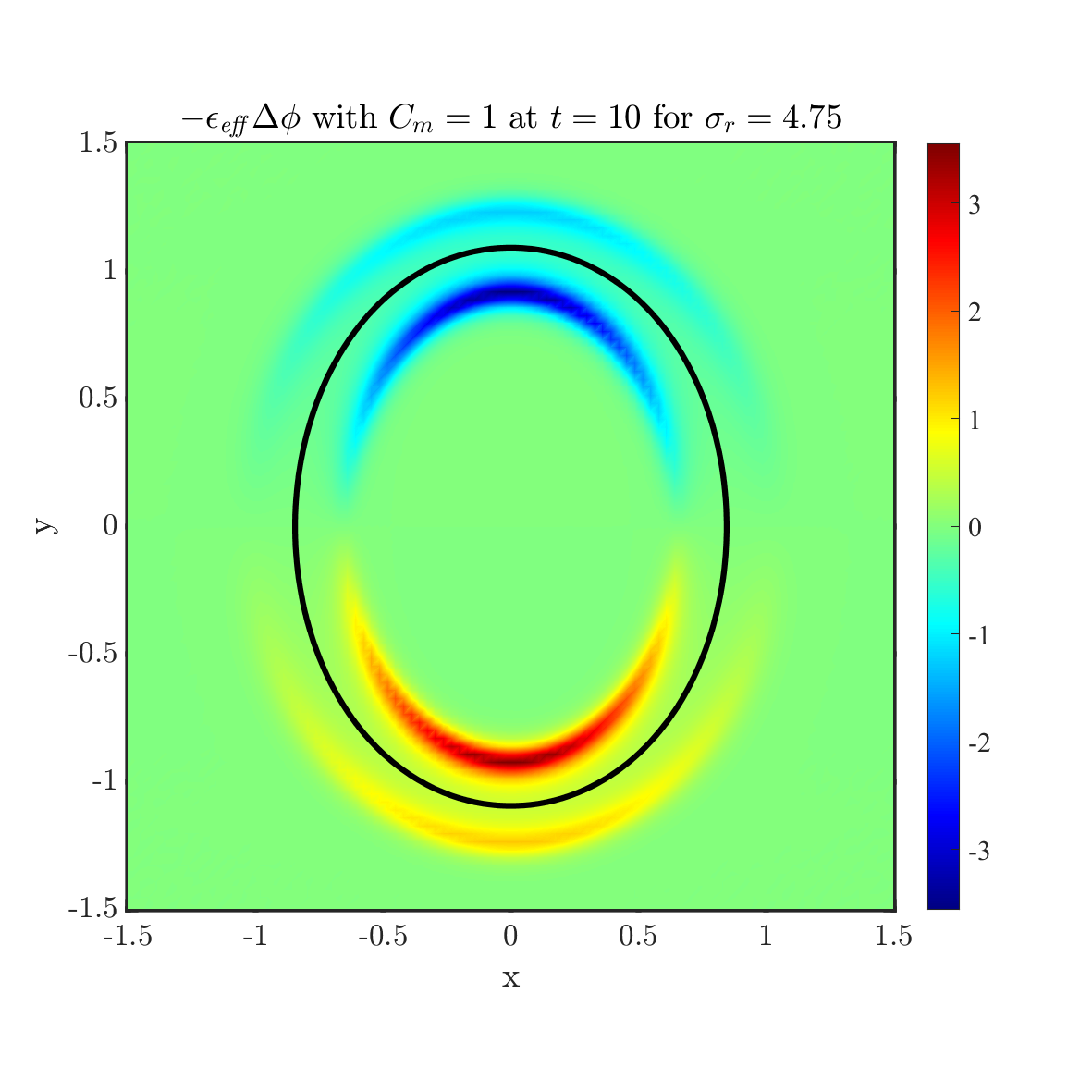}
\includegraphics[width=0.32\textwidth]{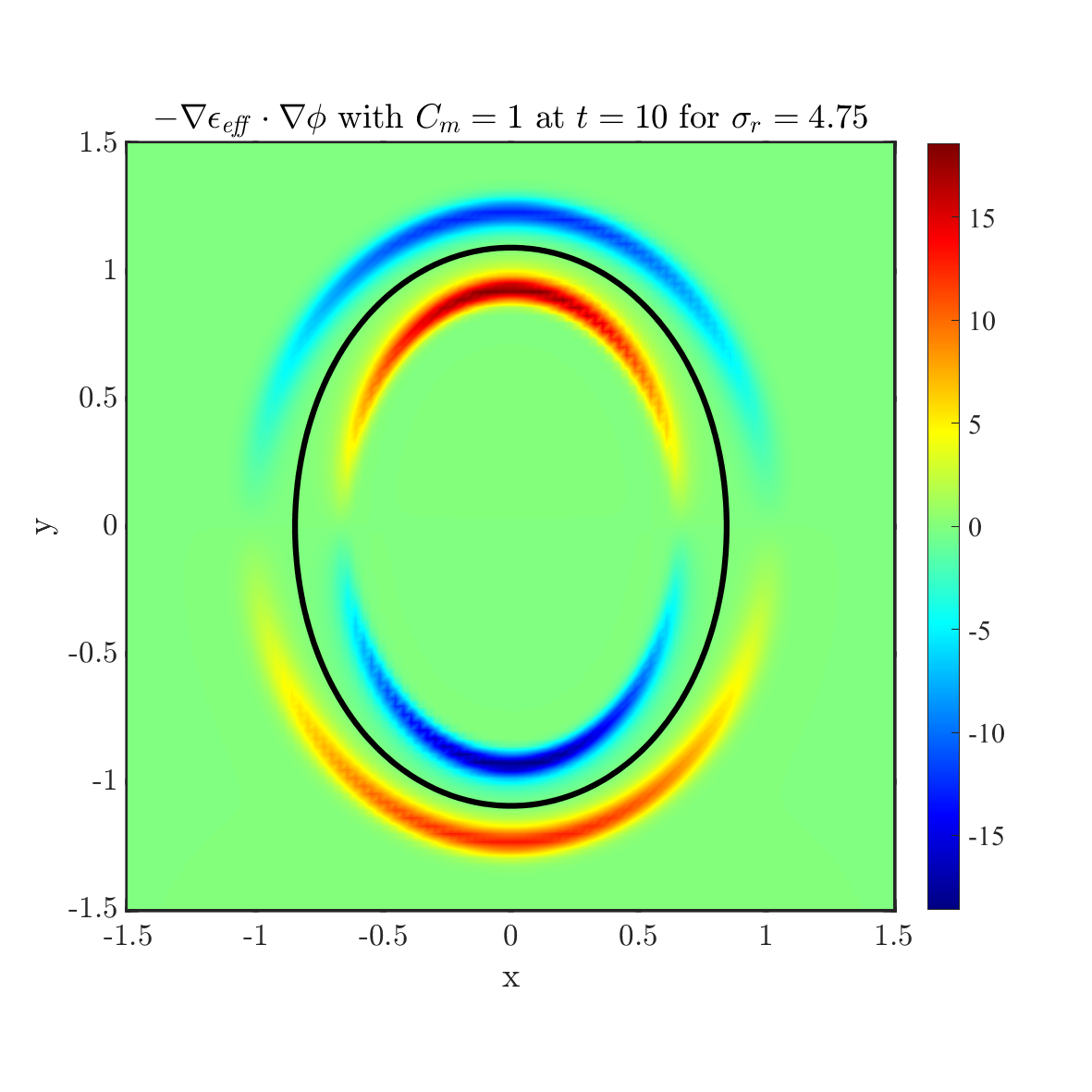}
\includegraphics[width=0.32\textwidth]{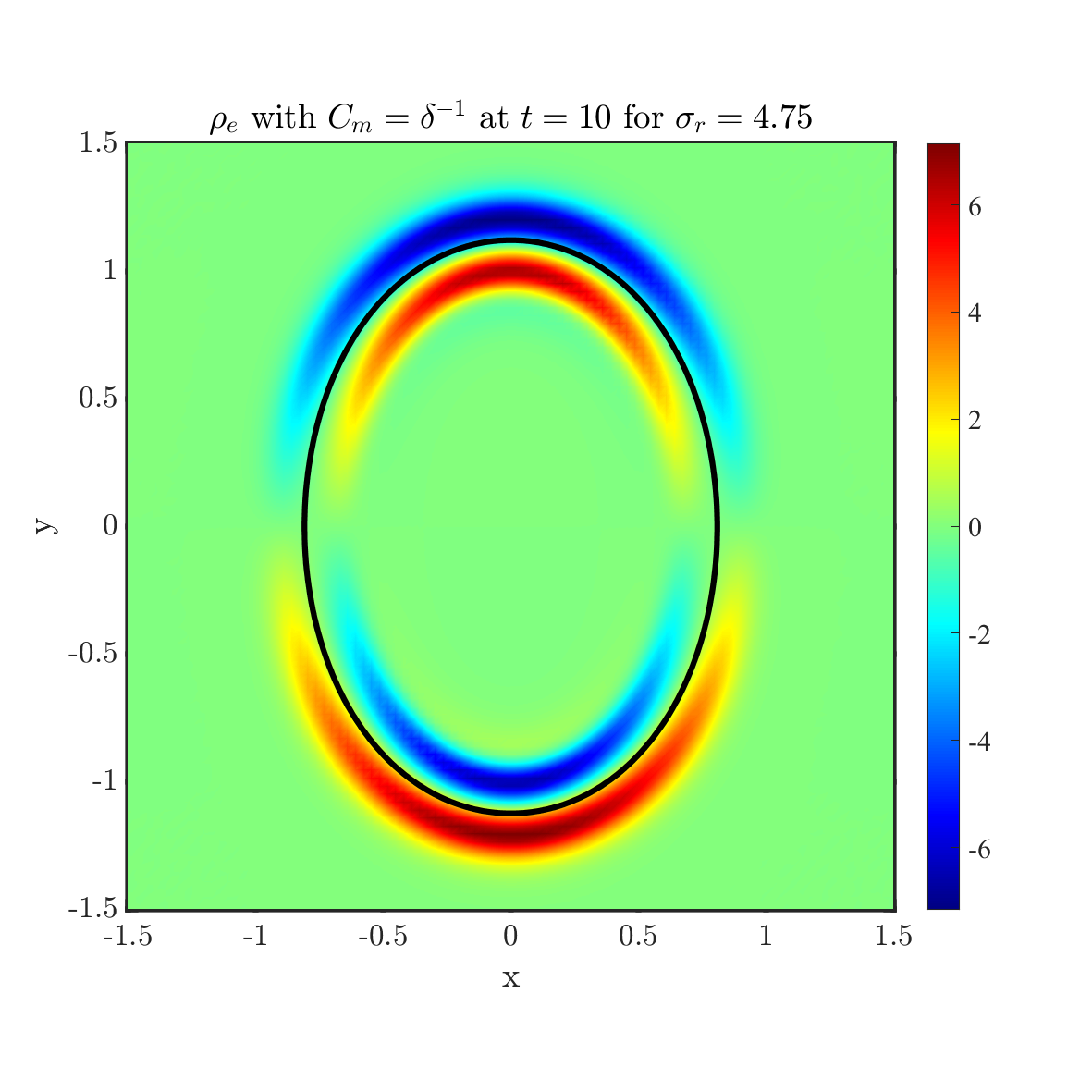}
\includegraphics[width=0.32\textwidth]{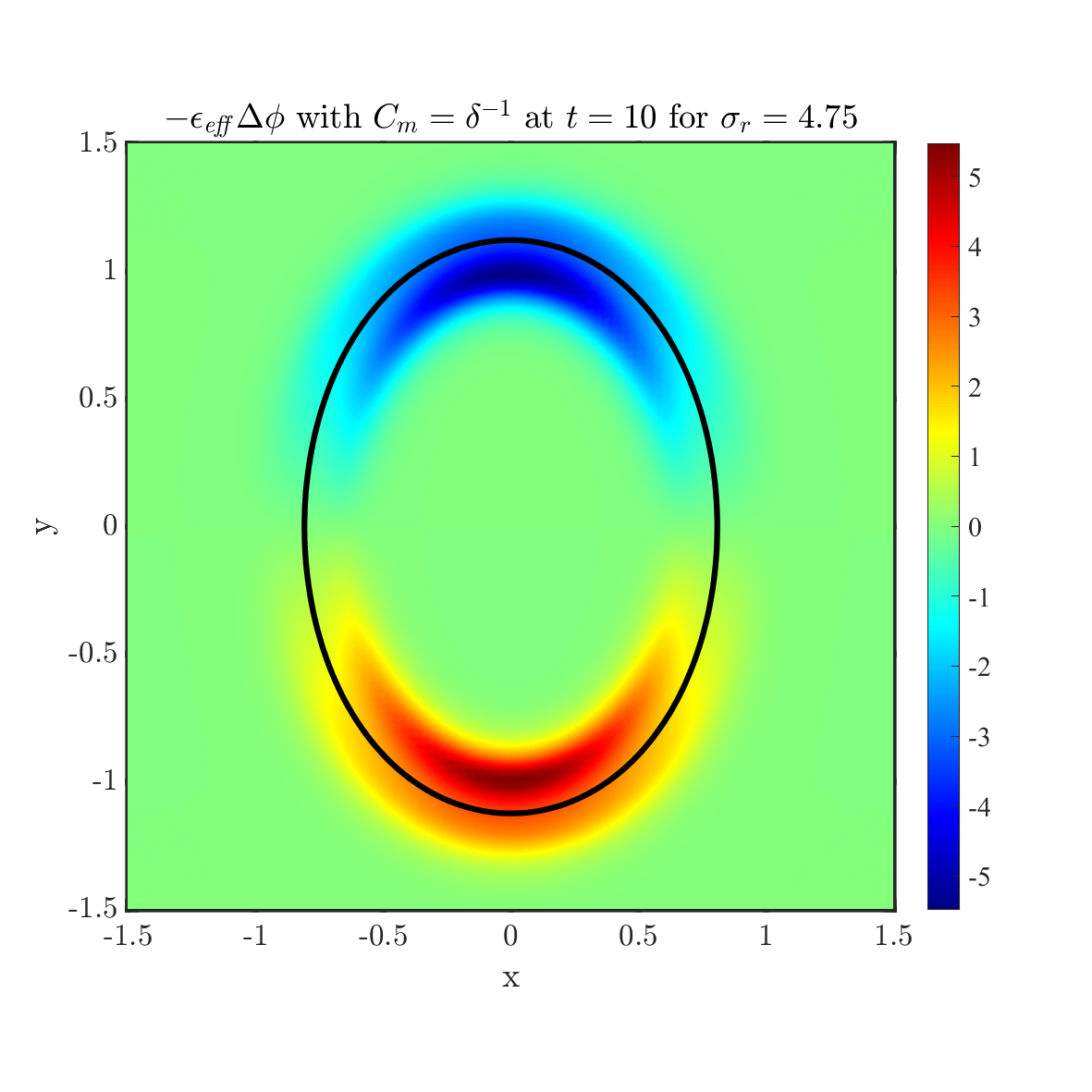}
\includegraphics[width=0.32\textwidth]{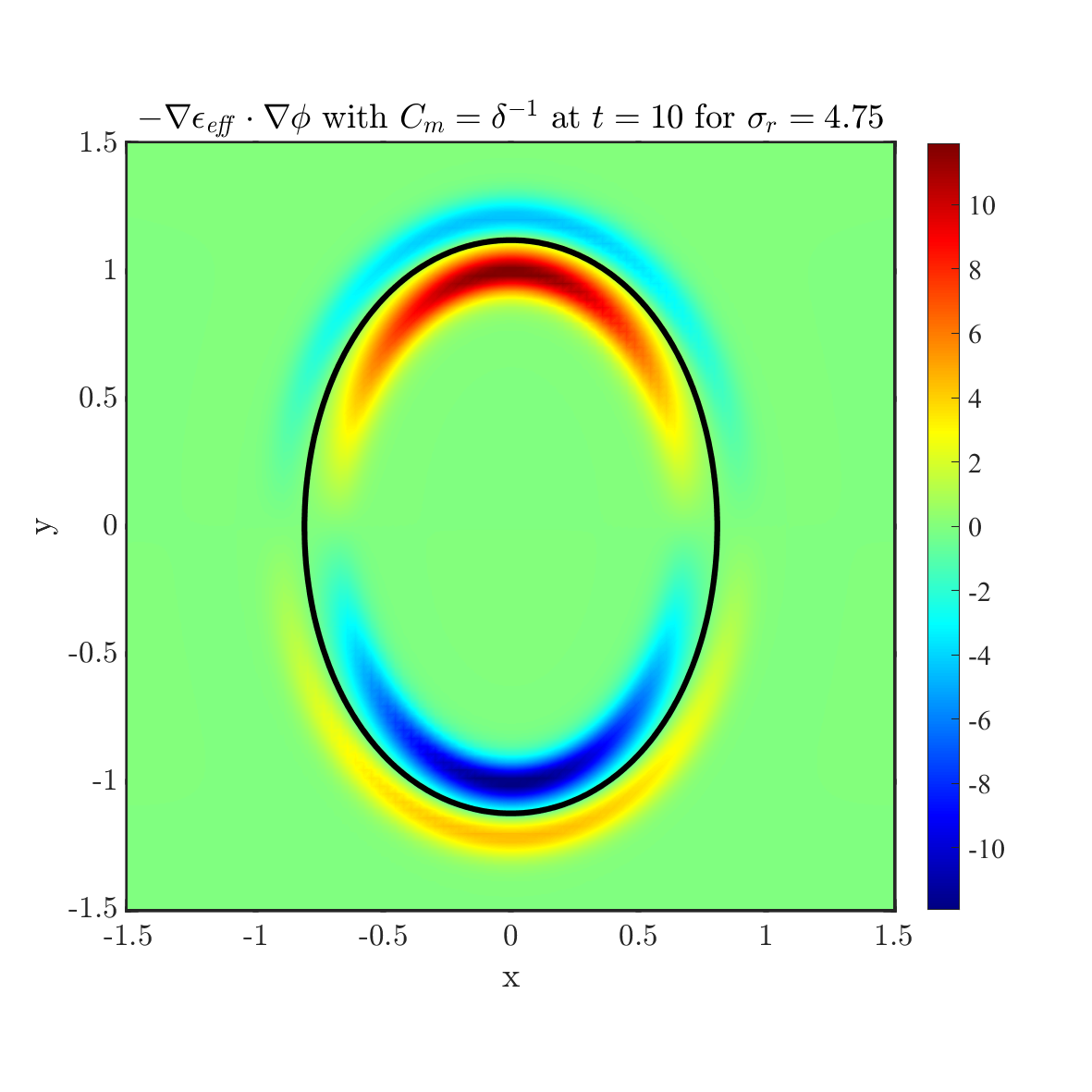}
\includegraphics[width=0.32\textwidth]{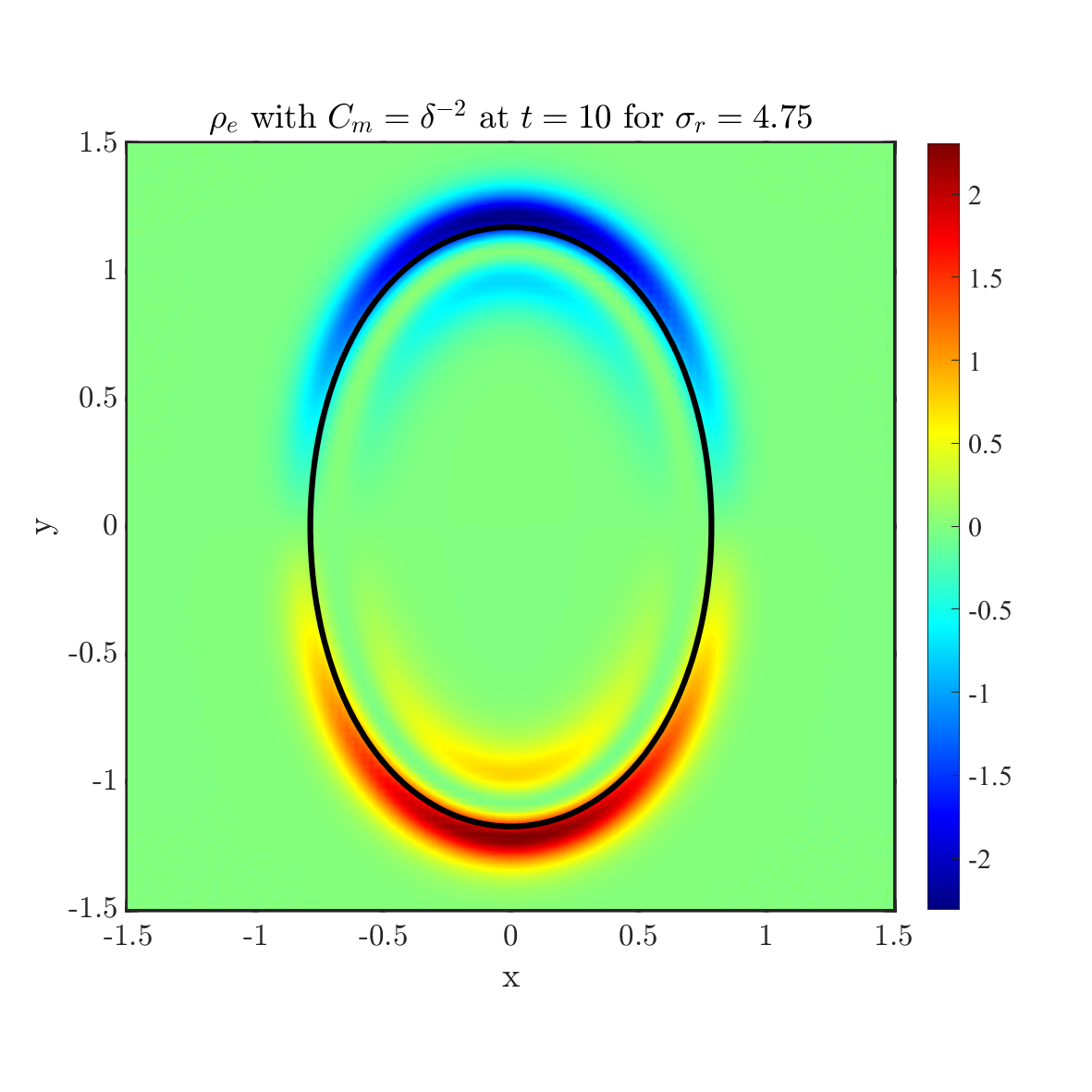}
\includegraphics[width=0.32\textwidth]{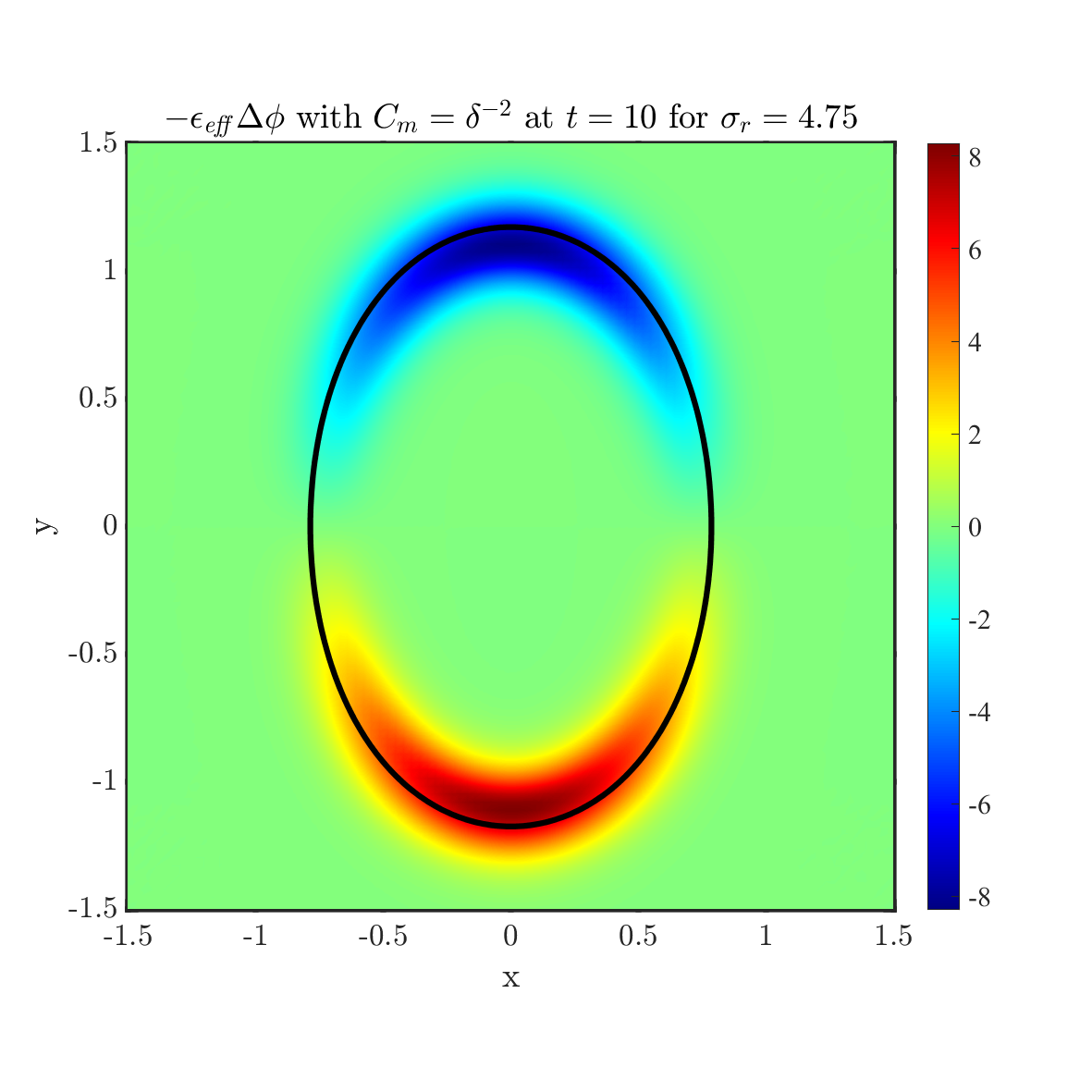}
\includegraphics[width=0.32\textwidth]{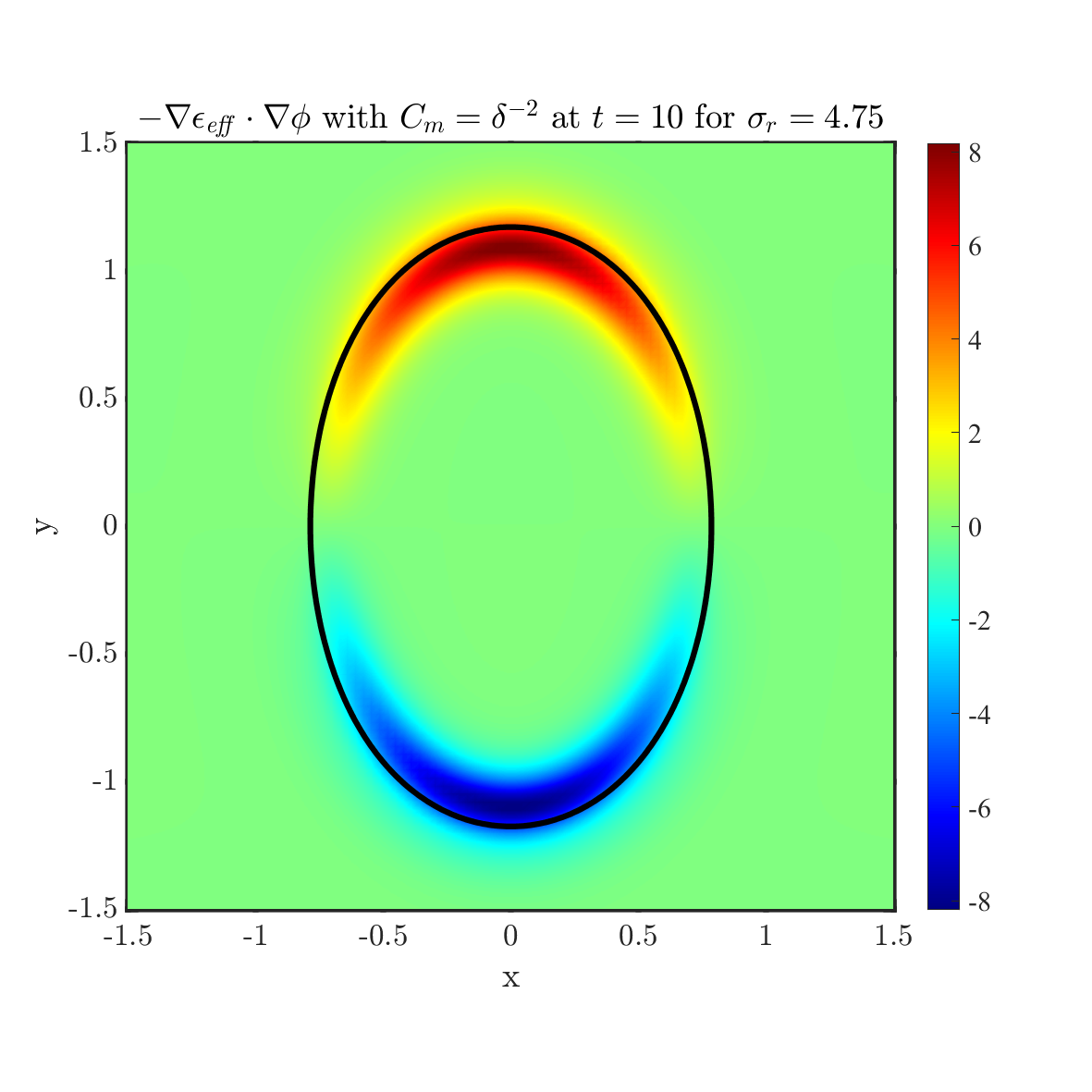}
\includegraphics[width=0.32\textwidth]{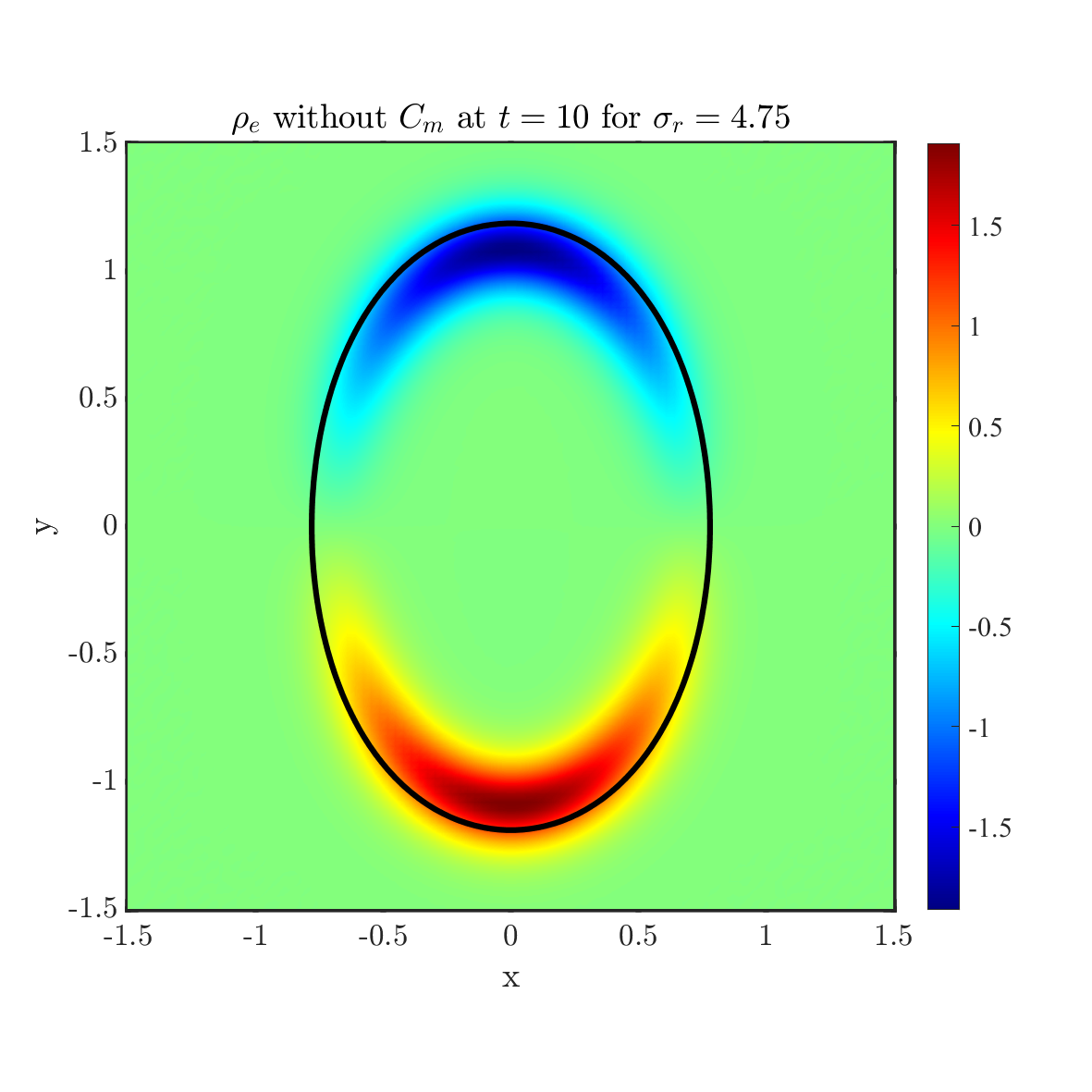}
\includegraphics[width=0.32\textwidth]{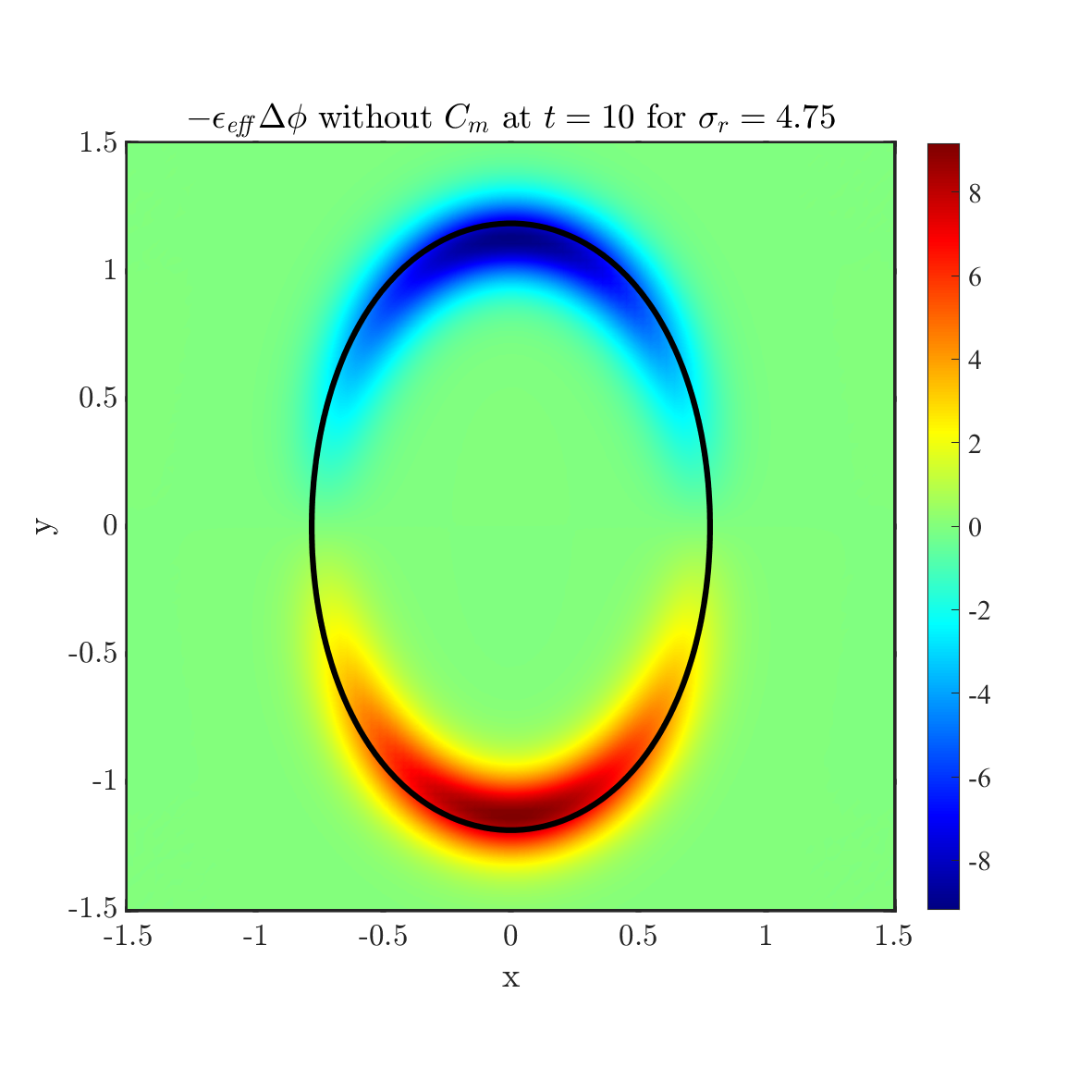}
\includegraphics[width=0.32\textwidth]{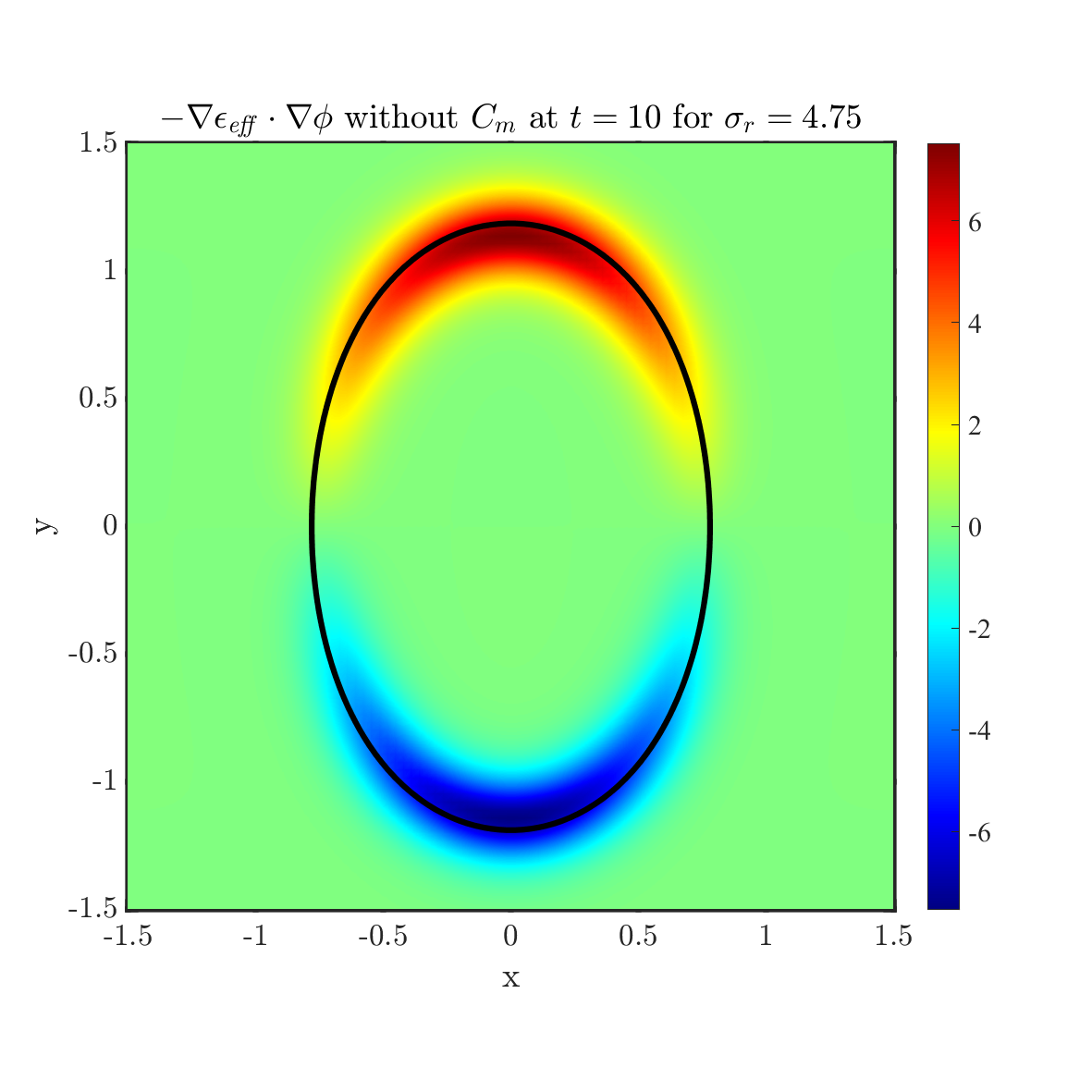}
\end{center} 
\caption{The net charge distribution for conductivity ratio $\sigma_{r} = 4.75$ 
by considering different capacitances $C_{m} = 1$, $C_{m} = \delta^{-1}$, $C_{m} = \delta^{-2}$ and no $C_{m}$ 
from top to bottom 	at time $t = 10$. In each figure, the solid line shows the zero level set ($\psi=0$).
The rest parameters are chosen as $\epsilon_{r} = 3.5$, $Ca_{E} = 1$.}
\label{fig: net charge for single drop with 3 cm sigma475}
\end{figure}

\begin{figure}
\begin{center}
\includegraphics[width=0.32\textwidth]{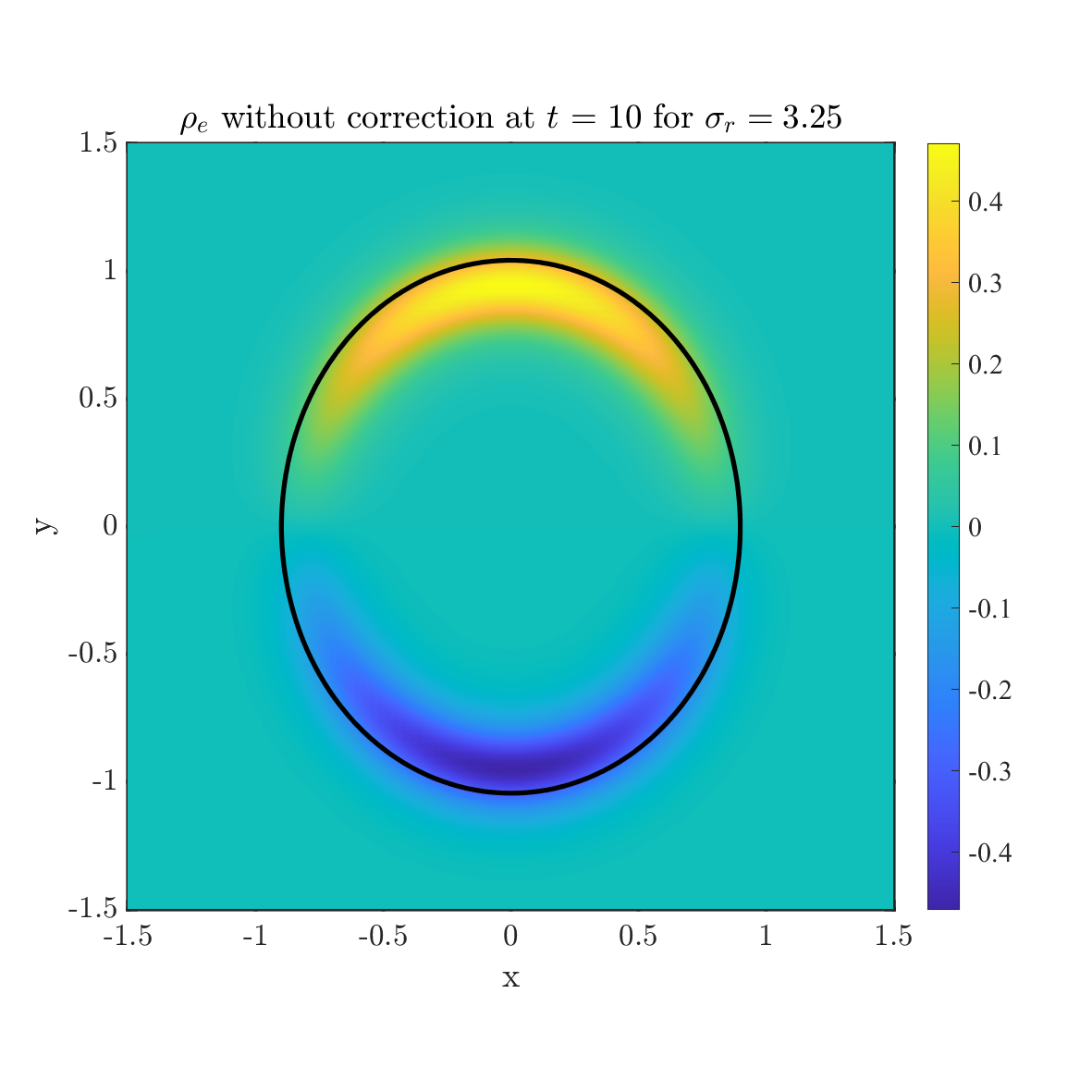}
\includegraphics[width=0.32\textwidth]{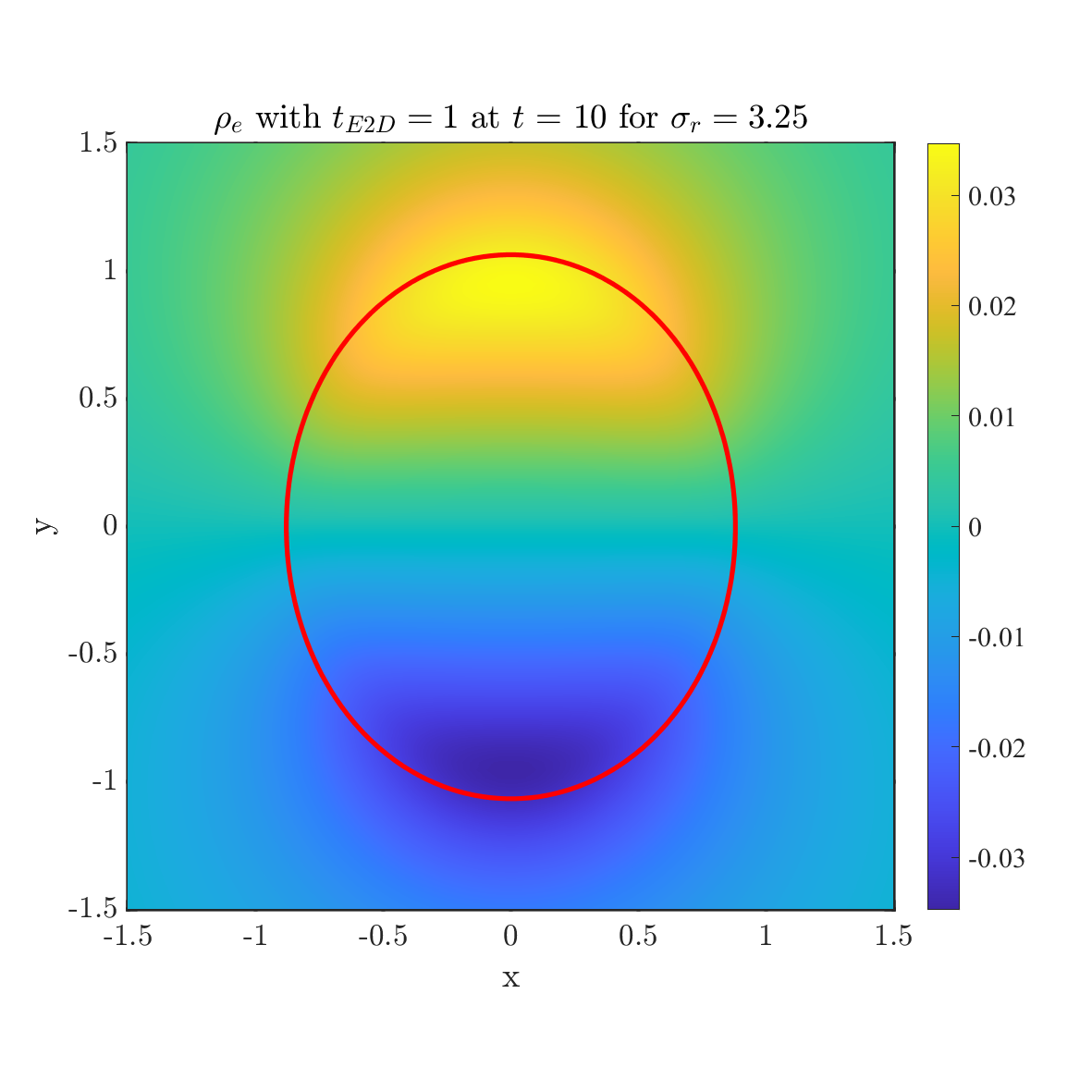}
\includegraphics[width=0.32\textwidth]{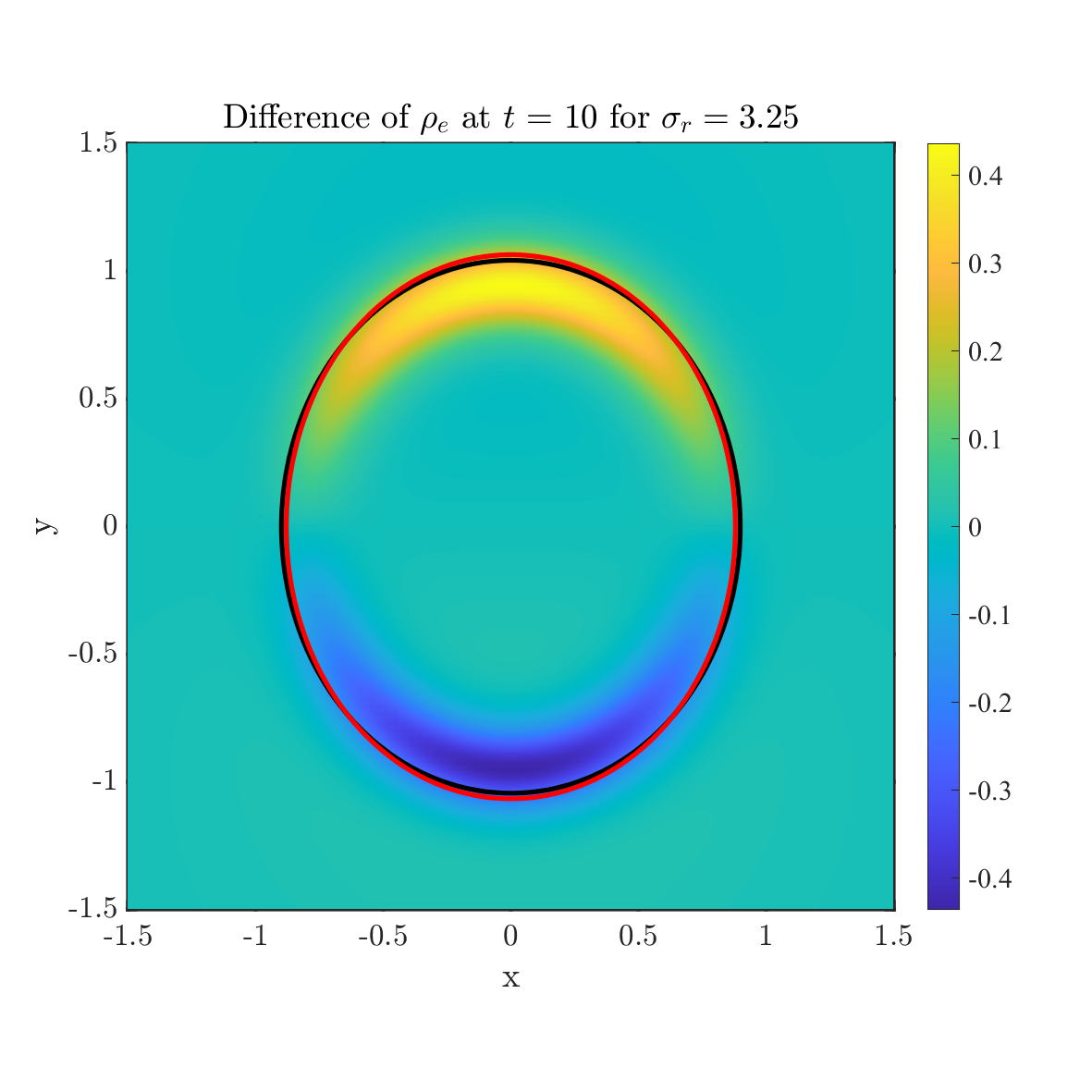}
\includegraphics[width=0.32\textwidth]{no_correction_325_no_Cm.png}
\includegraphics[width=0.32\textwidth]{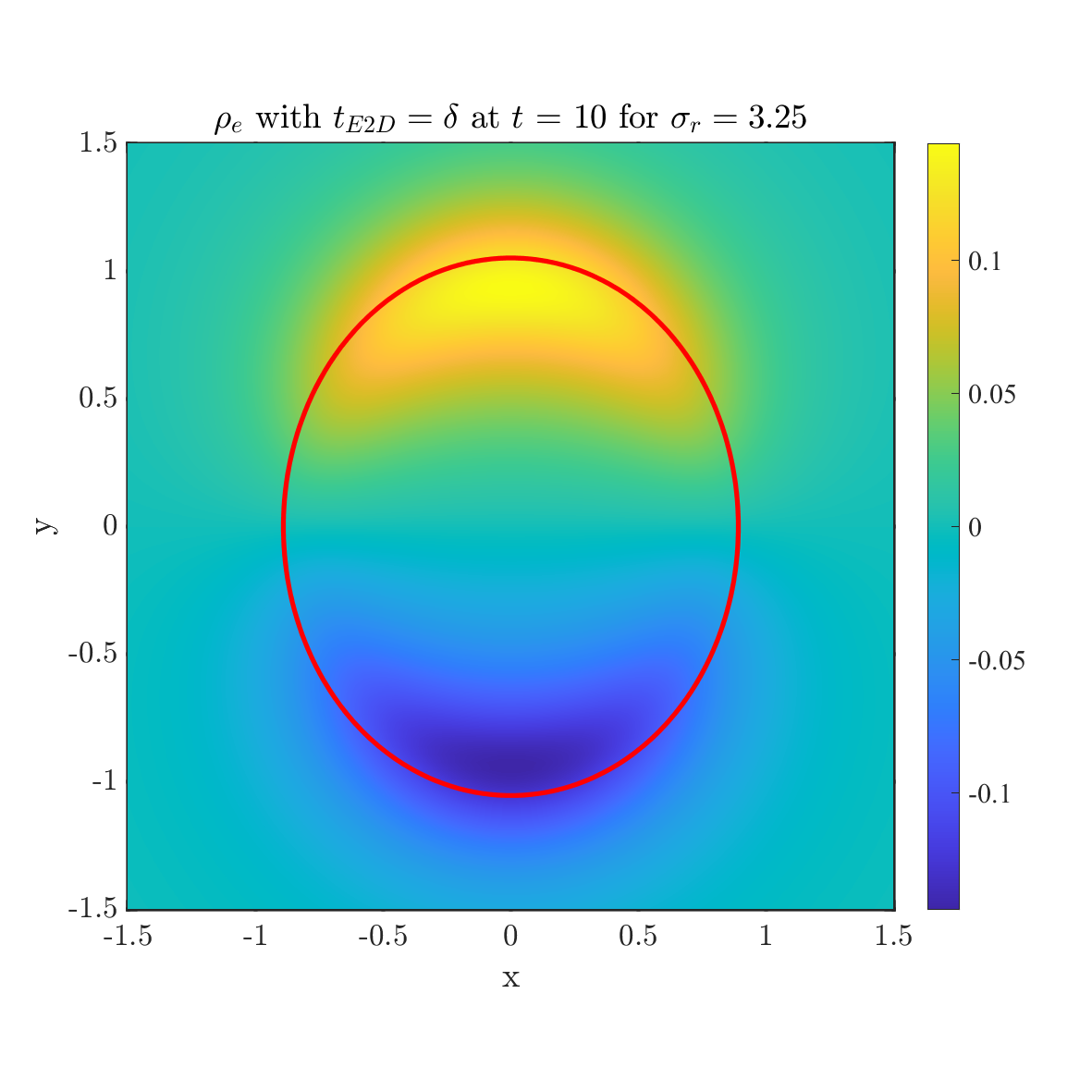}
\includegraphics[width=0.32\textwidth]{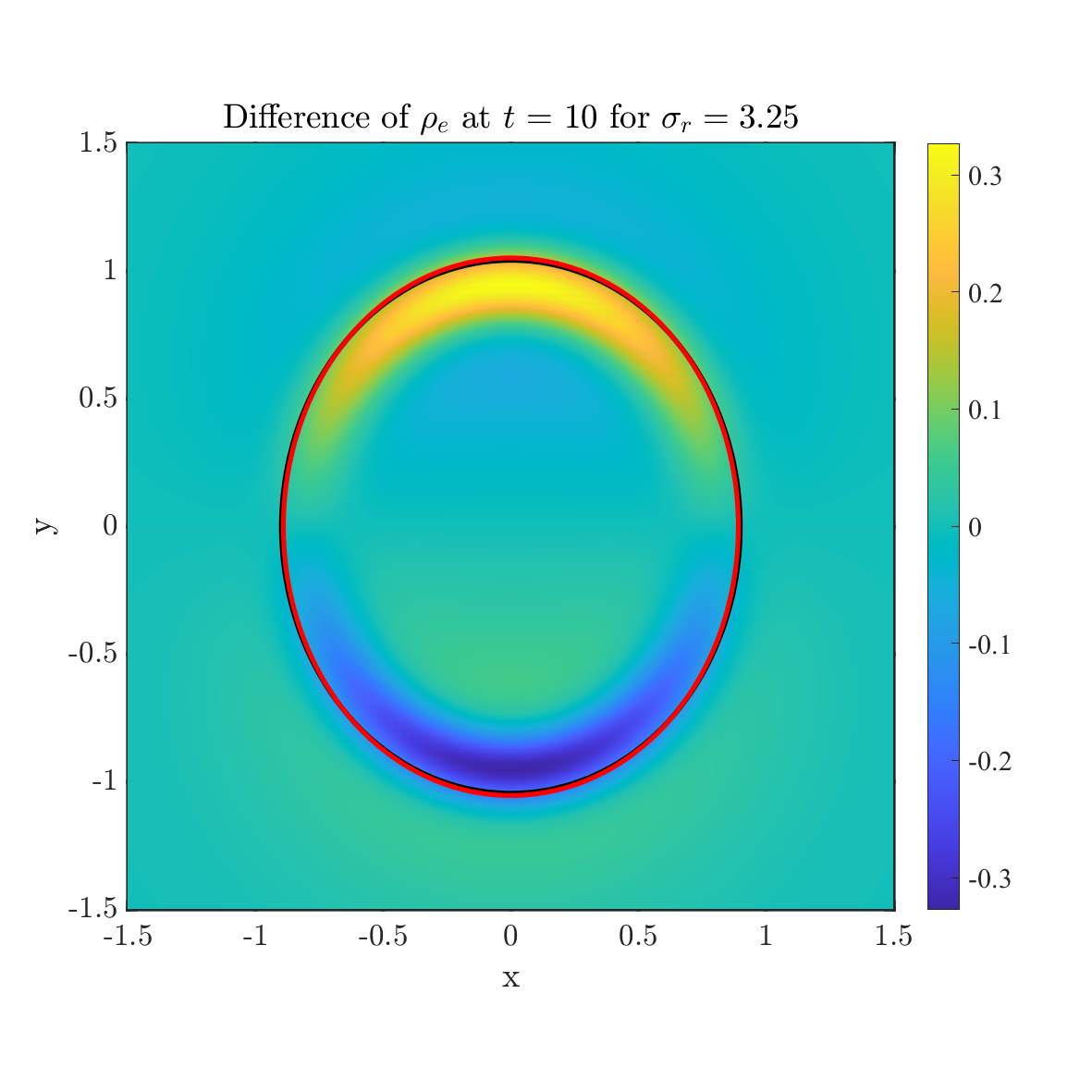}
\includegraphics[width=0.32\textwidth]{no_correction_325_no_Cm.png}
\includegraphics[width=0.32\textwidth]{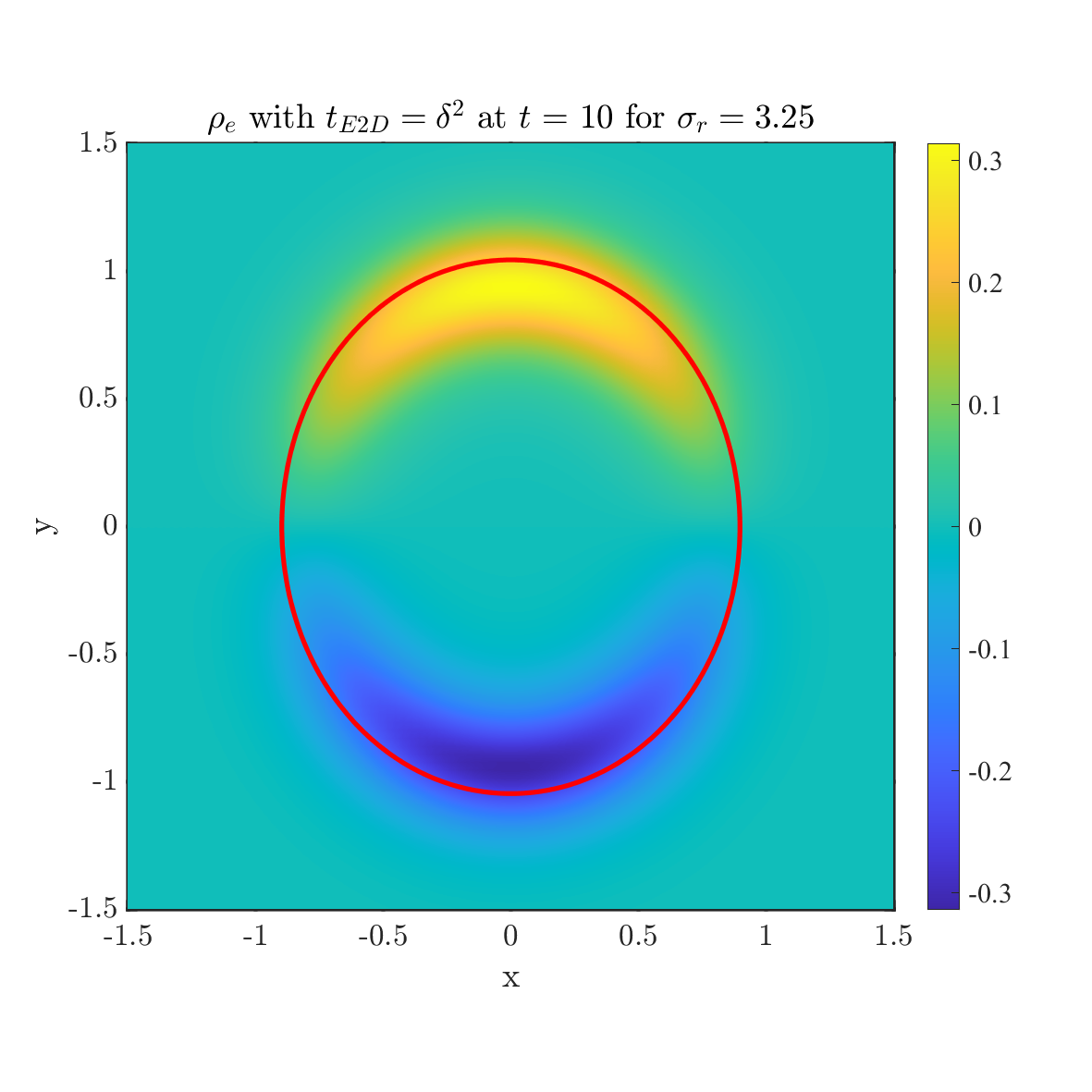}
\includegraphics[width=0.32\textwidth]{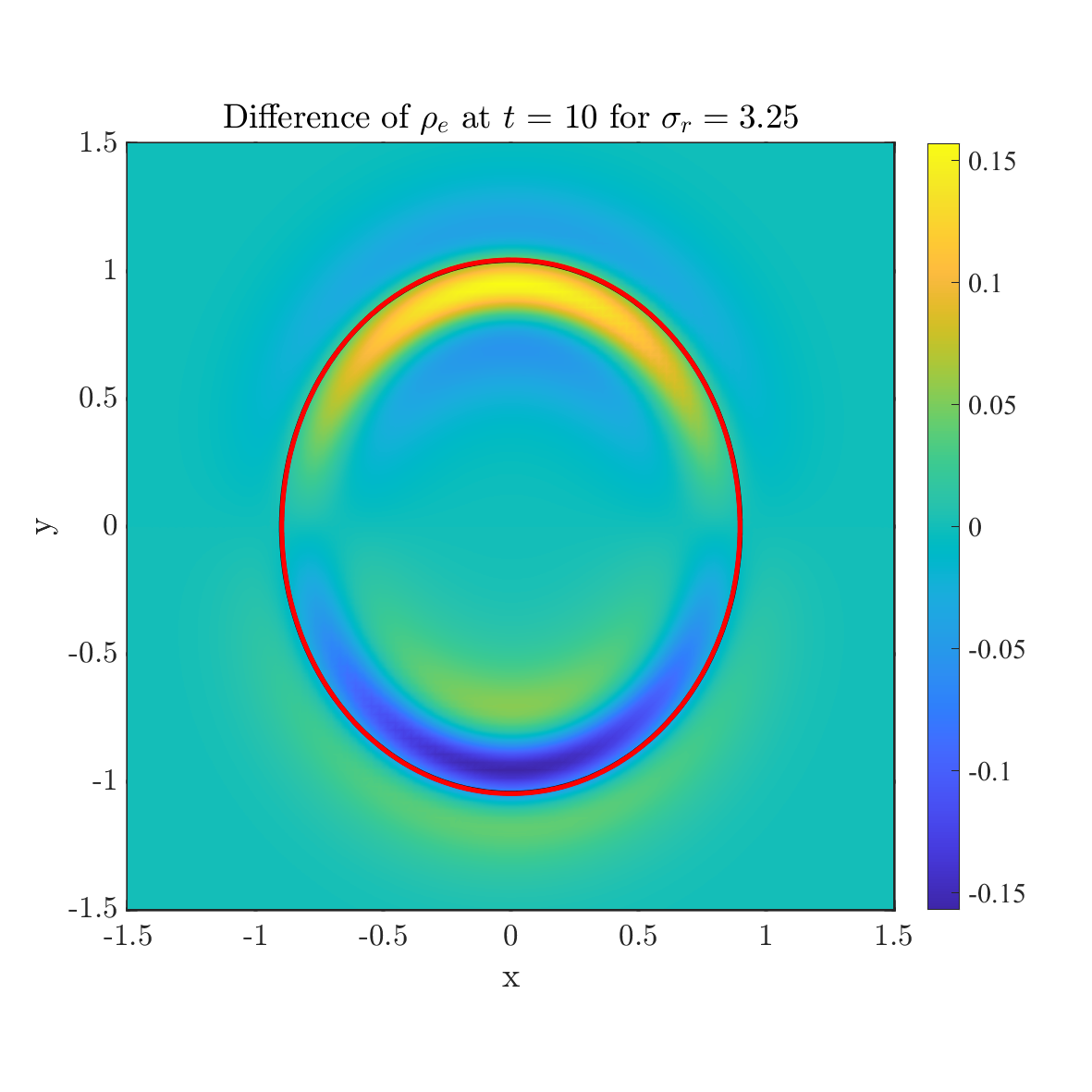}
\end{center}
\caption{The comparison between leaky dielectric model (left) and net charge model (middle) at $t = 10$. 
The conductivity ratio is $\sigma_{r} = 3.25$. 
And the difference between them is shown in the right column. 
relaxation time $t_{E2M} = t_{E2D} = 1$ (top), $t_{E2M} = t_{E2D} =\delta$ (middle) and $t_{E2M} = t_{E2D} = \delta^{2}$ (bottom) are considered here. 
In each figure, the solid line shows the zero level set ($\psi=0$) where the black line shows the drop shape without correction 
and the red line shows the drop shape with correction. 
The rest parameters are chosen as $\epsilon_{r} = 3.5$, $Ca_{E} = 1$.}
\label{fig: correction2_325}
\end{figure}
\begin{figure}
\begin{center}
\includegraphics[width=0.32\textwidth]{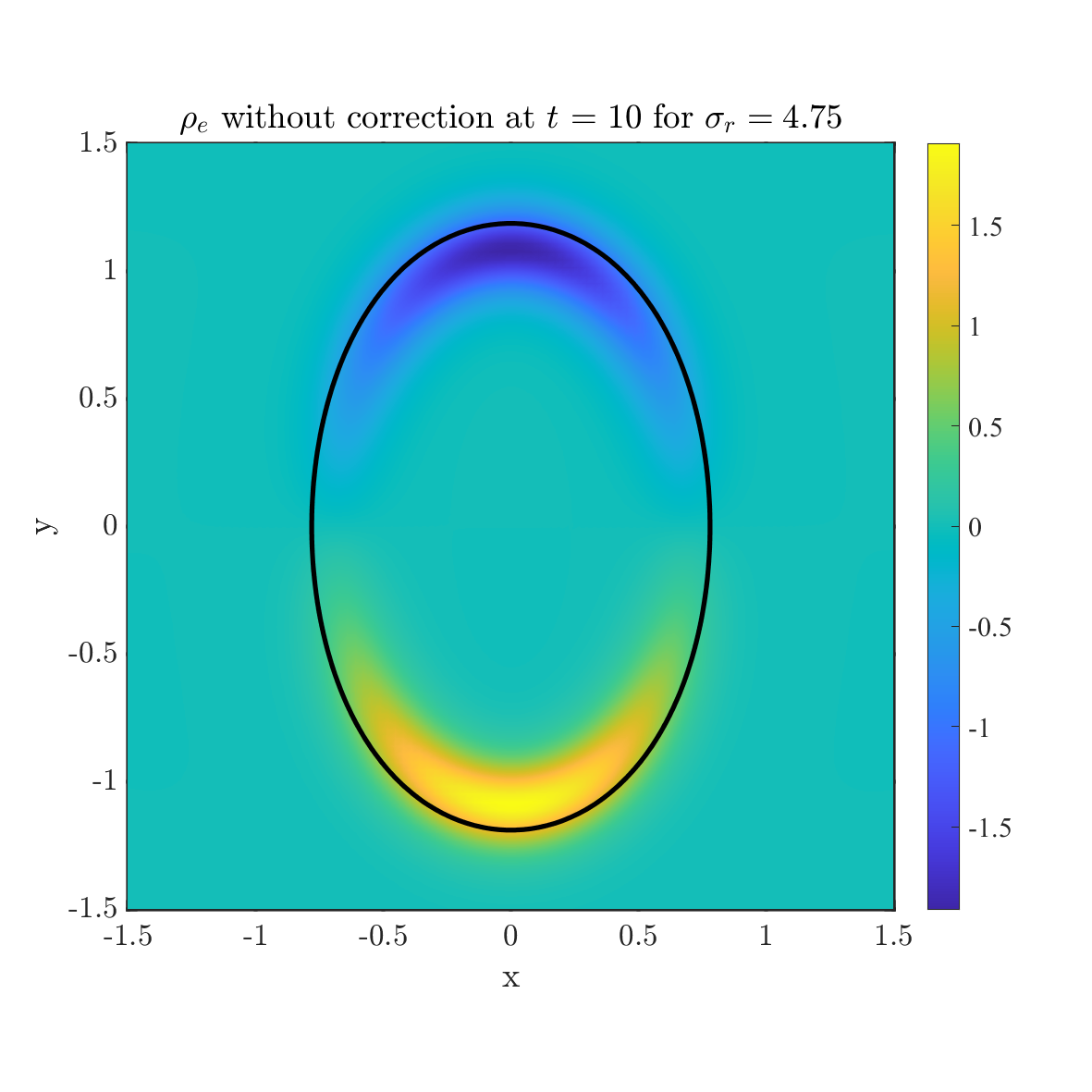}
\includegraphics[width=0.32\textwidth]{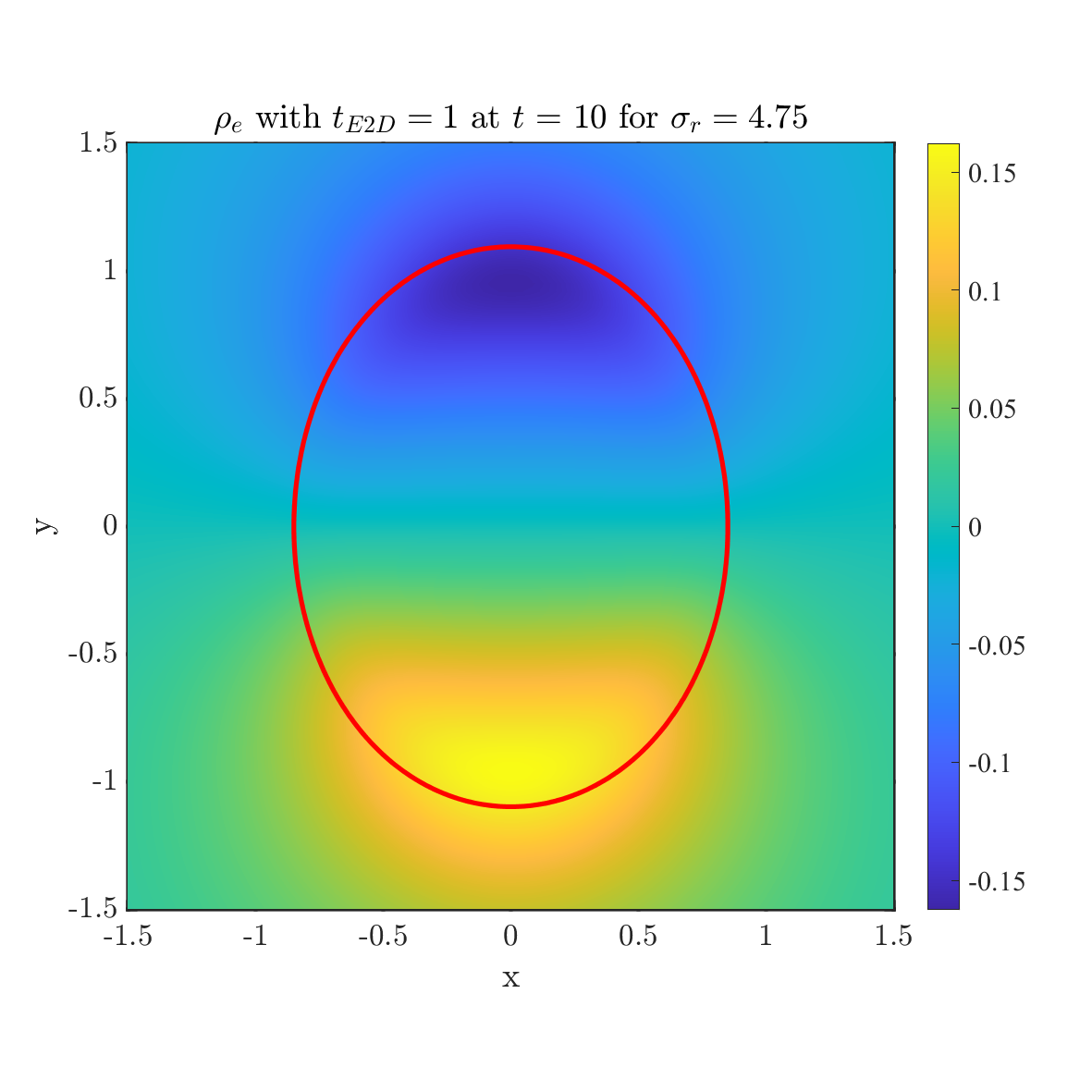}
\includegraphics[width=0.32\textwidth]{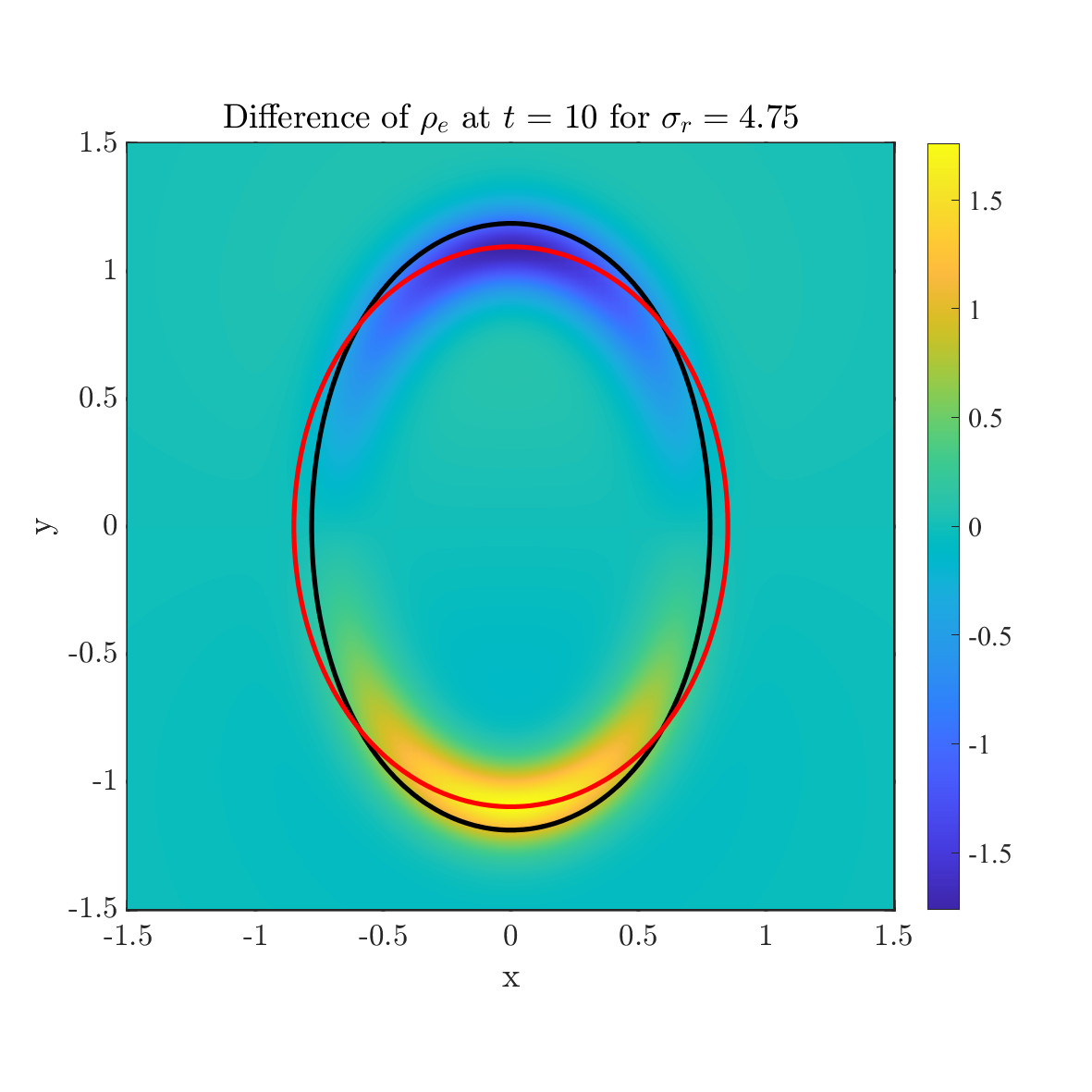}
\includegraphics[width=0.32\textwidth]{no_correction_475_no_Cm.png}
\includegraphics[width=0.32\textwidth]{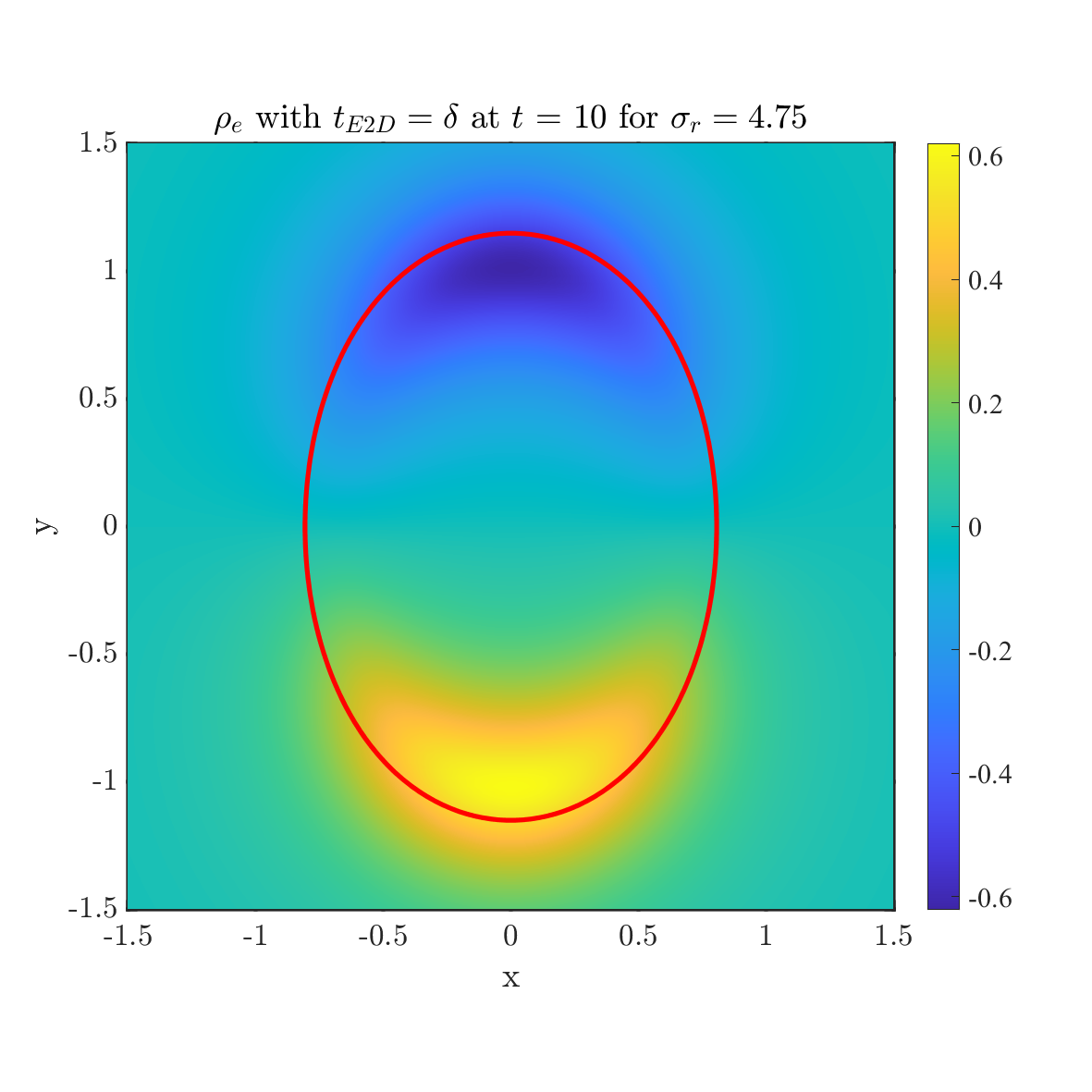}
\includegraphics[width=0.32\textwidth]{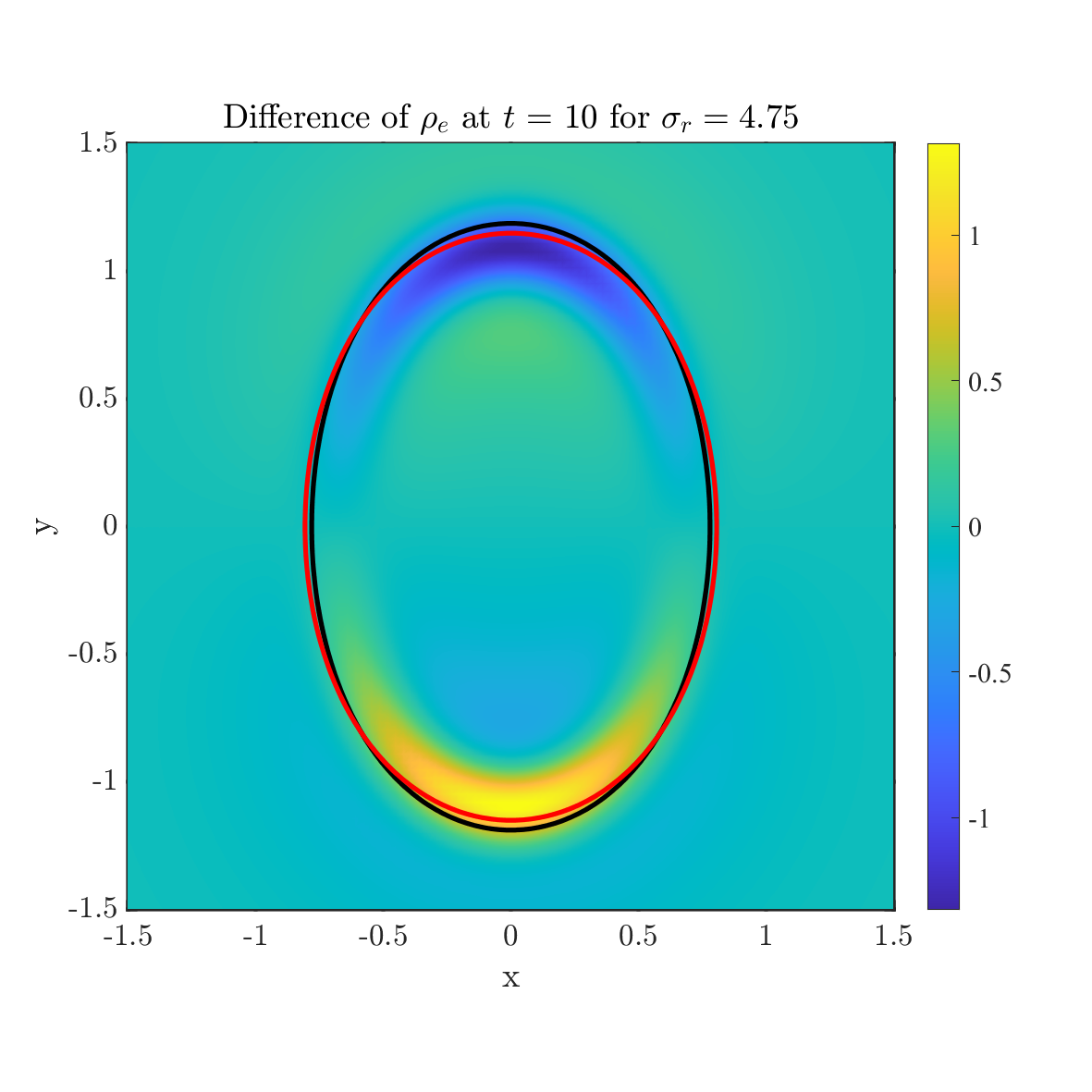}
\includegraphics[width=0.32\textwidth]{no_correction_475_no_Cm.png}
\includegraphics[width=0.32\textwidth]{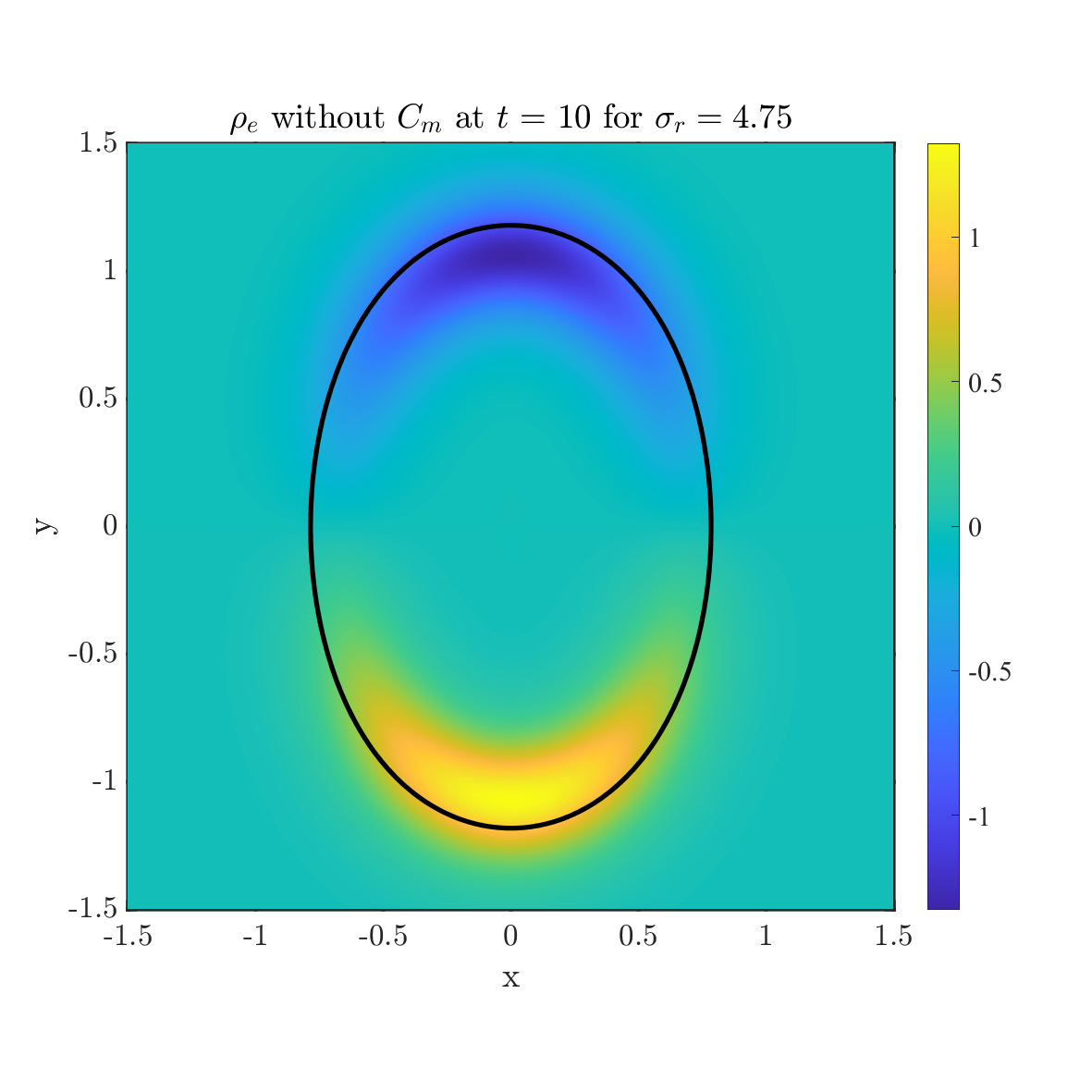}
\includegraphics[width=0.32\textwidth]{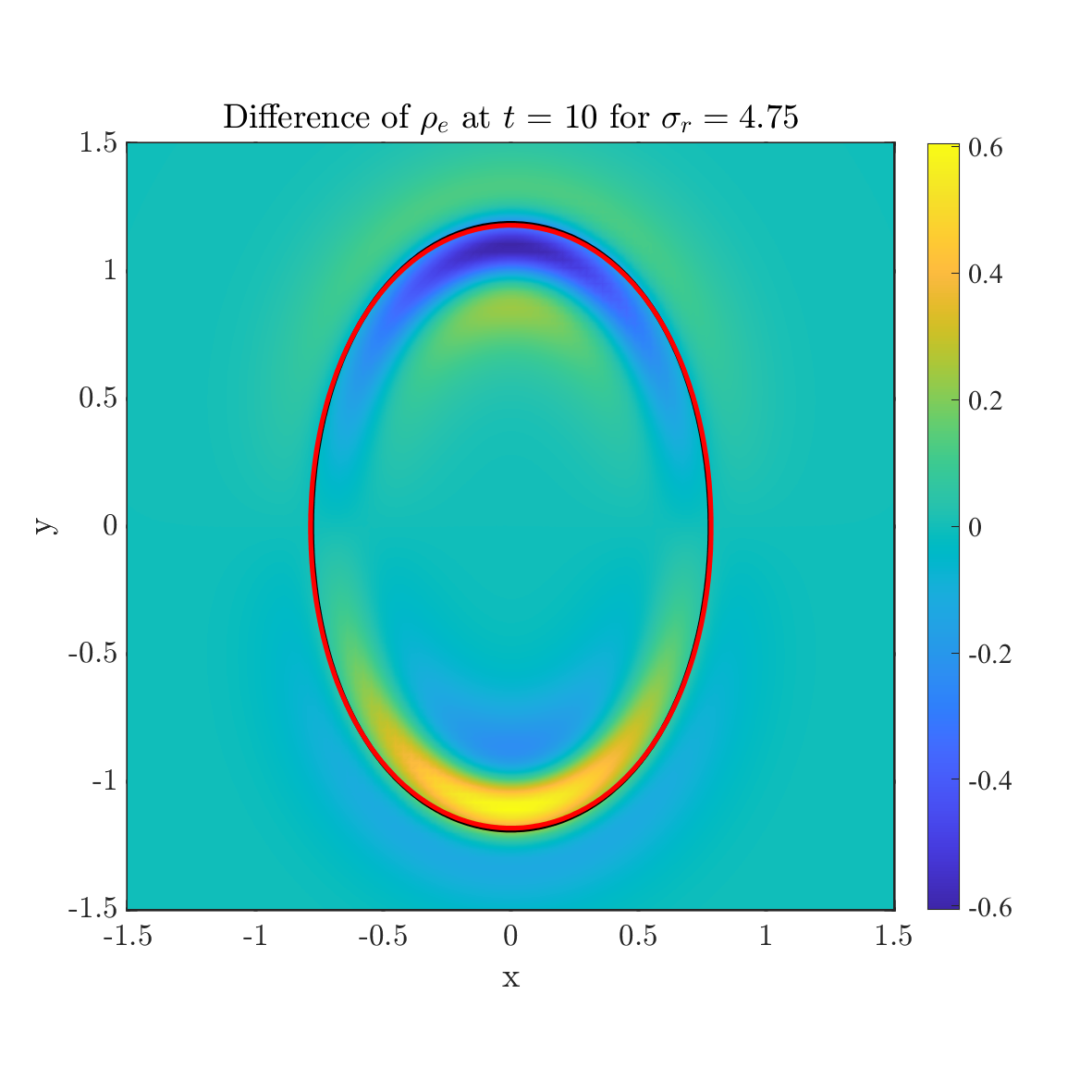}
\end{center}
\caption{The comparison between leaky dielectric model (left) and net charge model (middle) at $t = 10$. 
The conductivity ratio is $\sigma_{r} = 4.75$. 
And the difference between them is shown in the right column. 
relaxation time $t_{E2M} = t_{E2D} = 1$ (top), $t_{E2M} = t_{E2D} =\delta$ (middle) and $t_{E2M} = t_{E2D} = \delta^{2}$ (bottom) are considered here. 
In each figure, the solid line shows the zero level set ($\psi=0$) where the black line shows the drop shape without correction 
and the red line shows the drop shape with correction. 
The rest parameters are chosen as $\epsilon_{r} = 3.5$, $Ca_{E} = 1$.}
\label{fig: correction2_475}
\end{figure}
\begin{figure}
\begin{center}
\includegraphics[width=0.32\textwidth]{correction_175_no_Cm_tE_d+2.png}
\includegraphics[width=0.32\textwidth]{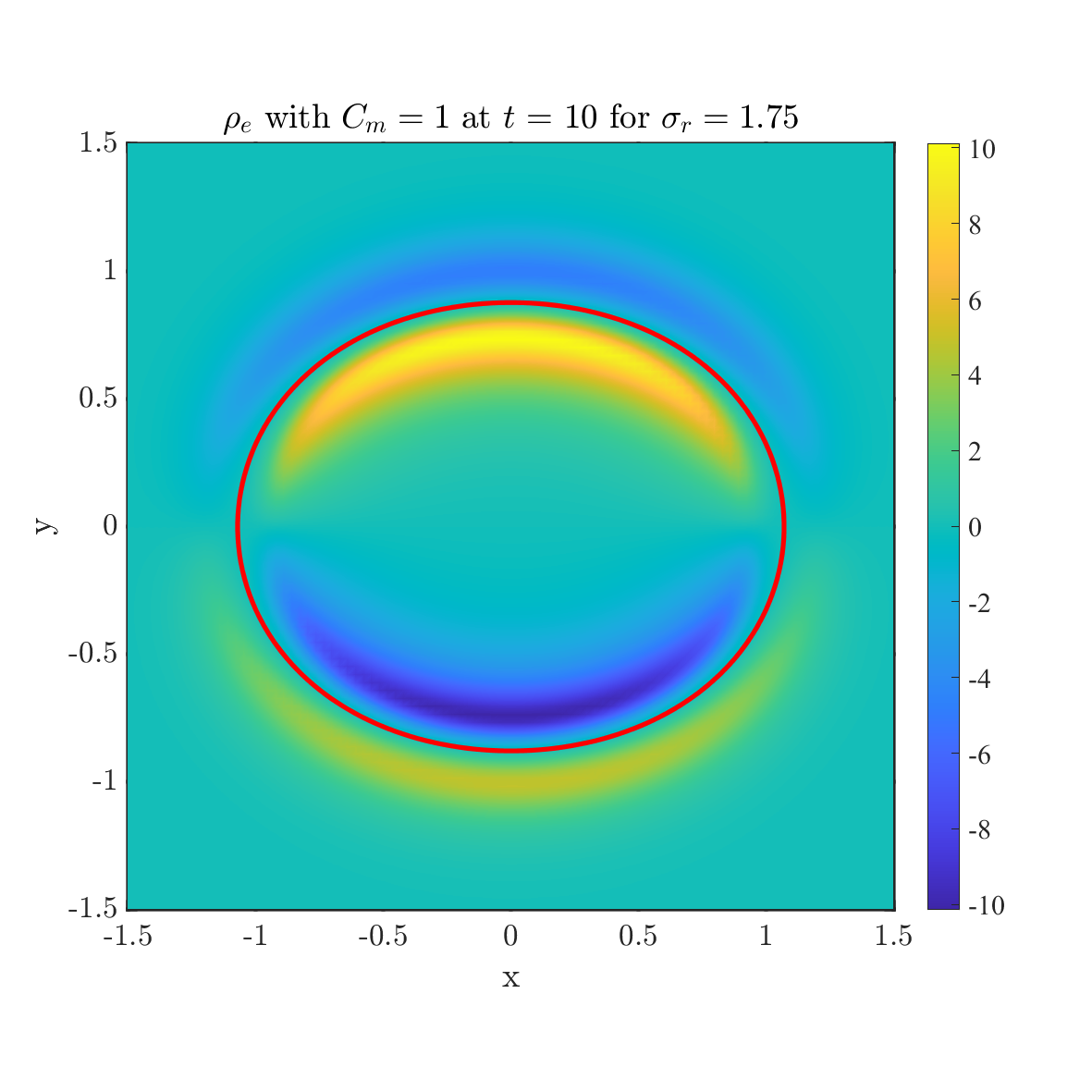}
\includegraphics[width=0.32\textwidth]{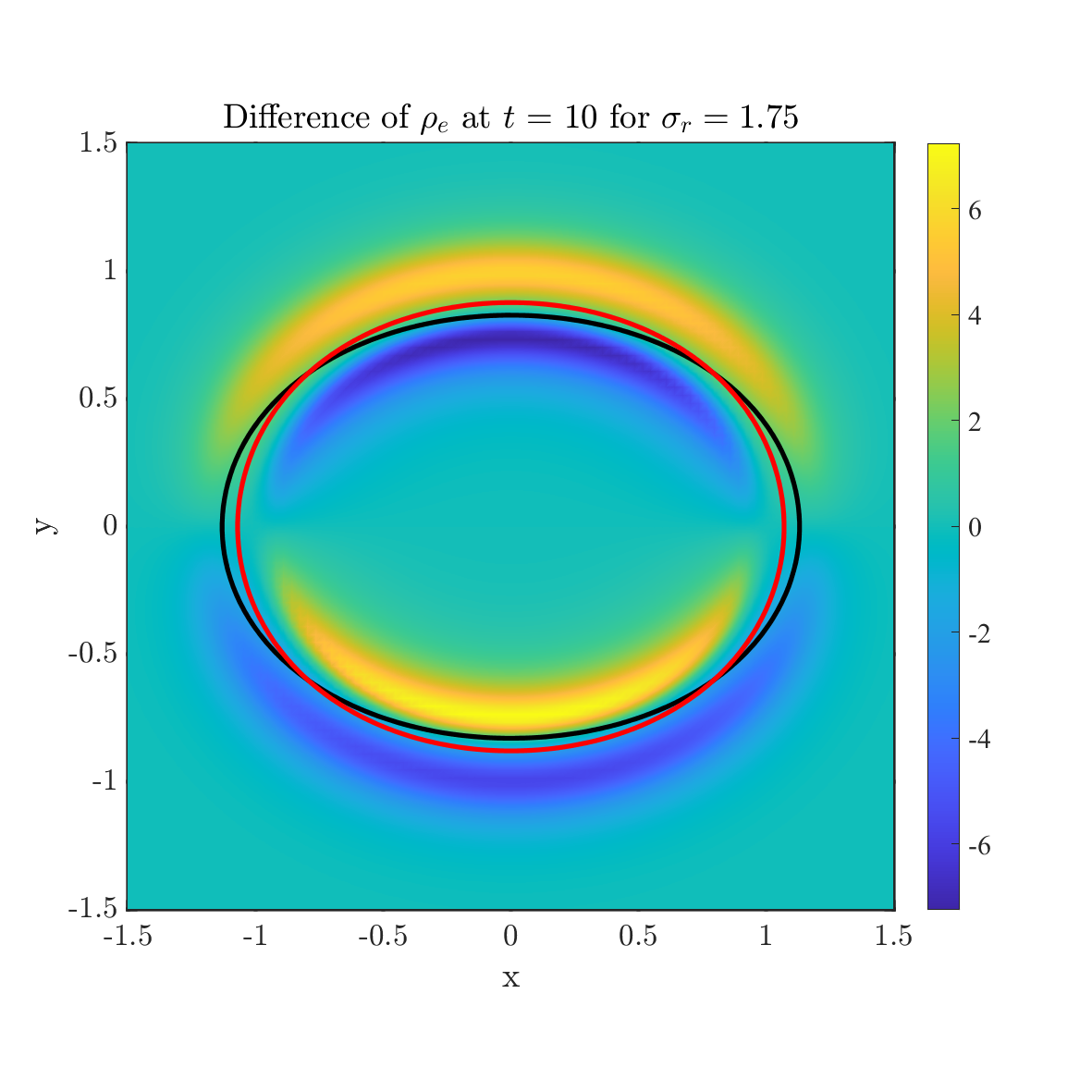}
\includegraphics[width=0.32\textwidth]{correction_175_no_Cm_tE_d+2.png}
\includegraphics[width=0.32\textwidth]{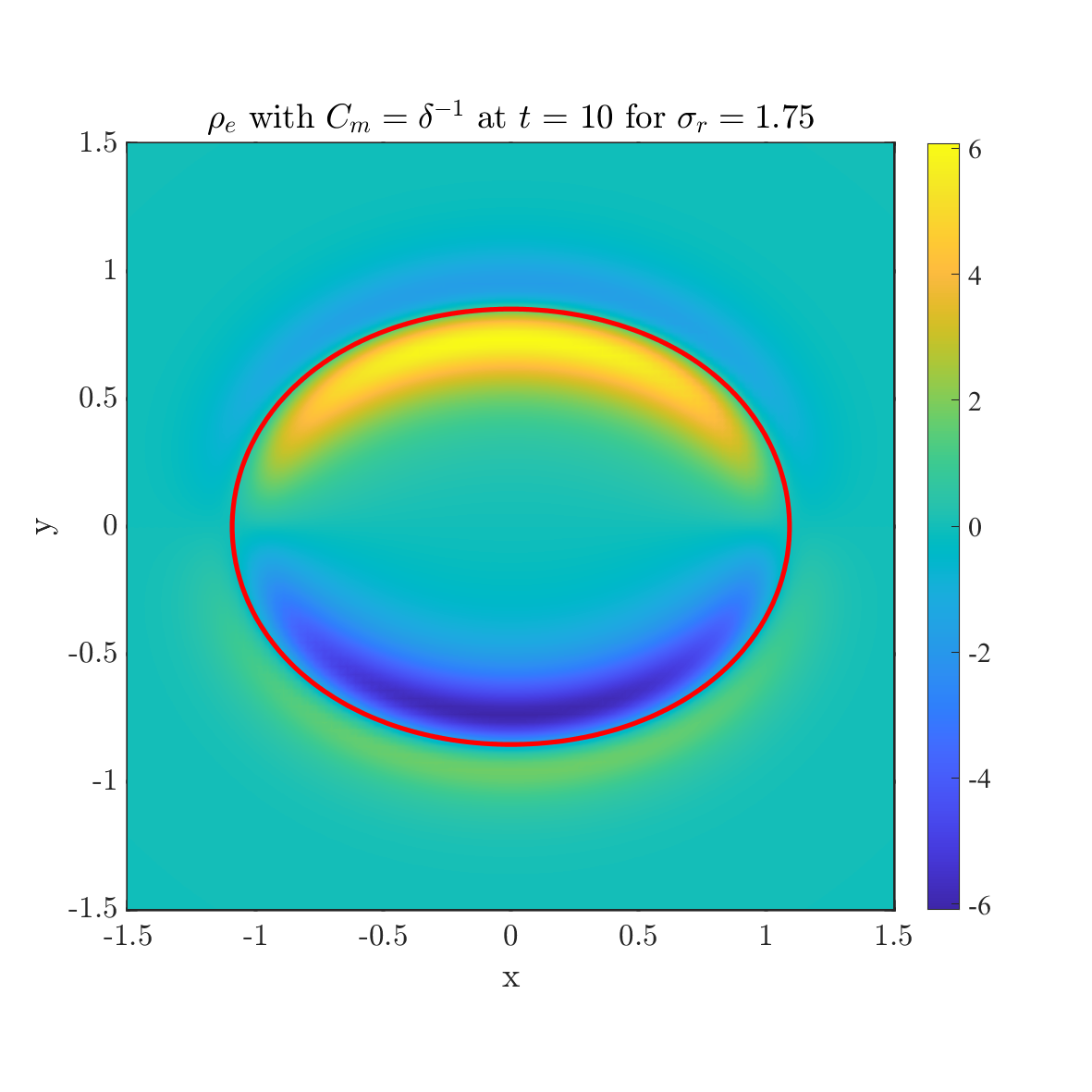}
\includegraphics[width=0.32\textwidth]{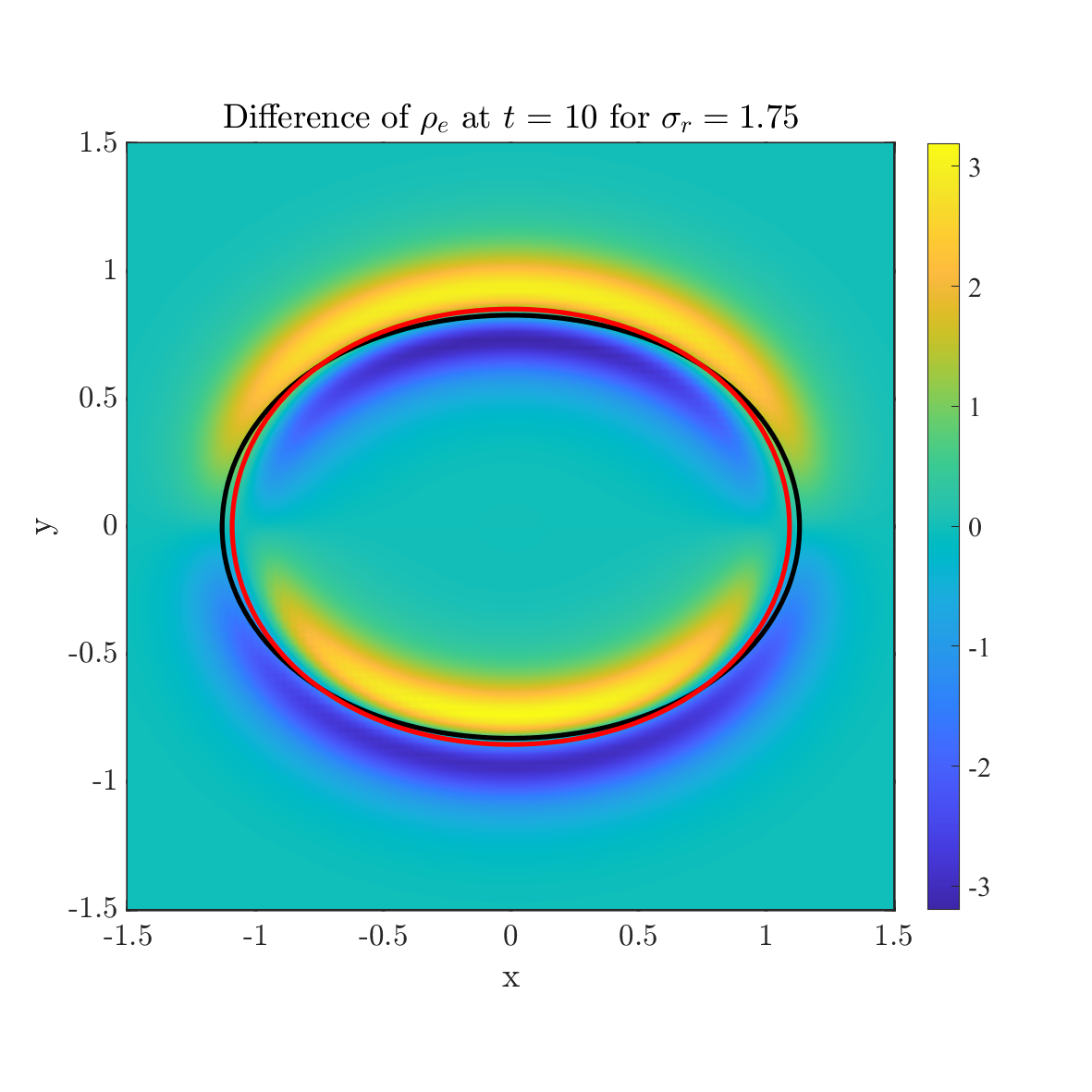}
\includegraphics[width=0.32\textwidth]{correction_175_no_Cm_tE_d+2.png}
\includegraphics[width=0.32\textwidth]{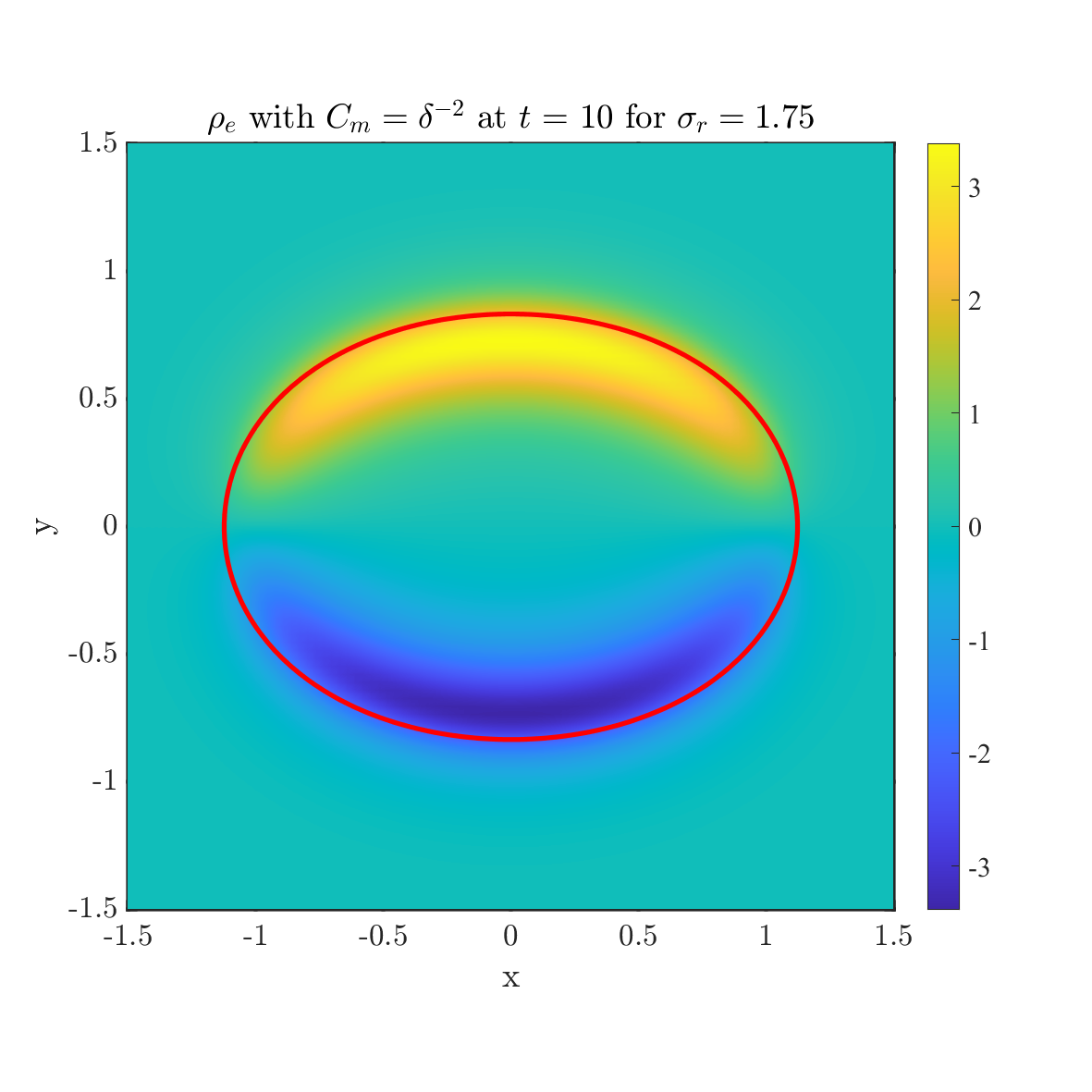}
\includegraphics[width=0.32\textwidth]{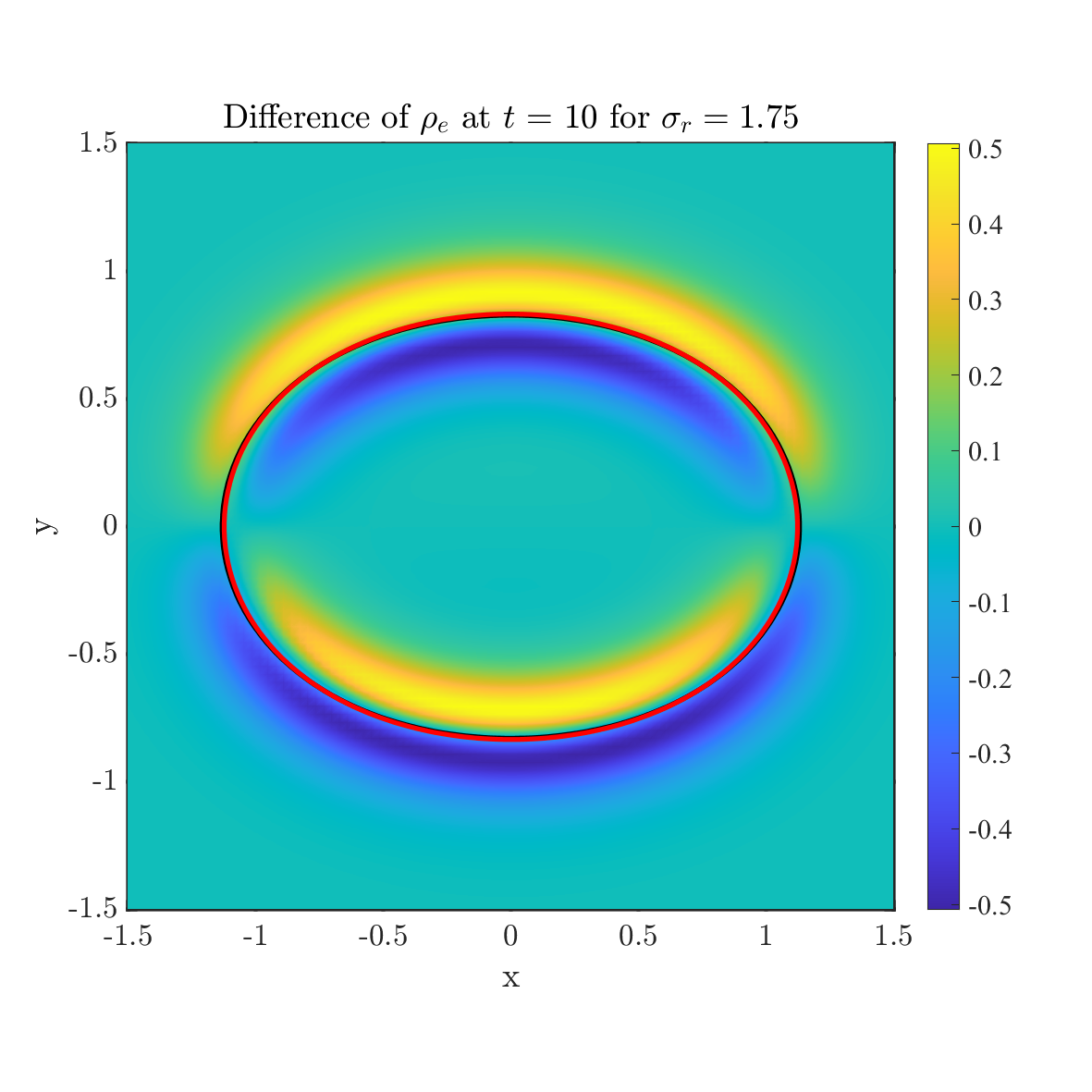}
\end{center}
\caption{The effect of capacitance for the net charge model at $t = 10$. 
Conductivity ratio $\sigma_{r} = 1.75$ is adopted here. 
We choose the situation without capacitance (left) as the reference. 
Three different capacitance $C_{m} = 1$ (top), $C_{m} = \delta^{-1}$ (middle) and $C_{m} = \delta^{-2}$ (bottom) is adopted 			in the middle column. 
In each figure, the solid line shows the zero level set ($\psi=0$) where the black line shows the drop shape without capacitance 
and the red line shows the drop shape with capacitance. 
The rest parameters are chosen as $\epsilon_{r} = 3.5$, $Ca_{E} = 1$ and $t_{E2D} = t_{E2M} = \delta^{2}$.}
\label{fig: correction2_175_with_cm}
\end{figure}

\begin{figure}
\begin{center}
\includegraphics[width=0.32\textwidth]{correction_325_no_Cm_tE_d+2.png}
\includegraphics[width=0.32\textwidth]{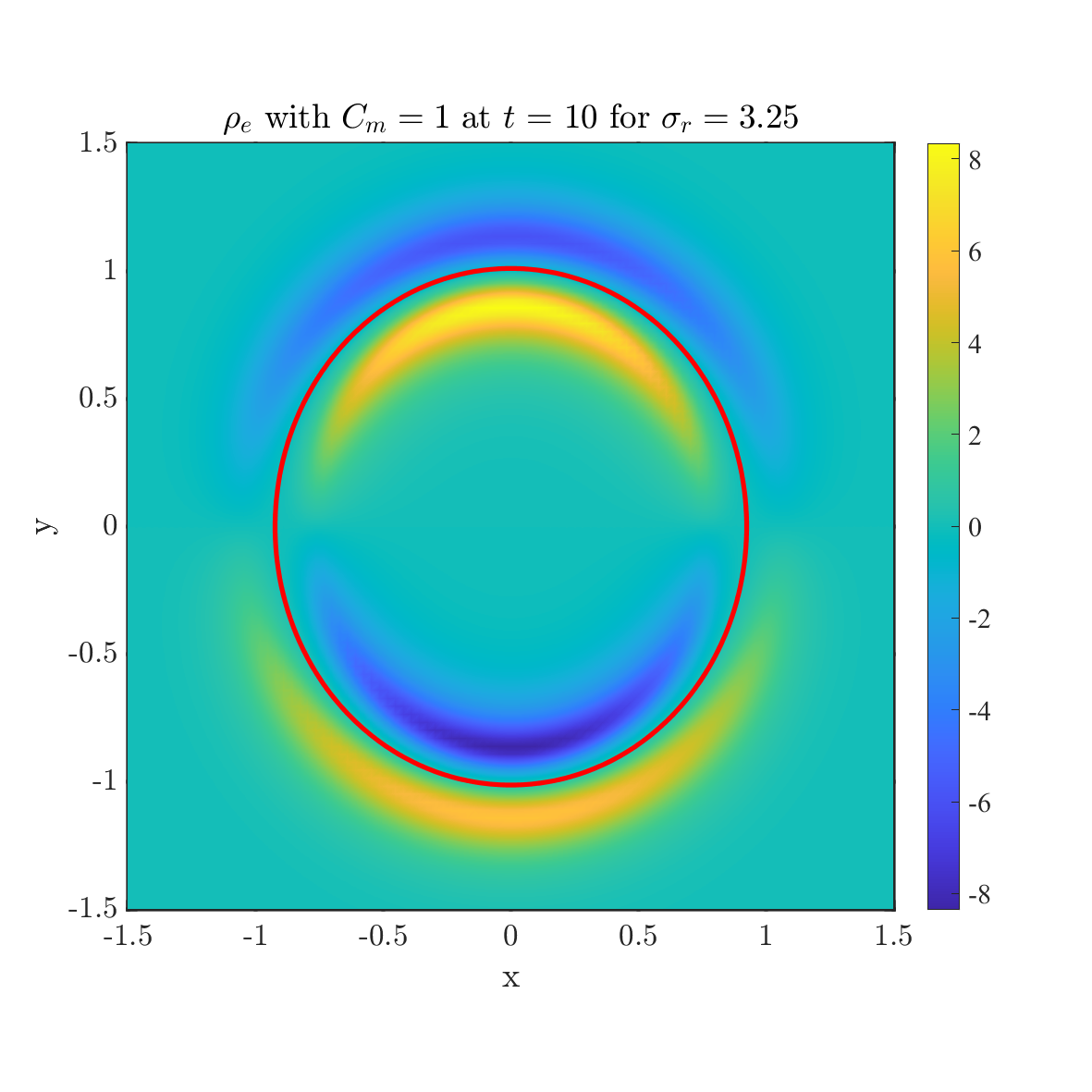}
\includegraphics[width=0.32\textwidth]{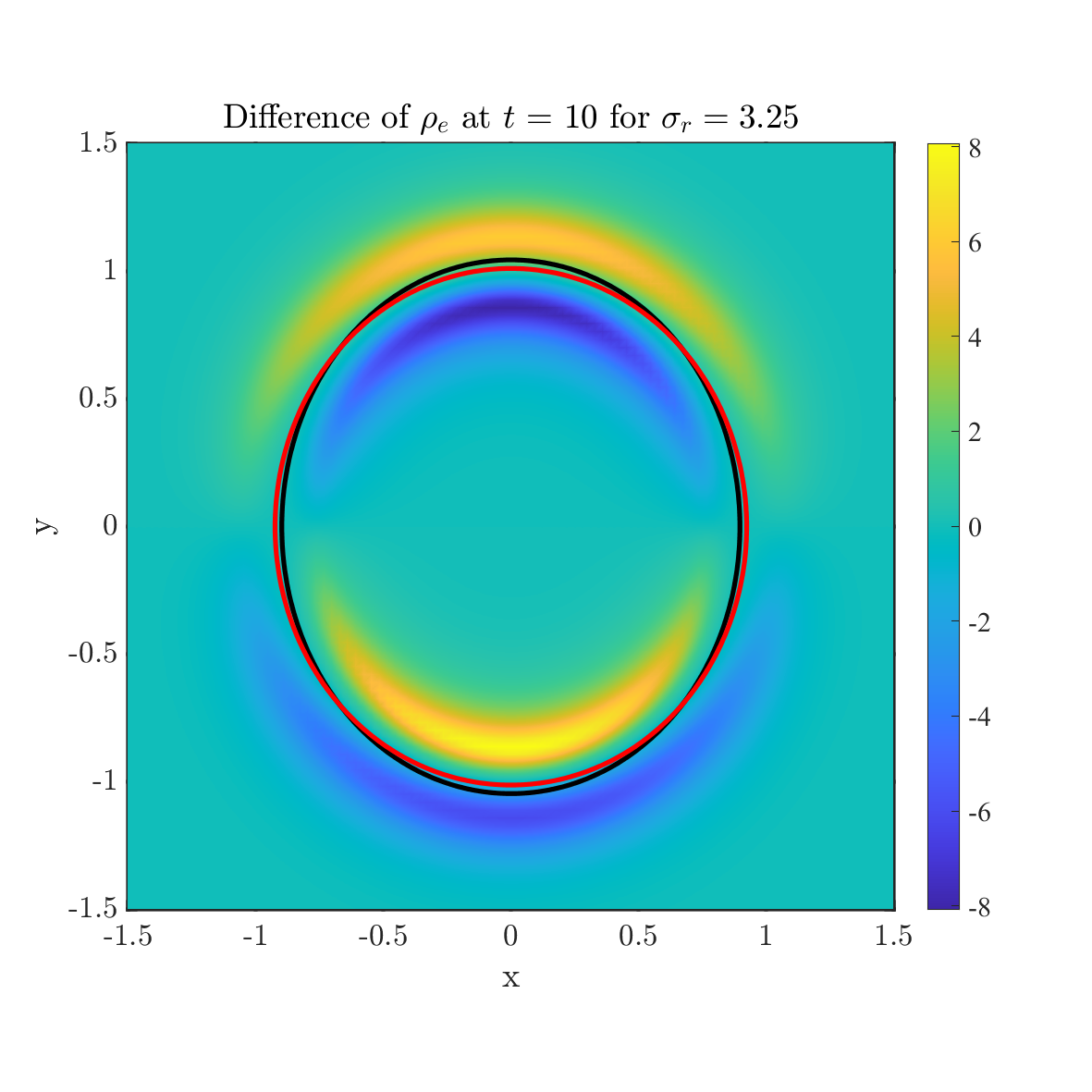}
\includegraphics[width=0.32\textwidth]{correction_325_no_Cm_tE_d+2.png}
\includegraphics[width=0.32\textwidth]{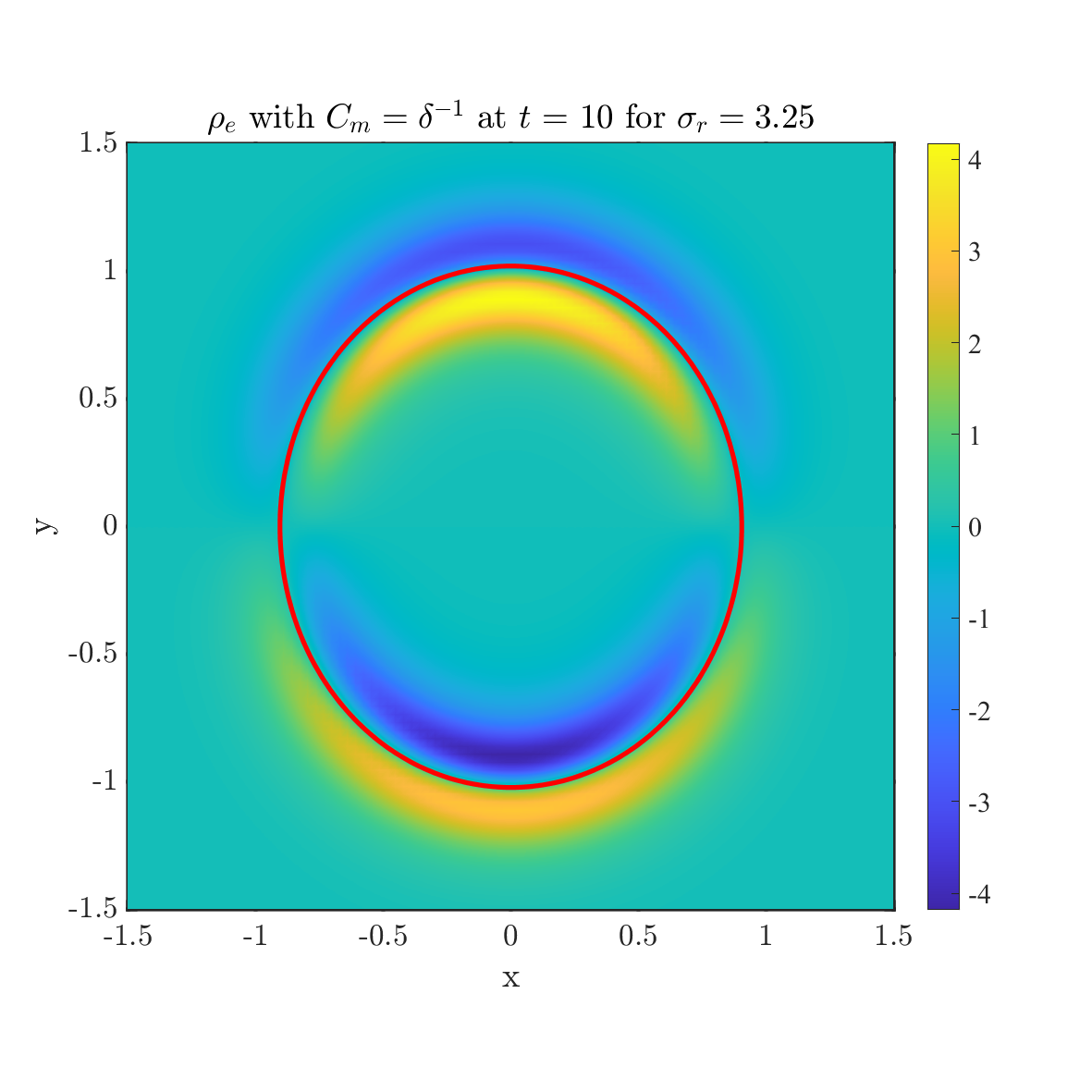}
\includegraphics[width=0.32\textwidth]{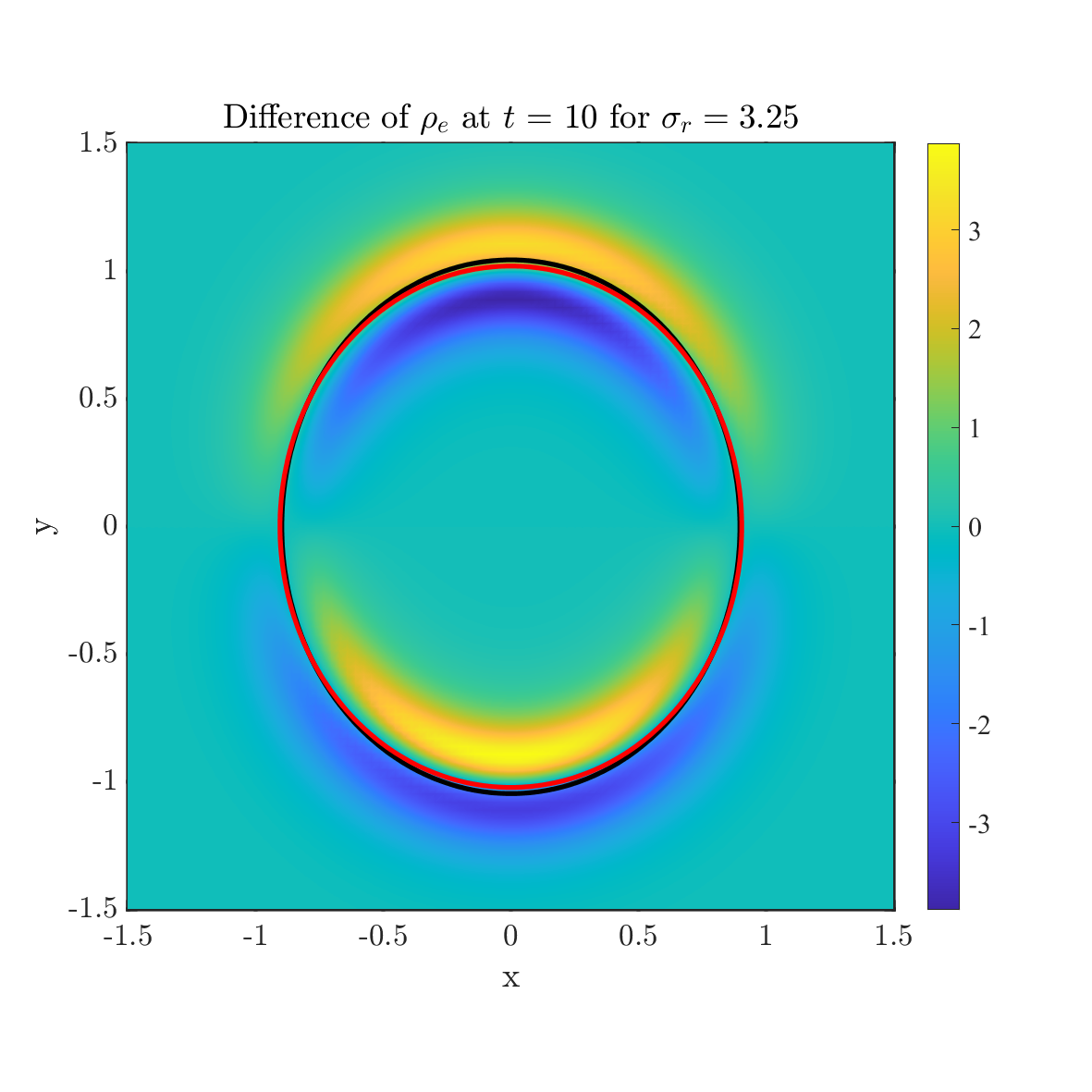}
\includegraphics[width=0.32\textwidth]{correction_325_no_Cm_tE_d+2.png}
\includegraphics[width=0.32\textwidth]{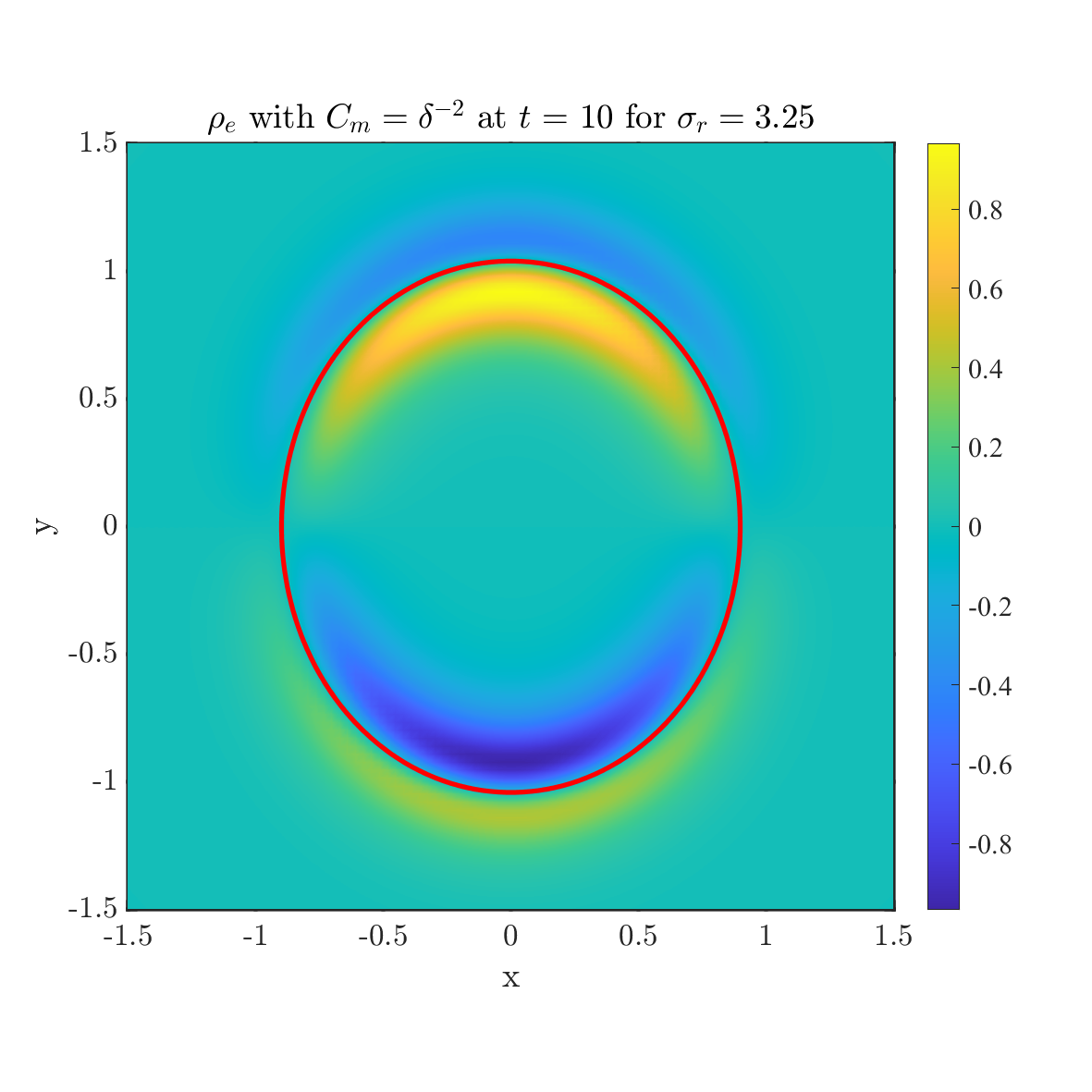}
\includegraphics[width=0.32\textwidth]{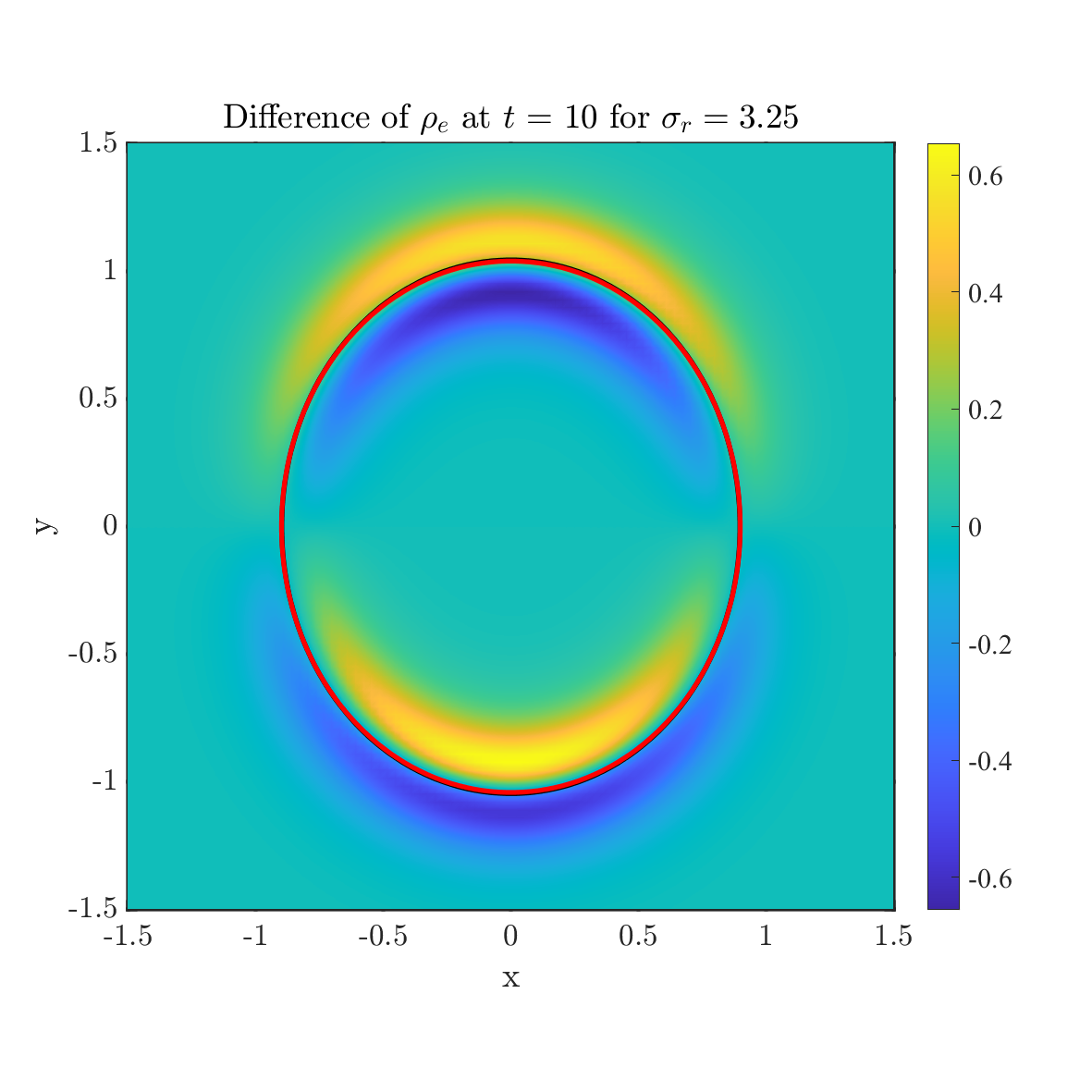}
\end{center}
\caption{The effect of capacitance for the net charge model at $t = 10$. 
Conductivity ratio $\sigma_{r} = 3.25$ is adopted here. 
We choose the situation without capacitance (left) as the reference. 
Three different capacitance $C_{m} = 1$ (top), $C_{m} = \delta^{-1}$ (middle) and $C_{m} = \delta^{-2}$ (bottom) is adopted 			in the middle column. 
In each figure, the solid line shows the zero level set ($\psi=0$) where the black line shows the drop shape without capacitance 
and the red line shows the drop shape with capacitance. 
The rest parameters are chosen as $\epsilon_{r} = 3.5$, $Ca_{E} = 1$ and $t_{E2D} = t_{E2M} = \delta^{2}$.}
\label{fig: correction2_325_with_cm}
\end{figure}

\begin{figure}
\begin{center}
\includegraphics[width=0.32\textwidth]{correction_475_no_Cm_tE_d+2.png}
\includegraphics[width=0.32\textwidth]{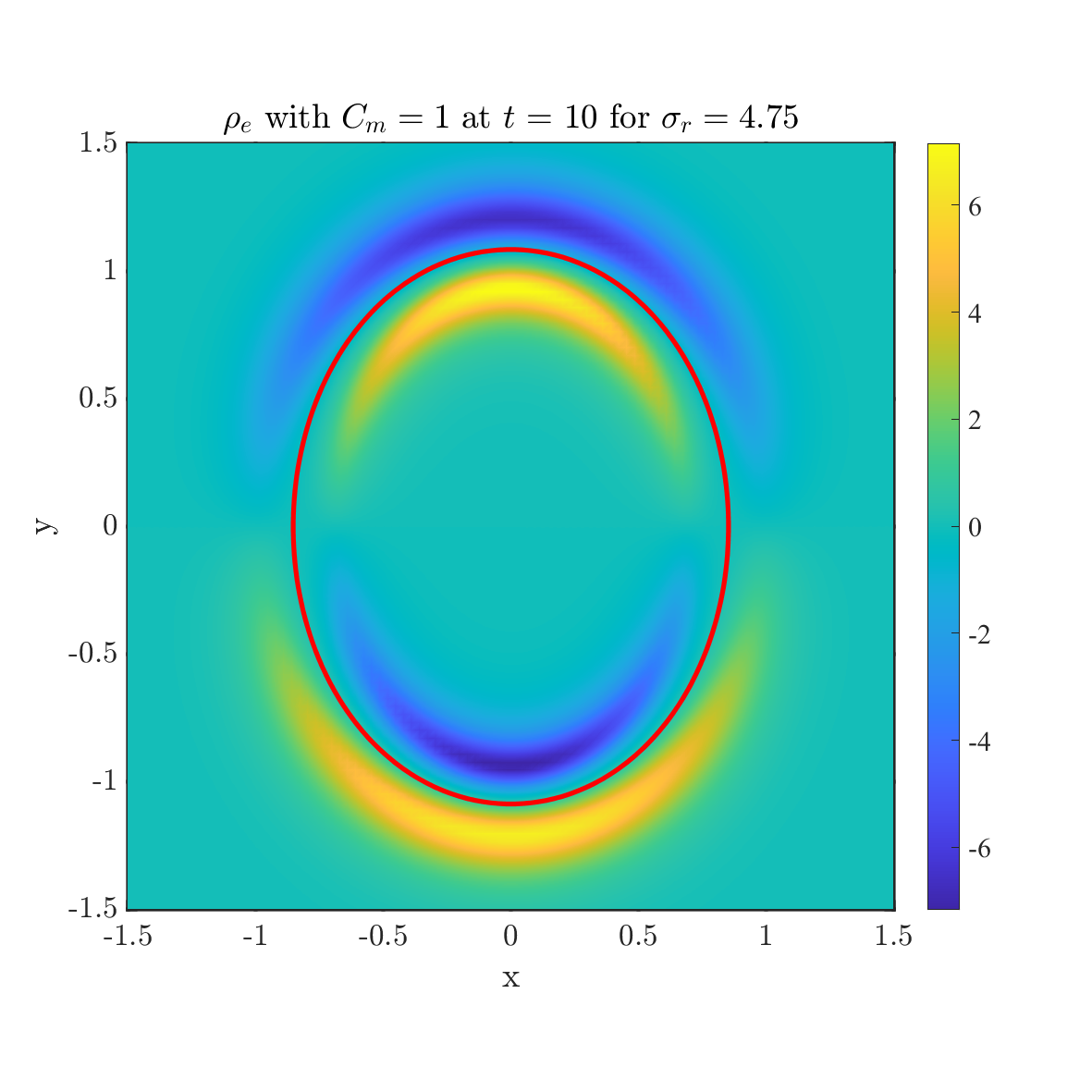}
\includegraphics[width=0.32\textwidth]{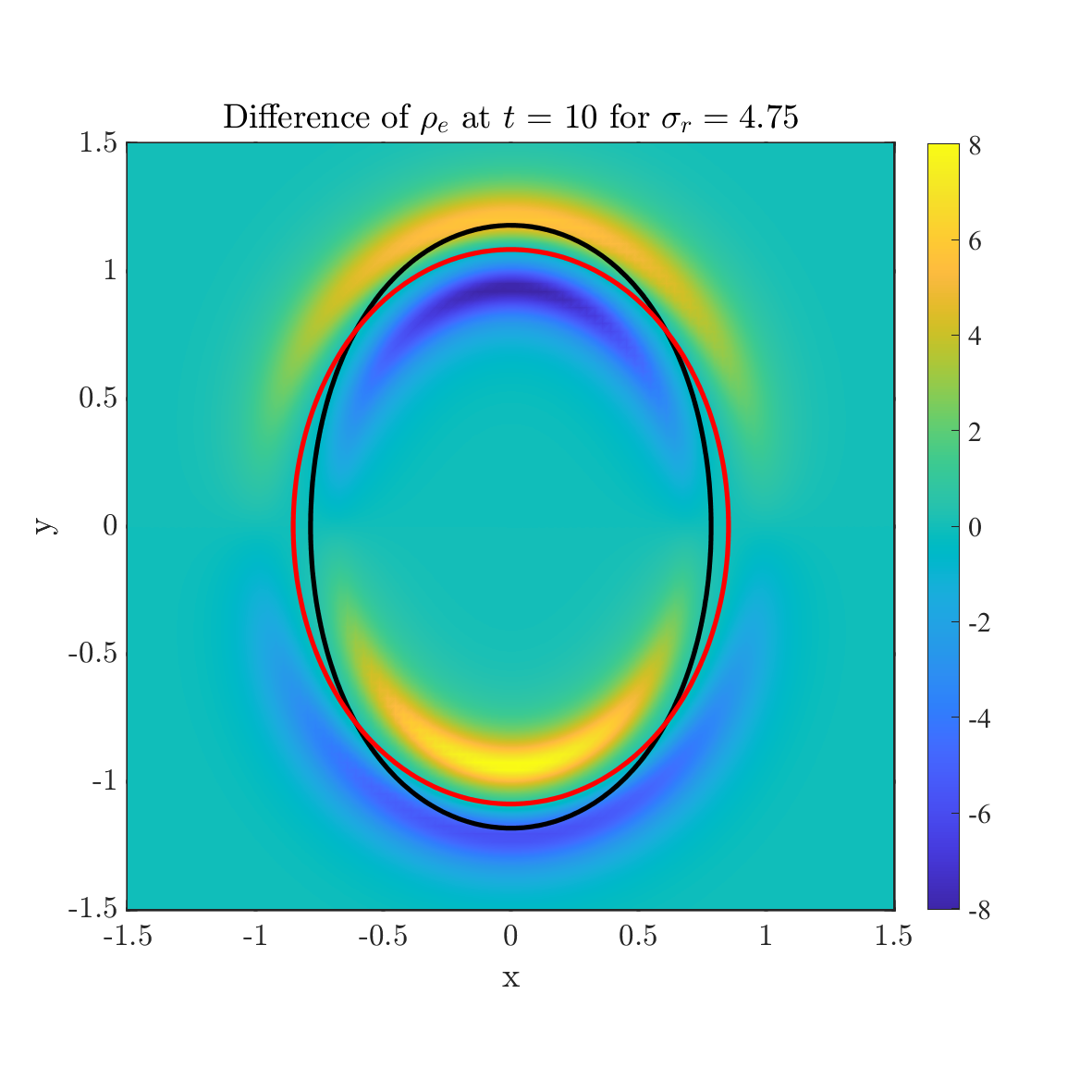}
\includegraphics[width=0.32\textwidth]{correction_475_no_Cm_tE_d+2.png}
\includegraphics[width=0.32\textwidth]{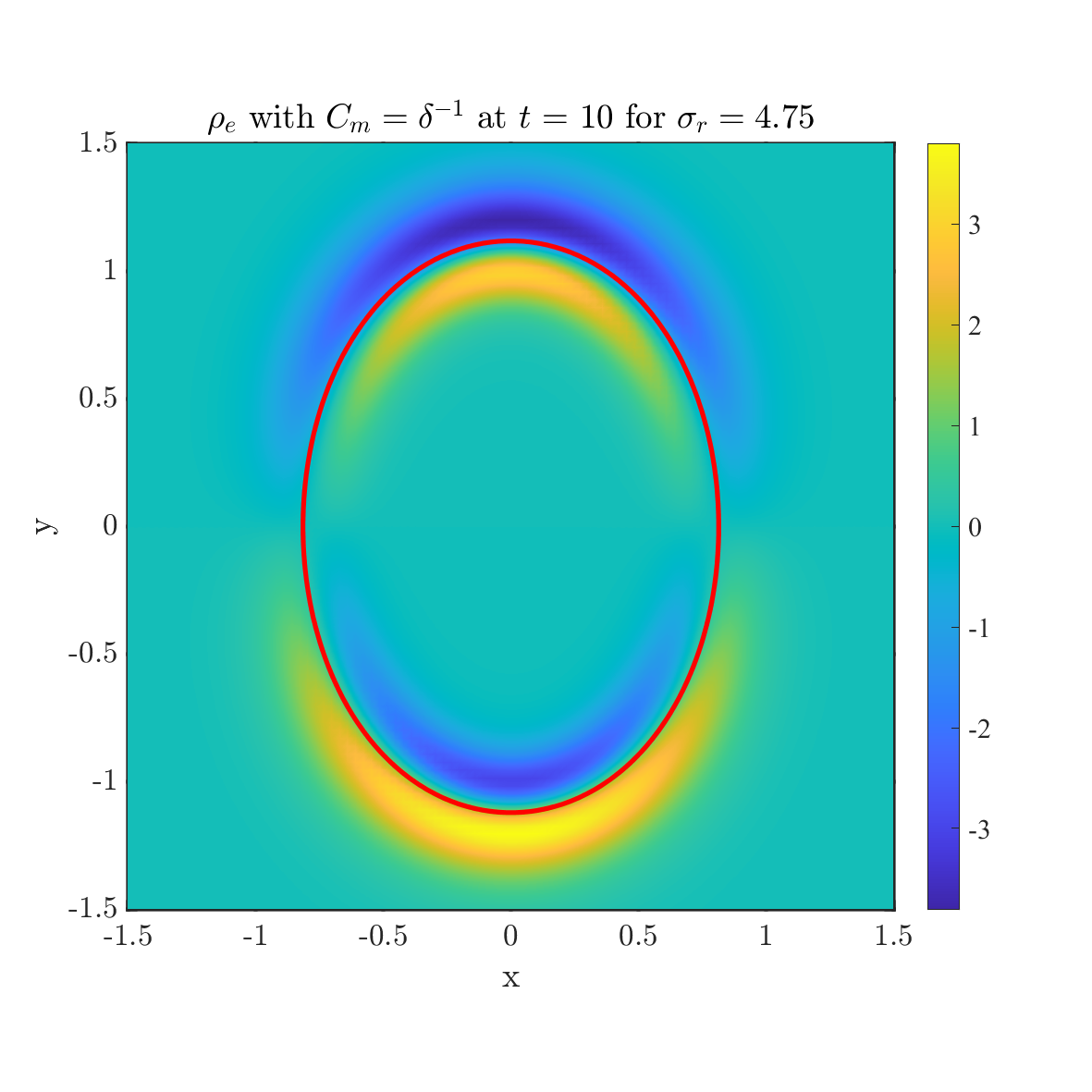}
\includegraphics[width=0.32\textwidth]{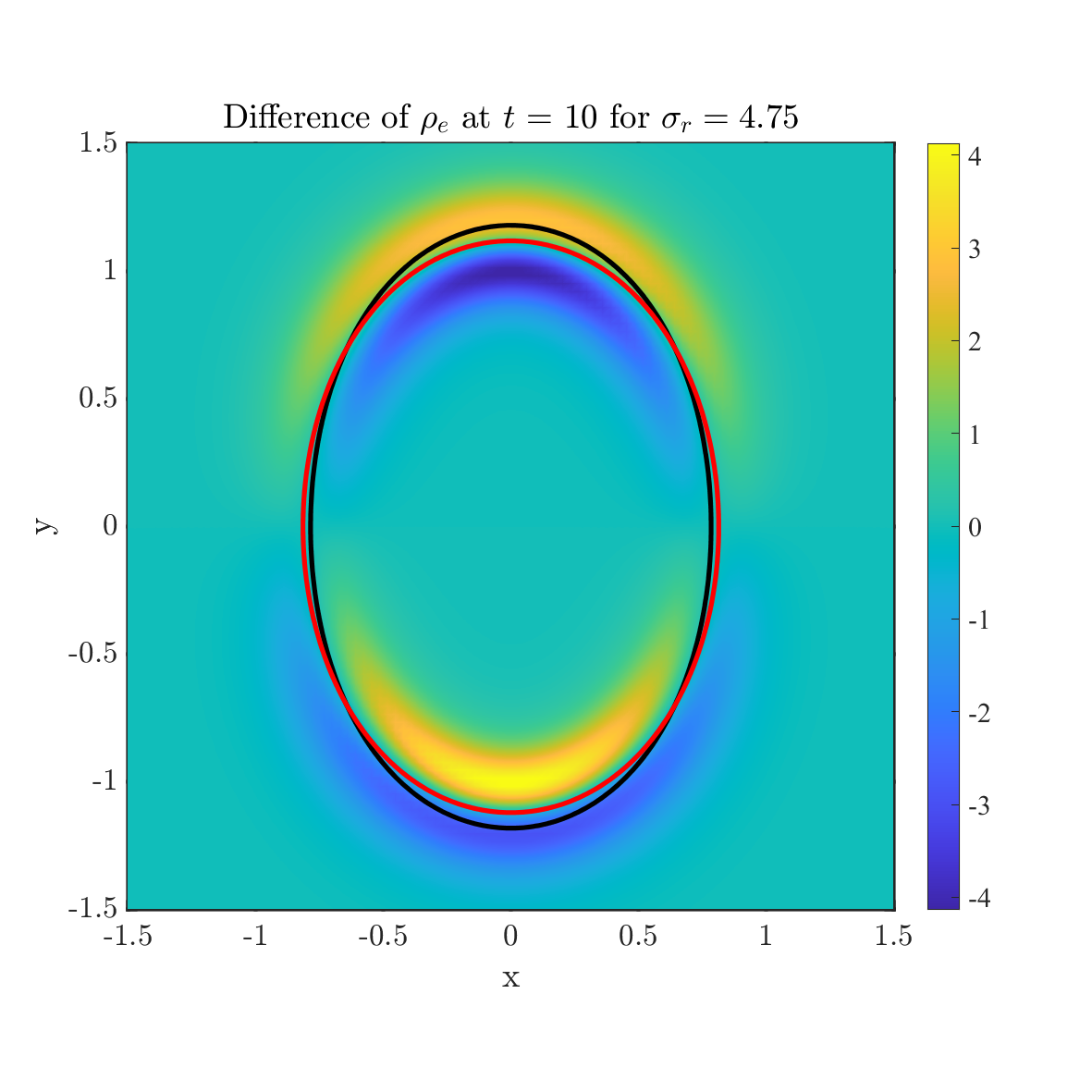}
\includegraphics[width=0.32\textwidth]{correction_475_no_Cm_tE_d+2.png}
\includegraphics[width=0.32\textwidth]{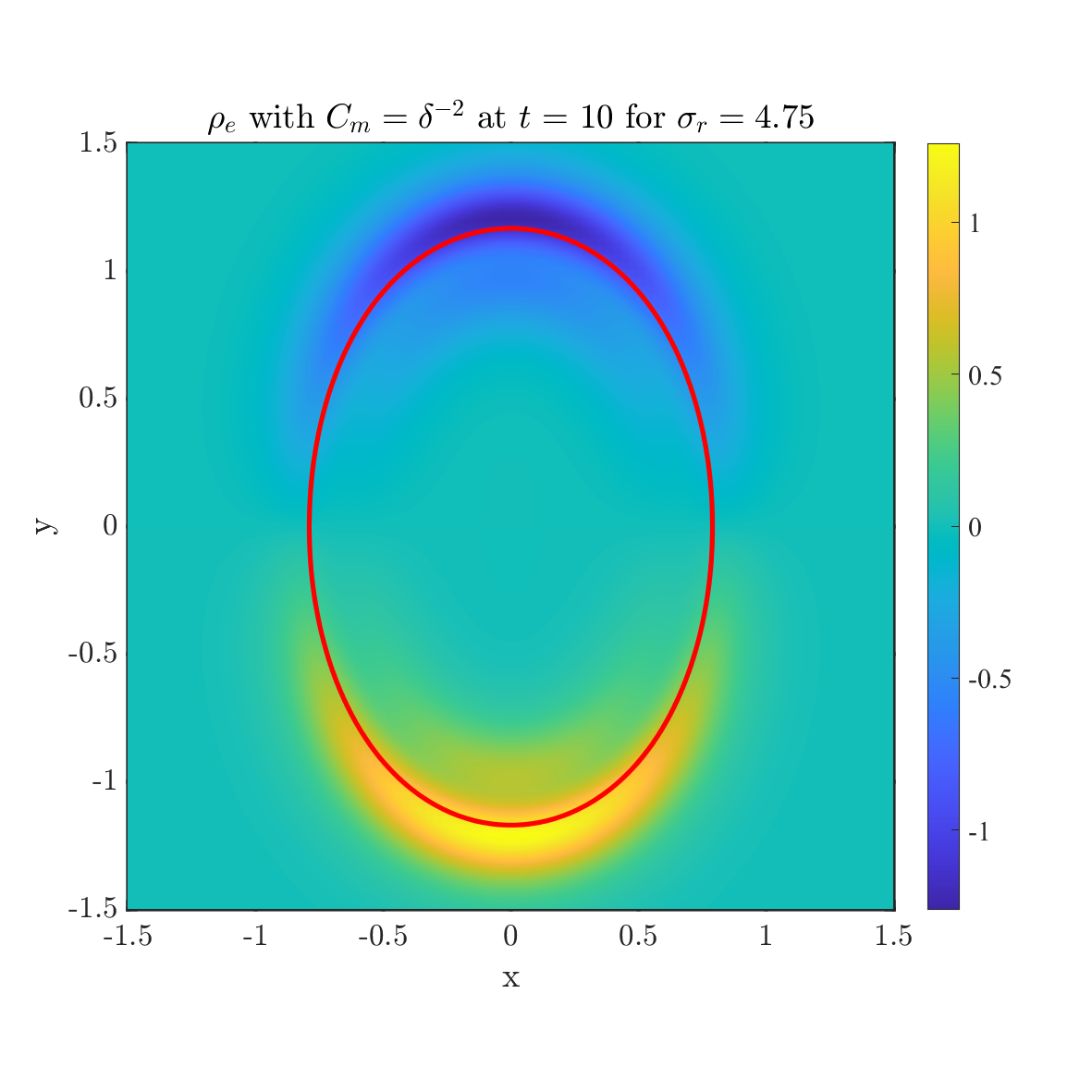}
\includegraphics[width=0.32\textwidth]{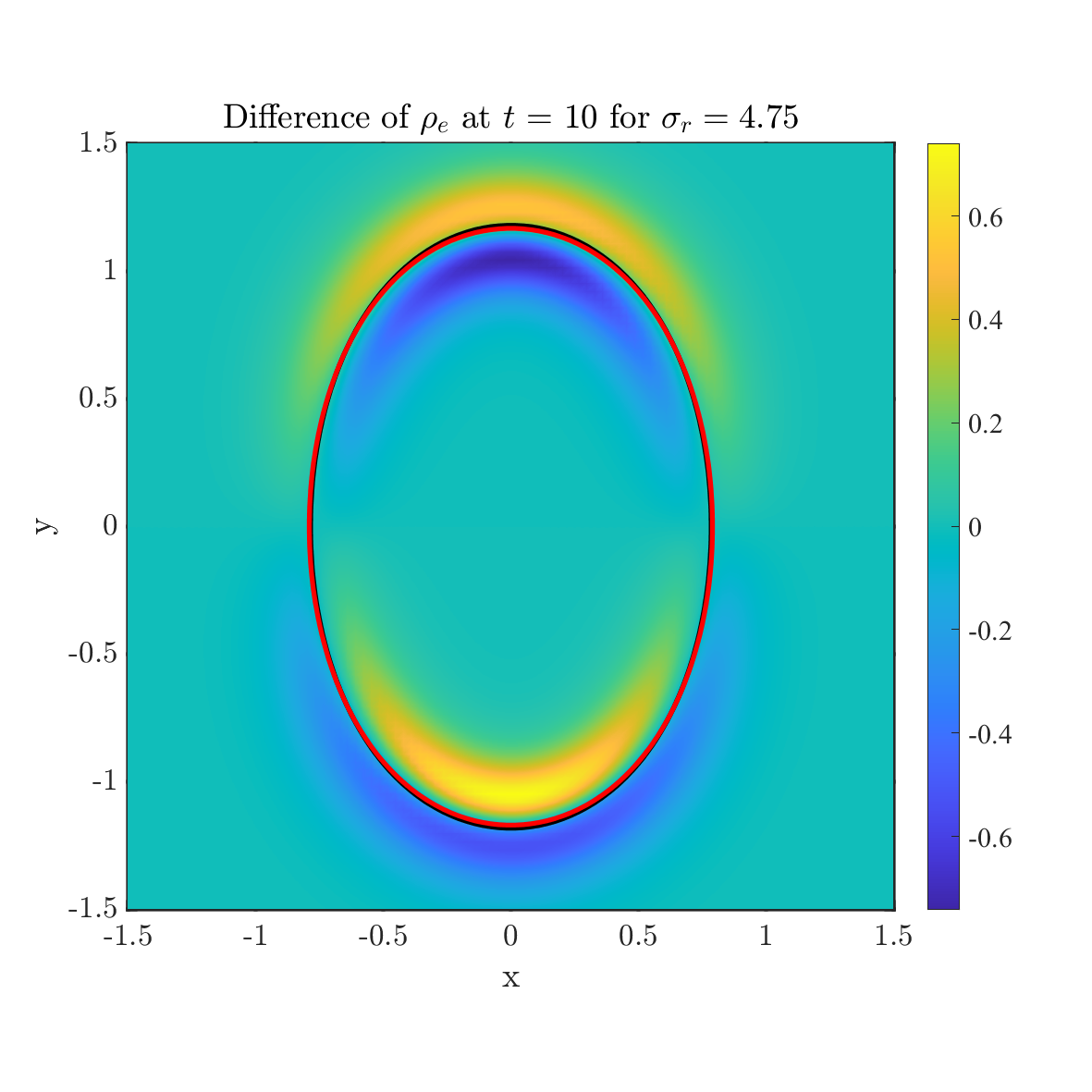}
\end{center}
\caption{The effect of capacitance for the net charge model at $t = 10$. 
Conductivity ratio $\sigma_{r} = 4.75$ is adopted here. 
We choose the situation without capacitance (left) as the reference. 
Three different capacitance $C_{m} = 1$ (top), $C_{m} = \delta^{-1}$ (middle) and $C_{m} = \delta^{-2}$ (bottom) is adopted 			in the middle column. 
In each figure, the solid line shows the zero level set ($\psi=0$) where the black line shows the drop shape without capacitance 
and the red line shows the drop shape with capacitance. 
The rest parameters are chosen as $\epsilon_{r} = 3.5$, $Ca_{E} = 1$ and $t_{E2D} = t_{E2M} = \delta^{2}$.}
\label{fig: correction2_475_with_cm}
\end{figure}

\begin{figure}
\begin{center}
\includegraphics[width=0.32\textwidth]{correction_175_Cm_d-0_tE_d+2.png}
\includegraphics[width=0.32\textwidth]{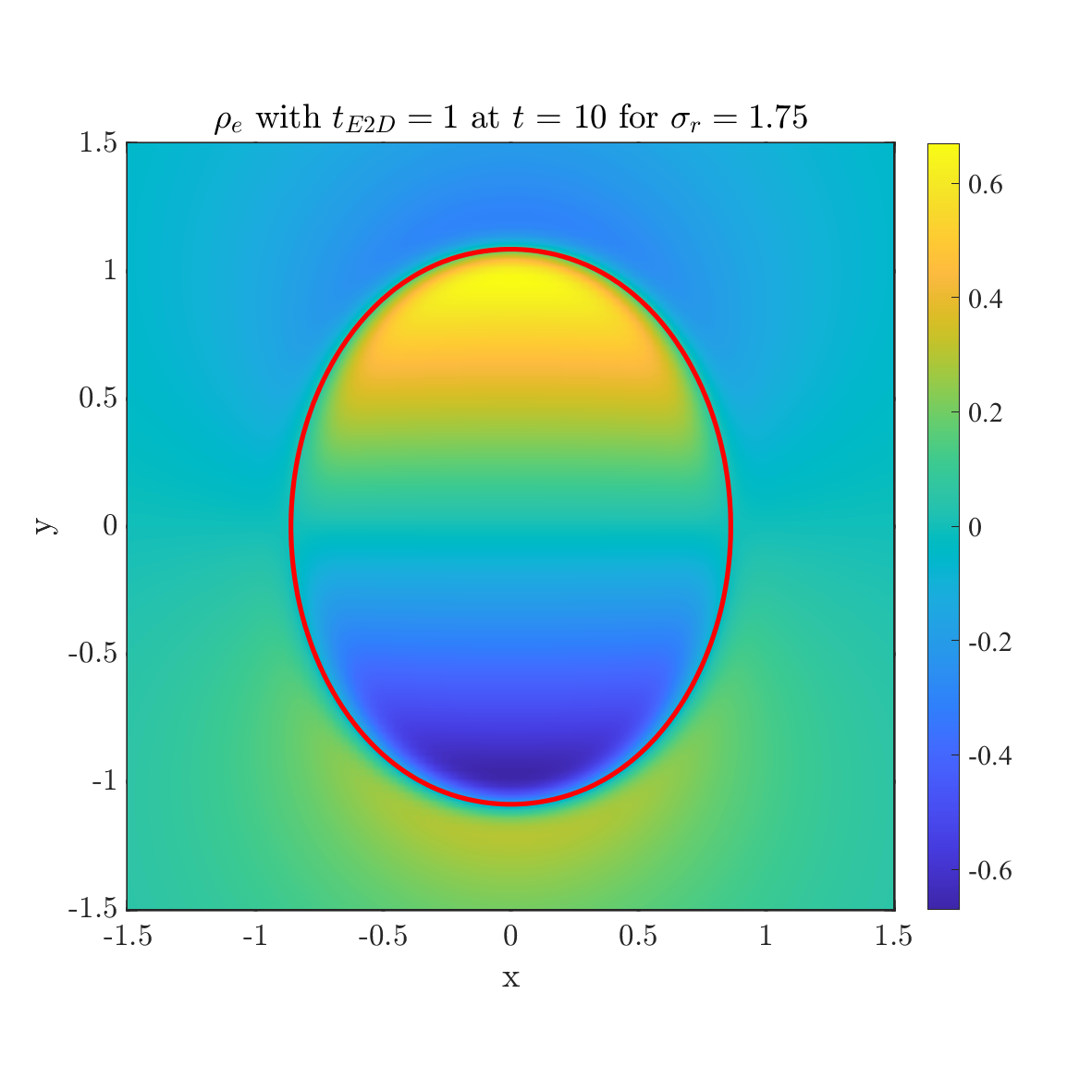}
\includegraphics[width=0.32\textwidth]{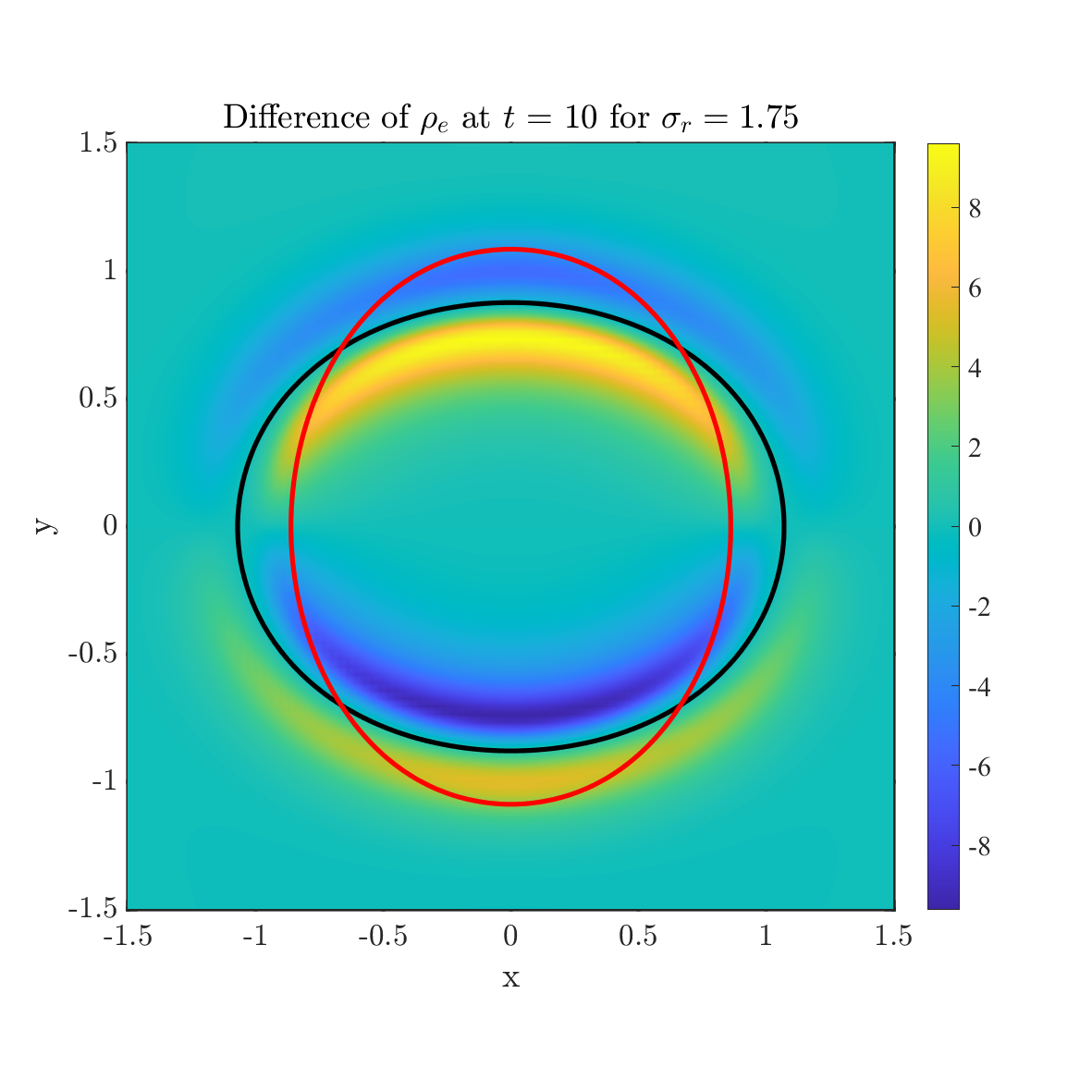}
\includegraphics[width=0.32\textwidth]{correction_175_Cm_d-0_tE_d+2.png}
\includegraphics[width=0.32\textwidth]{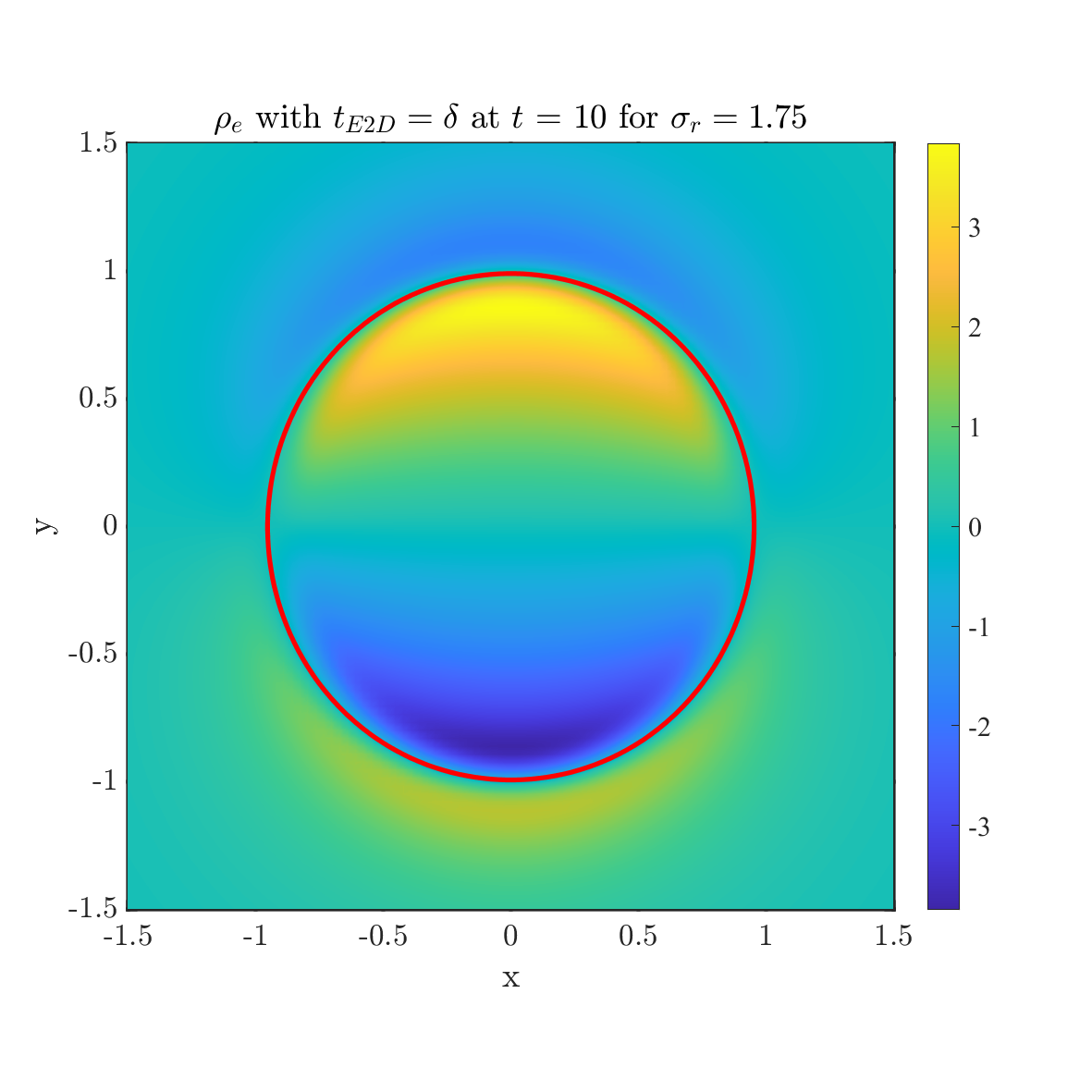}
\includegraphics[width=0.32\textwidth]{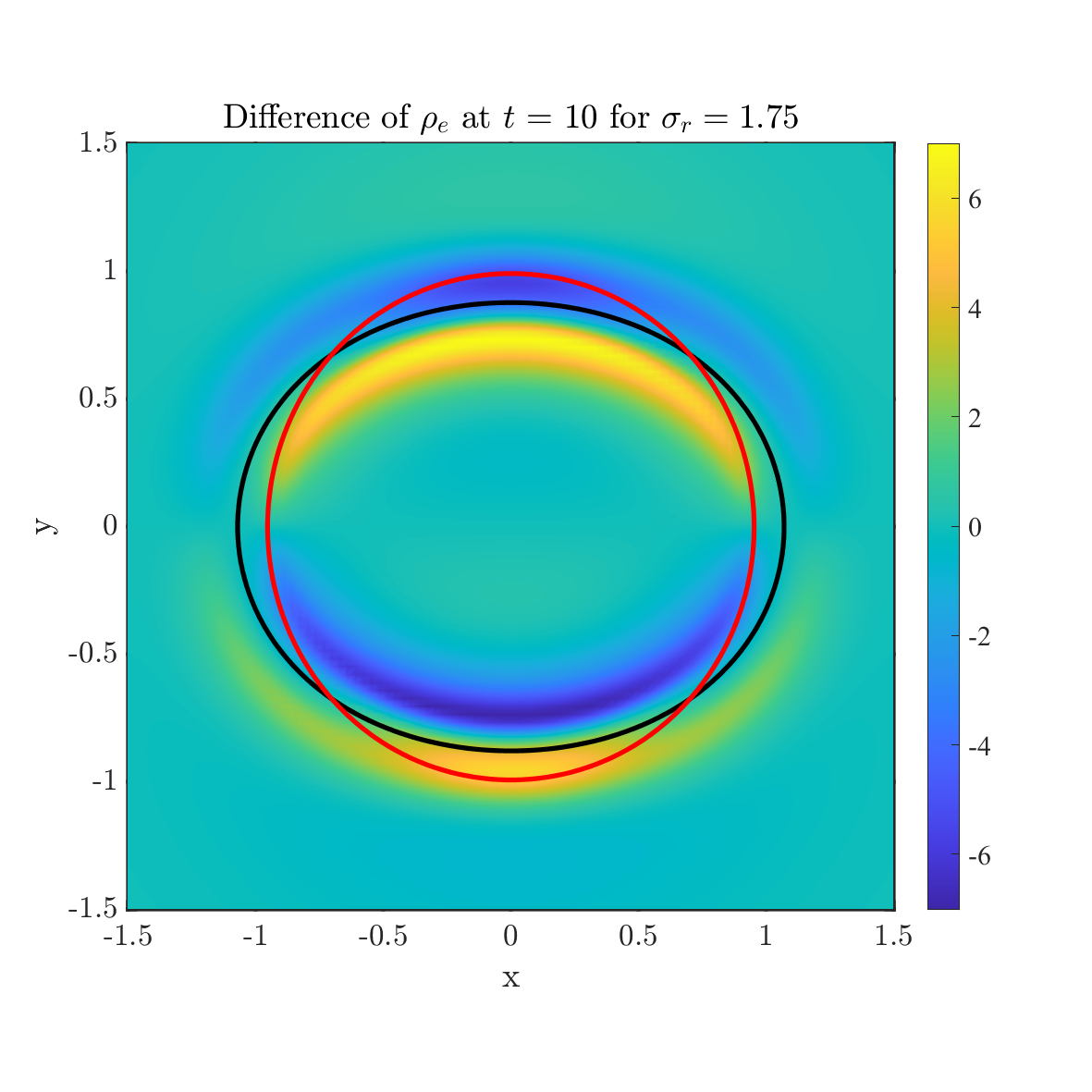}
\end{center}
\caption{The behavior of drop shapes and distributation of the net charge with conductivity ratio $\sigma_{r} = 1.75$ at $t = 10$ for the net charge model. 
We choose $t_{E2M} = \delta^{2}$ as the reference with black solid line and the results with $t_{E2M} = 1$ (top) and $t_{E2M} = \delta$ (bottom) with the red solid line. 
The rest parameters are chosen as $\epsilon_{r} = 3.5$, $Ca_{E} = 1$ and $C_{m} = 1$.}
\label{fig: correction2_175_Cm_1_tE}
\end{figure}

\begin{figure}
\begin{center}
\includegraphics[width=0.32\textwidth]{correction_175_Cm_d-2_tE_d+2.png}
\includegraphics[width=0.32\textwidth]{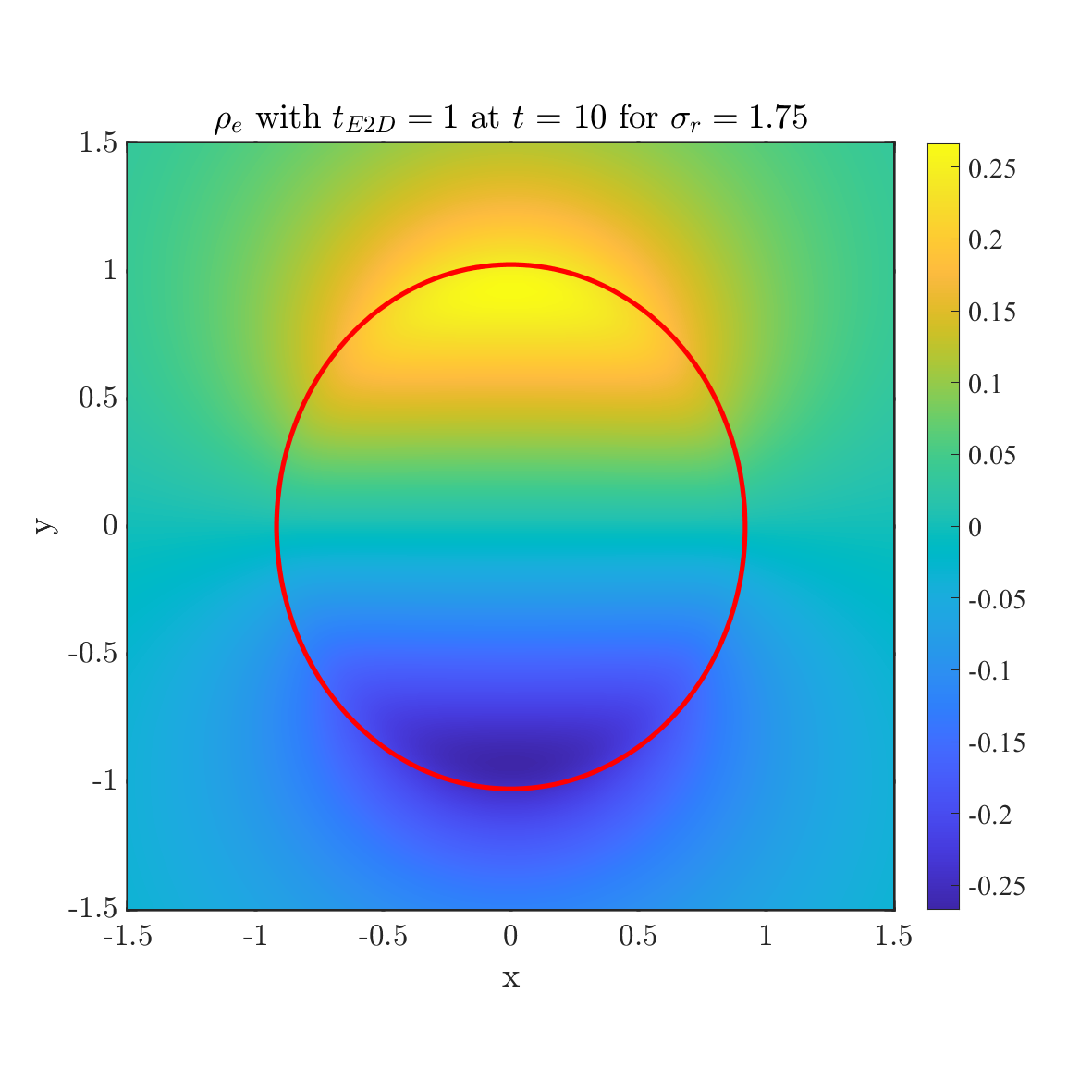}
\includegraphics[width=0.32\textwidth]{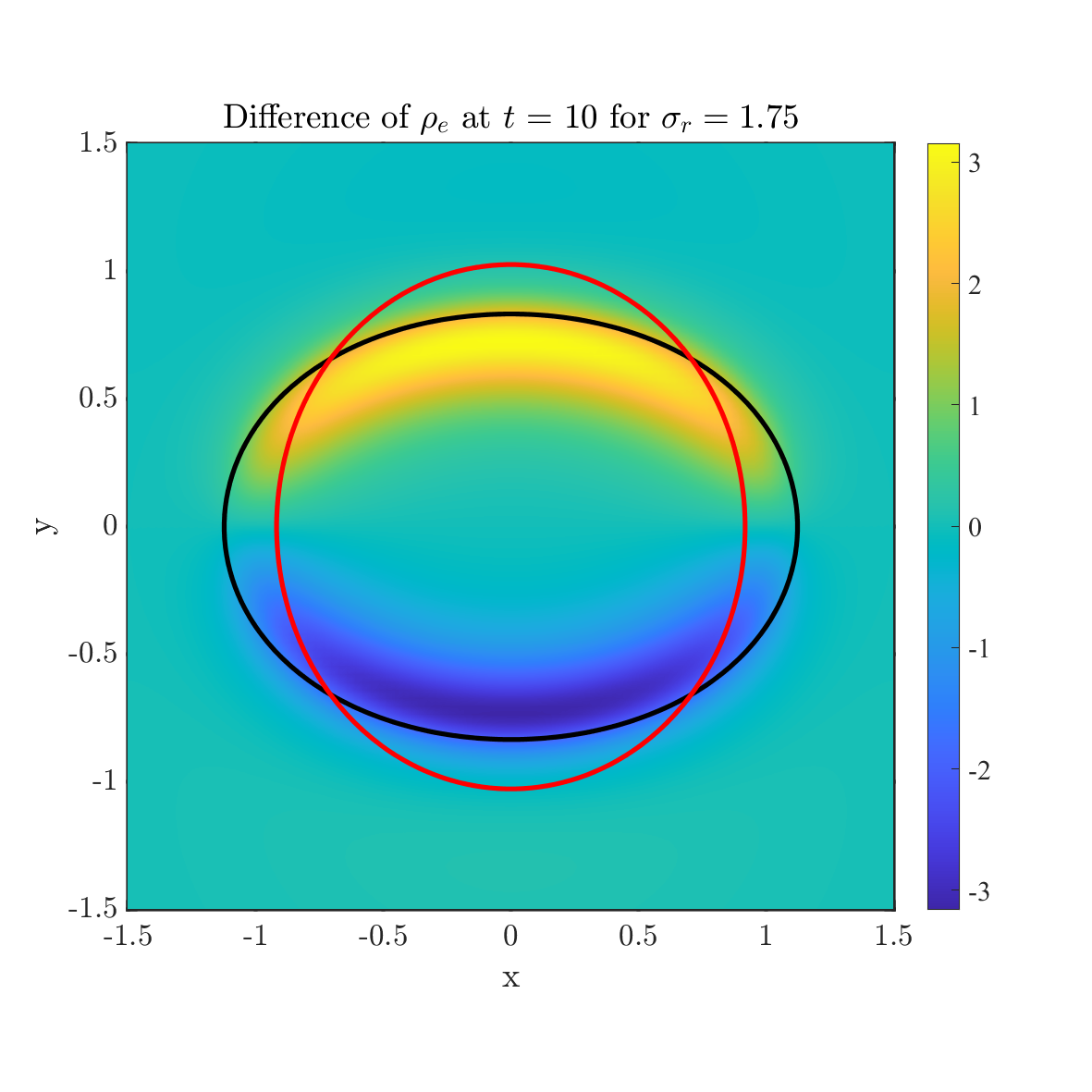}
\includegraphics[width=0.32\textwidth]{correction_175_Cm_d-2_tE_d+2.png}
\includegraphics[width=0.32\textwidth]{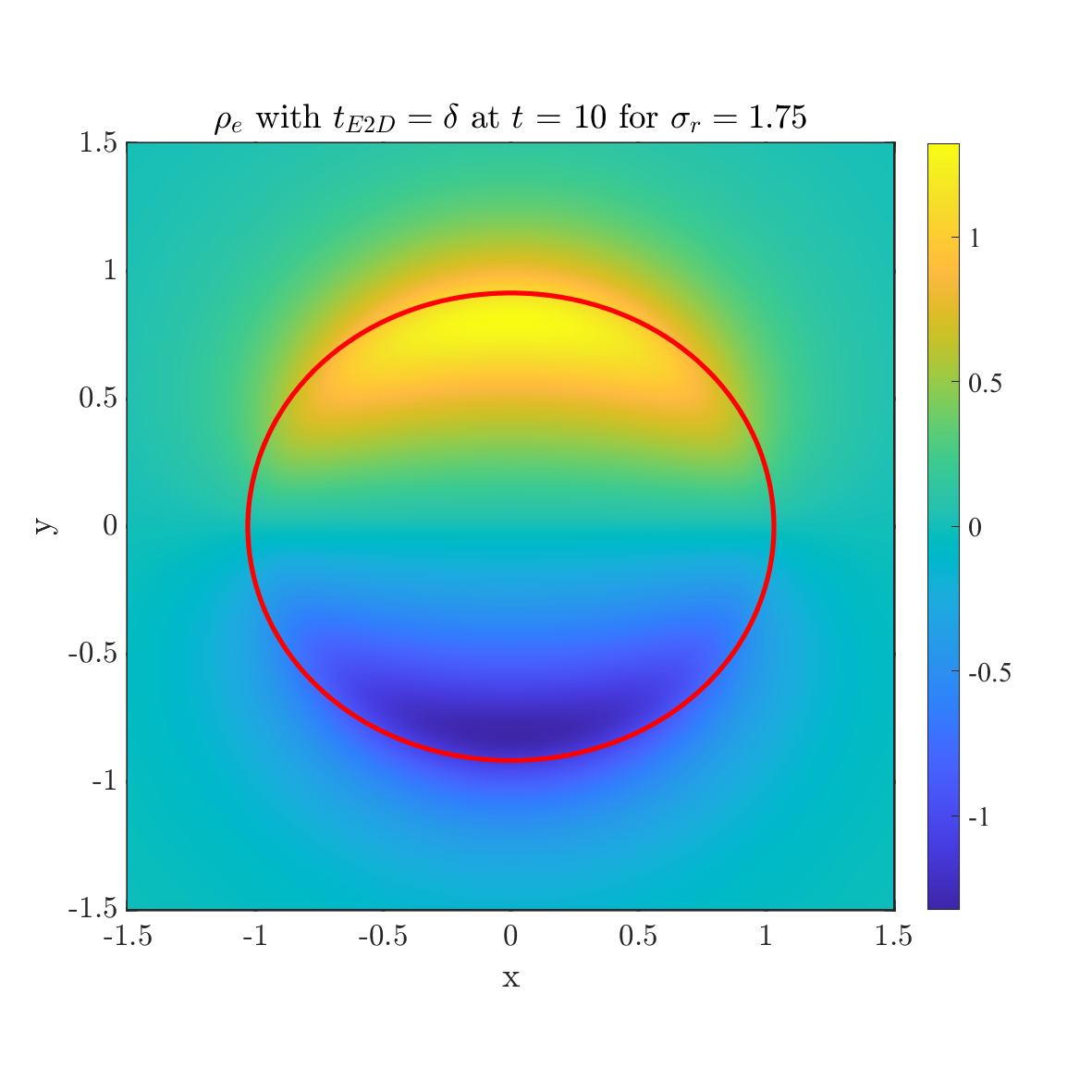}
\includegraphics[width=0.32\textwidth]{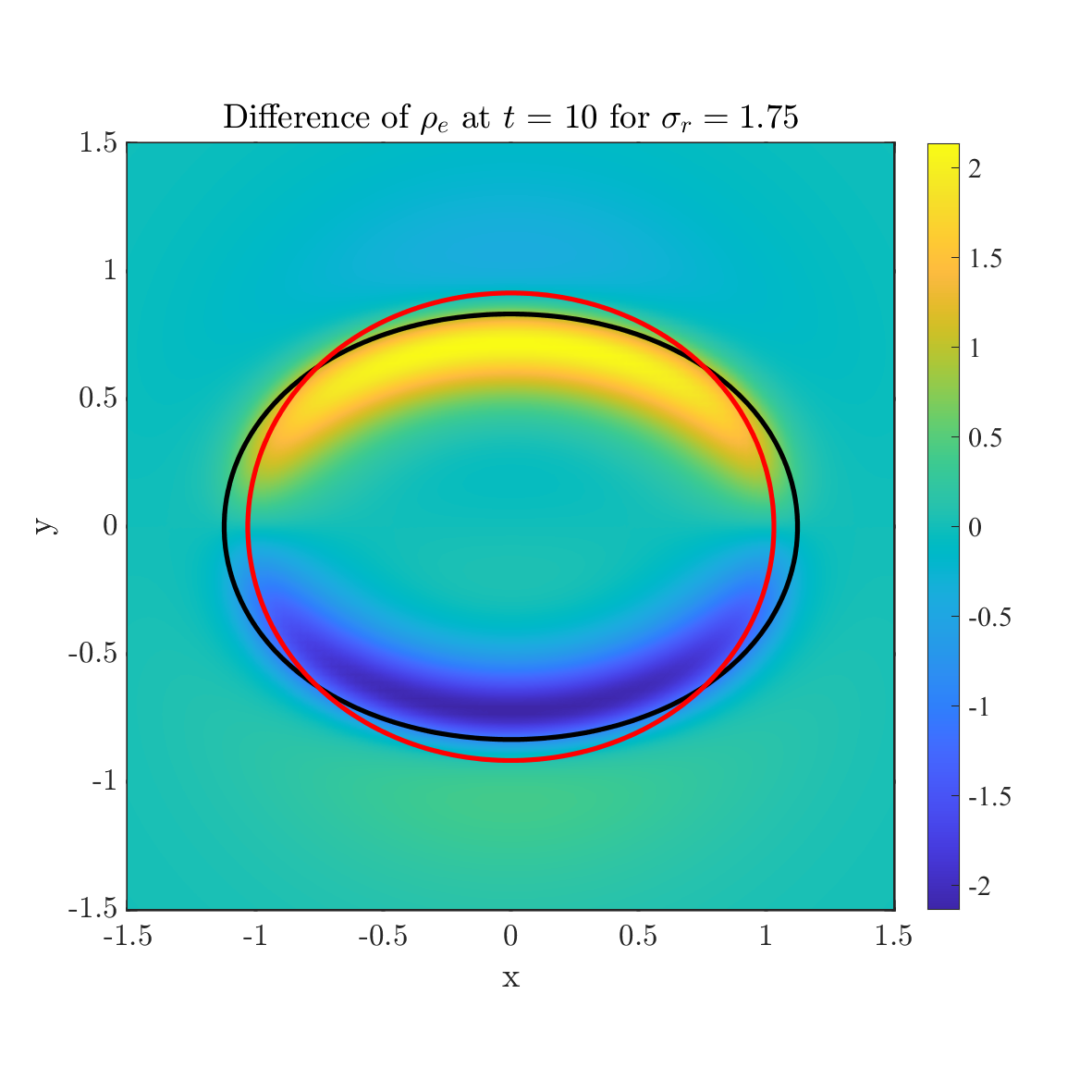}
\end{center}
\caption{The behavior of drop shapes and distributation of the net charge for conductivity ratio $\sigma_{r} = 1.75$ at $t = 10$ for the net charge model. 
We choose $t_{E2M} = \delta^{2}$ as the reference with black solid line and results with $t_{E2M} = 1$ (top) and $t_{E2M} = \delta$ (bottom) with the red solid line. 
The rest parameters are chosen as $\epsilon_{r} = 3.5$, $Ca_{E} = 1$ and $C_{m} = \delta^{-2}$.}
\label{fig: correction2_175_Cm_2_tE}
\end{figure}

\newpage
\bibliographystyle{plain}
\bibliography{jfm}

\end{document}